\documentclass[8pt]{article}
\usepackage[utf8]{inputenc}
\usepackage[flushleft]{threeparttable}
\usepackage{adjustbox, amssymb,amsmath,amsfonts,appendix, booktabs, comment, epigraph, eurosym,graphicx, geometry,ulem,caption, color,setspace,sectsty,comment,float,footmisc,multirow,pdflscape,array,hyperref, tabularx, subcaption, multirow, longtable, threeparttablex, gensymb, xcolor}
\newcolumntype{Y}{>{\centering\arraybackslash}X}
\usepackage[]{caption}
\usepackage{mathptmx}

\usepackage[]{natbib}

\usepackage{chapterbib}

\usepackage[paper=A4,pagesize]{typearea}
\usepackage{afterpage}
\usepackage{changepage}
\usepackage{pifont}%
\usepackage{csquotes}
\usepackage{booktabs} 
\usepackage{pdflscape}

\newcommand{\barroFixedCite}{%
  \begingroup
  \hypersetup{linkbordercolor={0 1 0},
  citebordercolor=green,}
  \hyperref[bib:barro]{Barro and Sala-i-Martin}
  (\hyperref[bib:barro]{2004})%
  \endgroup
}

\newcommand{\cmark}{\ding{51}}%
\newcommand{\xmark}{\ding{55}}%
\graphicspath{ {./plots/} }
\renewcommand{\thesection}{\Roman{section}} 
\renewcommand{\thesubsection}{\Alph{subsection}}
\def\sym#1{\ifmmode^{#1}\else\(^{#1}\)\fi}
\begin{document}

\begin{titlepage}
\title{Democracy and Growth in the 21st Century}
\author{Yusuke Narita\thanks{Narita: Department of Economics and Cowles Foundation, Yale University, email: \url{yusuke.narita@yale.edu}. I am grateful to Raj Chetty, Suresh Naidu, Rohini Pande, and many seminar participants for their feedback. I also thank exceptional research assistance by Ayumi Sudo, Richard Gong, Tra Nguyen, and Leonardo Fancello.}}
\date{\today}
\maketitle

\begin{abstract}
\noindent 
 We find that, in the 21st century, democracy has persistent \textit{negative} impacts on growth in GDP and night-time light intensity. This finding emerges from five different instrumental variable strategies that account for potential invalidity in some of the instruments.  
 Our analysis also suggests a key mechanism:
In this century, many electoral democracies shift toward populism and protectionism. These political changes weaken trade and investment, collectively dampening economic growth. Democracies also experienced lower growth in subjective life satisfaction among citizens. 
However, democracy causes slower growth in CO2 emissions and energy use, suggesting a trade-off between economic growth and socio-environmental performance.\\
\vspace{0in}\\
\noindent\textit{Keywords:} Political Institution, Economic Growth, Inequality, Happiness, Environment, Populism, Protectionism, Causality\\
\vspace{0in}\\

\bigskip
\end{abstract}

\setcounter{page}{0}
\thispagestyle{empty}
\end{titlepage}

\pagebreak \newpage


\section{Introduction} \label{sec:intro}
Is democracy good for economic growth? This long-standing question is becoming increasingly debatable. 
Over the past few decades, the remarkable economic rise of China, the setbacks of the Arab Spring, and rising populism in many democracies have fueled new doubts about the advantages of democracy.
This concern is well expressed by recent bestseller titles such as \textit{How Democracies Die} and \textit{How Democracy Ends}, and is reinforced by democratic backsliding  \citep{haggard2021backsliding}, where democratic institutions erode or disappear in many countries. 

This paper shows that democracy negatively impacts economic growth in the 21st century, a departure from previous findings using 20th-century data.
We start by looking at the cross-country correlation between national outcomes and a widely used electoral democracy index. The index quantifies the extent to which the ideal of electoral democracy is achieved by aggregating freedom of association and expression, clean elections, and suffrage. 
As reported in Figure \ref{fig:ols}, democracy is associated with lower growth in 2001-2022. This contrasts the 1980s and 1990s, for which periods we and the prior literature find no such negative correlation between democracy and economic growth. 

We aim to investigate whether this emerging association of democracy with lower growth has any causality. To identify the causal effect of democracy, we adopt the most influential IVs for current political institutions. 
\begin{itemize}
    \item Mortality of European colonial settlers \citep{acemogluColonialOriginsComparative2001}, which influenced the types of political institutions Europeans introduced to former  colonies
    \item Population density in the 1500s \citep{acemogluReversalFortuneGeography2002}\t, which also influenced the types of political institutions former European colonies adopted
    \item Legal origin, based on the impact of a European colonizer’s legal structure on the colony's eventual political regime \citep{portaLawFinance1998}
    \item Availability of crops and minerals, which reflects historical agricultural endowments and influences political organization through heterogeneous demand for slave labor \citep{easterlyTropicsGermsCrops2003} 
    \item Fraction of the population speaking English and a Western European language \citep{hallWhyCountriesProduce1999}
\end{itemize}
These IVs help identify the effects of political institutions by tracing their origins back to geographical and historical determinants. These determinants of today's democracy level capture the feasibility and incentives of colonial powers to invest in local institution-building and each country's cultural and industrial affinities with Western culture. 
Indeed, first-stage regressions show that these IVs are significant drivers of the cross-country variation in today's democracy levels. These IVs also pass the J test of overidentification, providing no evidence against the validity of the instruments.

All of these IVs produce similar estimates of democracy's impact, based on Two-Stage Least Squares (2SLS) and two-step efficient Generalized Method of Moment (GMM). 
These estimates indicate that democracy has a persistent negative impact on economic growth in this century. 
The median estimate among our five IV strategies is that a standard deviation increase in the democracy level causes a 2.2 percentage point GDP decrease per year in 2001-2019 (57.9\% of the outcome mean) and a 0.9 percentage point GDP decrease per year in 2020-2022 (47.3\% of the outcome mean). To facilitate interpretation of the findings, the political-regime difference between China and the US is equivalent to a 3 standard deviation difference in the democracy index in 2019. 
Put differently, a standard deviation change in the democracy index is equivalent to the political-regime difference between Mozambique and Mexico, or Mexico and Denmark. We also apply recent methodological advances to address the possibility that some of the instruments may be invalid. Allowing for potential IV invalidity does not substantially change our results. 

This finding suggests that the way democracy influences growth has fundamentally shifted in the 21st century. To allay concerns that the GDP data may be manipulated, especially by authoritarian governments,  we also use night-time light intensity growth as a nonmanipulable proxy for growth. 
 The results remain similar: a one standard deviation increase in the democracy level leads to a 2.3 percentage point decrease in night-time light intensity growth in 2001-13 (33.3\% of the outcome mean). Therefore, the adverse effect of democracy on growth is not due to misreporting or mismeasurement by authoritarian governments. 

Our finding is robust to various alternative specifications and measurement choices. 
Controlling for baseline total or per capita GDP has little effect on the estimates. Controlling for latitude, temperature, precipitation, population density, median age, continent dummies, and diabetes prevalence does not change the results. 
 The results change little with alternative indices for democracy or alternative weighting of countries. Moreover, the adverse effect of democracy is robust to excluding outlier nations from the sample. The results are not driven by the US and China alone, nor do the G7 nations drive them. 
The negative effect of democracy persists across different time periods before, during, and after the Great Recession.\footnote{We also analyze a complementary question of the effect of \textit{becoming} more democratic (in contrast to the effect of \textit{being} more democratic at the baseline). Consistent with the above main results, we find that positive changes in democracy level are negatively associated with GDP growth in 2001-2019 and night-time light intensity growth in 2001-2013. However, we cannot conclude a causal relationship due to lack of power.}
 
 We explore many potential mechanisms that underlie the perverse effect of democracy. 
 One critical factor appears to be the rise in xenophobia and myopia, manifested through both political and economic channels.
 Consistent with the widespread concern, we find that democracies experienced larger increases in threats to the quality of democracy, especially protectionism, political polarization, hate speech, and populism. These trends can be interpreted as political expressions of xenophobic and myopic tendencies. These political shifts may discourage long-term economic planning. For instance, democracies experienced stagnation in investment in the future (tangible capital formation, intangible R\&D and education) and trade with foreign countries (both imports and exports). 
 Democracy also depresses TFP growth. 
As a result of these factors, 2SLS estimates suggest that democracy slows value-added growth especially in manufacturing and services.\footnote{In contrast, other potentially important channels such as taxes,  child mortality, domestic conflict, and the number of new businesses appear to play a less significant role in explaining democracy’s adverse effect.} 
Together, these results suggest that since the turn of the twenty-first century, democracy may be increasingly influenced by short-termism and nationalist sentiment, undermining key drivers of long-term growth.

A possible interpretation of the negative effect of democracy is that democracies may try to achieve other objectives than average economic growth. We investigate this possibility by examining the impacts of democracy on additional outcomes. 
We find that democracy negatively affects growth in subjective life satisfaction between 2001 and 2022, which is consistent with our main results. At the same time, we also find that democracy alleviates income inequality between 2001 and 2022. Democracy also slows down the growth in CO2 emissions and energy consumption per capita. 
This result suggests that democracies may achieve more equality and environmental sustainability at the expense of average growth. 

Finally, we provide a supplementary analysis of the Covid-19 pandemic. 
From 2020 to 2022, the US and other major democracies faced historic recessions and death tolls due to the pandemic. 
During this period, democracy is associated with bigger shocks to GDP and more excess deaths.
These associations turn out to be causal. Our median estimate among our five IV strategies is that a standard deviation increase in a country’s democracy index results in 22.4 (s.e. = 8) more excess deaths per 100,000 people (25.5\% of the outcome mean). 
We also provide evidence that a significant channel for democracy's adverse effect appears to be weaker and narrower containment policies at the beginning of the pandemic, rather than the speed of policy implementation. 

\textbf{Related Literature.} 
Any cause of national outcomes is difficult to identify due to omitted variable biases, measurement errors, and limited data size.  
Classic cross-country regression studies claim that democracy's cumulative effect on economic growth may be negligible \citep{barroDeterminantsEconomicGrowth1997,przeworskiPoliticalRegimesEconomic,przeworskiDemocracyDevelopmentPolitical2000}. With more quasi-experimental research designs, however, later studies show that democracies experience more stable, long-term growth than non-democracies \citep{acemogluDemocracyDoesCause2018,alesinaDemocracyTechnologyGrowth2007,
papaioannouDemocratisationGrowth2008, perssonDemocracyDevelopmentDevil2006, quinnDemocracyNationalEconomic2001, rodrikDemocraticTransitionsProduce2005}. Similar findings exist for democracy's positive effects on health, such as child mortality \citep{besleyHealthDemocracy2006a,gerring2012democracy}. More broadly defined Western social institutions are also shown to positively affect economic growth \citep{acemogluColonialOriginsComparative2001, acemogluReversalFortuneGeography2002, easterlyTropicsGermsCrops2003, hallWhyCountriesProduce1999}.\footnote{Other studies inspect the micro mechanisms behind democracy’s effects. Some studies use regional differences in democratic representation to find that higher representation leads to greater investments in education and public health \citep{baumPoliticalEconomyGrowth2003, doucouliagosDemocracyEconomicGrowth2008, 
tavaresHowDemocracyAffects2001}. Studies such as \citet{besleyPoliticalInstitutionsPolicyChoices2003} and \citet{burgessValueDemocracyEvidence2015} focus on how different political processes within countries lead to different income redistributions and provisions of public goods.} 

The prior work chiefly focuses on the 20th and earlier centuries, while we analyze the 21st century. We deliberately use the same quasi-experimental research designs as prior work, so as to see the difference between the 21st and ealier centuries. 
Our results suggest that the role of democracy in economic growth may differ between this and previous centuries. 
This finding echoes a growing set of recent facts that challenge the conventional wisdom about economic growth. For example, as opposed to studies from the 1990s, \citet{convergingtoconvergence} and references therein note a trend towards convergence (poor countries catch up with rich) since 2000. See also \citet{acemoglu2021converging} for the causal interpretation of \citet{convergingtoconvergence}'s descriptive finding.  
\citet{EasterlyNewStylizedFactsOnPolicyAndGrowthOutcomes2019} reports that policy outcomes in inflation, black market premiums, currency overvaluation, real interest rates, and trade shares to GDP started improving in developing countries since the late 1990s.
 \citet{growinglikechina} document a series of facts about China's unprecedented economic transition and present a new growth model to explain the facts. 
 \citet{autor2016china} and references therein point out that American labor-market adjustments to China's trade shocks challenge much of the received empirical wisdom about economic growth. 

Our analysis of 2020-2022 also contributes to the literature on the economics of pandemics. 
Studies show that obedience to travel restrictions or social distancing differs by government communication and political systems \citep{allcott2020, 
freyDemocracyCultureContagion2020, grossman2020, Schmelze2016385118}. In contrast to the prior work, which is mostly correlational, this paper sheds light on democracy as a root cause of Covid-19-related outcomes. 

We integrate these strands of the literature to find that democracy causes worse economic and social outcomes since the beginning of the 21st century. To our knowledge, this paper seems to be the only study that shows any substantially adverse effect of democracy on any crucial national outcome. 

We organize this paper as follows. 
Section \ref{sec:ols} analyzes the correlation between democracy and national outcomes. Section \ref{sec:causal} presents our 2SLS estimates of the causal effect of democracy. 
Section \ref{sec:channel}
explores the channels behind democracy's effect. 
Section \ref{sec:discuss} discusses alternative specifications and measurement choices, nonlinear relationships, and placebo tests using 1980-2000. This section also studies the effect of becoming more democratic. Section \ref{sec:conclusion} concludes.


\section{Democracy is Associated with Lower Growth} \label{sec:ols}

Before turning to causal analysis, we first examine whether democracy is correlated with growth in 2001–2022, by using the following types of data. Table \ref{tab:descriptive-stats-main} provides descriptive statistics for our main variables.\footnote{Table \ref{tab:sources} provides details on data sources. Descriptive statistics for the remaining variables are in Table \ref{tab:descriptive-stats-appendix}.}

\textbf{Outcomes.} 
We examine two complementary measures of economic growth: GDP growth and night-time light intensity growth. 
GDP growth rates come from the World Bank.
For our sample of 159 countries, the mean annual GDP growth rate between 2001 and 2019 is 3.8\% with a standard deviation of 1.8\% (Table \ref{tab:descriptive-stats-main} row 1). We also look at the mean GDP growth rate between 2020 and 2022. Although 2020 was a disastrous year, countries' economies bounced back in 2021 and 2022, yielding an average growth rate of 1.8\% during this period. 

To address the 
widespread concern that the GDP growth data may be manipulated and exaggerated, especially in authoritarian countries, we also use mean night-time light intensity growth rate in 2001-2013, sourced from \cite{martinez2022much}. Nighttime light intensity, derived from satellite imagery capturing the Earth's surface after dark, serves as an alternative proxy for economic growth by measuring the luminosity associated with human activity. Nighttime light intensity is less prone to manipulation than GDP. For our sample of 155 countries, the mean is 6.8\% with a standard deviation of 6.0\% (Table \ref{tab:descriptive-stats-main} row 3), which is higher and more volatile than GDP growth. 

\textbf{Democracy indices.} 
Measuring the extent of democracy is tricky. Our baseline measure is the electoral democracy index from the \emph{Varieties of Democracy} (V-Dem) Project. It considers multiple facets of democracy, such as the freedom of association and expression, and clean elections. It is increasingly accepted in the economics and political science literature as a measure of democracy. 
As shown in Table \ref{tab:descriptive-stats-main}, the index captures our intuitive notion of democratic countries. According to the index, the most democratic countries are Sweden and Denmark, while the least democratic country is Saudi Arabia. 
As a further sanity check, Table \ref{tab:dem-ranking-country} ranks 30 nations with the largest GDP by their democracy levels. 
As robustness checks, we also use the polity index by the Center for Systemic Peace, the freedom index by Freedom House, and the democracy index by the Economist Intelligence Unit.\footnote{The polity index measures democratic and autocratic authority in governing institutions by evaluating executive recruitment, constraints on executive authority, and political competition. Meanwhile, the freedom index focuses more on citizens' political rights and civil liberties. The democracy index by the Economist Intelligence Unit rates democracy holistically by considering electoral processes, government functions, political participation, democratic culture, and civil liberties. Table \ref{tab:indices-correlation} shows that the indices are highly correlated with each other.}




\textbf{Country characteristics.} To control for country characteristics, we collect country-level data about GDP, absolute latitude, mean temperature, mean precipitation, population density, median age, and diabetes prevalence. We source data from the United Nations, the World Bank, and the International Diabetes Federation. 

With this data, we first examine the performance of democratic and authoritarian countries in the 21st century. Figure \ref{fig:ols-mean-gdp} shows that higher levels of democracy are associated with lower GDP growth rates from 2001 to 2019. For the period 2020–2022, Figure \ref{fig:ols-gdp-2020} indicates that more democratic nations experienced greater GDP losses due to the pandemic. A similar negative association with democracy is observed for nighttime light-intensity growth, as shown in Figure \ref{fig:ols-light}. Additionally, OLS estimates in Table \ref{tab:ols-gdp} confirm that democracy is strongly and significantly associated with worse performance in the 21st century. The baseline estimates in column 1 suggest that a one standard deviation increase in the democracy measure corresponds to a 1.6 percentage‐point decrease in the annual GDP growth rate over 2001–2019 (s.e. = 0.4), a 1.6 percentage‐point decrease in nighttime light-intensity growth rate between 2001 and 2013 (s.e. = 0.3), and, from 2020 onward, a 0.8 percentage‐point decrease in the mean annual GDP growth rate (s.e. = 0.2). These patterns hold after controlling for absolute latitude, mean temperature, mean precipitation, population density, and median age, along with baseline GDP per capita and baseline total GDP. For the pandemic period 2020-2022, we additionally control for diabetes prevalence.

\section{Causal Effects of Democracy in the 21st Century} \label{sec:causal}

\subsection{IVs for Political Regimes} \label{subsec:instruments}

We cannot interpret the above relationship as causal, however. Many omitted determinants of outcomes also correlate with the degree of democracy. To assess whether democracy causes slower growth, we use five established instruments for democracy that reflect historical and geographic determinants. 

\textbf{European settler mortality.} European settler mortality is the mortality rate (annualized deaths per thousand mean strength) of European soldiers, bishops, and sailors stationed in the colonies between the seventeenth and nineteenth centuries. 
Europeans used mortality rates to decide where to settle \citep{curtinDeathMigrationEurope1989}. In colonies with inhospitable germs, they did not settle and established extractive institutions that extracted local resources and lacked checks and balances against government expropriation. In colonies with hospitable disease environments, Europeans settled and established inclusive institutions that protected individual liberties. The effect of these institutions persists today. 
Consistent with this hypothesis by \citet{acemogluColonialOriginsComparative2001}, Figure \ref{fig:first-stage-european-settlers} shows that countries with higher European settler mortality have lower democracy levels today. Doubling European settler mortality leads to a 0.8 (s.e. = 0.2) standard deviation decrease in democracy levels (Table \ref{tab:first-stage} column 1). This fact motivates us to use European settler mortality as an IV among ex-European colonies. 
 
 \textbf{Past population density.} Population density in the 1500s is the number of inhabitants per square kilometer in the 16th century. Sparse populations in the 16th century induced Europeans to settle and develop Western-style institutions, while denser populations made extractive institutions more profitable. As a result, population density at the beginning of the colonial age determined colonial institutions' inclusiveness. \citet{acemogluReversalFortuneGeography2002} use this IV to show that European institutions positively affect economic growth.\footnote{They also use urbanization in the 1500s as an IV. Using this IV for our analysis produces similar estimates (available upon request).} Figure \ref{fig:first-stage-population-density} confirms that higher population density in the 16th century corresponds to lower democracy levels today. Doubling population density at the beginning of the colonial age is associated with a 0.5 (s.e. = 0.09) standard deviation decrease in democracy (Table \ref{tab:first-stage} column 3). Similar to the European settler mortality IV, we use this IV for ex-European colonies. 
 
\textbf{Legal origin.} This IV is a dummy variable for British legal origin that takes the value 1 if the country's legal origin is British (common law) and 0 if it is French, German, or Scandinavian (civil law). Many countries derive their legal systems from European colonization \citep{portaLawFinance1998}. Such legal origin determines how the law protects civil liberties and political rights. Indeed, first-stage regressions show that countries with a British legal origin are significantly more democratic today, exhibiting a 2.0 (s.e. = 0.6) standard deviation increase in democracy (Table \ref{tab:first-stage}, column 5). 

\textbf{Fraction speaking English or European.} This variable is the fraction of the population speaking English or a major Western European language (French, German, Portuguese, and Spanish) as a mother tongue in 1992. As \citet{hallWhyCountriesProduce1999} argue, an essential feature of modern world history is the spread of Western European influence, which created an institutional and cultural background conducive to democracy. The language variable is a proxy for such influence. Indeed, the fraction of the population speaking a major European language positively correlates with democracy (Figure \ref{fig:first-stage-fraction-european}).\footnote{The original specification also uses absolute latitude and the Frankel-Romer trade share as IVs. Our results remain similar with or without these variables as IVs (available upon request).} For example, a 100\% increase in the fraction of the population speaking a European language is associated with a 1.8 (s.e. = 0.6) standard deviation increase in democracy (line 5 in column 7 of Table \ref{tab:first-stage}). Like the original authors, we include all countries in the world in the sample definition. \footnote{Missing data restricts the actual sample to 132 countries.} 

\textbf{Availability of crops and minerals.} Bananas, coffee, maize, millet, rice, rubber, sugarcane, and wheat are dummy variables coded 1 if a country produced the crop in 1990. Copper and silver are coded 1 if a country mined the mineral in 1990.\footnote{The binary availability of crops and minerals as of 1990 is a good proxy for historical agricultural endowments \citep{easterlyTropicsGermsCrops2003}. The reason is that although the quantity produced would endogenously respond to price incentives, institutions, and other country characteristics, whether any of the commodity is produced is likely to reflect exogenous characteristics like soil and climate, which are stable over time.} According to \citet{sokoloffInstitutionsFactorEndowments2000}, certain commodities induced economies of scale and incentivized slave labor, leading to weaker liberty and rights protection for the broad population. Meanwhile, other commodities encouraged production by middle-class farmers, which induced inclusive institutions. The historical agricultural endowments thus influenced political regimes. Consistent with this narrative by  \citet{easterlyTropicsGermsCrops2003}, first-stage regressions confirm that several of these IVs are significant determinants of today’s democracy levels (Table \ref{tab:first-stage}). For example, mining silver in 1990 corresponds to a 1.1 (s.e. = 0.4) standard deviation increase in democracy (column 9, line 13). \ We include all countries in the world in the base sample.\footnote{Easterly and Levine's dataset only contains 71 countries. We extend their data-gathering process to cover 142 countries.} 


\textbf{Evaluating the Validity of the IVs.}
We are aware that these IVs are not ideal. Each IV is likely to be threatened by its own mix of potential measurement errors, omitted variables, and exclusion violations. 
At the very least, however, we provide suggestive evidence that the IVs satisfy the independence and monotonicity requirements.\footnote{We test whether the first-stage relationship between the univariate IVs and democracy is monotonic, i.e., of the same sign for different countries. Table \ref{tab:monotonicity-check} evaluates this assumption by estimating the first stage for different groups of countries (created by randomly dividing continents into groups). 
The first-stage estimates are mostly of the same sign and never have opposite signs with statistical significance. This result suggests that the first-stage relationship satisfies monotonicity. The IVs also achieve covariate balance, i.e., are not significantly correlated with covariates such as the length of the country's name.} 
The IVs also pass the overidentification test, as shown by J statistics in Table \ref{tab:2sls_tab1}. Except for one specification with marginally significant rejection (column 7), all the specifications have J-test p-values above 0.05, providing no evidence against the validity of their instruments.

Our strategy is to use these five different IVs as robustness checks with each other. Indeed, Table \ref{tab:ivs-corr} shows that the correlation among the IVs is limited, suggesting that the different IVs exploit different sources of variation to estimate democracy's effect. 
Importantly, we find no apparent reason to believe that potential exclusion violations by different IVs lead to biases of the same sign. For example, the European settler mortality IV may have excluded negative effects on growth since worse disease environments may directly hamper economic activities. On the other hand, the population density IV may have excluded \textit{positive} effects on growth thanks to returns to scale and agglomeration effects. These two exclusion violations would result in biases of opposite signs. Table \ref{tab:iv-bias-summary} summarizes the likely direction of potential bias for each IV.  Different IVs have expected bias of the opposite signs, providing support for the idea of using the different IVs as mutual robustness checks.
As a further robustness check, we also implement state-of-the-art methods to allow for potentially invalid instruments \citep{guo2018confidence, guo2023causal}. 




\subsection{IV Estimation} \label{subsec:equation}
This section presents our main results. 
With the above IVs, we estimate democracy's impact by the following 2SLS regressions: 
\begin{align}
    Y_i &= \alpha_2 + \beta_2 Democracy_i + \gamma_2X_i  + \epsilon_{2i} \label{eqn:2sls-second}\\
    Democracy_i &= \alpha_1 + \beta_1 Z_i  + \gamma_1X_i + \epsilon_{1i} \label{eqn:2sls-first}
\end{align}
\noindent 
The coefficient $\beta_2$ represents the effect of $Democracy_i$ on $Y_i$, the outcome variable, conditional on a varying vector of country characteristics $X_i$. Given that $Democracy_i$ is far from randomly assigned, we instrument for $Democracy_i$ by each vector of IVs, $Z_i$, in the first-stage equation (\ref{eqn:2sls-first}).  

Does democracy cause worse economic performance? Reduced-form figures using the IVs suggest so. Figures \ref{fig:reduced-mean-gdp-logem}, \ref{fig:reduced-gdp-2020-logem}, and \ref{fig:reduced-deaths-logem} show that lower European settler mortality causes higher democracy levels, which cause slower economic growth in 2001-2019, bigger shocks to GDP in 2020-2022, and lower nighttime light intensity growth between 2001 and 2013. A higher log of population density in the 1500s is also associated with lower growth outcomes (Figures \ref{fig:rf_appendix_a}, \ref{fig:rf_appendix_b}, and \ref{fig:rf_appendix_c}). Likewise, Figures \ref{fig:rf_appendix_d}, \ref{fig:rf_appendix_e}, and \ref{fig:rf_appendix_f} reveal that the British Legal Origin instrumental variable is also negatively related to these same outcomes.

Table \ref{tab:2sls_tab1}'s Panel A reports the 2SLS estimates of the effect of democracy, using each of the five IV strategies. They all indicate significant adverse effects of democracy. 
Columns 1 and 2 show our estimates using log European settler mortality as an IV for our base sample of ex-colonies. The corresponding 2SLS regression estimates in Panel A's column 1 in Table \ref{tab:2sls_tab1} show that a standard deviation increase in the democracy measure causes a -2.5\unskip (s.e. = 0.3\unskip) percentage-point decrease for the GDP growth rate in 2001-2019, and -2.8\unskip (s.e. = 0.7\unskip) percentage-point decrease for the night-time light intensity growth rate in 2001-2013. Democracy's effect persists in the next decade. We estimate that a standard deviation increase in democracy causes a -0.9\unskip (s.e. = 0.1\unskip percentage-point decrease for the GDP growth rate in 2020-2022. 
Our confidence in the plausibility of the IV estimates is bolstered by the fact that controlling for various potential sources of omitted variable bias has little impact on our estimates. In column 2, we control for climate, population density, population aging, and diabetes prevalence. The coefficients remain similar. 

To check whether the above results are sensitive to the choice of IVs, columns 3 and 4 use population density in the 1500s as an IV for a similar sample of ex-colonies. We continue to find a negative effect of democracy. 
The 2SLS estimates in Table \ref{tab:2sls_tab1} column 3 show that the effects of a standard deviation increase in the democracy measure are -2.2\unskip (s.e. = 0.4\unskip) percentage points decrease for the GDP growth rate per year in 2001-2019, 
-1.8\unskip (s.e. = 0.4\unskip) for the night-time light intensity growth rate in 2001-2013, and
-0.9\unskip (s.e. = 0.1\unskip) for the GDP growth rate in 2020-2022. 

The overall pattern remains the same for the legal origin IV in columns 5 and 6. The corresponding 2SLS estimates in Table \ref{tab:2sls_tab1}, Panel A, column 5 show that the effects of a standard deviation increase in the democracy measure are -1.8\unskip (s.e. = 0.5\unskip) percentage-points decrease for the GDP growth rate in 2001-2019, 
-2.5\unskip (s.e. = 0.6\unskip) for the night-time light intensity growth rate in 2001-2013, and
-0.8\unskip (s.e. = 0.2\unskip) for the GDP growth rate in 2020-2022. Adding controls in column 6 preserves the estimates. 


Columns 7 and 8 use the fraction of the population speaking English or a European language as IVs. Unlike the previous three IVs, the base sample definition is not limited to former European colonies. 
Yet, the results remain similar to the previous estimates. Column 7's estimates in Table \ref{tab:2sls_tab1} (Panel A) show that the effects of a standard deviation increase in the democracy measure are  -1.0\unskip (s.e. = 0.9\unskip) for the GDP growth rate in 2001-2019,
-1.6\unskip (s.e. = 1.1\unskip) for the night-time light intensity growth rate in 2001-2013, and
-0.7\unskip (s.e. = 0.3\unskip) for the GDP growth rate in 2020-2022. Controlling for baseline covariates in column 8 barely changes the estimate. 


Finally, we use dummies for the ability to grow crops and mine minerals as IVs, finding estimates consistent with our baseline results. 
The coefficients in Panel A's column 9 in Table \ref{tab:2sls_tab1} are -2.2\unskip (s.e. = 0.5\unskip) percentage-point decrease for the average GDP growth rates per year in 2001-2019, 
-2.3\unskip (s.e. = 0.5\unskip) for the night-time light intensity growth in 2001-2013, and
-1.0\unskip (s.e. = 0.2\unskip) for the GDP growth rate in 2020-2022. 
The regression with controls in column 10 produces similar results. 




In summary, the several different sources of variation in democracy from the historical democratization process lead to similar estimates of the negative impact of democracy. The estimates are also similar to the OLS estimates. It is particularly reassuring that the different IV strategies, which use different sources of variation in democracy, nonetheless produce similar estimates. A majority of these estimates also pass \citet{lee2020valid}’s 95\% confidence level test, which explicitly allows for the presence of potentially weak IVs. We also perform two-step efficient GMM estimation in Table \ref{tab:2sls_tab1} Panel B and obtain almost the same estimates as in 2SLS.\footnote{We use the ivreg2 implementation of 2SLS and GMM. For an exactly identified model, under the assumptions of conditional homoskedasticity and independence, the efficient GMM estimator coincides with the traditional 2SLS estimator. 
The two estimators differ in more general cases (though they share the same estimand). In two-step efficient GMM, the efficient or optimal weighting matrix is the inverse of an estimate of the covariance matrix of orthogonality conditions. The efficiency gains of this estimator relative to the traditional 2SLS estimator derive from the use of the optimal weighting matrix, the overidentifying restrictions of the model, and the relaxation of the i.i.d. assumption.  For further details, see \cite{hayashi2000econometrics}, pp. 206-13 and 226-27.} 

\subsubsection{Potentially Invalid Instruments \label{subsubsec:invalid-iv}}

The IVs we use for democracy may not fully satisfy instrument relevance or exclusion restriction. To deal with the concern, in Table \ref{tab:2sls_tab1} Panel C, we implement two recent methods \citep{guo2018confidence, guo2023causal} for IV estimation with potential invalidity in some of the instruments. 

The first method is two-stage hard thresholding with voting, which selects a set of valid instruments from a set of candidate instruments. This method uses a three step process to select valid instruments. The first hard thresholding keeps strong IVs based on their correlation with the treatment, and the second hard thresholding generates candidate sets of valid IVs based on satisfying the exclusion restriction and no unmeasured confounding. Majority or plurality voting rules then select the final set of valid IVs using the number of candidate sets each IV appears in as votes. This method selects the valid IVs with probability one as sample size grows large, so the estimator has the same asymptotic distribution as the oracle 2SLS estimator using only valid IVs (see Section 3.5 of \cite{guo2018confidence}).

The second method, searching-sampling, improves on a potential shortcoming of the first method: confidence intervals produced by two-stage hard thresholding with voting may not be robust to bias from locally invalid IVs, which may be hard to separate from valid IVs due to data limitations. The searching-sampling method constructs uniformly valid confidence intervals that are robust to mistakes in separating valid and invalid IVs. The method searches for a range of treatment effects that permit sufficiently many valid IVs and uses resampling to improve the precision of the confidence interval. Searching sampling produces confidence intervals instead of point estimates with standard errors.\footnote{In Table \ref{tab:2sls_tab1} Panel C, we report point estimates and standard errors derived from  95\% confidence intervals. We assume searching sampling produces an equal bandwidth confidence interval around a normally distributed point estimate.}

Table \ref{tab:2sls_tab1} Panel C shows the results from applying both methods to the 2SLS specifications with multiple instruments. Both methods may choose an empty set of valid IVs when candidate IVs are too weak conditional on controls. We are unable to report estimates for such cases. Overall, we find results consistent with those in Panels A and B, although standard errors tend to be larger.



\section{Mechanisms Behind Democracy's Adverse Effect}
\label{sec:channel}

We now turn to potential mechanisms. What explains democracy’s negative impact on growth? Our analysis documents that the effect of democracy is reversed between the 21st and earlier centuries. 
This reversal suggests that the key mechanisms cannot be stable factors like property rights institutions, which remained similar between the 21st and late 20th centuries.  
 We obtain data on other potential mechanisms that might experience key changes in the last few decades. 
 Our data cover different mechanisms, including political, economic, demographic, and educational channels, from 2001 to 2019.
 We list and summarize these variables in Table \ref{tab:descriptive-stats-main}. 

For the mean annual growth rate of each potential mechanism $M_i$ in 2001-2019, we estimate the following 2SLS equations:
\begin{align}
M_{i} = \alpha_2 + \beta_2 Democracy_i + \gamma_2X_i + \epsilon_{2i} \\
\text{First Stage: }Democracy_i = \alpha_1 + \beta_1Z_i  + \gamma_1 X_i + \epsilon_{1i}.
\end{align}
\noindent In the following, we report the coefficient $\beta_2$, which captures the percentage points increase in the mean annual growth rate of $M_i$ in response to a standard deviation (s.d.) increase in the democracy index. 
This specification mirrors our main 2SLS specification except that the second-stage dependent variable $M_i$ is a mechanism rather than an outcome. This approach is similar to Acemoglu et al. (2003)'s, which evaluates channels behind democratization's effects using a similar 2SLS. 
All of the following results are robust to controlling for the baseline level of the mechanism variable, absolute latitude, mean temperature, mean precipitation, population density, and median age (results are available upon request).

\textbf{Political Mechanisms.} We first find that in this century, democracies experience greater increases in threats to quality democracy. In Table \ref{tab:2sls-mechanisms-21st} Panel 1 (columns 1–4), we consider four political mechanisms: Protectionism (measured by a formula incorporating tariffs and other regulatory trade barriers), populism (the extent to which representatives of a political party use populist rhetoric), hate speech (how often major political parties use hate speech in their rhetoric), and polarization (how significant the differences of opinion are on major political issues among major political parties). The overidentified 2SLS estimates in Panel G indicate that for 2001–2019, a standard deviation increase in democracy increases protectionism by 0.4 s.d. (s.e. = 0.1 s.d.), populism by 0.6 s.d. (s.e. = 0.2 s.d.), hate speech by 0.5 s.d. (s.e. = 0.1 s.d.), and political polarization by 0.6 s.d. (s.e. = 0.08 s.d.). 
This result is consistent with the widespread concern that democracies become more myopic and xenophobic. Other 2SLS estimates are similar, though they tend to be less precisely estimated than the overidentified estimates. 
In the remainder of this section, we report the overidentified estimates in Panel G.


\textbf{Economic Mechanisms.} These political changes come with analogous economic changes. 
Consistent with the rising populism in democracies, democracy decreases growth in tangible and intangible investment in the future. 
For example, column 5 in Table \ref{tab:2sls-mechanisms-21st} shows that a standard deviation increase in democracy decreases the growth of capital stock formation by -0.6 percentage points (35.3\% of the mean, s.e. = 0.2 percentage points). 
Capital stock formation consists of outlays on additions to the fixed assets of the economy plus net changes in the level of inventories. 
A similar effect is observed for R\&D investments in column 6 of Table \ref{tab:2sls-mechanisms-21st}, where the mean annual growth rate decreases by 2.1 percentage points (36.8\% of the mean, s.e. = 0.5 percentage point) per a standard deviation increase in democracy. 

We observe similar negative impacts of democracy on education, which can also be interpreted as intangible social investment. 
As shown in columns 11 and 12 of Table \ref{tab:2sls-mechanisms-21st}, 
from 2001 to 2019, a standard deviation increase in democracy reduces the mean annual growth rate of primary school enrollment by 0.3 percentage points (60\% of the mean, s.e. = 0.1 percentage points) and that of secondary school enrollment by 1 percentage point (58.8\% of the mean, s.e. = 0.2 percentage points). 
Potentially as a result of stagnating investment, democracy also depresses TFP growth. As shown in Table \ref{tab:2sls-mechanisms-21st} column 8, the mean annual growth rate of TFP experiences a reduction of 0.4 percentage points (s.e. = 0.1 percentage points, 50.0\% of the mean) in response to a standard deviation increase in democracy.\footnote{The impact of democracy on labor force growth is unclear. As shown in column 7, the overidentified estimate is that a standard deviation increase in democracy insignificantly causes the mean annual growth rate of the labor force (the number of people aged 15 and older who supply labor for the production of goods and services) to decrease.} 
Consistent with increasing protectionism, democracies also experience stagnation in trade growth. 
Panel A column 9's overidentified estimate is -3.9 percentage points (s.e. = 0.4 percentage points, 40.6\% of the mean) decrease in the mean annual growth rate of the import value in 2001-2019. Column 10 exhibits similar estimates for exports, with the overidentified estimate being a 3.4 percentage point decrease in the mean annual growth rate (s.e. = 0.5 percentage points, 34.7\% of the mean) in response to a standard deviation democracy increase.
\footnote{A potential explanation for the dampening effect of democracy on trade is that electoral competition could lead to trade barriers \citep{anderson2013political}. We also check whether democratic nations are less likely to trade with China in Table \ref{tab:additional_mechanism}. We run 2SLS regressions with the share of the total value of imports from China or exports to China in GDP as the outcome variable. We find negative effects of democracy on imports from and exports to China.}


As a result of these factors, democracy slows value-added growth in manufacturing and services. 
Value added is the net output of a sector after adding up all the output and subtracting intermediate input. 
In columns 1-3 of Table \ref{tab:additional_mechanism}, we analyze the effect of democracy on the value added by sector. 
Column 1's overidentified estimate indicates that a standard deviation increase in democracy reduces the mean growth rate of the value added in agriculture by -1.3 percentage points (s.e. = 0.4 percentage points, 18.8\% of the mean). Similarly, columns 2-3 show that democracy significantly dampens value-added growth in manufacturing and services. The overidentified estimate in column 2 is that a standard deviation increase in democracy causes a -2 percentage points (s.e. = 0.7 percentage points, 23.8\% of the mean) decrease in the mean annual growth rate of manufacturing value added in 2001-2019. For services, column 3's overidentified estimate is -3.7 percentage points (s.e. = 0.4 percentage points, 40.7\% of the mean) decrease in the mean annual growth rate. 

Note that value-added growth can be decomposed into changes in capital input, labor input, and productivity.\footnote{Change in value-added in industry $i$ in period $t$ can be expressed in logarithms as $\Delta \ln V_{i,t} = \frac{\bar{v}_{K_{i,t}}}{\bar{v}_{V_{i,t}}} \Delta \ln K_{i,t} + \frac{\bar{v}_{L_{i,t}}}{\bar{v}_{V_{i,t}}} \Delta \ln L_{i,t} +  \frac{1}{\bar{v}_{V_{i,t}}} \Delta \ln A_{i,t} $. Here $\Delta \ln K_{i,t}$ $\Delta \ln L_{i,t}$ and $\Delta \ln A_{i,t}$ denote the log change in capital investments, labor inputs, and productivity of industry $i$ in period $t$, respectively. $v_{K_{i,t}}$ and $v_{L_{i,t}}$ are the shares of capital or labor compensation in the gross output of industry $i$. $v_{V_{i,t}}$ is the share of industry value added in industry gross output. The bars denote averages for the current and previous periods, $t$ and $t-1$. See \citet{jorgenson20083} for details.} 
The above results suggest that the slower value-added growth in manufacturing and services is
primarily caused by less capital investments and lower productivity, rather than less labor.
Ultimately, our results suggest that democracy might have stopped improving building blocks (investment in the future and trade with other countries) for growth, along with the degradation of democracy in the form of populism, protectionism, and hate speech, as summarized in the picture.\footnote{It is possible that the degradation of democracy was induced by the spread of the mobile web and the social media, though it is hard to empirically prove it due to the lack of data.} We also confirm that these mechanisms are not active during prior decades in the past century (results available upon request). 

\begin{figure}[htbp]
  \centering
  \label{fig:image_mech_effect}
  \includegraphics[width=0.99\linewidth]{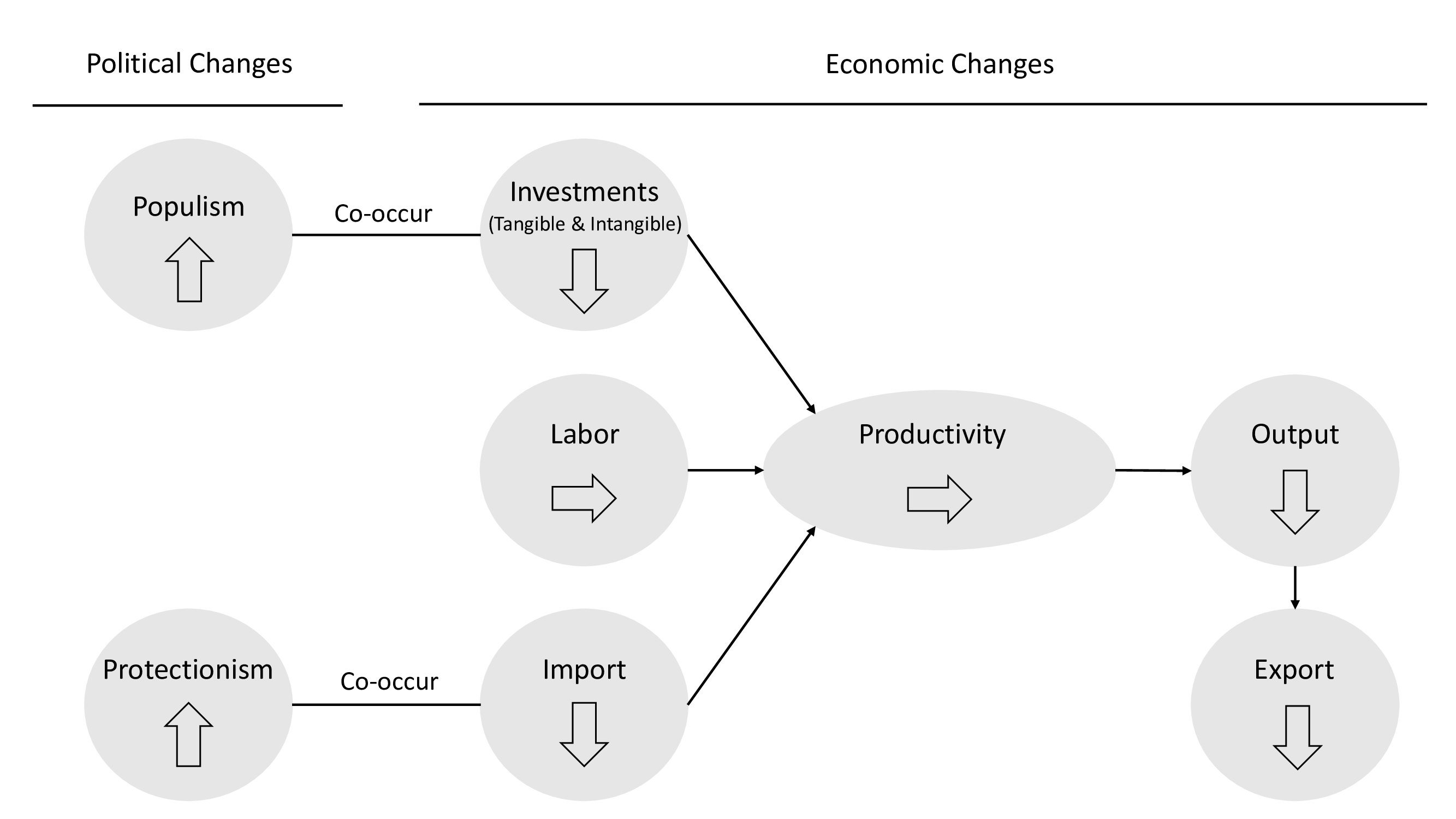}
\end{figure}

\textbf{Other Mechanisms.} In contrast, demographic channels appear less important in explaining the adverse effect of democracy. 
As shown in columns 13-14 of Table \ref{tab:2sls-mechanisms-21st}, democracy causes only insignificant effects on population growth. 
The effect on median age is even negative. 
In Table \ref{tab:additional_mechanism}, we also explore other potentially important channels, such as tax revenue growth, R\&D researchers, the number of new registered businesses, foreign direct investment (FDI), domestic conflict, and child mortality. In column 11, we find that, at the overidentified estimate (Panel G), a standard deviation increase in democracy slows the mean annual growth rate of tax revenue by 1 percentage point (s.e. = 0.4 percentage points, 71.4\% of the mean). Additionally, in column 11, we find that a standard deviation increase in democracy impacts child mortality, with its mean annual growth rate increasing by 1.9 percentage points (s.e. = 0.2 percentage points, 55.8\% of the mean) as the overidentified estimate (Panel G). For the other aforementioned mechanisms of Table \ref{tab:additional_mechanism}, we do not find a strong causal effect of democracy. 

Appendix \ref{subsubsec:mechanism-2020} provides a separate analysis of policy channels in the pandemic period (2020-22). A major channel for democracy's adverse effect in this period appears to be weaker and narrower containment policies at the beginning of the pandemic rather than the speed of containment policy implementation.

\section{Life Satisfaction, Inequality, and the Environment}\label{subsec:more}

The negative effect of democracy on average economic growth may be a side effect of democracies' efforts to achieve other objectives. To study this possibility, we analyze the impacts of democracy on additional outcomes, especially life satisfaction, economic inequality, and environmental burden. Consistent with the negative effect of democracy on economic growth, we also find a negative impact of democracy on life satisfaction, especially during the pandemic period. Figures \ref{fig:democracy_outcomes_lf2019} and \ref{fig:democracy_outcomes_lf2022} report a negative association between democracy and life satisfaction in 2001–2019 and a more pronounced negative association during 2020–2022. Causally, the median estimate among our five IV strategies in Panel A in Table \ref{tab:2sls} is that a standard-deviation increase in democracy reduces the mean growth rate of life satisfaction by 1.6 percentage points (s.e. = 0.2) during 2020–2022 (third line of column 9) and by 0.1 percentage points (s.e. = 0.4) during 2010–2019 (second line of column 9).\footnote{Democracy is also positively associated with excess deaths during the Covid-19 period, as shown in Figure \ref{fig:democracy_outcomes_deaths}. Furthermore, in Table \ref{tab:2sls}, Panel A, the median estimate (first line of column 3) suggests that a standard-deviation increase in democracy increases the average excess deaths per 100,000 people by 22.4 (s.e. = 8.1).} 
Therefore, even subjective well-being lag under democratic regimes in this period.

More nuanced views emerge from results on inequality and environmental burden. 
In Figures \ref{fig:democracy_outcomes_in2019} and \ref{fig:democracy_outcomes_in2022}, democracy is slightly negatively associated with the growth of the top 1\% income share in 2001–2019 but is positively associated with it in 2020–2022. Moreover, according to the median 2SLS estimates in Table \ref{tab:2sls}, a standard-deviation increase in democracy reduces the mean annual growth rate of the top 1\% income share by -0.4 percentage points (s.e. = 0.2) during 2001–2019 (fourth line of column 9) and increases it by 0.5 percentage points (s.e. = 0.1) during 2020–2022 (fifth line of column 9). 

In Figures \ref{fig:democracy_outcomes_co2019}, \ref{fig:democracy_outcomes_co2022}, \ref{fig:democracy_outcomes_en2019}, and \ref{fig:democracy_outcomes_en2022}, we find a strongly negative association between democracy and environmental burden. Additionally, the median 2SLS estimates in Table \ref{tab:2sls} suggest that a standard deviation increase in democracy slows the mean annual growth rate of CO\textsubscript{2} emissions per capita by 2.7 percentage points (s.e. = 0.5) during 2001–2019 (sixth line of column 7) and by 1.6 percentage points (s.e. = 0.4) during 2020–2022 (seventh line of column 5). Similarly, a one-standard-deviation increase in democracy decreases per-capita energy consumption by 2.5 percentage points (s.e. = 0.4) during 2001–2019 (eighth line of column 5) and by 1.5 percentage points (s.e. = 0.3) during 2020–2022 (ninth line of column 7).

These results suggest that democracies' economic growth may slow as they pursue more complex objectives, such as economic equality and environmental sustainability. 
This creates a trade-off that must be weighed by policymakers.

\section{Discussion} \label{sec:discuss}

\subsection{Alternative Specifications and Measurements}
Our analysis may be sensitive to measurement and modeling choices, such as how to measure economic performance, how to measure democracy, which countries and periods to analyze, whether to control for baseline GDP and other important characteristics, and how to weight countries. Extreme nations may also be driving our results. Below we check whether these concerns threaten our findings. 


\textbf{GDP per capita and light intensity per capita growth as the outcome.} We check whether our results hold for the per-capita GDP and per-capita light intensity growth rates in Table \ref{tab:2sls-gdppc}. We continue to observe a negative democracy effect.

\textbf{Alternative democracy indices.} 
We adopt alternative democracy indices by Polity (the Center for Systemic Peace), Freedom House, and the Economist Intelligence Unit. These indices are modestly correlated with each other (Table \ref{tab:indices-correlation}). Importantly, Table \ref{tab:2sls-compare-indices} confirms that our results stay similar regardless of which democracy index is used. Our results are thus not dependent on the particular V-Dem index.

\textbf{Separating the Great Recession.} We check if the Great Recession drives our results. Table \ref{tab:2sls-recession} conducts the same analysis separately for growth rates before, during, and after the recession period (2008-9). We find negative effects of democracy in every one of the periods.

\textbf{Alternative sample definitions.} To check if our results are driven by a few countries, such as the US and China, we show our results without the two countries in Table \ref{tab:2sls-compare-samples}. We re-estimate our preferred specification without outlier countries with a standardized residual above 1.96 or below -1.96 in Table \ref{tab:2sls-outliers}. Furthermore, we remove G7 countries from the sample in Table \ref{tab:2sls-remove-g7}. In all cases, we continue to estimate democracy's adverse effects. Limiting the sample to G20 countries also produces similar results. Thus, the negative impact of democracy is a global phenomenon and not driven by a handful of countries. 
%

\textbf{Alternative weightings.} Our 2SLS results so far weight countries by baseline GDP. We believe that baseline GDP weighting is reasonable, especially when the outcomes are GDP growth rates. Nonetheless, we compare our results with weighting by population or no weighting in Table \ref{tab:2sls-compare-weighting}. The qualitative pattern is the same among the three ways to weight countries. 

\textbf{Control for baseline GDP.} We test whether our results are due to the mechanical reason that more developed countries tend to grow slower. Table \ref{tab:2sls-control-gdp} runs regressions with baseline total GDP or GDP per capita as controls. 
The resulting estimates continue to find democracy's negative effect, confirming that baseline GDP conditions do not drive our results. 

\textbf{Control for continents.} We additionally control for dummy variables for each continent in  Table \ref{tab:2sls-control-continent}. Although the estimates are less precisely estimated, we continue to observe democracy's negative effect on economic growth and public health. This suggests that the democracy treatment is significant within each continent. 


\subsection{21st vs 20th Centuries.} 

It is natural to ask whether our finding is unique to the 21st century. Additional evidence suggests so. 
We apply the same descriptive and IV analyses to data from the 20th century. 
The resulting estimates show that the negative association between democracy and economic growth did not exist in 1981-1990 or 1991-2000 (Figure \ref{fig:ols-decade}). More importantly, for the same period, we do not observe a significantly negative causal effect of democracy (Table \ref{tab:2sls-by-decade}). The reduced-form relationships between the IVs and economic growth in 1981-2000 are either insignificant or of the opposite sign to those in 2001-2020. 
These results suggest that democracy’s growth effects worsened specifically in the 21st century.
Thus, how democracy matters for economic growth might have changed around the turn of the 21st century. 

\citet{barro2004economic} find an inverse U-shaped relationship between democracy (electoral rights) and economic growth using data from 1965–1994. That is, for the early period, democracy is positively associated with growth for non-democratic countries, but is negatively associated with growth for democratic countries. We also find the same inverse U-shaped relationship in 1981–2000. However, the relationship changes and becomes monotonically downwards sloping in 2001–2020, i.e., democracy becomes globally negatively associated with growth (Figures \ref{fig:ols-decade_quad} and \ref{fig:ols-decade_quad_res}). This finding further supports our results and suggests that the way democracy matters for growth has changed around the turn of the 21st century. 
We also test whether the causal relationship between democracy and growth is nonlinear. We find no conclusive evidence, mainly due to the small data size. Results are available upon request.

\subsection{Effects of Becoming More Democratic} \label{subsec:change}

We have focused on studying the adverse effect of being more democratic at the baseline. A related problem is the effect of \textit{becoming} more democratic (i.e., democratization) in this century. 
This question is especially relevant due to the recent concern about democratic backsliding. In the past decade, several countries have experienced changes towards less democratic institutions and cultures. 
The effect of democratization could be positive, especially because democracy change in 2001–2019 is slightly negatively associated with baseline democracy level in 2000 (Figure \ref{fig:ols_change3}).
However, Figures \ref{fig:ols_change1} and \ref{fig:ols_change2} show that the correlation between economic growth and democracy change (the average annual change in the democracy index) is negative. Democracy change is associated with lower GDP growth in 2001–2019 and lower night-time light intensity growth in 2001–2013.\footnote{Standard errors are large for the correlation between night-time light intensity growth and democracy change. The association between the variables could be neutral or positive.} 
Table \ref{tab:2sls_change} panel B provides OLS estimates of the negative correlations. 
We also find similar negative correlations when using baseline democracy change between 1999 and 2000 instead. 
Table \ref{tab:2sls_change}  presents 2SLS estimates of the effects of democracy change on growth outcomes using the same set of instruments from before. Due to a weak first stage, however, we cannot detect a significant causal effect. 

In Appendix Section \ref{subsubsec:mechanism-2020}, we further explore the causal effect of democratization on growth under the dynamic panel-data framework of \citet{acemogluDemocracyDoesCause2018}. We only find inconclusive evidence, yet the points estimates suggest that the effect of democratization may be weaker after 2000 than before that. 
The causal effects of democracy change on economic growth in the 21st century are therefore not conclusive but are largely consistent with our main results. 

\section{Conclusion} \label{sec:conclusion}
Skepticism about the performance of democratic political regimes is as old as the invention of democracy: 


\begin{quote}
``\textit{having them [the multitude of the citizens] take part in the greatest offices is not safe: through injustice and imprudence they would act unjustly in some respects and err in others.}" (Aristotle, \textit{Politics}, 1281b25)
\end{quote}

We bring data and five different IVs to find that democracy has a significant negative effect on economic growth in the 21st century — in sharp contrast to 20th-century findings.
The channels likely behind the negative effect of democracy are rising myopic and xenophobic tendencies in democracies, which are associated with stagnant investment and trade growth. The negative impact of democracy remains strong during the pandemic, in which democracy causes not only worse GDP shocks, but also more excess deaths. 
Overall, political institutions still matter for economic growth, but how they matter might have changed between the prior and current centuries. 

We do not argue that democracy is undesirable overall, or that authoritarianism is preferable, for at least two reasons. First, democracy has normative and procedural virtues, regardless of whether they result in good economic outcomes. Second, despite our findings on democracy's negative impacts in the 21st century, democracies may produce better outcomes in the long run or other aspects. Indeed, we find that democracy causes improvements in environmental burden for the same period, pointing to a trade-off between economic growth and social or environmental outcomes. 
Understanding how democracy can adapt to meet both economic and social goals is a central question for scholars and policymakers alike.


\newpage

\newgeometry{left=0.3cm, right = 0.3cm, top = 1cm, bottom=1in}
\begin{figure}
\centering
\caption{Correlation Between Democracy and Growth}\label{fig:ols}
\captionsetup{width=0.99\textwidth}

\begin{subfigure}[c]{.49\linewidth}
    \centering
    \caption{Mean GDP Growth Rate in 2001-2019}\label{fig:ols-mean-gdp}
    \includegraphics[width=0.99\textwidth]{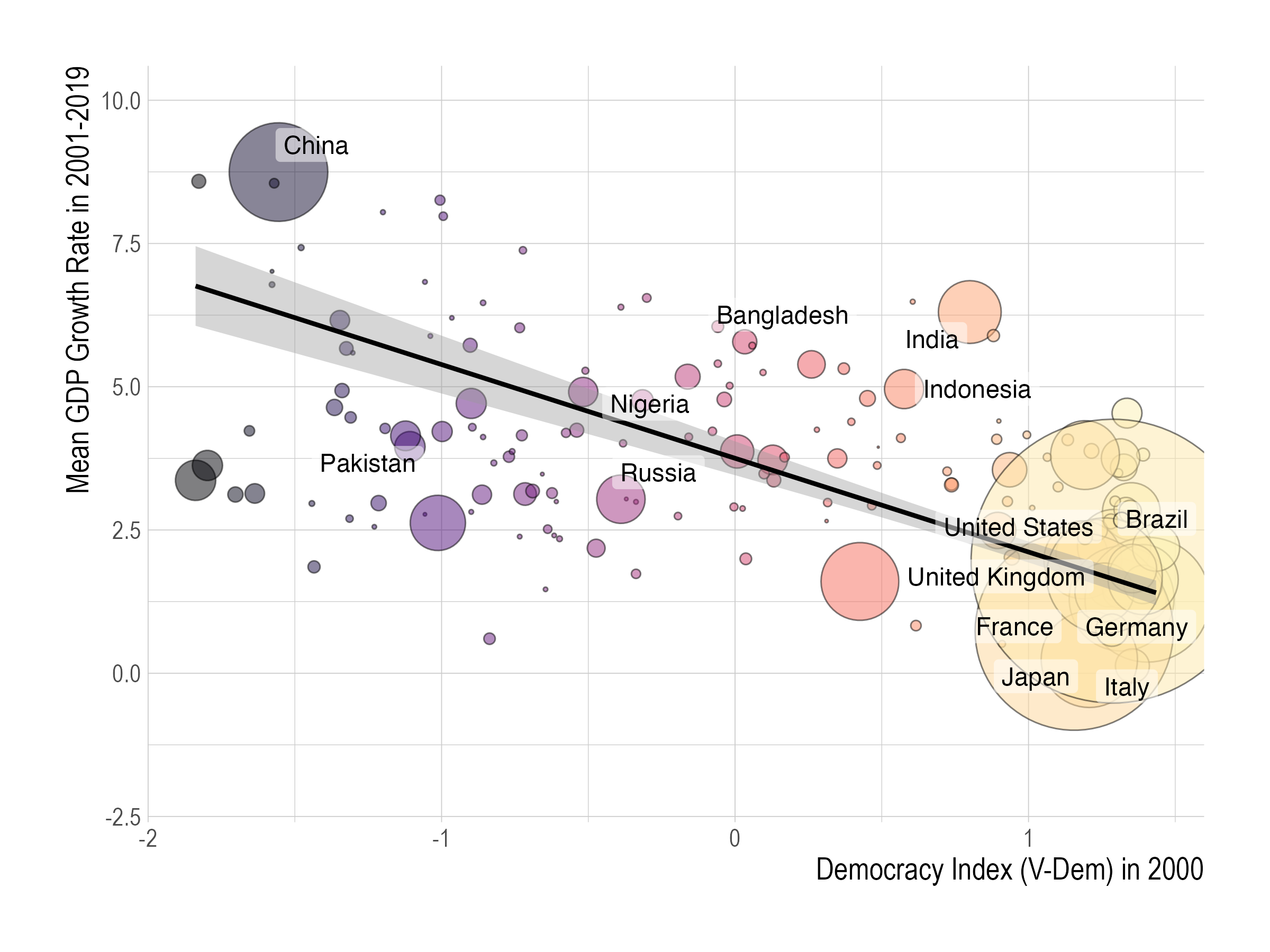}
    
    \caption{Mean GDP Growth Rate in 2020-2022}\label{fig:ols-gdp-2020}
    \includegraphics[width=0.99\textwidth]{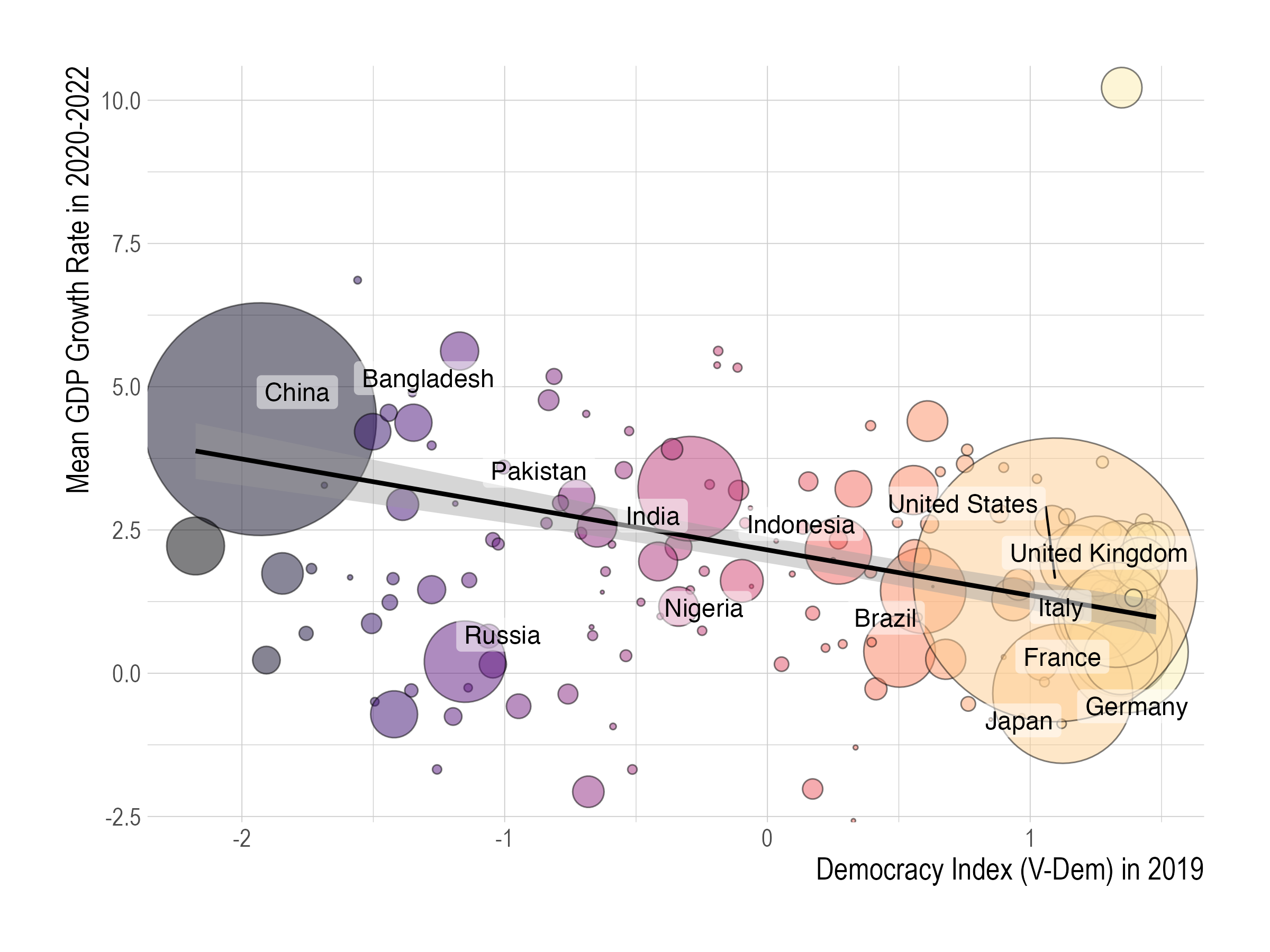}

    \caption{Mean Night-time Light Intensity Growth Rate in 2001-2013}\label{fig:ols-light}
    \includegraphics[width=0.99\textwidth]{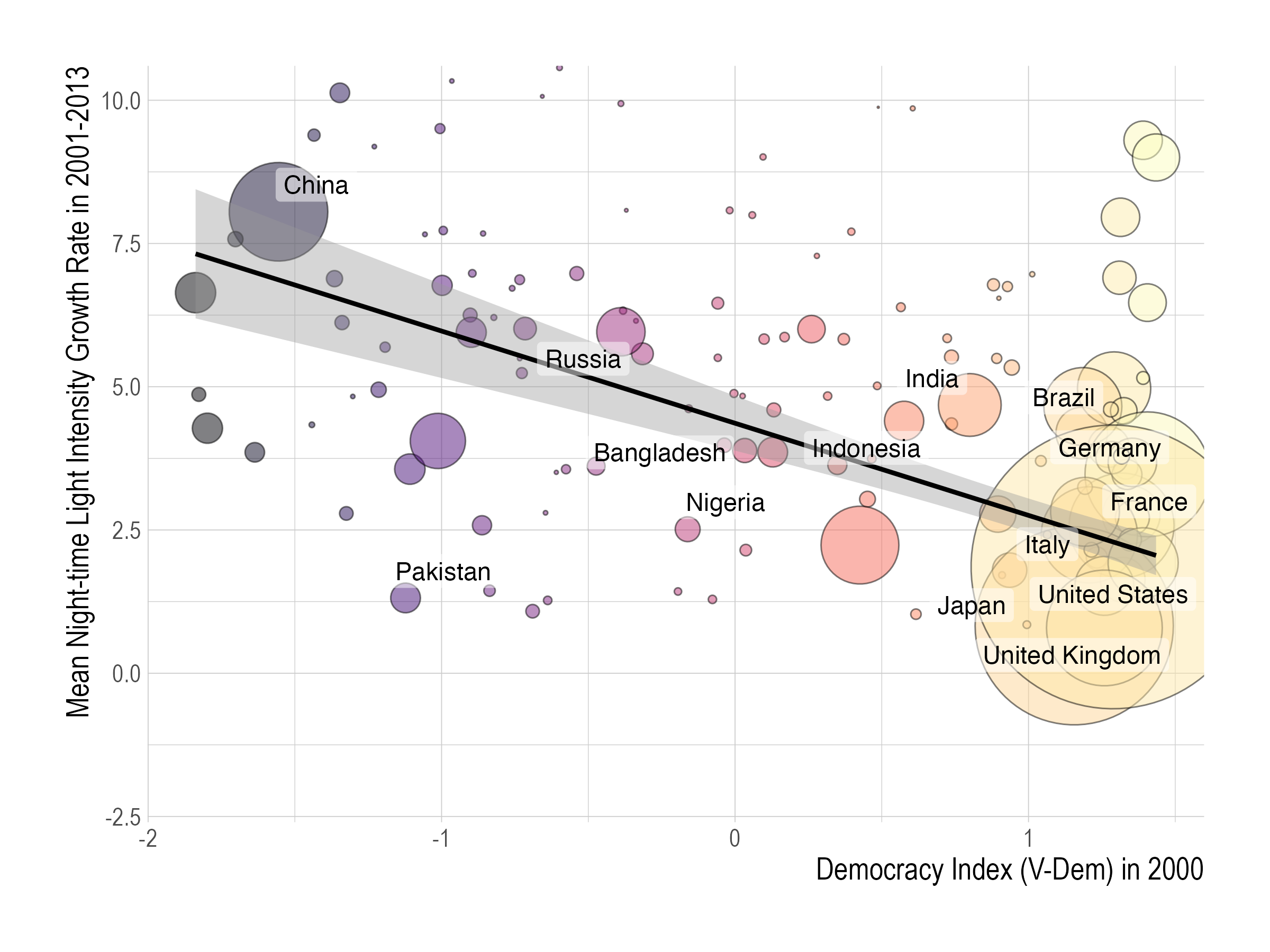}
\end{subfigure}

\caption*{\footnotesize{\textit{Notes:} This figure shows the correlation between democracy and growth outcomes. Panels (a) and (b) display the mean GDP growth rates in 2001-19 and 2020-22, respectively. Panel (c) examines the mean night-time light intensity growth rate in 2001-13. The size of each circle (country) is proportional to its baseline GDP. The colors represent the level of democracy, with warmer colors indicating higher levels of democracy. The line represents the OLS regression fitted line without controls, weighted by countries' baseline GDP. The shaded area corresponds to the 95\% confidence interval. Variable definitions and data sources are in Appendix Table \ref{tab:sources}.}}

\end{figure}
\restoregeometry
\restoregeometry

\newpage
\newgeometry{left=0.3cm, right = 0.3cm, top = 1cm, bottom=1in}
\begin{figure}
\centering
\caption{Causal Effects of Democracy: First Look}\label{fig:first-stage}
\captionsetup{width=0.99\textwidth}
\begin{subfigure}[c]{.49\linewidth}
    \centering
    \caption{First-stage: Log European Settler Mortality IV}\label{fig:first-stage-european-settlers}
    \includegraphics[width=.99\textwidth]{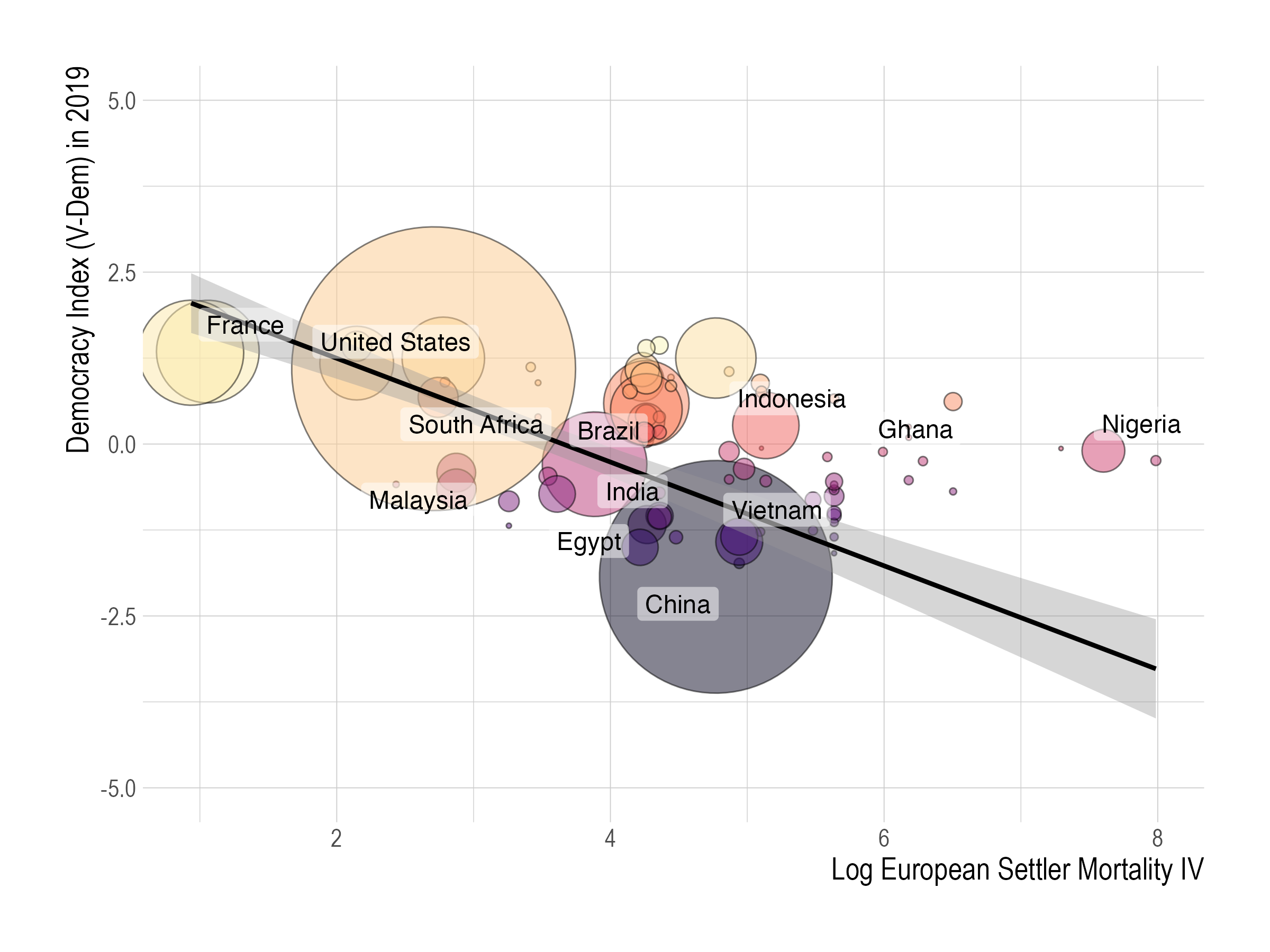}
    
    \caption{First-stage: Log Population Density in 1500s IV}\label{fig:first-stage-population-density}
    \includegraphics[width=.99\textwidth]{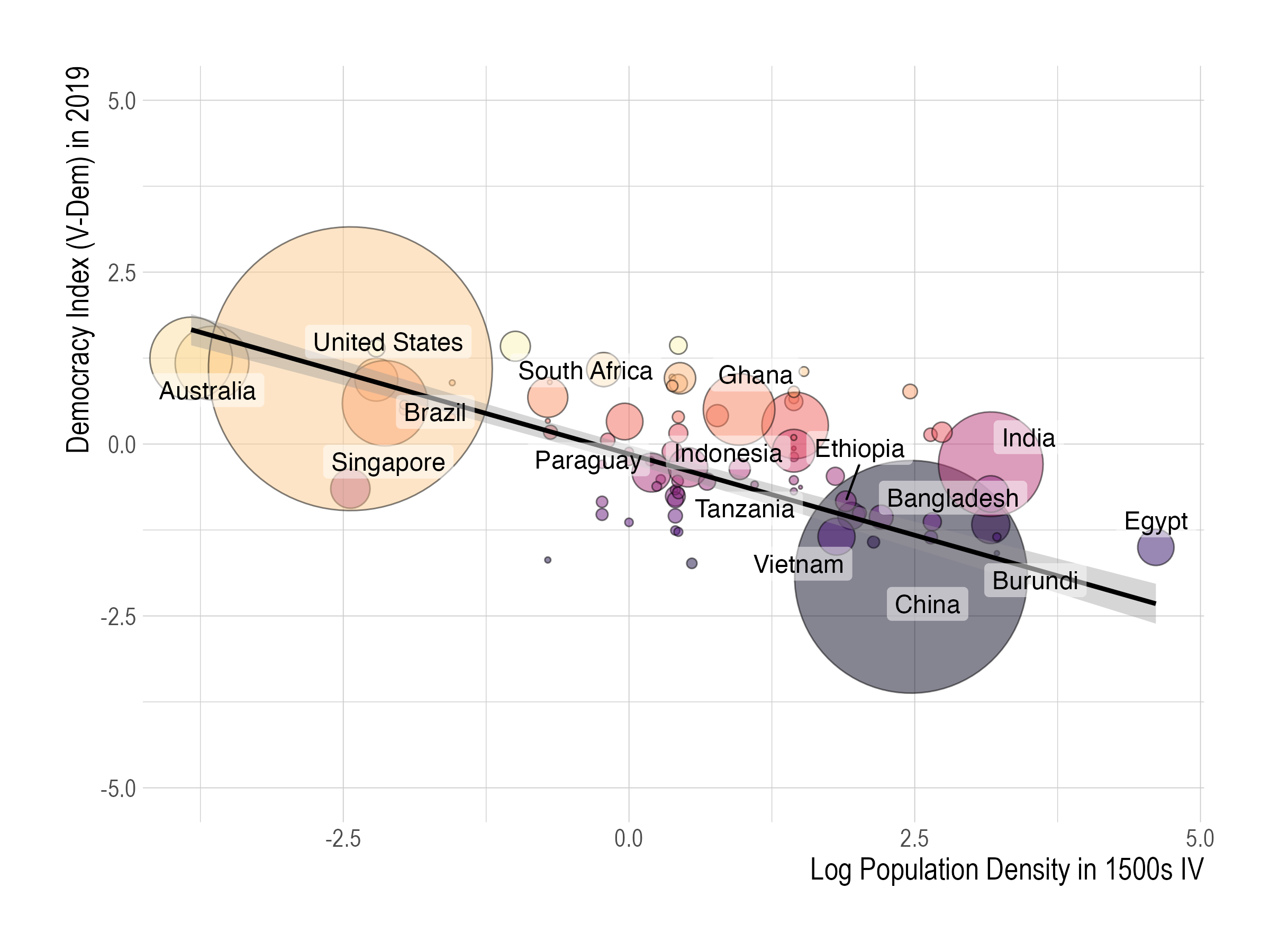}

     \caption{First-stage: Fraction Speaking European IV}\label{fig:first-stage-fraction-european}
    \includegraphics[width=.99\textwidth]{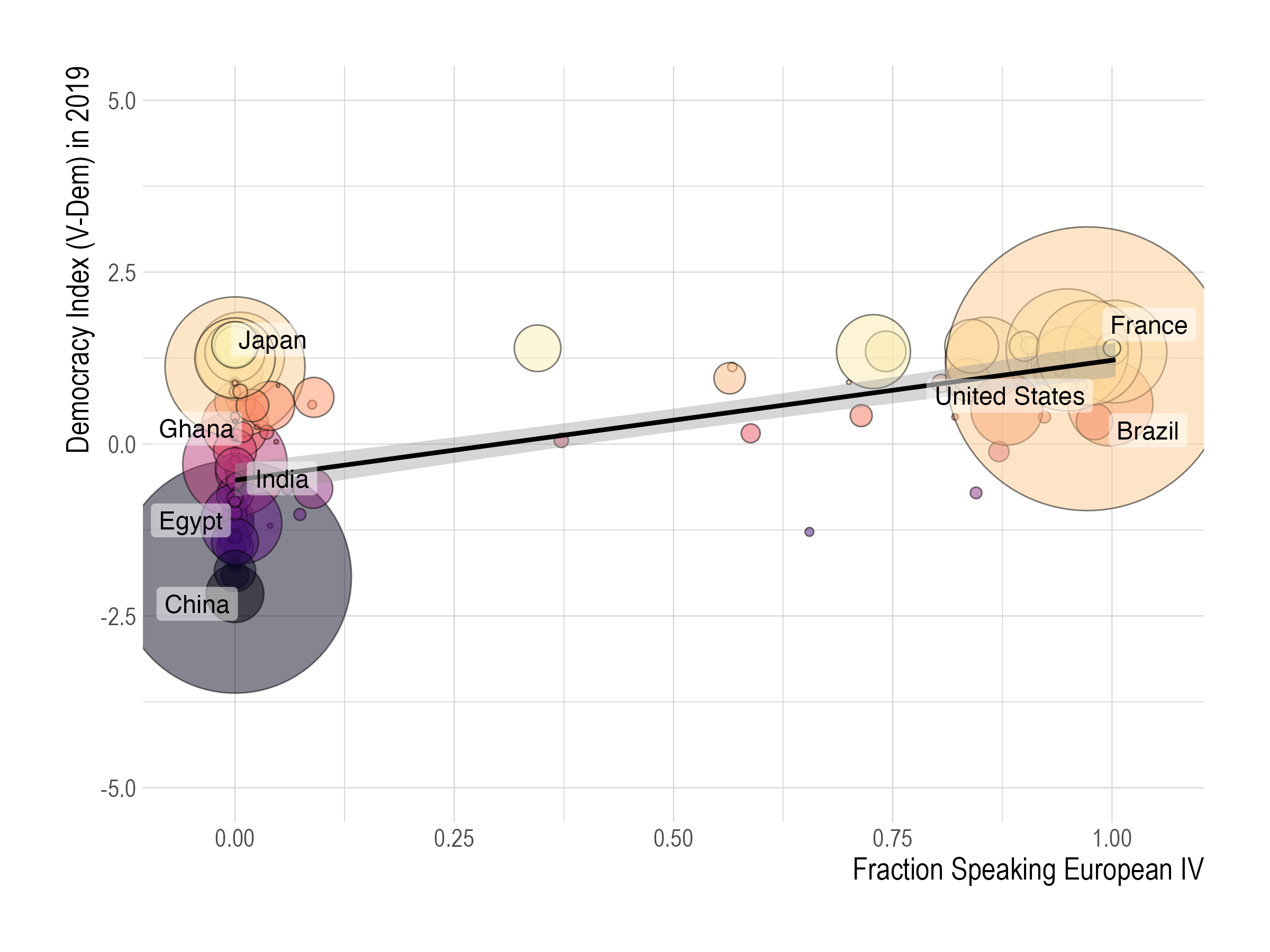}

\end{subfigure}
\begin{subfigure}[c]{.49\linewidth}
    \centering
    
    \caption{Reduced form: Mean GDP Growth Rate in 2001–2019}\label{fig:reduced-mean-gdp-logem}
    \includegraphics[width=.99\textwidth]{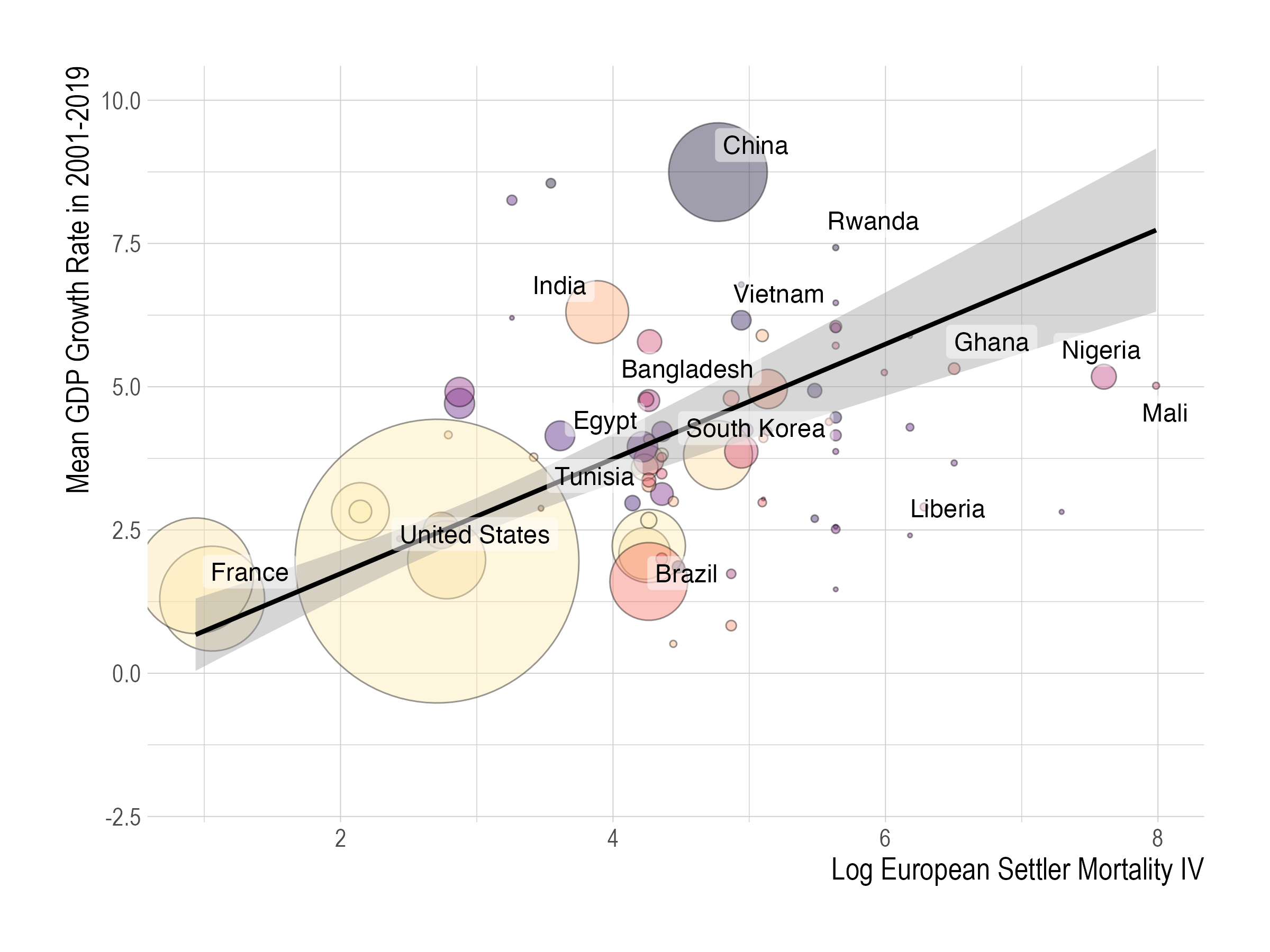}
    
    \caption{Reduced form: Mean GDP Growth Rate in 2020–2022}\label{fig:reduced-gdp-2020-logem}
    \includegraphics[width=.99\textwidth]{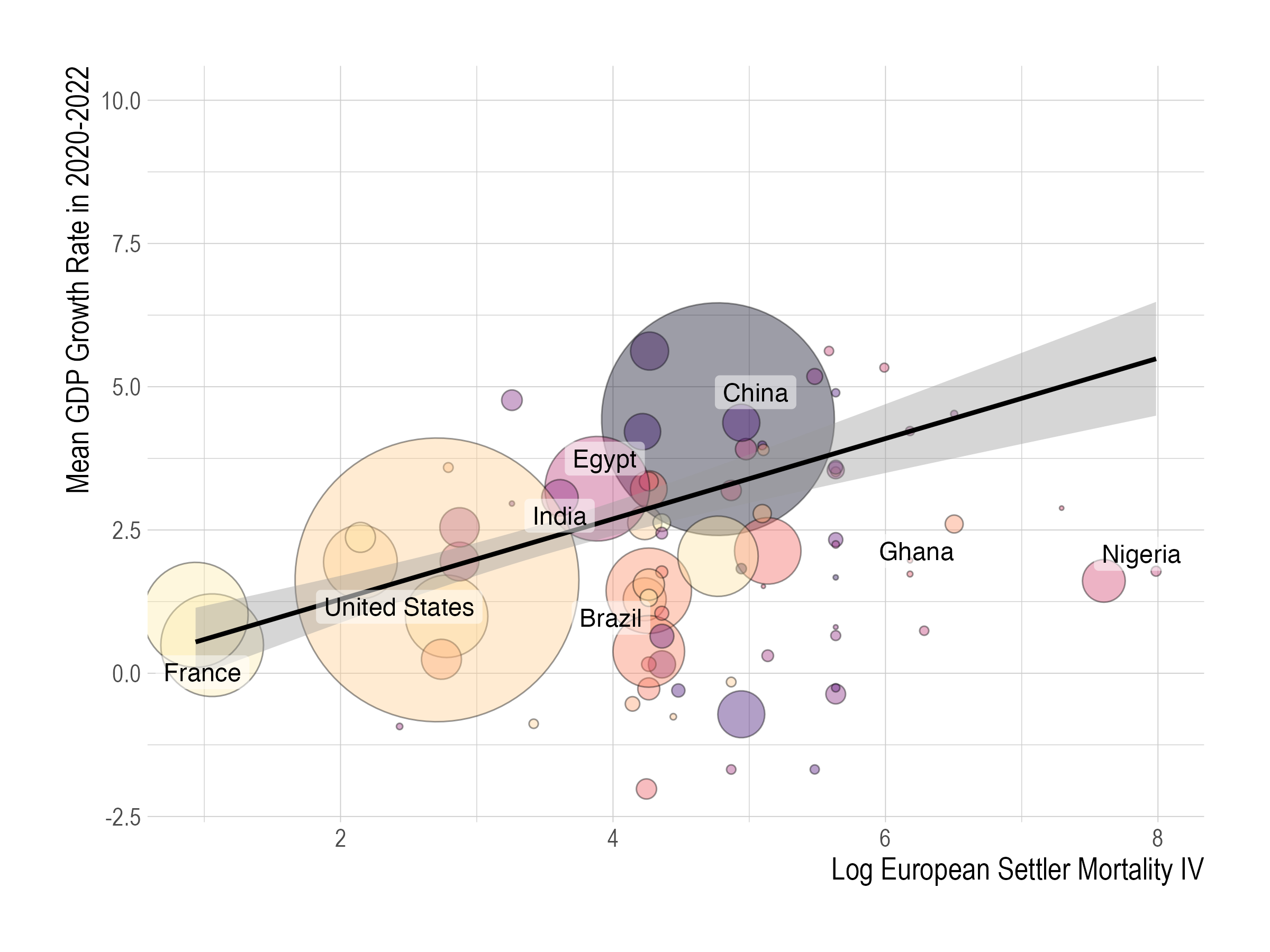}
 
     \caption{Reduced form: Mean Nighttime Light Intensity Growth Rate in 2001-2013}\label{fig:reduced-deaths-logem}
    \includegraphics[width=.99\textwidth]{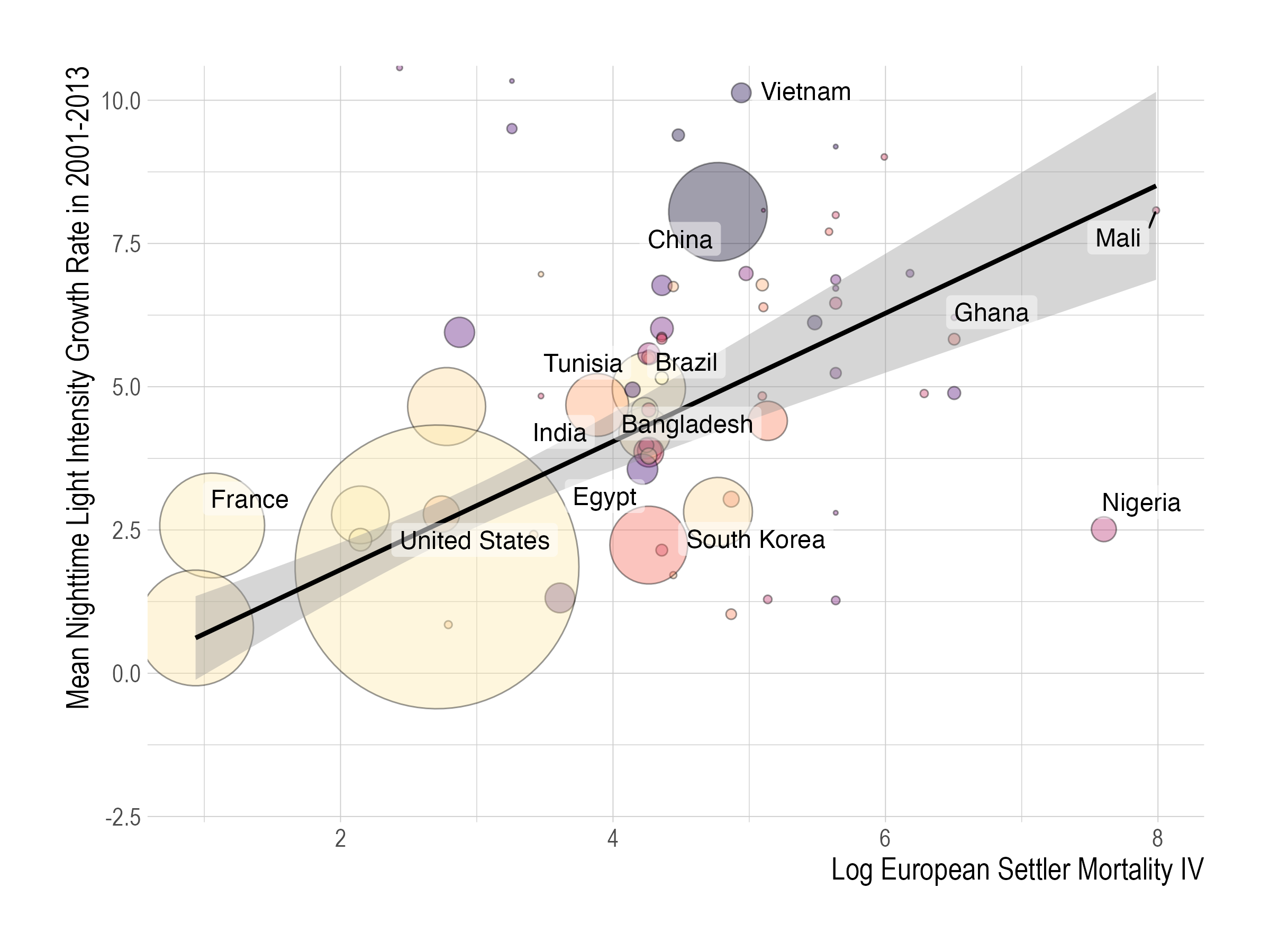}
\end{subfigure}


\caption*{\footnotesize{\textit{Notes:} Panels (a), (b), and (c) show the first-stage relationships between democracy in 2019 and three univariate IVs: the log European settler mortality IV, the log population density in 1500s IV, and the fraction speaking European IV. Panels (d), (e), and (f) show the reduced-form relationship between the log European settler mortality IV and three outcomes: the mean
GDP growth rate in 2001-2019, the mean GDP growth rate in 2020-2022, and the mean night-time light intensity growth rate in 2001-2013. \input{supporting_files/coefs/explanation_dem} The size of each circle (country) is proportional to its baseline GDP. The colors depend on the level of the democracy index (warmer colors for democracy and darker colors for autocracies). The line is the OLS regression fitted line without controls and weights countries by baseline GDP. The shaded area corresponds to the 95\% confidence interval. \input{supporting_files/coefs/explanation_vars}}}
\end{figure}

\restoregeometry

\newgeometry{left=0.3cm, right = 0.3cm, bottom=1in, top = 1in}

\begin{center}
\begin{table}  \centering
  \caption{2SLS and GMM Estimates of Democracy's Effects}\label{tab:2sls_tab1} 
  \footnotesize
  \begin{threeparttable}
  
  \begin{tabular}{@{\extracolsep{0pt}}lcccccccccccc} 
\hline 
\hline 
 & (1) & (2) & (3) & (4) & (5) & (6) & (7) & (8) & (9) & (10) & (11) & (12)\\ 
\hline 
\multicolumn{13}{l}{\textbf{Panel A: Two-Stage Least Squares}} \\ 
& \multicolumn{12}{c}{Dependent Variable is Mean GDP Growth Rate in 2001-2019} \\\cline{2-13}\\[-1.8ex]
Democracy Index (V-Dem, 2000)&        -2.5&        -4.2&        -2.2&        -3.1&        -1.8&        -1.8&        -1.0&        -1.1&        -2.2&        -1.6&        -2.1&        -2.6\\
&       (0.3)&       (2.7)&       (0.4)&       (0.6)&       (0.5)&       (1.3)&       (0.9)&       (0.6)&       (0.5)&       (0.6)&       (0.2)&       (0.2)\\
&        0.00&        0.13&        0.00&        0.00&        0.00&        0.18&        0.28&        0.06&        0.00&        0.01&        0.00&        0.00\\

& \multicolumn{12}{c}{Dependent Variable is Mean Nighttime Light Intensity Growth Rate in 2001-2013} \\\cline{2-13}\\[-1.8ex]
Democracy Index (V-Dem, 2000)&        -2.8&        -5.3&        -1.8&        -1.3&        -2.5&        -3.2&        -1.6&        -3.0&        -2.3&        -2.7&        -2.0&        -2.0\\
&       (0.7)&       (3.8)&       (0.4)&       (0.6)&       (0.6)&       (1.0)&       (1.1)&       (1.0)&       (0.5)&       (0.5)&       (0.2)&       (0.4)\\
&        0.00&        0.17&        0.00&        0.02&        0.00&        0.00&        0.16&        0.00&        0.00&        0.00&        0.00&        0.00\\

& \multicolumn{12}{c}{Dependent Variable is Mean GDP Growth Rate in 2020-2022} \\\cline{2-13}\\[-1.8ex]
Democracy Index (V-Dem, 2019)&        -0.9&        -0.8&        -0.9&        -1.0&        -0.8&        -0.7&        -0.7&        -0.6&        -1.0&        -0.8&        -0.9&        -1.0\\
&       (0.1)&       (0.2)&       (0.1)&       (0.1)&       (0.2)&       (0.1)&       (0.3)&       (0.2)&       (0.2)&       (0.2)&       (0.1)&       (0.1)\\
&        0.00&        0.00&        0.00&        0.00&        0.00&        0.00&        0.02&        0.02&        0.00&        0.00&        0.00&        0.00\\
IVs & \multicolumn{2}{c}{settler mortality} &  \multicolumn{2}{c}{population density} & \multicolumn{2}{c}{legal origin} & \multicolumn{2}{c}{language} & \multicolumn{2}{c}{crops \& minerals} & \multicolumn{2}{c}{all IVs} \\
Number of IVs & 1 & 1 & 1 & 1 & 1 & 1 & 2 & 2 & 10 & 10 & 15 & 15 \\
F-Statistic (First stage)&         9.7&         7.3&        27.2&       135.2&        12.0&        16.8&         4.3&        14.0&         6.4&         6.0&        57.5&       351.8\\
J-Statistic (p-value)&           .&           .&           .&           .&           .&           .&        0.03&        0.22&        0.48&        0.12&        0.45&        0.31\\
N                   &          81&          81&          86&          86&          90&          90&         132&         132&         138&         138&          71&          71\\

 \\[-1.8ex] 
\hline 
\multicolumn{13}{l}{\textbf{Panel B: Two-Step Efficient GMM}} \\ 
& \multicolumn{12}{c}{Dependent Variable is Mean GDP Growth Rate in 2001-2019} \\\cline{2-13}\\[-1.8ex]
Democracy Index (V-Dem, 2000)&        -2.5&        -4.2&        -2.2&        -3.1&        -1.8&        -1.8&        -0.5&        -1.2&        -1.9&        -1.6&        -2.1&        -2.6\\
&       (0.3)&       (2.7)&       (0.4)&       (0.6)&       (0.5)&       (1.3)&       (0.9)&       (0.6)&       (0.3)&       (0.5)&       (0.1)&       (0.2)\\
&        0.00&        0.13&        0.00&        0.00&        0.00&        0.18&        0.54&        0.04&        0.00&        0.00&        0.00&        0.00\\

& \multicolumn{12}{c}{Dependent Variable is Mean Nighttime Light Intensity Growth Rate in 2001-2013} \\\cline{2-13}\\[-1.8ex]
Democracy Index (V-Dem, 2000)&        -2.8&        -5.3&        -1.8&        -1.3&        -2.5&        -3.2&        -1.7&        -2.6&        -2.8&        -2.7&        -2.1&        -2.2\\
&       (0.7)&       (3.8)&       (0.4)&       (0.6)&       (0.6)&       (1.0)&       (1.1)&       (0.9)&       (0.3)&       (0.4)&       (0.1)&       (0.3)\\
&        0.00&        0.17&        0.00&        0.02&        0.00&        0.00&        0.13&        0.00&        0.00&        0.00&        0.00&        0.00\\

& \multicolumn{12}{c}{Dependent Variable is Mean GDP Growth Rate in 2020-2022} \\\cline{2-13}\\[-1.8ex]
Democracy Index (V-Dem, 2019)&        -0.9&        -0.8&        -0.9&        -1.0&        -0.8&        -0.7&        -0.7&        -0.6&        -1.0&        -0.9&        -0.9&        -0.9\\
&       (0.1)&       (0.2)&       (0.1)&       (0.1)&       (0.2)&       (0.1)&       (0.3)&       (0.2)&       (0.1)&       (0.2)&       (0.1)&       (0.1)\\
&        0.00&        0.00&        0.00&        0.00&        0.00&        0.00&        0.02&        0.01&        0.00&        0.00&        0.00&        0.00\\
IVs & \multicolumn{2}{c}{settler mortality} &  \multicolumn{2}{c}{population density} & \multicolumn{2}{c}{legal origin} & \multicolumn{2}{c}{language} & \multicolumn{2}{c}{crops \& minerals} & \multicolumn{2}{c}{all IVs} \\

 \\[-1.8ex] 
\hline 
\multicolumn{13}{l}{\textbf{Panel C: IV Estimation with Potentially Invalid IVs}} \\ 
& \multicolumn{6}{c}{\textbf{I) Two-Stage Hard Thresholding}} & \multicolumn{6}{c}{\textbf{II) Searching-Sampling}}\\
\cmidrule(lr){2-7} \cmidrule(lr){8-13}
& \multicolumn{12}{c}{Dependent Variable is Mean GDP Growth Rate in 2001-2019} \\\cline{2-13}\\[-1.8ex]
Democracy Index (V-Dem, 2000)&. &-1.3 &-0.8 &-0.5 &-2.8 &-1.0 &. &. &. &. &-2.2 &-4.6 \\
&. &(1.0) &(1.0) &(1.0) &(0.9) &(1.9) &. &. &. &. &(1.3) &(1.8) \\
&. &0.21 &0.43 &0.61 &0.00 &0.62 &. &. &. &. &0.10 &0.01 \\

& \multicolumn{12}{c}{Dependent Variable is Mean Nighttime Light Intensity Growth Rate in 2001-2013} \\\cline{2-13}\\[-1.8ex]
Democracy Index (V-Dem, 2019)&. &-1.3 &-1.4 &-1.0 &-3.2 &-1.3 &. &. &. &-1.3 &-3.4 &-3.6 \\
&. &(1.5) &(2.2) &(1.3) &(1.1) &(1.1) &. &. &. &(1.8) &(1.0) &(1.8) \\
&. &0.41 &0.52 &0.45 &0.00 &0.22 &. &. &. &0.47 &0.00 &0.05 \\

& \multicolumn{12}{c}{Dependent Variable is Mean GDP Growth Rate in 2020-2022} \\\cline{2-13}\\[-1.8ex]
Democracy Index (V-Dem, 2000)&-1.0 &-0.9 &-0.9 &-0.8 &-0.9 &-0.8 &. &. &-0.7 &-0.1 &-3.7 &-1.6\\
&(0.3) &(0.4) &(0.2) &(0.2) &(0.3) &(0.7) &. &. &(0.2) &(0.7) &(1.9) &(1.2)\\
&0.00 &0.02 &0.00 &0.00 &0.01 &0.23 &. &. &0.00 &0.93 &0.05 &0.20 \\
IVs & \multicolumn{2}{c}{language} &  \multicolumn{2}{c}{crops \& minerals} & \multicolumn{2}{c}{all IVs} & \multicolumn{2}{c}{language} & \multicolumn{2}{c}{crops \& minerals} & \multicolumn{2}{c}{all IVs} \\
Number of IVs & 2 & 2 & 10 & 10 & 15 & 15 & 2 & 2 & 10 & 10 & 15 & 15 \\
 \\[-1.8ex] 
\hline 
\hline 
\end{tabular} 
\begin{tablenotes} 
\item {\footnotesize {\textit{Notes:} Panel A reports the 2SLS estimates of democracy's effect on the mean GDP growth rate in 2001-2019, the mean nighttime light intensity growth rate in 2001-2013, and the mean GDP growth rate in 2020-2022, using five different IV strategies.  
 The Democracy Index (V-Dem) is normalized to have mean zero and standard deviation one in each year.\unskip 
For IVs, columns 1 and 2 use log European settler mortality, columns 3 and 4 use log population density in the 1500s, columns 5 and 6 use British legal origin, columns 7 and 8 use the fraction speaking English and the fraction speaking European, columns 9 and 10 use the ability to grow crops and mine minerals, and columns 11 and 12 use all the IVs together. \unskip 
Robust standard errors are in parentheses. \unskip 
The p-values, presented under the standard errors, are displayed as 0.00 if they are strictly smaller than the 0.005 threshold. The F-statistics are from the first-stage regressions of the IVs against the democracy index in 2019. The corresponding first-stage coefficients are in Appendix Table \ref{tab:first-stage}. 
The J-statistic (p-value) shows the p-value from the overidentification J test of the null hypothesis that the instruments are valid. A low p-value suggests that some of the instruments may be invalid. 
Columns 1, 3, 5, 7, 9, and 11 have no controls, while columns 2, 4, 6, 8,  10, and 12 have the following baseline controls: absolute latitude, mean temperature, mean precipitation, population density, and median age. For the mean GDP growth rate in 2020-2022, we also control for diabetes prevalence. \unskip 
N refers to the number of countries for which IVs, controls, and outcomes are available. \unskip
Panel B reports the two-step efficient GMM estimates. Panel C reports the estimates of democracy's effect using two methods to allow for potentially invalid IVs. Columns 1–6 use two-stage hard thresholding to select valid IVs, and columns 7-12 use searching-sampling to select valid IVs. For potential IVs, columns 1–2 and 7–8 use the fraction speaking English and the fraction speaking European, columns 3–4 and 9–10 use the ability to grow crops and mine minerals, and columns 5–6 and 11–12 use all the IVs together.
Variable definitions and data sources are in Appendix Table \ref{tab:sources}. \unskip}}
\end{tablenotes}
\end{threeparttable}
\end{table} 
\end{center}

\begin{landscape}
\begin{table}[htbp]
\centering
\begin{threeparttable}
  \caption{Mechanisms Behind Democracy’s Effect in 2001-2019}\label{tab:2sls-mechanisms-21st}
\scriptsize
\setlength{\tabcolsep}{4pt} 
\renewcommand{\arraystretch}{1.1} 

\hspace*{-4cm} 
\vspace*{1.1cm} 
\begin{tabular}{l@{}*{14}{c@{}}} 
\hline\hline \\[-1.8ex]

& \multicolumn{4}{c}{\textbf{1) Political Channels}} & \multicolumn{6}{c}{\textbf{2) Economic Channels}}
& \multicolumn{2}{c}{\textbf{3) Education Channels}}
& \multicolumn{2}{c}{\textbf{4) Demographic Channels}}\\

\cmidrule(lr){2-5} \cmidrule(lr){6-11} \cmidrule(lr){12-13} \cmidrule(lr){14-15}

& \multicolumn{1}{c}{\shortstack{Protectionism\\Growth}} 
& \multicolumn{1}{c}{\shortstack{Populism\\Growth}} 
& \multicolumn{1}{c}{\shortstack{Hate Speech\\Growth}} 
& \multicolumn{1}{c}{\shortstack{Polarization\\Growth}} 
& \multicolumn{1}{c}{\shortstack{Capital\\Investment\\Growth}} 
& \multicolumn{1}{c}{\shortstack{R\&D\\Investment\\Growth}} 
& \multicolumn{1}{c}{\shortstack{Labor Force\\Growth}}
& \multicolumn{1}{c}{\shortstack{TFP\\Growth}}
& \multicolumn{1}{c}{\shortstack{Imports\\Growth}} 
& \multicolumn{1}{c}{\shortstack{Exports\\Growth}} 
& \multicolumn{1}{c}{\shortstack{Primary\\Education\\Growth}}
& \multicolumn{1}{c}{\shortstack{Secondary\\Education\\Growth}}
& \multicolumn{1}{c}{\shortstack{Population\\Growth}} 
& \multicolumn{1}{c}{\shortstack{Median Age\\Growth}}\\

&\multicolumn{1}{c}{(1)} 
&\multicolumn{1}{c}{(2)} 
&\multicolumn{1}{c}{(3)} 
&\multicolumn{1}{c}{(4)}  
&\multicolumn{1}{c}{(5)} 
&\multicolumn{1}{c}{(6)}  
&\multicolumn{1}{c}{(7)} 
&\multicolumn{1}{c}{(8)} 
&\multicolumn{1}{c}{(9)} 
&\multicolumn{1}{c}{(10)} 
&\multicolumn{1}{c}{(11)} 
&\multicolumn{1}{c}{(12)} 
&\multicolumn{1}{c}{(13)}
&\multicolumn{1}{c}{(14)} 
\\

\hline

& \multicolumn{13}{c}{\textbf{Panel A: OLS}}\\
Democracy Index (V-Dem, 2000)&         0.2&         0.4&         0.4&         0.4&        -0.5&        -3.7&        -0.5&        -0.2&        -3.2&        -3.0&        -0.1&        -0.6&        -0.4&        -0.3\\
&       (0.1)&       (0.2)&       (0.1)&       (0.1)&      (0.10)&       (2.0)&       (0.3)&       (0.3)&       (0.6)&       (0.6)&      (0.05)&       (0.1)&       (0.2)&      (0.06)\\
N                   &         115&         148&         160&         160&         125&         125&         159&         113&         115&         115&         146&         138&         154&         154\\
& \multicolumn{13}{c}{\textbf{Panel B: Instrument for Democracy by Settler Mortality}}\\
Democracy Index (V-Dem, 2000)&         0.7&         0.6&        -0.3&       -0.08&        -0.6&        -2.9&        -0.4&        -0.6&        -5.0&        -4.5&        -0.1&        -1.6&        -0.3&        -0.6\\
&       (0.3)&       (0.3)&       (0.6)&       (0.4)&       (0.2)&       (1.0)&       (0.4)&       (0.2)&       (0.8)&       (0.7)&       (0.1)&       (0.3)&       (0.2)&       (0.2)\\
N                   &          69&          77&          82&          82&          71&          65&          82&          59&          61&          61&          78&          71&          82&          82\\
& \multicolumn{13}{c}{\textbf{Panel C: Instrument for Democracy by Population Density in 1500s}}\\
Democracy Index (V-Dem, 2000)&         0.6&         0.7&         0.2&         0.4&        -0.6&        -2.2&        -0.2&        -0.4&        -4.3&        -3.8&        -0.4&        -0.9&        -0.1&        -0.4\\
&       (0.2)&       (0.3)&       (0.3)&       (0.2)&       (0.2)&       (0.6)&       (0.3)&       (0.2)&       (0.8)&       (0.8)&       (0.2)&       (0.3)&       (0.2)&       (0.1)\\
N                   &          69&          80&          87&          87&          71&          67&          87&          60&          62&          62&          81&          73&          86&          86\\
& \multicolumn{13}{c}{\textbf{Panel D: Instrument for Democracy by Legal Origin}}\\
Democracy Index (V-Dem, 2000)&         0.8&         0.8&         0.3&         0.3&        -0.6&        -2.7&        -0.4&        0.05&        -4.4&        -3.8&       -0.03&        -2.3&       -0.01&        -0.6\\
&       (0.4)&       (0.4)&       (0.4)&       (0.3)&       (0.2)&       (1.0)&       (0.4)&       (0.5)&       (0.8)&       (0.7)&       (0.2)&       (0.6)&       (0.1)&       (0.2)\\
N                   &          71&          84&          91&          91&          73&          68&          91&          62&          65&          65&          85&          77&          90&          90\\
& \multicolumn{13}{c}{\textbf{Panel E: Instrument for Democracy by Language}}\\
Democracy Index (V-Dem, 2000)&         0.8&         2.3&         0.7&         1.4&        -0.7&        -4.0&        0.05&        -0.2&        -4.0&        -2.8&        -0.2&        -0.5&         0.3&        -0.6\\
&       (0.4)&       (1.4)&       (0.5)&       (0.8)&       (0.2)&       (1.7)&       (0.5)&       (0.4)&       (0.9)&       (1.3)&       (0.1)&       (0.2)&       (0.5)&       (0.2)\\
N                   &         107&         122&         133&         133&         106&         106&         132&         101&          97&          97&         125&         118&         130&         130\\
& \multicolumn{13}{c}{\textbf{Panel F: Instrument for Democracy by Crops and Minerals}}\\
Democracy Index (V-Dem, 2000)&         0.6&         0.7&       -0.08&         0.2&        -0.6&        -2.4&        -0.5&        -0.3&        -4.7&        -4.3&        -0.2&        -0.5&        -0.3&        -0.5\\
&       (0.3)&       (0.4)&       (0.4)&       (0.3)&       (0.2)&       (0.9)&       (0.3)&       (0.3)&       (0.8)&       (0.8)&      (0.07)&       (0.3)&       (0.2)&       (0.2)\\
N                   &         107&         127&         139&         139&         109&         108&         138&          99&         101&         101&         130&         122&         137&         137\\
& \multicolumn{13}{c}{\textbf{Panel G: Use all IVs}}\\
Democracy Index (V-Dem, 2000)&         0.4&         0.6&         0.5&         0.6&        -0.6&        -2.1&        -0.1&        -0.4&        -3.9&        -3.4&        -0.3&        -1.0&       -0.05&        -0.3\\
&       (0.1)&       (0.2)&       (0.1)&      (0.08)&       (0.2)&       (0.5)&       (0.3)&       (0.1)&       (0.4)&       (0.5)&       (0.1)&       (0.2)&       (0.1)&      (0.06)\\
N                   &          62&          67&          72&          72&          63&          58&          72&          52&          52&          52&          68&          62&          72&          72\\
\\ Outcome Mean & N/A & N/A & N/A & N/A & 1.7 & 5.7 & 1.9 & 0.8 & 9.6 & 9.8 & 0.5 & 1.7 & 1.6 & 0.9\\
\\[-1.8ex] 

\hline\hline
\end{tabular}

\begin{tablenotes}
\item {\footnotesize {\textit{Notes:} This table reports the OLS (Panel A) and 2SLS regression (Panels B-G) estimates of democracy's effect on potential mechanisms in 2001-2019.  The Democracy Index (V-Dem) is normalized to have mean zero and standard deviation one in each year.\unskip 
Protectionism measures a country's aversion to international trade; Populism measures the extent to which representatives of a political party use populist rhetoric; Hate Speech measures how often major political parties use hate speech in their rhetoric; Polarization indicates how significant the differences of opinions are on major political issues among major political parties in a society; Capital Investment is the gross capital formation for a specific country; R\&D Investment is the R\&D expenditure expressed as a share of GDP; Labor Force is the total number of workers above the age of 15 in a specific country; Exports is the total value of exports of goods and services in US dollars; Imports is the total value of imports of goods and services in US dollars; Primary Education and Secondary Education represent the enrollment rates for primary education and secondary education, respectively. \unskip
 For IVs, Panel B uses log European settler mortality, Panel C uses log population density in the 1500s, Panel C uses British legal origin, Panel E uses the fraction speaking English and the fraction speaking European, Panel F uses the ability to grow crops and mine minerals, and Panel G uses all the IVs together. \unskip 
Robust standard errors are in parentheses. \unskip 
We report the global mean of the dependent variables in the bottom row. The specifications do not include controls. N refers to the number of countries for which IVs and outcomes are available.
Variable definitions and data sources are in Appendix Table \ref{tab:sources}. \unskip}}
\end{tablenotes}
\end{threeparttable}
\end{table}
\end{landscape}

\restoregeometry

\clearpage
\section*{References}

\bibliographystyle{chicago}
{\footnotesize

}

\newpage
\appendix
\setcounter{table}{0}
\renewcommand{\thetable}{\thesubsection\arabic{table}}
\setcounter{figure}{0}
\renewcommand{\thefigure}{\thesubsection\arabic{figure}}
\subsection{Appendix}
\setcounter{page}{1} 

\renewcommand{\thepage}{A\arabic{page}}


\subsubsection{Extending \citet{acemogluDemocracyDoesCause2018}'s Analysis of the Effect of Democratization}\label{subsubsec:acemoglu-ext}

\citet{acemogluDemocracyDoesCause2018} document that transitions to democracy cause per-capita GDP growth from 1960 to 2010. They use dynamic panel strategies that control for country fixed effects and utilize regional waves of democratization as an IV. While our finding that democracy has had a negative cross-country impact on growth since the beginning of the century neither denies nor corroborates \citet{acemogluDemocracyDoesCause2018}'s finding, it is interesting nevertheless to explore whether their positive effect of democratization persists or weakens in recent years. We do so by extending their analysis to include 2010-2020. We examine how the impact of democratization on growth has evolved over time. 

 For the extended period of 1960 to 2020, we confirm that democratization positively impacts GDP per capita. 
Specifically, using \citet{acemogluDemocracyDoesCause2018}'s preferred model — which includes country fixed effects and four lags of GDP — our analysis in Column 3 in Table \ref{tab:Table1_ext} (Panel (a), first line) suggests a 1.16\% immediate increase in GDP per capita following a permanent shift to democracy. Over the long term, this effect is substantially larger, resulting in a 28.63\% (s.e. = 7.87\%) increase in GDP per capita, as detailed in the fourth line of Panel (a) in Table \ref{tab:Table1_ext}. 
In Panel (b), where we limit the analysis to the years 1980 to 2020, the long-term effect remains comparable at 27.03\% (s.e. = 7.15\%). 
This result is similar to Table 2 in \citet{acemogluDemocracyDoesCause2018}

For the more recent periods 1995-2020 and 2001-2020, however, the estimates are statistically insignificant and exhibit larger standard errors (Panels (c) and (d)). Notably, in Panel (d), the sign of the democratization coefficient reverses, suggesting a potential shift in the relationship in recent years. Table \ref{tab:Table2_ext} provides additional robustness checks and finds a similar pattern to Table \ref{tab:Table1_ext}. Figure \ref{fig:Figure1_ext} presents corroborating evidence graphically. Overall, the impact of democratization on GDP per capita is inconclusive for 2001-2020.


\subsubsection{Policy Mechanisms Behind Democracy's Effect during the Pandemic}\label{subsubsec:mechanism-2020}


Why does having a stronger democracy cause worse economic and public health outcomes during the Covid-19 pandemic? Media and policy discussions point to the speed, coverage, and severity of containment policies as potential proximate mechanisms. For example, Paul Krugman blames ``\textit{catastrophically slow and inadequate}" responses by the US government for its failure.\footnote{Krugman, Paul. 2020. ``3 Rules for the Trump Pandemic." \emph{New York Times.} March 19. \url{https://www.nytimes.com/2020/03/19/opinion/trump-coronavirus.html}} We explore whether this differential in policy responses explains democracy's negative effect. Our findings suggest that a key channel for the negative impact of democracy is weaker and narrower containment policies at the beginning of the outbreak. In contrast, the speed of containment policies appears to be less important. 

To measure the severity of the policy, we use the Containment Health Index for the 10th confirmed case of Covid-19. The Containment Health Index is a measure of how strict the government's response to containing Covid-19 was \citep{OxCGRT}.\footnote{We get similar results when we use the index at the 100th confirmed case or the index's mean during 2020.} To quantify how widely initial responses cover aspects of civilian life, we look at the percentage of 13 domains in which the government introduced containment measures at the 10th Covid-19 case. The domains are schools, workplaces, public events, gatherings, public transport, stay-at-home requirements, domestic travel, international travel, public information campaigns, testing, contact tracing, facial coverings, and vaccinations. To assess policy speed, we consider the number of days between the 10th confirmed case and the introduction of any containment policy.\footnote{We get similar results with the 100th confirmed case and January 1st, 2020 as the start date. The introduction date of any containment policy is the date when the Containment Health Index becomes positive.} 



For each policy response mechanism $M$ (severity, coverage, or speed of containment response), we estimate the following 2SLS equations:
\begin{align}
M_{i} = \alpha_2 + \beta_2 Democracy_i + \gamma_2X_i + \epsilon_{2i} \\
\text{First Stage: }Democracy_i = \alpha_1 + \beta_1Z_i  + \gamma_1X_i + \epsilon_{1i}.
\end{align}

Table \ref{tab:channels} summarizes the results of this analysis.\footnote{We get similar results with alternative democracy indices, weighting, and sample definitions.} Panel A shows that democracy causes less severe responses at the 10th confirmed case of Covid-19. The median estimate is that a standard deviation increase in democracy causes the Containment Health Index to decrease by 0.4 standard deviations, which corresponds to 20\% of the mean. Democracy also narrows the scope of containment policies. The median estimate in Panel B suggests that a standard deviation increase in democracy causes a 9.3 percentage-point decrease in the coverage of the initial policy. On the other hand, democracy does not appear to cause slower responses. In fact, in Panel C, all columns predict that democracy causes \textit{faster} responses. This leads to the bottom line that the severity and coverage of initial containment policies is a more important mechanism for the adverse effect of democracy than their speed.

\setcounter{table}{0}
\clearpage
\subsubsection{Additional Results}
\begin{footnotesize}
\begin{center}
\begin{spacing}{.7}
\begin{longtable}{p{0.07\linewidth} p{0.2\linewidth} p{0.27\linewidth} p{0.36\linewidth}} 
    \caption{Data Sources and Description} 
    \label{tab:sources}
    \scriptsize

    \\\hline  \\
    & \textbf{Variable}  &  \textbf{Data Source} & \textbf{Short Description}\\ 
    \hline \\ \endhead
    Outcomes 
    
    & Mean GDP Growth Rate &
    \citet{wdi2021} & Annual percentage change in real GDP. It is calculated using GDP and GDP deflator (base year varies by country) from the World Bank's WDI dataset. \\
    

    & Mean Nighttime Light Intensity Growth Rate &
    \cite{martinez2022much} &  Nighttime light intensity as measured by satellite images \\ 
    
    
    \hline \\
    
    Additional Outcomes
    
    & Mean Excess Deaths Per One Hundred Thousand & \citet{theEconomist} & Number of deaths per one hundred thousand between 2020/01/01 and 2022/12/31 in excess of the the baseline number of deaths we might normally have expected in this period.\\

    & Life Satisfaction & \citet{helliwell2023happiness} & Life Satisfaction (life ladder) measures the well-being by asking respondents to imagine a ladder from 0 to 10, with 10 being their best possible life and 0 their worst. Respondents then rate their current life on this scale, capturing their overall life satisfaction. \\

    & Top 1\% Income Share & \citet{wid} & Percentage of total income held by the wealthiest 1\% of the population. \\

    & CO2 Emissions Per-Capita & \citet{co2emissions} & Carbon dioxide (CO2) emissions from fossil fuels and industry, measured in tonnes per person. \\

    & Energy Consumption Per-Capita & \citet{energy} & Energy consumption is expressed in kilowatt-hours per person. It refers to primary energy—that is, the energy required by the end user for electricity, transportation, and heating—plus the energy lost due to inefficiencies during the conversion of raw resources into a usable form.  \\
    
    \hline \\
    
    Treatments 
    
    & Democracy Index (V-Dem) &
    \citet{coppedge2021a} &  Index  aggregating indices measuring freedom of association, clean elections, freedom of expression, elected officials, and suffrage.\\
    
    & Democracy Index (Polity) & \citet{centerforsystemicpeacePolity5AnnualTime2018} &  Index measuring the level of democracy by having the lower extreme being "hereditary monarchy" and the upper "consolidated democracy".\\

    & Democracy Index (Freedom House) & \citet{freedomhouseFreedomWorld20202020} &  Index measuring the degree of democratic freedom by considering political rights and civil liberties.\\

     & Democracy Index (Economist Intelligence Unit) & \citet{DemocracyIndex2020} & Index measuring the state of democracy. Ranges from least democratic to most democratic.\\

\hline \\

    Weightings \& Controls 
    
    
    
    & Population (Millions) & \citet{unitednationsdepartmentofeconomicandsocialaffairspopulationdivisionWorldPopulationProspects2019} & \\

    & Absolute Latitude & \citet{GooglePublicData} & Latitude of the centroid of each country (i.e., a measure of distance from the equator).\\
    
    & Mean Temperature & \citet{theworldbankgroupClimateChangeKnowledge2020} & Average monthly temperature in degrees Celcius. \\
    
    & Mean Precipitation & \citet{theworldbankgroupClimateChangeKnowledge2020} & Average monthly precipitation in millimeters. \\
    
    & Population Density & \citet{unitednationsdepartmentofeconomicandsocialaffairspopulationdivisionWorldPopulationProspects2019} & The number of people divided by land area, measured in square kilometers. \\
    
    & Median Age & \citet{unitednationsdepartmentofeconomicandsocialaffairspopulationdivisionWorldPopulationProspects2019} & United Nations projections of the median age of the population. \\
    
    & Diabetes Prevalence & \citet{internationaldiabetesfederationIDFDiabetesAtlas2019} & \% of population with diabetes aged 20 to 79. \\ \hline \\

    IVs & 
    Log European Settler Mortality & \citet{ColonialOriginsReplicationData} & Log annualized deaths per thousand mean strength of European settlers between the seventeenth and nineteenth century.\\ 
    
    & Log Population Density in 1500s & \citet{UnbundlingInstitutionsReplicationData} & Log population density in the 1500s measured as the number of inhabitants per square kilometer. \\
    
    & British Legal Origin & \citet{portaLawFinanceReplicationData} & Binary indicator that the country's legal origin is British. \\

    & Fraction Speaking English & \citet{hallWhyCountriesProduceReplicationData} & The fraction of the population speaking English as a mother tongue in 1992.\\ 
    
    & Fraction Speaking European & \citet{hallWhyCountriesProduceReplicationData} & The fraction of the population speaking English, French, German, Portuguese or Spanish as a mother tongue in 1992.\\ 
    
    

    
    & Bananas, Coffee, Maize, Millet, Rice, Sugarcane, Rubber, Wheat & \citet{easterlyReplicationData, foodandagricultureassociationoftheunitednationsFAOGlobalStatistical2020} & Binary indicator that the country produced any of the particular commodity in 1990. \\
    
    & Copper, Silver & \citet{easterlyReplicationData, CopperStatistics, SilverStatistics} & Binary indicator that the country mined any of the particular commodity in 1990. \\ 

    \\\hline \\

    Mechanisms
    
    & Protectionism Growth & \citet{ftti} & A measure of a country’s aversion to trade internationally. \\
    
    & Populism Growth & \citet{populism} & A measure of the extent to which members and representatives of a political party use populist rhetoric. \\
    
    & Hate Speech Growth & \citet{digitalSociety} & A measure of the extent to which hate speech is used in the rhetoric of major political parties. \\
    
    & Polarization Growth & \citet{digitalSociety} & A measure of how significantly opinions differ on major political issues among major political parties in a society. \\
    
    & Capital Investment Growth & \citet{wdi2021} & Annual growth rate of gross capital formation expressed as a \% of the GDP. Gross capital formation (formerly gross domestic investment) consists of outlays on additions to the fixed assets of the economy plus net changes in the level of inventories. Fixed assets include land improvements (fences, ditches, drains, and so on); plant, machinery, and equipment purchases; and the construction of roads, railways, and the like, including schools, offices, hospitals, private residential dwellings, and commercial and industrial buildings. Inventories are stocks of goods held by firms to meet temporary or unexpected fluctuations in production or sales, and ``work in progress." According to the 1993 SNA, net acquisitions of valuables are also considered capital formation.\\
    
    & R\&D Expenditure Growth & \citet{UNESCOdata}&Annual growth rate of gross domestic expenditures on R\&D, expressed as a percent of GDP. They include both capital and current expenditures in the four main sectors: business enterprise, government, higher education and private non-profit. R\&D covers basic research, applied research, and experimental development. \\
    
     & Labor Force Growth & \citet{wdi2021} & Annual growth rate of the labor force, which comprises people ages 15 and older who supply labor for the production of goods and services during a specified period. It includes people currently employed and people unemployed but seeking work as well as first-time job-seekers. Not everyone who works is included, however. Unpaid workers, family workers, and students are often omitted. Some countries do not count members of the armed forces. \\
     
     & TFP Growth & \citet{pennworldtables} &  Annual growth rate of total factor productivity calculated using real GDP and factor input growth rates obtained from national accounts data. \\
     
     & Import Value Growth & \citet{wdi2021} & Annual growth rate of the imports of goods and services (BoP, in current US\$).  \\
     
    & Export Value Growth & \citet{wdi2021} & Annual growth rate of the exports of goods and services (BoP, in current US\$).\\
    
    & Population Growth & \citet{popGrowth} & Annual growth rate of population. \\
    
    & Median Age Growth & \citet{medianAge} & Annual growth rate of median age. \\
    
    & Primary Education Growth & \citet{wdi2021} & Annual growth rate of the ratio of children of official school age enrolled in primary school to the total population of children in the corresponding official school age group. \\
    
    & Secondary Education Growth & \citet{wdi2021} & Annual growth rate of the ratio of children of official school age enrolled in secondary school to the total population of children in the corresponding official school age group.\\

    \\\hline \\
    
    Additional Mechanisms 
    
    & Value Added, Agriculture Growth & \citet{wdi2021} & Annual growth rate for agricultural value added expressed in current US dollars. Agriculture corresponds to ISIC divisions 1-5 and includes forestry, hunting, and fishing, as well as cultivation of crops and livestock production. Value added is the net output of a sector after adding up all outputs and subtracting intermediate inputs. It is calculated without making deductions for depreciation of fabricated assets or depletion and degradation of natural resources. The origin of value added is determined by the International Standard Industrial Classification (ISIC), revision 3. \\
    
    & Value Added, Manufacturing Growth & \citet{wdi2021} & Annual growth rate for manufacturing value added expressed in current US dollars. 
    Manufacturing refers to industries belonging to ISIC divisions 15-37.\\ 
    
    & Value Added, Services Growth & \citet{worldbankVAService} & Annual growth rate for value added in services expressed in current US dollars. 
    Services correspond to ISIC divisions 50-99. They include value added in wholesale and retail trade (including hotels and restaurants), transport, and government, financial, professional, and personal services such as education, health care, and real estate services. Also included are imputed bank service charges, import duties, and any statistical discrepancies noted by national compilers as well as discrepancies arising from rescaling. \\
    
    & Tex Revenue Growth & \citet{taxRevenue} & Annual growth rate of the ratio of tax revenues to GDP. It includes both direct and indirect taxes, as well as social contributions. Taxes are defined as compulsory, unrequited payments to the government, while social contributions encompass both compulsory and voluntary social insurance contributions from employers, employees, and the self-employed.\\
    
    & R\&D Researchers Growth & \citet{UNESCOdata}& Annual growth rate of the number of researchers engaged in R\&D per million. Researchers are professionals who conduct research and improve or develop concepts, theories, models techniques instrumentation, software of operational methods. \\
    
    & New Business Registrations Growth & \citet{WorldBankEntrepreneur}& Annual growth rate of new businesses per 1,000 people. New businesses registered are the number of new limited liability corporations registered in the calendar year. For cross-country comparability, only limited liability corporations that operate in the formal sector are included.\\
    
    & FDI Growth & \citet{fdi} & Annual growth rate of foreign direct investments (FDI). FDI are defined as the net inflows of investment used to acquire a lasting management interest (10 percent or more of the voting stock) in an enterprise operating in a foreign economy. It comprises equity capital, reinvested earnings, other long-term capital, and short-term capital, as recorded in the balance of payments. The series represents the net inflows from foreign investors in the reporting economy, as a share of GDP. \\
    
    & Conflict Index Growth & \citet{banks2021a} & Weighted average of indices measuring assassinations, strikes, guerilla warfare, government crises, purges, riots, revolutions, and anti-government demonstrations. \\
    & Child Mortality Growth &\citet{wdi2021} & Annual growth rate of the number of infants who die before one year of age, per 1,000 live births.\\
    
    \\\hline \\
    Policy Responses in 2020 &
    
    Containment Health Index at 10th Covid-19 Case & \citet{OxCGRT} & Average of 13 sub-scores which record severity of government responses in a specific domain on an ordinal scale (for example, the school sub-index is on a 0 (no measure) to 4 (require closing) scale) and subtracts 0.5 if it is targeted. It is scaled to take a value between 0 and 100. The domains are schools, workplaces, public events, gatherings, public transport, stay-at-home requirements, domestic travel, international travel, public information campaigns, testing, contact tracing, facial coverings, and vaccinations. We use the index at the date when the 10th case of Covid-19 is confirmed. \\
    
    & Coverage of Containment Measures at 10th Covid-19 Case & \citet{OxCGRT} & The percentage of the 13 domains in which the data records any policy introduction at the date when the 10th case of Covid-19 is confirmed. \\
    
    & Days between 10th Covid-19 Case and Any Containment Measure & \citet{OxCGRT} & The number of days between the date when the 10th Covid-19 case is confirmed and the date when the containment health index becomes positive. 
    
\\\hline

\end{longtable}

\end{spacing}
\end{center}
\end{footnotesize}

\clearpage
\newgeometry{left=0.3cm, right = 0.3cm, top = 1cm, bottom=1in}
\begin{center}
\begin{spacing}{1}
\begin{table}[H]  
  \centering
  \caption{Descriptive Statistics}
  \label{tab:descriptive-stats-main} 
  \scriptsize
  \begin{threeparttable}

\begin{tabular}{@{\extracolsep{0pt}}lp{5cm}cccccc}
\\[-1.8ex]\hline 
\hline \\ 
& Variable & \multicolumn{1}{c}{N} & \multicolumn{1}{c}{Mean} & \multicolumn{1}{c}{St. Dev.} & \multicolumn{1}{c}{Min} & \multicolumn{1}{c}{Median} & \multicolumn{1}{c}{Max} \\\hline \\

Outcomes 
& Mean GDP Growth Rate in 2001-2019 & 160 & 3.8 & 1.8 & 0.1 & 3.6 & 8.7 \\ 
& & & & & (Greece) & (Chile) & (China) \\
& Mean GDP Growth Rate in 2020-2022 & 160 & 1.9 & 4.1 & $-$7.3 & 1.7 & 44.0 \\ 
& & & & & (Ukraine) & (Burundi) & (Guyana) \\
& Mean Nighttime Light Intensity Growth Rate in 2001-2013 & 156 & 6.8 & 6.0 & 0.8 & 5.5 & 48.1 \\ 
& & & & & (United Kingdom) & (Bolivia) & (Cambodia) \\
\\[-1.8ex] 
\hline \\

Treatments 
& Democracy Index (V-Dem, 2000) & 160 & 0 & 1.0 & $-$1.8 & $-$0.01 & 1.4 \\ 
& & & & & (Saudi Arabia) & (Madagascar) & (Sweden) \\
& Democracy Index (V-Dem, 2019) & 160 & 0 & 1.0 & $-$2.2 & 0.02 & 1.5 \\ 
& & & & & (Saudi Arabia) & (Bhutan) & (Denmark) \\
\\[-1.8ex] 
\hline \\ 

Controls 
& GDP (Current USD, Billions, 2000) & 160 & 206.7 & 931.2 & 0.1 & 11.7 & 10,251.0 \\ 
& & & & & (Sao Tome and Principe) & (El Salvador) & (United States) \\
& GDP (Current USD, Billions, 2019) & 160 & 532.3 & 2,116.4 & 0.4 & 54.6 & 21,381.0 \\ 
& & & & & (Sao Tome and Principe) & (Slovenia) & (United States) \\
& Absolute Latitude & 160 & 26.4 & 17.5 & 0.0 & 23.5 & 65.0 \\ 
& & & & & (Democratic Republic of the Congo) & (Bangladesh) & (Iceland) \\
& Mean Temperature (\degree c, 1991-2000) & 160 & 18.5 & 8.4 & $-$6.2 & 22.0 & 28.6 \\ 
& & & & & (Canada) & (Angola) & (Mali) \\
& Mean Temperature (\degree c, 1991-2016) & 160 & 18.7 & 8.3 & $-$6.0 & 22.3 & 28.9 \\ 
& & & & & (Canada) & (Botswana) & (Mali) \\
& Mean Precipitation (mm per Month, 1991-2000) & 160 & 92.1 & 63.7 & 2.7 & 78.6 & 252.7 \\ 
& & & & & (Egypt) & (Albania) & (Malaysia) \\
& Mean Precipitation (mm per Month, 1991-2016) & 160 & 93.4 & 64.6 & 2.5 & 79.8 & 259.1 \\ 
& & & & & (Egypt) & (Angola) & (Malaysia) \\
& Population Density (No. of People per km$^2$, 2000) & 160 & 155.3 & 481.2 & 1.5 & 61.7 & 5,755.5 \\ 
& & & & & (Mongolia) & (Benin) & (Singapore) \\
& Population Density (No. of People per km$^2$, 2019) & 160 & 213.3 & 701.0 & 2.1 & 81.5 & 8,291.9 \\ 
& & & & & (Mongolia) & (Greece) & (Singapore) \\
& Median Age (2000) & 160 & 25.7 & 8.0 & 15.0 & 22.7 & 41.2 \\ 
& & & & & (Burundi) & (Guyana) & (Japan) \\
& Median Age (2019) & 160 & 30.5 & 9.3 & 15.2 & 29.6 & 48.4 \\ 
& & & & & (Niger) & (Lebanon) & (Japan) \\
& Diabetes Prevalence (\%, 2019) & 160 & 7.5 & 4.0 & 1.0 & 6.4 & 22.1 \\ 
& & & & & (Benin) & (Cambodia) & (Sudan) \\
\\[-1.8ex] 
\hline \\

IVs
& Log European Settler Mortality (Annual No. of Deaths per Thousand) & 82 & 4.6 & 1.3 & 0.9 & 4.5 & 8.0 \\ 
& & & & & (United Kingdom) & (Barbados) & (Mali) \\
& Log Population Density in 1500s (No. of Inhabitants per km$^2$) & 87 & 0.6 & 1.6 & $-$3.8 & 0.4 & 4.6 \\ 
& & & & & (Canada) & (Costa Rica) & (Egypt) \\
& British Legal Origin & 91 & 0.4 & 0.5 & 0 & 0 & 1 \\ 
& & & & & (Algeria) & (Algeria) & (Australia) \\
& Fraction Speaking English & 133 & 0.1 & 0.2 & 0 & 0 & 1 \\ 
& & & & & (Algeria) & (Algeria) & (Barbados) \\
& Fraction Speaking European & 133 & 0.2 & 0.4 & 0 & 0 & 1 \\ 
& & & & & (Angola) & (Angola) & (France) \\
& Bananas & 139 & 0.7 & 0.5 & 0 & 1 & 1 \\ 
& & & & & (Albania) & (Angola) & (Angola) \\
& Coffee & 139 & 0.5 & 0.5 & 0 & 0 & 1 \\ 
& & & & & (Albania) & (Albania) & (Angola) \\
& Copper & 147 & 0.3 & 0.5 & 0 & 0 & 1 \\ 
& & & & & (Algeria) & (Algeria) & (Albania) \\
& Maize & 139 & 0.9 & 0.3 & 0 & 1 & 1 \\ 
& & & & & (Bahrain) & (Albania) & (Albania) \\
& Millet & 139 & 0.5 & 0.5 & 0 & 0 & 1 \\ 
& & & & & (Albania) & (Albania) & (Angola) \\
& Rice & 139 & 0.7 & 0.5 & 0 & 1 & 1 \\ 
& & & & & (Austria) & (Albania) & (Albania) \\
& Rubber & 139 & 0.2 & 0.4 & 0 & 0 & 1 \\ 
& & & & & (Albania) & (Albania) & (Bangladesh) \\
& Silver & 145 & 0.4 & 0.5 & 0 & 0 & 1 \\ 
& & & & & (Albania) & (Albania) & (Algeria) \\
& Sugarcane & 139 & 0.6 & 0.5 & 0 & 1 & 1 \\ 
& & & & & (Albania) & (Angola) & (Angola) \\
& Wheat & 139 & 0.7 & 0.5 & 0 & 1 & 1 \\ 
& & & & & (Bahrain) & (Albania) & (Albania) \\
\\[-1.8ex] 
\hline \\

\end{tabular}
\end{threeparttable}
\end{table}
\end{spacing}
\end{center}

\begin{center}
\begin{spacing}{1}
\begin{table}[H]  
  \centering
  \addtocounter{table}{0} 
  \caption*{Table \thetable{} (Continued)}
  \scriptsize
  \begin{threeparttable}

\begin{tabular}{@{\extracolsep{0pt}}lp{5cm}cccccc}
\\[-1.8ex]\hline 
\hline \\ 
& Variable & \multicolumn{1}{c}{N} & \multicolumn{1}{c}{Mean} & \multicolumn{1}{c}{St. Dev.} & \multicolumn{1}{c}{Min} & \multicolumn{1}{c}{Median} & \multicolumn{1}{c}{Max} \\\hline \\

Mechanisms
\\in 2001--2019
& Capital Investments (Mean Annual \% Growth) & 125 & 1.7 & 3.2 & $-$6.1 & 0.8 & 14.3 \\ 
& & & & & (Equatorial Guinea) & (South Africa) & (Ivory Coast) \\
& R\&D Expenditure (Mean Annual \% Growth) & 125 & 5.7 & 13.1 & $-$7.9 & 1.9 & 74.0 \\ 
& & & & & (Madagascar) & (Spain) & (Bosnia and Herzegovina) \\
& Total Labor Force (Mean Annual \% Growth) & 159 & 1.9 & 1.6 & $-$1.4 & 2.0 & 10.4 \\ 
& & & & & (Romania) & (Colombia) & (Qatar) \\
& TFP (Mean Annual \% Growth) & 113 & 0.8 & 1.6 & $-$4.7 & 0.5 & 8.0 \\ 
& & & & & (Qatar) & (Kenya) & (Tajikistan) \\
& Total Import Value (Mean Annual \% Growth) & 115 & 9.6 & 3.7 & 4.3 & 9.1 & 33.0 \\ 
& & & & & (Jamaica) & (Colombia) & (Djibouti) \\
& Total Export Value (Mean Annual \% Growth) & 115 & 9.8 & 4.6 & 3.7 & 8.8 & 41.3 \\ 
& & & & & (Jamaica) & (Paraguay) & (Djibouti) \\
& Median Age (Mean Annual \% Growth) & 154 & 1.6 & 1.4 & $-$1.6 & 1.4 & 8.2 \\ 
& & & & & (Moldova) & (Ireland) & (Qatar) \\
& Population (Mean Annual \% Growth) & 154 & 0.9 & 0.5 & $-$0.6 & 0.8 & 2.5 \\ 
& & & & & (Central African Republic) & (Barbados) & (Maldives) \\
& Primary School Enrollment (Mean Annual \% Growth) & 146 & 0.5 & 1.1 & $-$2.5 & 0.1 & 4.8 \\ 
& & & & & (Equatorial Guinea) & (Mauritius) & (Djibouti) \\
& Secondary School Enrollment (Mean Annual \% Growth) & 138 & 1.7 & 2.2 & $-$1.5 & 0.9 & 11.1 \\ 
& & & & & (Liberia) & (Mali) & (Mozambique) \\
\\[-1.8ex] 
\hline\hline 
\end{tabular}

\begin{tablenotes} 
\item{\textit{Notes:} Parentheses contain country names corresponding to the minimum, median and maximum values of each variable. When we observe multiple countries corresponding to the same minimum, median or maximum, we choose the first country in alphabetical order. When we do not find a country that corresponds exactly to the median, we choose the country with the closest value. Variable definitions and data sources are in Appendix Table \ref{tab:sources}. \unskip}
\end{tablenotes}
\end{threeparttable}
\end{table} 
\end{spacing}
\end{center}

\nopagebreak
\begin{center}
\begin{spacing}{1}
\begin{table}[H]  
  \centering
  \caption{Additional Descriptive Statistics}
  \label{tab:descriptive-stats-appendix} 
  \scriptsize
  \begin{threeparttable}

\begin{tabular}{@{\extracolsep{0pt}}lp{5cm}cccccc}
\\[-1.8ex]\hline 
\hline \\ 
& Variable & \multicolumn{1}{c}{N} & \multicolumn{1}{c}{Mean} & \multicolumn{1}{c}{St. Dev.} & \multicolumn{1}{c}{Min} & \multicolumn{1}{c}{Median} & \multicolumn{1}{c}{Max} \\\hline \\ 

Outcomes
& Mean GDP Growth Rate in 1981-1990 & 130 & 2.9 & 2.3 & $-$2.5 & 2.6 & 9.9 \\ 
& & & & & (Guyana) & (Jamaica) & (Botswana) \\
& Mean GDP Growth Rate in 1991-2000 & 155 & 3.7 & 4.2 & $-$1.8 & 3.1 & 38.4 \\ 
& & & & & (Moldova) & (New Zealand) & (Equatorial Guinea) \\
& Mean GDP Growth Rate in 2001-2010 & 159 & 4.1 & 2.5 & $-$3.0 & 3.8 & 18.0 \\ 
& & & & & (Zimbabwe) & (Benin) & (Equatorial Guinea) \\
& Mean GDP Growth Rate in 2011-2019 & 160 & 3.3 & 1.9 & $-$3.0 & 3.3 & 8.5 \\ 
& & & & & (Equatorial Guinea) & (Egypt) & (Ethiopia) \\
& Mean GDP Per Capita Growth Rate in 2001-2019 & 160 & 2.4 & 1.9 & $-$1.5 & 2.1 & 8.7 \\ 
& & & & & (United Arab Emirates) & (Libya) & (Burma) \\
& Mean GDP Per Capita Growth Rate in 2020-2022 & 160 & 0.8 & 4.2 & $-$12.8 & 0.9 & 41.7 \\ 
& & & & & (Lebanon) & (Burkina Faso) & (Guyana) \\
& Mean Excess Deaths per 100k People 2020-2022 & 154 & 87.9 & 64.7 & $-$33.8 & 71.7 & 363.9 \\ 
& & & & & (Bhutan) & (Kenya) & (Bulgaria) \\
& Life Satisfaction Growth Rate 2010-2019 & 144 & 0.5 & 2.5 & $-$11.8 & 0.5 & 12.2 \\ 
& & & & & (Angola) & (Croatia) & (Gambia) \\
& Life Satisfaction Growth Rate 2020-2022 & 129 & $-$0.9 & 5.4 & $-$25.6 & $-$0.4 & 13.8 \\ 
& & & & & (Democratic Republic of the Congo) & (Iceland) & (Mauritania) \\
& Top 1\% Income Share Growth Rate in 2001-2019 & 158 & $-$0.1 & 1.2 & $-$3.1 & $-$0.01 & 6.1 \\ 
& & & & & (Maldives) & (Nigeria) & (Cyprus) \\
& Top 1\% Income Share Growth Rate in 2020-2022 & 158 & $-$0.4 & 3.1 & $-$16.3 & 0.0 & 16.0 \\ 
& & & & & (Costa Rica) & (Albania) & (Ivory Coast) \\
& CO2 Emissions Growth Rate in 2001-2019 & 154 & 1.9 & 3.2 & $-$3.4 & 1.8 & 17.8 \\ 
& & & & & (Gabon) & (Latvia) & (Laos) \\
& CO2 Emissions Growth Rate in 2020-2022 & 154 & $-$0.5 & 4.1 & $-$10.0 & $-$0.9 & 19.1 \\ 
& & & & & (Ukraine) & (Egypt) & (Singapore) \\
& Energy Consumption Growth Rate in 2001-2019 & 154 & 2.2 & 6.5 & $-$2.2 & 1.5 & 76.1 \\ 
& & & & & (Zimbabwe) & (Belarus) & (Equatorial Guinea) \\
& Energy Consumption Growth Rate in 2020-2022 & 154 & $-$1.6 & 4.0 & $-$16.7 & $-$1.4 & 11.3 \\ 
& & & & & (Papua New Guinea) & (Comoros) & (Albania) \\
 \\[-1.8ex] 
\hline \\

Controls
& GDP (Current USD, Billions, 1980) & 128 & 85.7 & 293.4 & 0.03 & 7.4 & 2,857.3 \\ 
& & & & & (Equatorial Guinea) & (Ethiopia) & (United States) \\
& GDP (Current USD, Billions, 1990) & 135 & 163.5 & 619.2 & 0.1 & 8.0 & 5,963.1 \\ 
& & & & & (Sao Tome and Principe) & (Dominican Republic) & (United States) \\
& GDP (Current USD, Billions, 2000) & 160 & 206.7 & 931.2 & 0.1 & 11.7 & 10,251.0 \\ 
& & & & & (Sao Tome and Principe) & (El Salvador) & (United States) \\
& GDP (Current USD, Billions, 2010) & 160 & 402.8 & 1,427.1 & 0.2 & 37.4 & 15,049.0 \\ 
& & & & & (Sao Tome and Principe) & (Costa Rica) & (United States) \\
& GDP Per Capita (Current USD, 1980) & 124 & 4,241.3 & 6,214.0 & 123.4 & 1,431.8 & 41,311.9 \\ 
& & & & & (Equatorial Guinea) & (Tunisia) & (United Arab Emirates) \\
& GDP Per Capita (Current USD, 1990) & 133 & 5,848.5 & 8,769.9 & 87.2 & 1,310.4 & 39,842.8 \\ 
& & & & & (Sudan) & (Republic of the Congo) & (Switzerland) \\
& GDP Per Capita (Current USD, 2000) & 159 & 6,655.8 & 10,252.9 & 130.2 & 1,675.8 & 48,984.2 \\ 
& & & & & (Burundi) & (Paraguay) & (Luxembourg) \\
& GDP Per Capita (Current USD, 2010) & 160 & 13,182.9 & 18,936.0 & 231.5 & 4,604.7 & 112,049.1 \\ 
& & & & & (Burundi) & (Ecuador) & (Luxembourg) \\
& GDP Per Capita (Current USD, 2019) & 160 & 14,861.5 & 20,346.0 & 261.3 & 5,906.3 & 113,860.5 \\ 
& & & & & (Burundi) & (Bosnia and Herzegovina) & (Luxembourg) \\
& Population (Millions, 2000) & 159 & 37.1 & 135.4 & 0.1 & 8.1 & 1,290.6 \\ 
& & & & & (Seychelles) & (Azerbaijan) & (China) \\
& Population (Millions, 2019) & 159 & 47.0 & 162.4 & 0.1 & 10.1 & 1,439.3 \\ 
& & & & & (Seychelles) & (Azerbaijan) & (China) \\
& Mean Temperature (\degree c, 1971-1980) & 160 & 17.9 & 8.5 & $-$7.4 & 21.4 & 28.2 \\ 
& & & & & (Canada) & (Angola) & (Mali) \\
& Mean Temperature (\degree c, 1981-1990) & 160 & 18.2 & 8.5 & $-$7.0 & 21.8 & 28.6 \\ 
& & & & & (Canada) & (Angola) & (Mali) \\
& Mean Temperature (\degree c, 1991-2000) & 160 & 18.5 & 8.4 & $-$6.2 & 22.0 & 28.6 \\ 
& & & & & (Canada) & (Angola) & (Mali) \\
& Mean Temperature (\degree c, 2001-2010) & 160 & 18.9 & 8.4 & $-$5.8 & 22.5 & 29.1 \\ 
& & & & & (Canada) & (Botswana) & (Mali) \\
& Mean Precipitation (mm per Month, 1971-1980) & 160 & 93.4 & 64.5 & 3.0 & 83.8 & 260.3 \\ 
& & & & & (Egypt) & (Angola) & (Costa Rica) \\
& Mean Precipitation (mm per Month, 1981-1990) & 160 & 92.1 & 64.0 & 3.1 & 79.8 & 256.5 \\ 
& & & & & (Egypt) & (Albania) & (Papua New Guinea) \\
& Mean Precipitation (mm per Month, 1991-2000) & 160 & 92.1 & 63.7 & 2.7 & 78.6 & 252.7 \\ 
& & & & & (Egypt) & (Albania) & (Malaysia) \\
& Mean Precipitation (mm per Month, 2001-2010) & 160 & 94.9 & 66.3 & 2.2 & 82.5 & 265.7 \\ 
& & & & & (Egypt) & (Angola) & (Malaysia) \\
& Population Density (No. of People per km$^2$, 1980) & 160 & 110.7 & 297.8 & 1.1 & 42.6 & 3,445.3 \\ 
& & & & & (Mongolia) & (Malaysia) & (Singapore) \\
& Population Density (No. of People per km$^2$, 1990) & 160 & 131.4 & 367.2 & 1.4 & 50.9 & 4,304.2 \\ 
& & & & & (Mongolia) & (Cambodia) & (Singapore) \\
& Population Density (No. of People Per km$^2$, 2000) & 160 & 155.3 & 481.2 & 1.5 & 61.7 & 5,755.5 \\ 
& & & & & (Mongolia) & (Benin) & (Singapore) \\
& Population Density (No. of People Per km$^2$, 2010) & 160 & 186.1 & 612.3 & 1.8 & 73.3 & 7,330.2 \\ 
& & & & & (Mongolia) & (Bosnia and Herzegovina) & (Singapore) \\
& Median Age (1980) & 160 & 22.4 & 6.3 & 15.0 & 19.3 & 36.5 \\ 
& & & & & (Kenya) & (Haiti) & (Germany) \\
& Median Age (1990) & 160 & 23.7 & 7.2 & 15.3 & 20.8 & 38.4 \\ 
& & & & & (Rwanda) & (Lebanon) & (Sweden) \\
& Median Age (2000) & 160 & 25.7 & 8.0 & 15.0 & 22.7 & 41.2 \\ 
& & & & & (Burundi) & (Guyana) & (Japan) \\
& Median Age (2010) & 160 & 28.1 & 8.7 & 15.0 & 26.2 & 44.7 \\ 
& & & & & (Niger) & (Mexico) & (Japan) \\
 \\[-1.8ex] 
\hline \\

\end{tabular}
\end{threeparttable}
\end{table}
\end{spacing}
\end{center}

\begin{center}
\begin{spacing}{1}
\begin{table}[H]  
  \centering
  \addtocounter{table}{0} 
  \caption*{Table \thetable{} (Continued)}
  \scriptsize
  \begin{threeparttable}

\begin{tabular}{@{\extracolsep{5pt}}lp{6cm}cccccc}
\\[-1.8ex]\hline 
\hline \\ 
& Variable & \multicolumn{1}{c}{N} & \multicolumn{1}{c}{Mean} & \multicolumn{1}{c}{St. Dev.} & \multicolumn{1}{c}{Min} & \multicolumn{1}{c}{Median} & \multicolumn{1}{c}{Max} \\\hline \\ 

Mechanisms 
\\in 2001--2019
& Value Added Agriculture (Mean Annual \% Growth) & 135 & 6.9 & 3.5 & 0.5 & 6.8 & 18.9 \\ 
& & & & & (Sudan) & (Cameroon) & (Chad) \\
& Value Added Manufacturing (Mean Annual \% Growth) & 124 & 8.4 & 6.0 & 1.2 & 6.9 & 43.7 \\ 
& & & & & (United Kingdom) & (Egypt) & (Gabon) \\
& Value Added Services (Mean Annual \% Growth) & 138 & 9.1 & 3.6 & 1.4 & 8.7 & 23.1 \\ 
& & & & & (Japan) & (Cape Verde) & (Ghana) \\
& Tax Revenue Share (Mean Annual \% Growth) & 125 & 1.4 & 2.2 & $-$2.6 & 1.0 & 16.4 \\ 
& & & & & (Angola) & (Cape Verde) & (Burma) \\
& R\&D Researchers (Mean Annual \% Growth) & 118 & 5.0 & 11.3 & $-$7.9 & 1.9 & 74.0 \\ 
& & & & & (Mauritius) & (Bulgaria) & (Brazil) \\
& No. of New Business Registrations (Mean Annual \% Growth) & 124 & 5.7 & 12.1 & $-$7.9 & 2.0 & 74.0 \\ 
& & & & & (Luxembourg) & (Paraguay) & (Bahrain) \\
& Foreign Direct Investments (Mean Annual \% Growth) & 99 & 5.5 & 395.5 & $-$3,205.0 & 31.4 & 890.5 \\ 
& & & & & (Belarus) & (Cyprus) & (Paraguay) \\
& Value of Imports from China (Mean Annual \% Growth) & 134 & 23.2 & 9.9 & 6.7 & 21.2 & 56.7 \\ 
& & & & & (Japan) & (Egypt) & (Armenia) \\
& Value of Exports to China (Mean Annual \% Growth) & 104 & 39.9 & 33.7 & 7.8 & 26.6 & 149.1 \\ 
& & & & & (Finland) & (Barbados) & (Jamaica) \\
& Conflict Index (Mean Annual \% Growth) & 155 & 0.1 & 0.8 & $-$0.1 & $-$0.1 & 7.7 \\ 
& & & & & (Laos) & (Senegal) & (Iraq) \\
& Child Mortality Rate (Mean Annual \% Growth) & 153 & $-$3.4 & 1.8 & $-$9.5 & $-$3.1 & 0.7 \\ 
& & & & & (Montenegro) & (Burma) & (Botswana) \\
 \\[-1.8ex] 
\hline \\

Policy Responses 
\\in 2020
& Containment Health Index at 10th Covid-19 Case & 151 & 1.8 & 1.0 & 0.0 & 1.7 & 3.9 \\ 
& & & & & (Algeria) & (Burkina Faso) & (Djibouti) \\
& Coverage of Containment Policy at 10th Covid-19 Case & 152 & 48.4 & 23.7 & 0.0 & 46.2 & 92.3 \\ 
& & & & & (Algeria) & (Azerbaijan) & (Bhutan) \\
& Days Between 10th Covid-19 Case Until Any Containment Measure & 152 & $-$42.8 & 33.3 & $-$270 & $-$40 & 34 \\ 
& & & & & (Solomon Islands) & (Azerbaijan) & (Thailand) \\
 \\[-1.8ex] 
\hline \hline

\end{tabular}
\begin{tablenotes} 
\item{\textit{Notes:} Parentheses contain country names corresponding to the minimum, median, and maximum values of each variable. When we observe multiple countries corresponding to the same minimum, median, or maximum, we choose the first country in alphabetical order. When we do not find a country that corresponds exactly to the median, we choose the country with the closest value. Variable definitions and data sources are in Appendix Table \ref{tab:sources}. \unskip}
\end{tablenotes}
\end{threeparttable}
\end{table}
\end{spacing}
\end{center}

\restoregeometry

\clearpage
\begin{table}\centering
        \caption{Democracy Indices for 30 Countries with Largest Total GDP in 2019}\label{tab:dem-ranking-country}
\begin{threeparttable}

\begin{tabular}{l*{1}{cccc}}
\hline\hline
            &\shortstack{Democracy Index \\ (V-Dem, 2000)}& Rank & \shortstack{Democracy Index\\ (V-Dem, 2019)} & Rank\\
\hline

United States&         1.3&          22&         1.1&          32\\
China       &        -1.6&         150&        -1.9&         158\\
Japan       &         1.2&          35&         1.1&          30\\
Germany     &         1.4&           2&         1.4&          11\\
United Kingdom&         1.3&          26&         1.3&          17\\
India       &         0.8&          50&        -0.3&          94\\
France      &         1.3&          16&         1.3&          13\\
Italy       &         1.2&          30&         1.3&          21\\
Brazil      &         1.3&          21&         0.6&          55\\
Canada      &         1.2&          33&         1.2&          24\\
Russia      &        -0.4&          96&        -1.2&         133\\
South Korea &         1.2&          31&         1.3&          23\\
Spain       &         1.4&           5&         1.3&          14\\
Australia   &         1.4&           9&         1.2&          26\\
Mexico      &         0.4&          62&         0.5&          59\\
Indonesia   &         0.6&          56&         0.3&          70\\
Netherlands &         1.3&          27&         1.3&          16\\
Saudi Arabia&        -1.8&         159&        -2.2&         159\\
Switzerland &         1.4&           7&         1.4&           4\\
Poland      &         1.3&          17&         0.6&          58\\
Thailand    &         0.0&          79&        -1.4&         144\\
Belgium     &         1.3&          14&         1.4&           8\\
Sweden      &         1.4&           1&         1.4&           2\\
Argentina   &         1.2&          34&         0.9&          40\\
Nigeria     &        -0.2&          87&        -0.1&          85\\
Austria     &         1.3&          25&         1.2&          25\\
United Arab Emirates&        -1.8&         157&        -1.8&         156\\
Norway      &         1.4&           6&         1.4&           6\\
Israel      &         0.9&          43&         0.6&          54\\
Ireland     &         1.3&          11&         1.3&          12\\

\\[-1.8ex] 
\hline\hline
\end{tabular}
\begin{tablenotes} 
\item {\footnotesize {\textit{Notes:} This table reports the democracy index in 2000 and 2019 and the corresponding rank in the dataset (N=159) for 30 countries with the largest total GDP in 2019. The countries are ordered by GDP size. 
 The Democracy Index (V-Dem) is normalized to have mean zero and standard deviation one in each year.\unskip}}
\end{tablenotes}
\end{threeparttable}
\end{table}

\clearpage

\begin{table}
\centering
\caption{Correlation Among Democracy Indices}
        \label{tab:indices-correlation}
\begin{threeparttable}
    \begin{tabularx}{0.9\textwidth}{{l}*{1}{Y} {c}*{4}{Y}}
\hline\hline
                & V-Dem & Polity   & Freedom House & Economist Intelligence Unit\\
\hline
\textbf{Panel A: Democracy Index for 2019} &&&&\\
V-Dem  (2019) &        1&         &         &         \\
Polity (2018) &    0.858&        1&         &         \\
Freedom House (2019) &    0.944&    0.838&        1&         \\
Economist Intelligence Unit (2019) &    0.891&    0.776&    0.946&        1\\ \hline \\[-1.8ex]

\textbf{Panel B: Democracy Index for 2000} &&&&\\
V-Dem (2000)&        1&         &         &         \\
Polity (2000) &    0.898&        1&         &         \\
Freedom House (2003) &    0.935&    0.889&        1&         \\
Economist Intelligence Unit (2006)&    0.909&    0.853&    0.917&        1\\

\hline\hline \\[-1.8ex]
\end{tabularx}
\begin{tablenotes} 
\item {\footnotesize {\textit{Notes:} This table reports the pairwise correlations among the V-Dem, Polity, Freedom House, and Economist Intelligence Unit's democracy indices in 2019 (Panel A) and 2000 (Panel B). The publication year of each index is in parentheses. When data for democracy levels in 2019 or 2000 are unavailable, we use the index from the nearest available year.
Variable definitions and data sources are in Appendix Table \ref{tab:sources}. \unskip}}
\end{tablenotes}
\end{threeparttable}
\end{table}

\clearpage

\begin{table} \centering
\begin{threeparttable}
\def\sym#1{\ifmmode^{#1}\else\(^{#1}\)\fi}
\caption{Correlation Between Democracy and Economic Growth With Control for Baseline GDP}\label{tab:ols-gdp}
\setlength{\tabcolsep}{5pt} 
\renewcommand{\arraystretch}{1.2} 

\begin{tabular}{lcccccccc}
\hline\hline
& (1) & (2) & (3) & (4) & (5) & (6) & (7) & (8) \\
\hline
\multicolumn{9}{c}{Dependent Variable is Mean GDP Growth Rate in 2001-2019} \\
\hline
Democracy Index (V-Dem, 2000) & -1.6 & -1.2 & -1.3 & -0.9 & -1.7 & -1.0 & -1.1 & -1.0 \\
                              & (0.4) & (0.6) & (0.5) & (0.4) & (0.4) & (0.5) & (0.4) & (0.4) \\
\hline
\multicolumn{9}{c}{Dependent Variable is Mean GDP Growth Rate in 2020-2022} \\
\hline
Democracy Index (V-Dem, 2019) & -0.8 & -0.7 & -1.0 & -0.9 & -0.8 & -0.8 & -0.8 & -0.9 \\
                              & (0.2) & (0.2) & (0.2) & (0.3) & (0.2) & (0.2) & (0.2) & (0.3) \\
\hline
\multicolumn{9}{c}{Dependent Variable is Mean Nighttime Light Intensity Growth Rate in 2001-2013} \\
\hline
Democracy Index (V-Dem, 2000) & -1.6 & -1.5 & -0.8 & -1.0 & -1.3 & -0.9 & -0.8 & -0.9 \\
                              & (0.3) & (0.4) & (0.3) & (0.2) & (0.3) & (0.3) & (0.3) & (0.3) \\
\hline
Baseline Controls Other Than Baseline GDP & & \checkmark & & \checkmark & & \checkmark & & \checkmark \\
Baseline GDP Per Capita Control           & & & \checkmark & \checkmark & & & \checkmark & \checkmark \\
Baseline Total GDP Control                & & & & & \checkmark & \checkmark & \checkmark & \checkmark \\
N                                         & 150 & 150 & 150 & 150 & 150 & 150 & 150 & 150 \\
\hline\hline
\end{tabular}

\begin{tablenotes}
\item {\footnotesize {\textit{Notes:} This table reports the results of OLS regressions of GDP and nightime light intensity growth rates on the democracy index with additional controls.
 The Democracy Index (V-Dem) is normalized to have mean zero and standard deviation one in each year.\unskip Columns 2, 4, 6, 8 have the following baseline controls: absolute latitude, mean temperature, mean precipitation, population density, and median age. For the mean GDP growth rate in 2020-2022, we also control for diabetes prevalence. Columns 3, 4, 7, and 8 additionally control for baseline GDP per capita. Columns 5, 6, 7, 8 additionally control for baseline total GDP. 
N refers to the number of countries for which IVs, controls, and outcomes are available. \unskip
Robust standard errors are in parentheses. \unskip Variable definitions and data sources are in Appendix Table \ref{tab:sources}. \unskip}}
\end{tablenotes}
\end{threeparttable}
\end{table}

\clearpage

\begin{table}\centering
  \caption{ First-stage Regression Estimates of IVs' Effects on Democracy}
  \label{tab:first-stage} 
  \footnotesize
  \begin{threeparttable}
  
\begin{tabular}{l*{12}{c}}
\hline\hline
                    &\multicolumn{1}{c}{(1)}&\multicolumn{1}{c}{(2)}&\multicolumn{1}{c}{(3)}&\multicolumn{1}{c}{(4)}&\multicolumn{1}{c}{(5)}&\multicolumn{1}{c}{(6)}&\multicolumn{1}{c}{(7)}&\multicolumn{1}{c}{(8)}&\multicolumn{1}{c}{(9)}&\multicolumn{1}{c}{(10)}&\multicolumn{1}{c}{(11)}&\multicolumn{1}{c}{(12)}\\
\hline &\multicolumn{10}{c}{ Dependent Variable is Democracy Index (V-Dem, 2019)}\\\cline{2-13}\\[-1.8ex]
Log European Settler Mortality&        -0.8&        -0.9&            &            &            &            &            &            &            &            &        0.09&        -0.3\\
                    &       (0.2)&       (0.3)&            &            &            &            &            &            &            &            &       (0.2)&       (0.1)\\
Log Population Density in 1500s&            &            &        -0.5&        -0.5&            &            &            &            &            &            &       0.008&      -0.004\\
                    &            &            &      (0.09)&      (0.04)&            &            &            &            &            &            &      (0.08)&      (0.06)\\
British Legal Origin&            &            &            &            &         2.0&         2.0&            &            &            &            &         1.3&         0.7\\
                    &            &            &            &            &       (0.6)&       (0.5)&            &            &            &            &       (0.3)&       (0.2)\\
Fraction Speaking English&            &            &            &            &            &            &       -0.05&         0.7&            &            &        -1.0&        -0.4\\
                    &            &            &            &            &            &            &       (0.2)&       (0.5)&            &            &       (0.5)&       (0.5)\\
Fraction Speaking European&            &            &            &            &            &            &         1.8&         1.2&            &            &         2.3&         2.1\\
                    &            &            &            &            &            &            &       (0.6)&       (0.3)&            &            &       (0.5)&       (0.3)\\
Bananas             &            &            &            &            &            &            &            &            &       -0.08&         0.3&        -0.7&        -0.6\\
                    &            &            &            &            &            &            &            &            &       (0.5)&       (0.4)&       (0.2)&       (0.2)\\
Coffee              &            &            &            &            &            &            &            &            &       -0.08&         0.8&        -0.3&       0.002\\
                    &            &            &            &            &            &            &            &            &       (0.3)&       (0.3)&       (0.2)&       (0.2)\\
Copper              &            &            &            &            &            &            &            &            &        -0.6&       -0.01&       -0.06&         0.5\\
                    &            &            &            &            &            &            &            &            &       (0.4)&       (0.4)&       (0.3)&       (0.4)\\
Maize               &            &            &            &            &            &            &            &            &         0.7&         1.2&         1.0&        -2.6\\
                    &            &            &            &            &            &            &            &            &       (0.4)&       (0.4)&       (0.8)&       (0.4)\\
Millet              &            &            &            &            &            &            &            &            &        -0.6&        -0.3&        -0.2&        -0.2\\
                    &            &            &            &            &            &            &            &            &       (0.4)&       (0.3)&       (0.3)&       (0.2)\\
Rice                &            &            &            &            &            &            &            &            &        -0.8&        -0.8&       -0.07&         0.5\\
                    &            &            &            &            &            &            &            &            &       (0.6)&       (0.5)&       (0.4)&       (0.3)\\
Rubber              &            &            &            &            &            &            &            &            &        -2.2&        -2.2&        -0.4&        0.03\\
                    &            &            &            &            &            &            &            &            &       (0.5)&       (0.3)&       (0.2)&       (0.3)\\
Silver              &            &            &            &            &            &            &            &            &         1.1&         0.6&         0.6&         0.1\\
                    &            &            &            &            &            &            &            &            &       (0.4)&       (0.4)&       (0.5)&       (0.4)\\
Sugarcane           &            &            &            &            &            &            &            &            &         1.0&         0.4&         1.0&         0.4\\
                    &            &            &            &            &            &            &            &            &       (0.6)&       (0.6)&       (0.4)&       (0.4)\\
Wheat               &            &            &            &            &            &            &            &            &        -0.4&         0.9&        -1.1&        -1.0\\
                    &            &            &            &            &            &            &            &            &       (0.5)&       (0.6)&       (0.5)&       (0.3)\\
F-Statistic (First stage) & 9.7 & 7.3 & 27.2 & 135.2 & 12.0 & 16.8 & 4.3 & 14.0 & 6.4 & 6.0 & 57.5 & 351.8\\ \hline \\[-1.8ex] Baseline Controls & \xmark & \cmark & \xmark & \cmark & \xmark & \cmark & \xmark & \cmark & \xmark & \cmark & \xmark & \cmark\\ 
N                   &          81&          81&          86&          86&          90&          90&         132&         132&         138&         138&          71&          71\\
\hline\hline
\end{tabular}

\begin{tablenotes} 
\item {\footnotesize {\textit{Notes:} This table reports the first-stage regression estimates of the effect of the five different sets of IVs on democracy levels in 2019.  The Democracy Index (V-Dem) is normalized to have mean zero and standard deviation one in each year.\unskip 
Columns 1, 3, 5, 7, 9, and 11 have no controls, while columns 2, 4, 6, 8, 10, and 12 have the following baseline controls: absolute latitude, mean temperature, mean precipitation, population density, median age, and diabetes prevalence.\unskip
N refers to the number of countries for which IVs, controls, and outcomes are available. \unskip
Robust standard errors are in parentheses. \unskip
Variable definitions and data sources are in Appendix Table \ref{tab:sources}. \unskip}}
\end{tablenotes}
\end{threeparttable}
\end{table}

\clearpage
\begin{table}[]    \centering
 \begin{threeparttable} 
    \scriptsize
    \caption{First-stage Monotonicity Check: By Continent}\label{tab:monotonicity-check} 
    \begin{tabular}{l*{6}{c}}
\hline\hline
                                       &\multicolumn{1}{c}{(1)}&\multicolumn{1}{c}{(2)}&\multicolumn{1}{c}{(3)}&\multicolumn{1}{c}{(4)}&\multicolumn{1}{c}{(5)}&\multicolumn{1}{c}{(6)}\\
\hline
\multicolumn{7}{l}{\textbf{Panel A: Africa \& Asia vs. Americas \& Oceania}} \\ &\multicolumn{6}{c}{ Dependent Variable is Democracy Index (V-Dem, 2019)}\\\cline{2-7}\\[-1.8ex]
Log European Settler Mortality&        -0.2&        -0.3&            &            &            &            \\
                    &       (0.3)&       (0.3)&            &            &            &            \\
Log Population Density in 1500s&            &            &        -0.3&        -0.2&            &            \\
                    &            &            &       (0.2)&       (0.1)&            &            \\
British Legal Origin&            &            &            &            &         1.4&         0.5\\
                    &            &            &            &            &       (0.3)&       (0.2)\\
\hline
N                   &          49&          49&          58&          58&          61&          61\\

 &\multicolumn{6}{c}{ Dependent Variable is Democracy Index (V-Dem, 2019)}\\\cline{2-7}\\[-1.8ex]
Log European Settler Mortality&        -0.3&       -0.04&            &            &            &            \\
                    &      (0.04)&      (0.08)&            &            &            &            \\
Log Population Density in 1500s&            &            &        -0.2&         0.1&            &            \\
                    &            &            &      (0.02)&      (0.04)&            &            \\
British Legal Origin&            &            &            &            &         0.5&        0.03\\
                    &            &            &            &            &      (0.06)&       (0.1)\\
\hline
N                   &          29&          29&          28&          28&          29&          29\\

Baseline Controls & \xmark & \cmark & \xmark & \cmark & \xmark & \cmark\\

\hline
\multicolumn{7}{l}{\textbf{Panel B: Africa \& Oceania vs. Americas \& Asia}} \\ &\multicolumn{6}{c}{ Dependent Variable is Democracy Index (V-Dem, 2019)}\\\cline{2-7}\\[-1.8ex]
Log European Settler Mortality&        -0.3&        -0.4&            &            &            &            \\
                    &       (0.1)&       (0.3)&            &            &            &            \\
Log Population Density in 1500s&            &            &        -0.3&        -0.3&            &            \\
                    &            &            &      (0.02)&      (0.07)&            &            \\
British Legal Origin&            &            &            &            &         1.6&         1.0\\
                    &            &            &            &            &       (0.3)&       (0.3)\\
\hline
N                   &          40&          40&          47&          47&          50&          50\\

&\multicolumn{6}{c}{ Dependent Variable is Democracy Index (V-Dem, 2019)}\\\cline{2-7}\\[-1.8ex]
Log European Settler Mortality&        -1.1&        -1.2&            &            &            &            \\
                    &       (0.3)&       (0.3)&            &            &            &            \\
Log Population Density in 1500s&            &            &        -0.5&        -0.6&            &            \\
                    &            &            &      (0.10)&      (0.05)&            &            \\
British Legal Origin&            &            &            &            &         2.0&         2.1\\
                    &            &            &            &            &       (0.6)&       (0.8)\\
\hline
N                   &          38&          38&          39&          39&          40&          40\\

Baseline Controls & \xmark & \cmark & \xmark & \cmark & \xmark & \cmark\\

\hline
\multicolumn{7}{l}{\textbf{Panel C: Africa \& S. America vs.  N. America, Asia, \& Oceania}}\\ &\multicolumn{6}{c}{ Dependent Variable is Democracy Index (V-Dem, 2019)}\\\cline{2-7}\\[-1.8ex]
Log European Settler Mortality&        -0.2&        0.02&            &            &            &            \\
                    &      (0.07)&       (0.2)&            &            &            &            \\
Log Population Density in 1500s&            &            &        -0.3&        -0.3&            &            \\
                    &            &            &      (0.04)&      (0.06)&            &            \\
British Legal Origin&            &            &            &            &        -0.2&         0.9\\
                    &            &            &            &            &       (0.3)&       (0.2)\\
\hline
N                   &          47&          47&          55&          55&          57&          57\\

&\multicolumn{6}{c}{ Dependent Variable is Democracy Index (V-Dem, 2019)}\\\cline{2-7}\\[-1.8ex]
Log European Settler Mortality&        -1.1&        -0.9&            &            &            &            \\
                    &       (0.3)&       (0.3)&            &            &            &            \\
Log Population Density in 1500s&            &            &        -0.5&        -0.5&            &            \\
                    &            &            &      (0.09)&      (0.06)&            &            \\
British Legal Origin&            &            &            &            &         2.4&         1.8\\
                    &            &            &            &            &       (0.4)&       (1.0)\\
\hline
N                   &          31&          31&          31&          31&          33&          33\\

Baseline Controls & \xmark & \cmark & \xmark & \cmark & \xmark & \cmark\\

 \hline
\multicolumn{7}{l}{\textbf{Panel D: Africa \& N. America vs.  S. America, Asia, \& Oceania}} \\
&\multicolumn{6}{c}{ Dependent Variable is Democracy Index (V-Dem, 2019)}\\\cline{2-7}\\[-1.8ex]
Log European Settler Mortality&        -0.4&        0.08&            &            &            &            \\
                    &      (0.10)&       (0.2)&            &            &            &            \\
Log Population Density in 1500s&            &            &        -0.3&        -0.4&            &            \\
                    &            &            &      (0.04)&      (0.06)&            &            \\
British Legal Origin&            &            &            &            &         1.2&         0.5\\
                    &            &            &            &            &       (0.3)&       (0.3)\\
\hline
N                   &          50&          50&          58&          58&          60&          60\\

&\multicolumn{6}{c}{ Dependent Variable is Democracy Index (V-Dem, 2019)}\\\cline{2-7}\\[-1.8ex]
Log European Settler Mortality&        -0.9&        -0.7&            &            &            &            \\
                    &       (0.3)&       (0.3)&            &            &            &            \\
Log Population Density in 1500s&            &            &        -0.4&        -0.4&            &            \\
                    &            &            &       (0.1)&      (0.07)&            &            \\
British Legal Origin&            &            &            &            &         1.3&         0.4\\
                    &            &            &            &            &       (0.6)&       (0.7)\\
\hline
N                   &          28&          28&          28&          28&          30&          30\\
\hline
Baseline Controls & \xmark & \cmark & \xmark & \cmark & \xmark & \cmark\\
\hline\hline
\end{tabular}
\begin{tablenotes}
\item {\scriptsize {\textit{Notes:} This table conducts monotonicity checks for the relationship between the univariate IVs and democracy by dividing the sample into two by random combinations of continents. Panel A has Africa and Asia as the sample for the regressions in the first three rows and Americas and Oceania as the sample in the following three rows. Similarly, Panel B compares Africa and Oceania with the Americas and Asia; Panel C compares Africa and South America with North America, Asia, and Oceania; Panel D compares Africa and North America with South America, Asia, and Oceania.
 The Democracy Index (V-Dem) is normalized to have mean zero and standard deviation one in each year.\unskip Columns 1, 3, 5 have no controls, while columns 2, 4, 6 have the following baseline controls: absolute latitude, mean temperature, mean precipitation, population density, median age, and diabetes prevalence. 
N refers to the number of countries for which IVs, controls, and outcomes are available. \unskip
Robust standard errors are in parentheses. \unskip
Variable definitions and data sources are in Appendix Table \ref{tab:sources}. \unskip}}
\end{tablenotes}
\end{threeparttable} 
\end{table}

\clearpage
\begin{landscape}
\begin{table}\centering
\caption{Correlation Among IVs}
        \label{tab:ivs-corr}
\begin{threeparttable}
\footnotesize
\resizebox{\columnwidth}{!}{
\begin{tabular}{l*{15}{c}}
\hline\hline
                    & \shortstack[c]{Log \\European \\Settler \\Mortality}&\shortstack[c]{Log \\Population \\ Density \\in 1500s}&\shortstack[c]{British \\Legal \\Origin} & \shortstack[c]{Fraction \\Speaking \\English}&\shortstack[c]{Fraction \\Speaking \\ European}&  Bananas&   Coffee&   Copper&    Maize&   Millet&     Rice&   Rubber&   Silver&Sugarcane&    Wheat\\ 
                    \hline \\
Log European Settler Mortality&        1&         &         &         &         &         &         &         &         &         &         &         &         &         &         \\
Log Population Density in 1500s&    0.367&        1&         &         &         &         &         &         &         &         &         &         &         &         &         \\
British Legal Origin&   -0.233&  -0.0732&        1&         &         &         &         &         &         &         &         &         &         &         &         \\
Fraction Speaking English&   -0.418&   -0.407&    0.504&        1&         &         &         &         &         &         &         &         &         &         &         \\
Fraction Speaking European&   -0.437&   -0.545&   0.0458&    0.564&        1&         &         &         &         &         &         &         &         &         &         \\
Bananas         &   -0.112&   0.0502&  -0.0388&   0.0422&   0.0913&        1&         &         &         &         &         &         &         &         &         \\
Coffee          &    0.163&   0.0814&   -0.119&   -0.121&   0.0233&    0.442&        1&         &         &         &         &         &         &         &         \\
Copper          &   -0.388&   -0.227&   0.0388&   0.0650&    0.163&   0.0779&   0.0957&        1&         &         &         &         &         &         &         \\
Maize           &    0.181&   0.0677&  -0.0583&   0.0364&    0.102&    0.282&    0.238&    0.103&        1&         &         &         &         &         &         \\
Millet          &    0.325&    0.325&    0.149&   -0.194&   -0.488&  -0.0700&   -0.149&  -0.0574&    0.178&        1&         &         &         &         &         \\
Rice            &    0.385&    0.203&   -0.211&   -0.358&   -0.138&    0.274&    0.318&   0.0693&    0.560&    0.216&        1&         &         &         &         \\
Rubber          &   0.0642&    0.154& -0.00643&   -0.239&   -0.188&    0.183&    0.333&    0.305&    0.110&   0.0653&    0.197&        1&         &         &         \\
Silver          &   -0.459&   -0.247&  -0.0487&   0.0788&    0.299& -0.00279&  -0.0131&    0.795&  -0.0473&   -0.208&  -0.0844&    0.148&        1&         &         \\
Sugarcane       &    0.102&    0.130&  -0.0857&   -0.100&  -0.0177&    0.535&    0.444&   0.0391&    0.447&    0.143&    0.493&    0.247&   -0.150&        1&         \\
Wheat           &   -0.185&  0.00169&  -0.0725&  -0.0964&   0.0214&   -0.179&   -0.230&    0.243&   0.0317&    0.236&  -0.0464&  -0.0265&    0.275&  -0.0582&        1\\
\hline\hline
\end{tabular}
}
\begin{tablenotes} 
\item {\footnotesize {\textit{Notes:} This table reports the pairwise correlations among the IVs.
Variable definitions and data sources are in Appendix Table \ref{tab:sources}. \unskip}}
\end{tablenotes}
\end{threeparttable}
\end{table}
\end{landscape}

\clearpage

\begin{landscape}
\begin{table} \centering
    \caption{Directions of Potential Bias in the IV Estimates}\label{tab:iv-bias-summary}
    \begin{tabular}{p{0.1\linewidth}
    p{0.15\linewidth}
    p{0.2\linewidth}p{0.2\linewidth} p{0.15\linewidth}} 
    \hline \hline \\[-1.8ex]
        IV &  Base Sample
        & $Cov(Z, Democracy)$ 
        & $Cov(Z, Potential Omitted Var)$  
        & \shortstack[c]{Likely Direction of Bias \\ $\frac{Cov(Z, Potential Omitted Var)}{Cov(Z, Democracy)} $}\\ \hline \\
        
        European settler mortality IV 
        & N = 81 (countries formerly under European rule with data available) 
        & Negative (Higher settler mortality led settlers to establish extractive institutions, resulting in lower levels of democracy) 
        & Likely negative (Worse disease environments may directly hamper growth) 
        & Likely positive \\
        
        Population density in 1500s IV 
        & N = 86 (countries formerly under European rule with data available) 
        & Negative (Higher population density at the beginning of colonial rule led European colonizers to establish extractive institutions, resulting in lower levels of democracy) & 
        Likely positive (Higher population density may positively affect growth through higher returns to scale and agglomeration effects) 
        & Likely negative \\
        
        British legal origin IV
        & N = 90 (countries formerly under European rule with data available) 
        & Positive (British colonial rule led to the establishment of a common-law legal system, which is correlated with less restrictions on individual freedoms and higher levels of democracy) 
        & Likely positive (Being formerly subjected to British rule may lead to greater advantages in an Anglo-centric world economy through linguistic or cultural influence)
        & Likely positive \\
        
        Fraction speaking English or European IVs 
        & N = 132 (all countries with data available)
        & Positive (The fraction of the population speaking English or European corresponds to the extent of Western influence, which is positively related to higher levels of democracy)
        & Likely positive (Higher fractions of the population speaking English or a European language may result in more globally competitive human capital)
        & Likely positive\\
        
        Crops and minerals IVs
        & N = 138 (all countries with data available)
        & Depends on the commodity
        & Depends on the commodity 
        & Depends on the commodity \\
        
        \hline \hline 
        
    \end{tabular}

\end{table}
\end{landscape}

\clearpage
\newgeometry{left=0.3cm, right = 0.3cm, top = 1cm, bottom=1in}
\begin{figure}
\centering
\caption{Reduced Forms for Causal Effects of Democracy}\label{fig:first-stage}
\captionsetup{width=0.99\textwidth}
\begin{subfigure}[c]{.49\linewidth}
    \centering
    \caption{Reduced form: Mean GDP Growth Rate in 2001–2019}\label{fig:rf_appendix_a}
    \includegraphics[width=.99\textwidth]{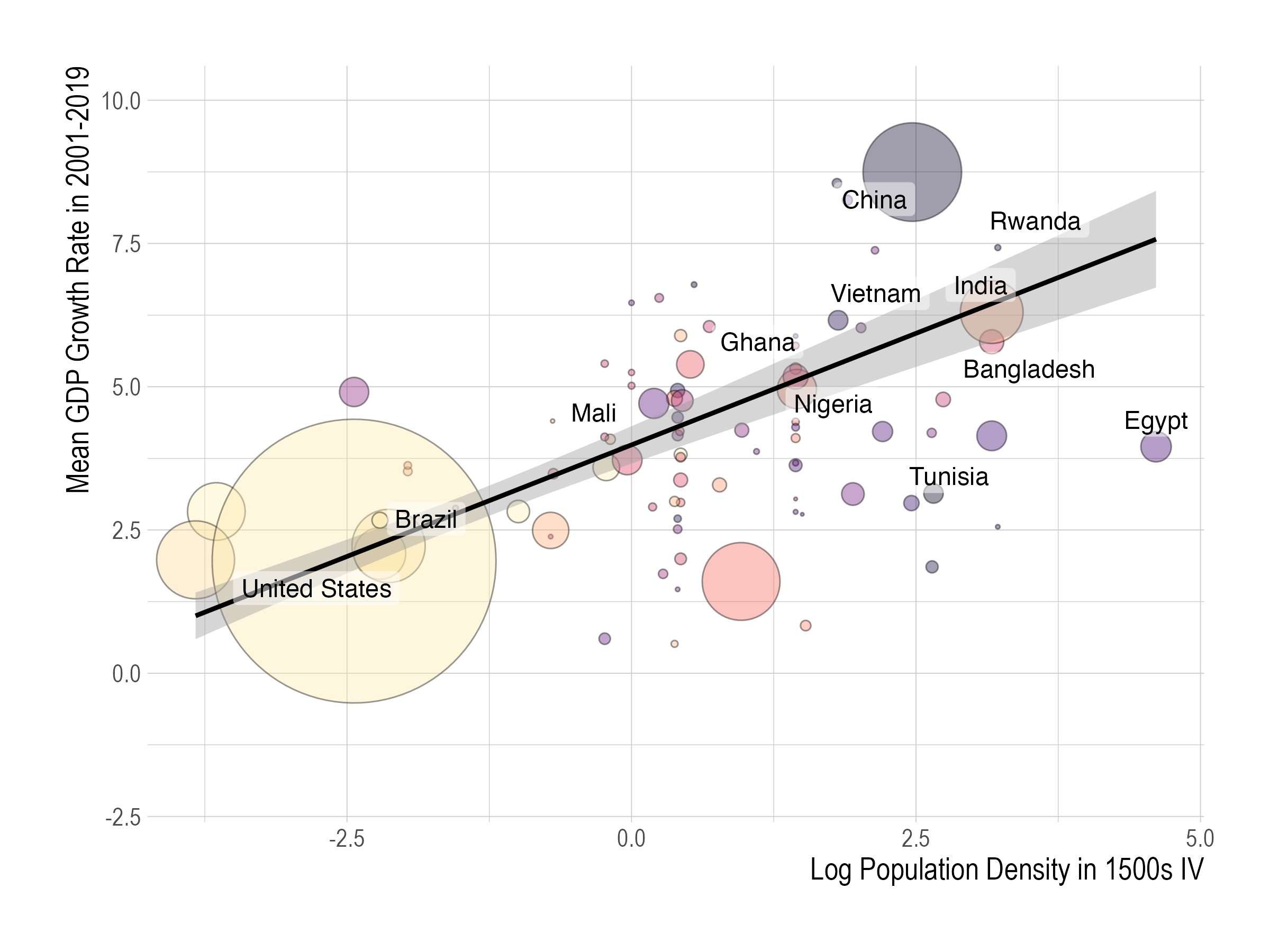}
    
    \caption{Reduced form: Mean GDP Growth Rate in 2020–2022}\label{fig:rf_appendix_b}
    \includegraphics[width=.99\textwidth]{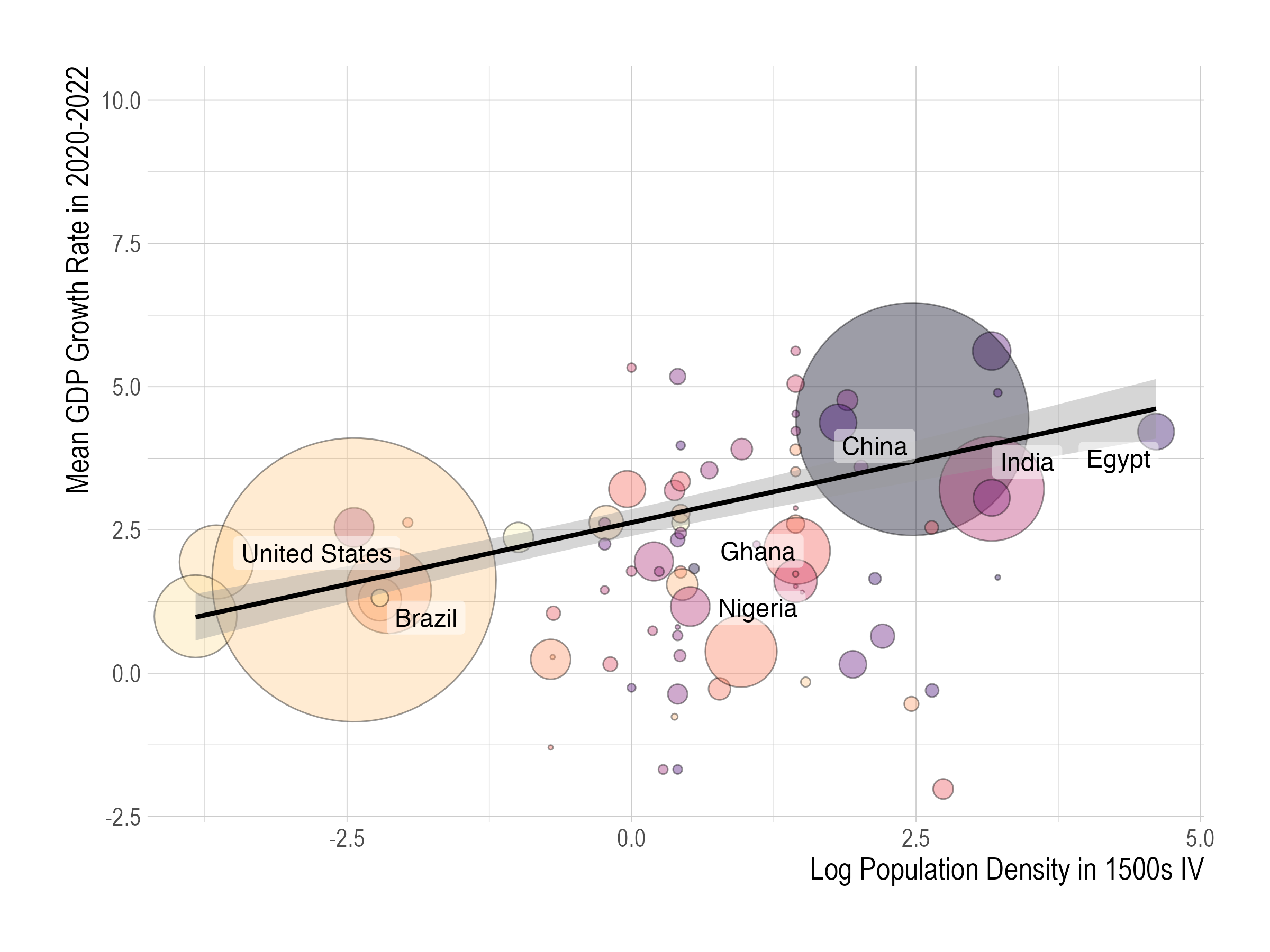}

     \caption{Reduced form: Mean Nighttime Light Intensity Growth Rate in 2001-2013}\label{fig:rf_appendix_c}
    \includegraphics[width=.99\textwidth]{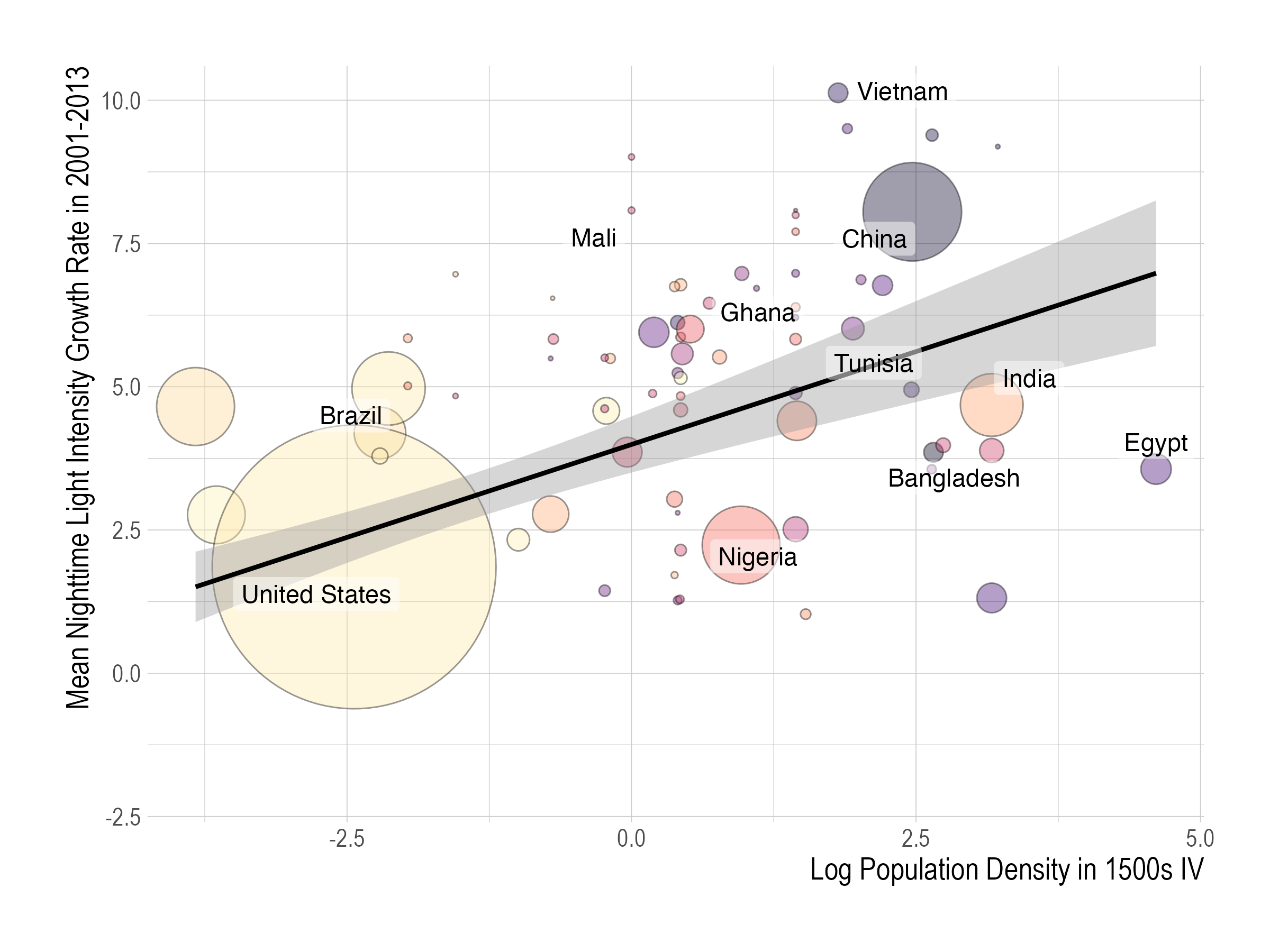}

\end{subfigure}
\begin{subfigure}[c]{.49\linewidth}
    \centering
    
    \caption{Reduced form: Mean GDP Growth Rate in 2001–2019}\label{fig:rf_appendix_d}
    \includegraphics[width=.99\textwidth]{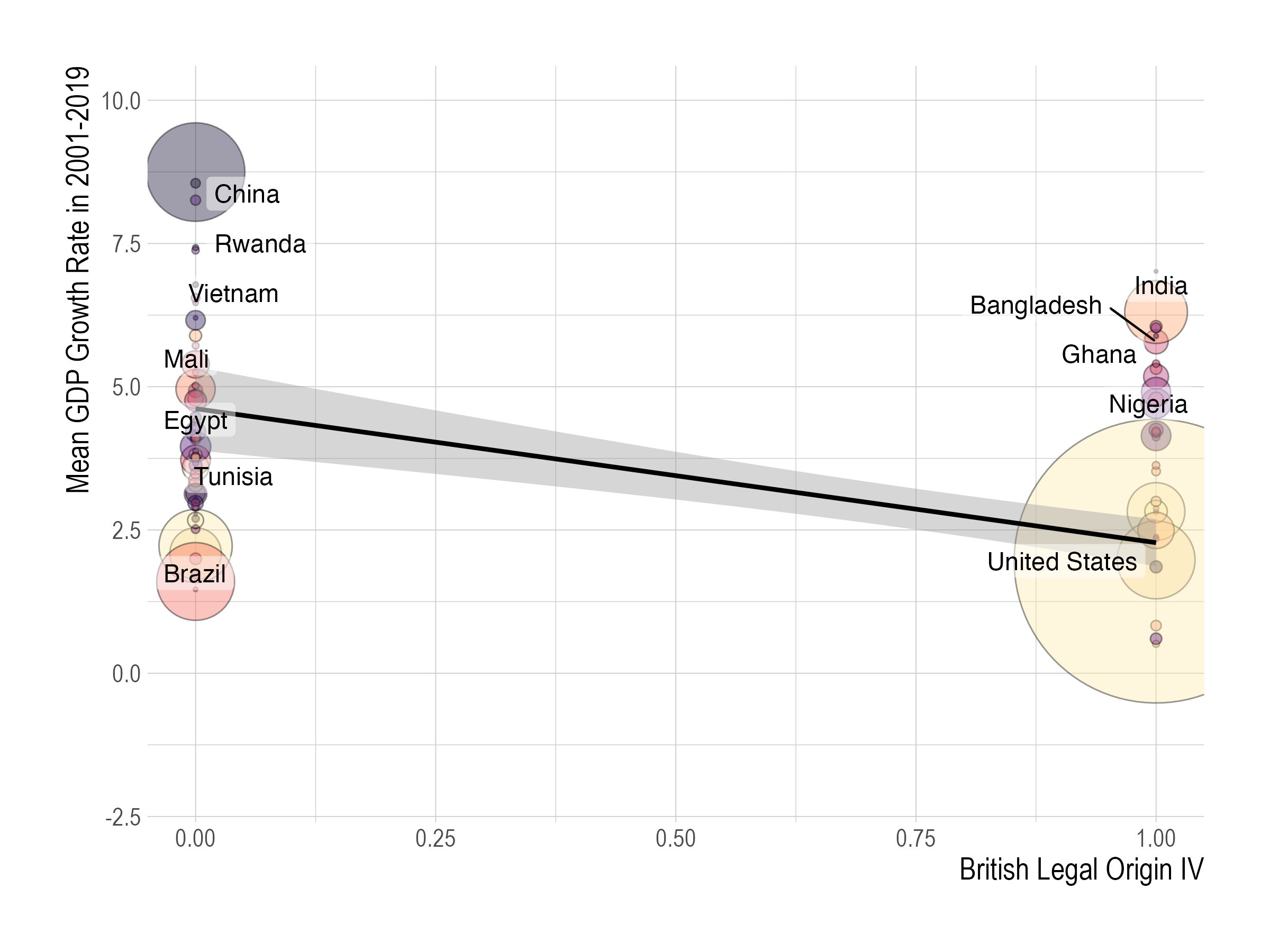}
    
    \caption{Reduced form: Mean GDP Growth Rate in 2020–2022}\label{fig:rf_appendix_e}
    \includegraphics[width=.99\textwidth]{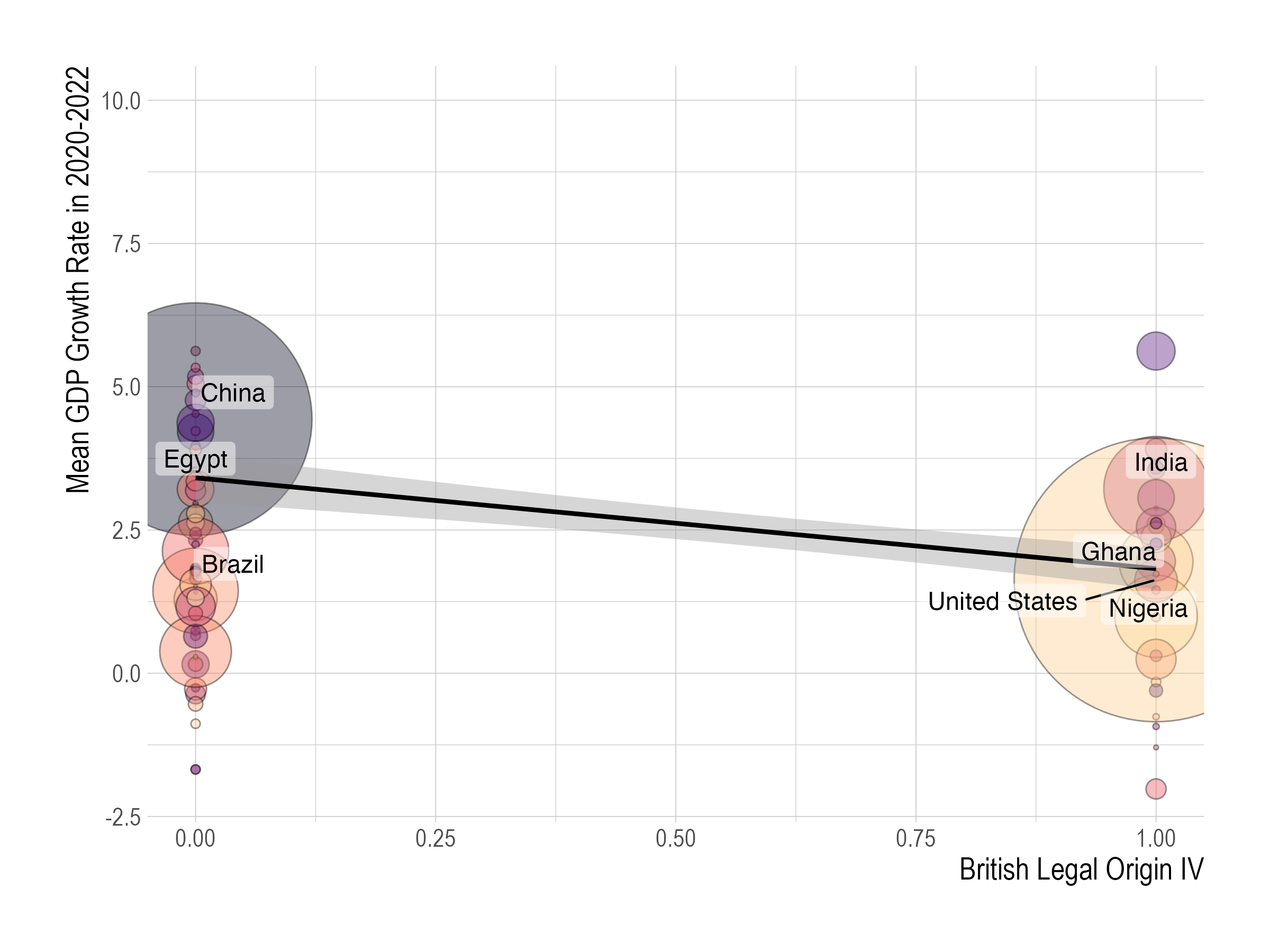}
 
     \caption{Reduced form: Mean Nighttime Light Intensity Growth Rate in 2001-2013}\label{fig:rf_appendix_f}
    \includegraphics[width=.99\textwidth]{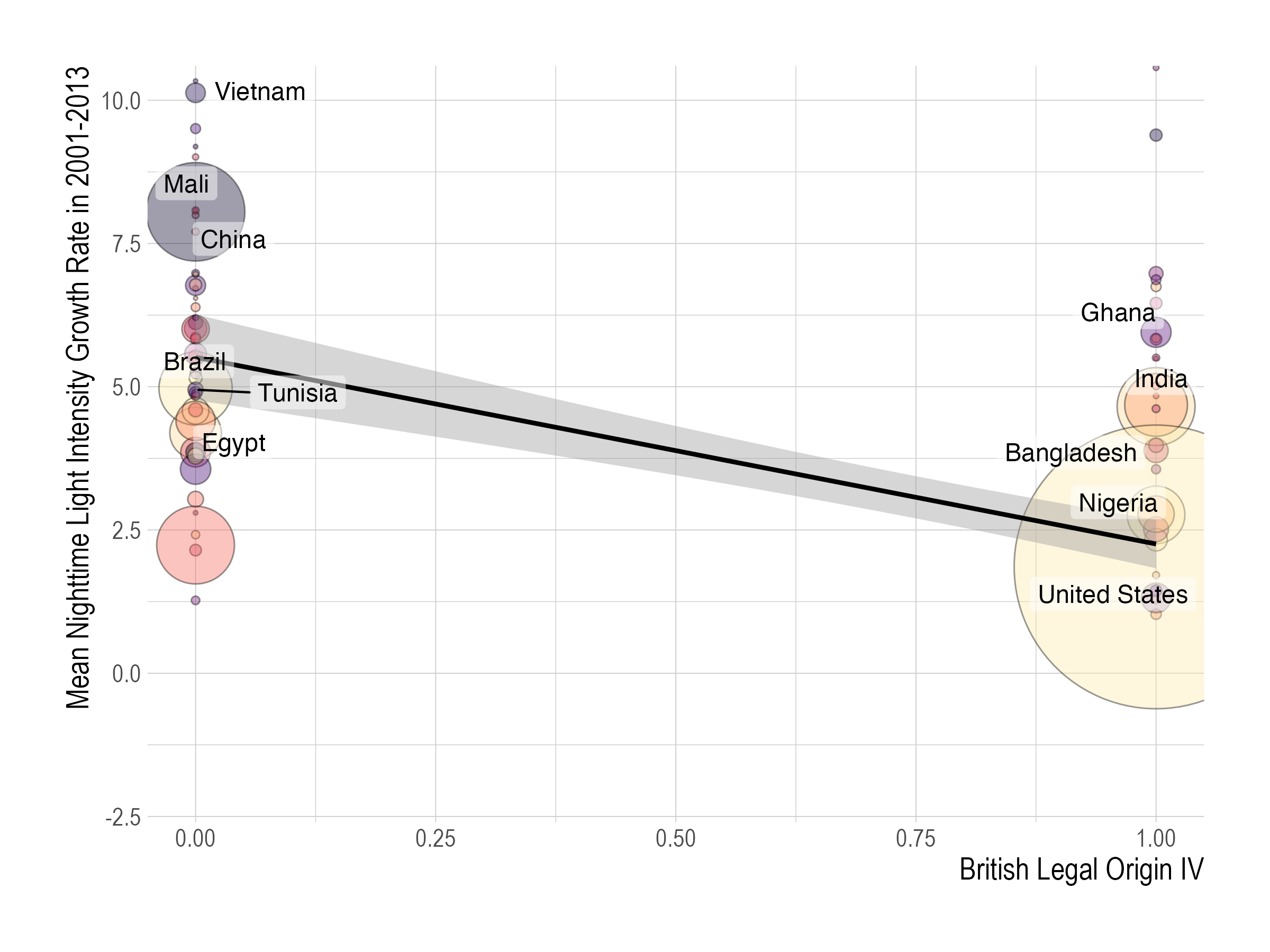}
\end{subfigure}


\caption*{\footnotesize{\textit{Notes:} Panels (a), (b), and (c) illustrate the reduced-form relationship between the log of Population Density in the 1500s (IV) and three outcomes: the mean GDP growth rate from 2001 to 2019, mean GDP growth rate from 2020 to 2022, and mean night-time light intensity growth rate from 2001 to 2013. Panels (d), (e), and (f) illustrate the reduced-form relationship between the same outcomes and the British Legal Origin (IV). 
\input{supporting_files/coefs/explanation_dem} The size of each circle (country) is proportional to its baseline GDP. The colors depend on the level of the democracy index (warmer colors for democracy and darker colors for autocracies). The line is the OLS regression fitted line without controls and weights countries by baseline GDP. The shaded area corresponds to the 95\% confidence interval. 
\input{supporting_files/coefs/explanation_vars}}}
\end{figure}

\restoregeometry

\clearpage
\begin{landscape}
\begin{table}[htbp]
\centering
\begin{threeparttable}
  \caption{Additional Mechanisms in 2001-2019}\label{tab:additional_mechanism}
\footnotesize
\setlength{\tabcolsep}{5pt} 

\hspace*{-4cm}  
\begin{tabular}{l@{}*{11}{c@{}}} 
\hline\hline \\[-1.8ex]

& \multicolumn{3}{c}{\textbf{1) Value Added Channels}}
& \multicolumn{6}{c}{\textbf{2) Economic Channels}} 
& \multicolumn{2}{c}{\textbf{3) Other Channels}}\\

\cmidrule(lr){2-4} \cmidrule(lr){5-10} \cmidrule(lr){11-12} \\

& \multicolumn{1}{c}{\shortstack{Value Added\\Growth\\Agriculture}} 
& \multicolumn{1}{c}{\shortstack{Value Added\\Growth\\Manufacturing}} 
& \multicolumn{1}{c}{\shortstack{Value Added\\Growth\\Services}} 
& \multicolumn{1}{c}{\shortstack{Tax Revenue\\Growth}}
& \multicolumn{1}{c}{\shortstack{R\&D\\Researchers\\Growth}}
& \multicolumn{1}{c}{\shortstack{New Business\\Registrations\\Growth}}  
& \multicolumn{1}{c}{\shortstack{FDI\\Growth}} 
& \multicolumn{1}{c}{\shortstack{Imports\\from China\\Growth}} 
& \multicolumn{1}{c}{\shortstack{Exports\\from China\\Growth}}
& \multicolumn{1}{c}{\shortstack{Conflict\\Index\\Growth}} 
& \multicolumn{1}{c}{\shortstack{Child Mortality\\Growth}} \\

&\multicolumn{1}{c}{(1)} 
&\multicolumn{1}{c}{(2)} 
&\multicolumn{1}{c}{(3)} 
&\multicolumn{1}{c}{(4)}  
&\multicolumn{1}{c}{(5)} 
&\multicolumn{1}{c}{(6)}  
&\multicolumn{1}{c}{(7)} 
&\multicolumn{1}{c}{(8)} 
&\multicolumn{1}{c}{(9)} 
&\multicolumn{1}{c}{(10)} 
&\multicolumn{1}{c}{(11)} 
\\

\hline

& \multicolumn{10}{c}{\textbf{Panel A: OLS}}\\
Democracy Index (V-Dem, 2000)&        -1.8&        -2.1&        -3.1&        -0.7&        -0.3&         1.8&       -10.7&        -4.8&       -10.0&        -0.2&         1.4\\
&       (0.4)&       (0.3)&       (0.6)&       (0.3)&       (1.0)&       (3.6)&       (9.8)&       (0.7)&       (1.8)&      (0.09)&       (0.3)\\
N                   &         135&         124&         138&         125&         118&         124&          99&         134&         104&         155&         153\\
& \multicolumn{10}{c}{\textbf{Panel B: Instrument for Democracy by Settler Mortality}}\\
Democracy Index (V-Dem, 2000)&        -3.8&        -7.3&        -4.6&        -1.7&        -5.1&        -8.5&       -55.0&        -9.0&       -14.8&        -0.4&         2.6\\
&       (0.9)&       (2.0)&       (0.8)&       (0.7)&       (6.9)&       (9.9)&      (42.4)&       (3.2)&       (4.9)&       (0.3)&       (0.5)\\
N                   &          74&          68&          76&          69&          62&          60&          50&          70&          53&          78&          81\\
& \multicolumn{10}{c}{\textbf{Panel C: Instrument for Democracy by Population Density in 1500s}}\\
Democracy Index (V-Dem, 2000)&        -1.3&        -2.6&        -3.7&        -1.5&        -0.9&         7.8&       -21.8&        -6.6&       -15.1&        -0.6&         2.0\\
&       (0.6)&       (1.7)&       (0.7)&       (0.5)&       (2.2)&       (8.4)&      (19.4)&       (2.0)&       (3.5)&       (0.4)&       (0.3)\\
N                   &          77&          69&          80&          71&          66&          63&          48&          74&          53&          83&          85\\
& \multicolumn{10}{c}{\textbf{Panel D: Instrument for Democracy by Legal Origin}}\\
Democracy Index (V-Dem, 2000)&        -0.8&       -38.9&        -3.8&        -2.5&       -12.3&      -112.9&       -58.6&        -8.6&       -21.5&         0.2&         2.6\\
&       (2.1)&     (251.9)&       (1.0)&       (0.9)&      (12.4)&     (535.6)&      (56.4)&       (3.6)&       (7.5)&       (0.2)&       (0.5)\\
N                   &          80&          72&          83&          75&          67&          65&          50&          77&          55&          87&          89\\
& \multicolumn{10}{c}{\textbf{Panel E: Instrument for Democracy by Language}}\\
Democracy Index (V-Dem, 2000)&        -2.2&        -4.4&        -1.8&        -3.6&         3.6&         9.8&        -3.2&        -3.4&        -4.9&        0.09&         2.9\\
&       (0.8)&       (1.7)&       (2.2)&       (1.7)&       (5.7)&      (13.9)&      (21.2)&       (3.5)&       (7.9)&       (0.3)&       (0.8)\\
N                   &         114&         104&         120&         105&         103&         104&          86&         113&          89&         129&         129\\
& \multicolumn{10}{c}{\textbf{Panel F: Instrument for Democracy by Crops and Minerals}}\\
Democracy Index (V-Dem, 2000)&        -3.0&        -4.0&        -4.2&        -0.9&        -6.3&        -1.3&       -52.0&        -8.0&       -15.3&        -0.2&         2.0\\
&       (0.7)&       (1.0)&       (0.9)&       (0.6)&       (6.0)&       (5.0)&      (34.2)&       (2.5)&       (3.3)&       (0.1)&       (0.4)\\
N                   &         119&         109&         122&         110&         105&         107&          87&         118&          92&         135&         136\\
& \multicolumn{10}{c}{\textbf{Panel G: Use all IVs}}\\
Democracy Index (V-Dem, 2000)&        -1.3&        -2.0&        -3.7&        -1.0&        0.02&         8.2&        -7.3&        -5.1&       -11.8&        -0.4&         1.9\\
&       (0.4)&       (0.7)&       (0.4)&       (0.4)&       (2.1)&       (7.7)&      (11.7)&       (1.4)&       (3.2)&       (0.2)&       (0.2)\\
N                   &          65&          59&          68&          60&          55&          54&          42&          62&          45&          68&          71\\
\\ Outcome Mean & 6.9 & 8.4 & 9.1 & 1.4 & 5.0 & 5.7 & 5.5 & 23.2 & 39.9 & 0.1 & -3.4\\\\[-1.8ex]

\hline\hline
\end{tabular}
\begin{tablenotes}
\item {\footnotesize {\textit{Notes:} This table reports the OLS (Panel A) and 2SLS (Panels B-G) estimates of democracy's effect on potential mechanisms in 2001-2019.
 The Democracy Index (V-Dem) is normalized to have mean zero and standard deviation one in each year.\unskip 
Tax Revenue represents the total tax revenues as a percentage of GDP; R\&D Researchers represents the number of researchers in R\&D per million people; New Business Registrations indicates the density of new business registrations per 1,000 people; Foreign Direct Investment (FDI) represents net outflows as a percentage of GDP; Import from China represents the total value of imports from China in US dollars; Export to China represents the total value of exports from China in US dollars; Conflict Index measures the intensity of internal conflict within a country; Child Mortality represents the mortality rate of infants per 1,000 live births. \unskip
 For IVs, Panel B uses log European settler mortality, Panel C uses log population density in the 1500s, Panel C uses British legal origin, Panel E uses the fraction speaking English and the fraction speaking European, Panel F uses the ability to grow crops and mine minerals, and Panel G uses all the IVs together. \unskip 
We report the global mean of the dependent variables in the bottom row. The specifications do not include controls.
N refers to the number of countries for which IVs, controls, and outcomes are available. \unskip 
Robust standard errors are in parentheses. \unskip
Variable definitions and data sources are in Appendix Table \ref{tab:sources}. \unskip}}
\end{tablenotes}
\end{threeparttable}
\end{table}
\end{landscape}

\clearpage
\newgeometry{left=0.3cm, right = 0.3cm, top = 1cm, bottom=1in}
\begin{figure}[p]
  \centering
  \captionsetup{width=0.99\textwidth}
  \caption{Correlation Between Democracy and Additional Outcomes}
  \label{fig:democracy_outcomes}
  
  \begin{subfigure}[c]{0.49\textwidth}
    \centering
    \caption{Mean Excess Deaths per 100k in 2020-2022}
    \label{fig:democracy_outcomes_deaths}
    \includegraphics[width=0.99\linewidth]{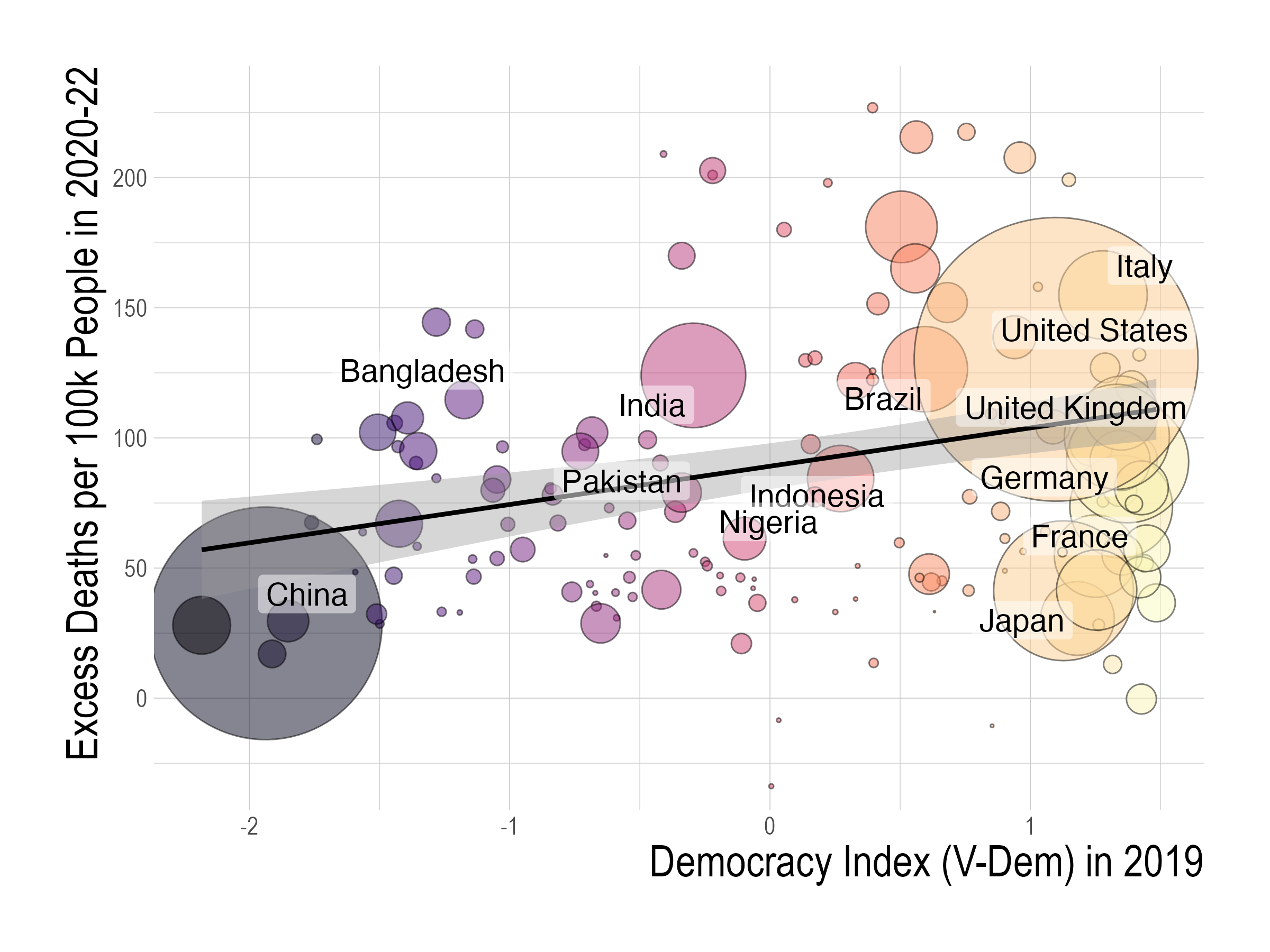}
  \end{subfigure}
  
  
  \begin{subfigure}[c]{0.49\textwidth}
    \centering
    \caption{Mean Life Satisfaction Growth Rates in 2010-2019}
    \label{fig:democracy_outcomes_lf2019}
    \includegraphics[width=0.99\linewidth]{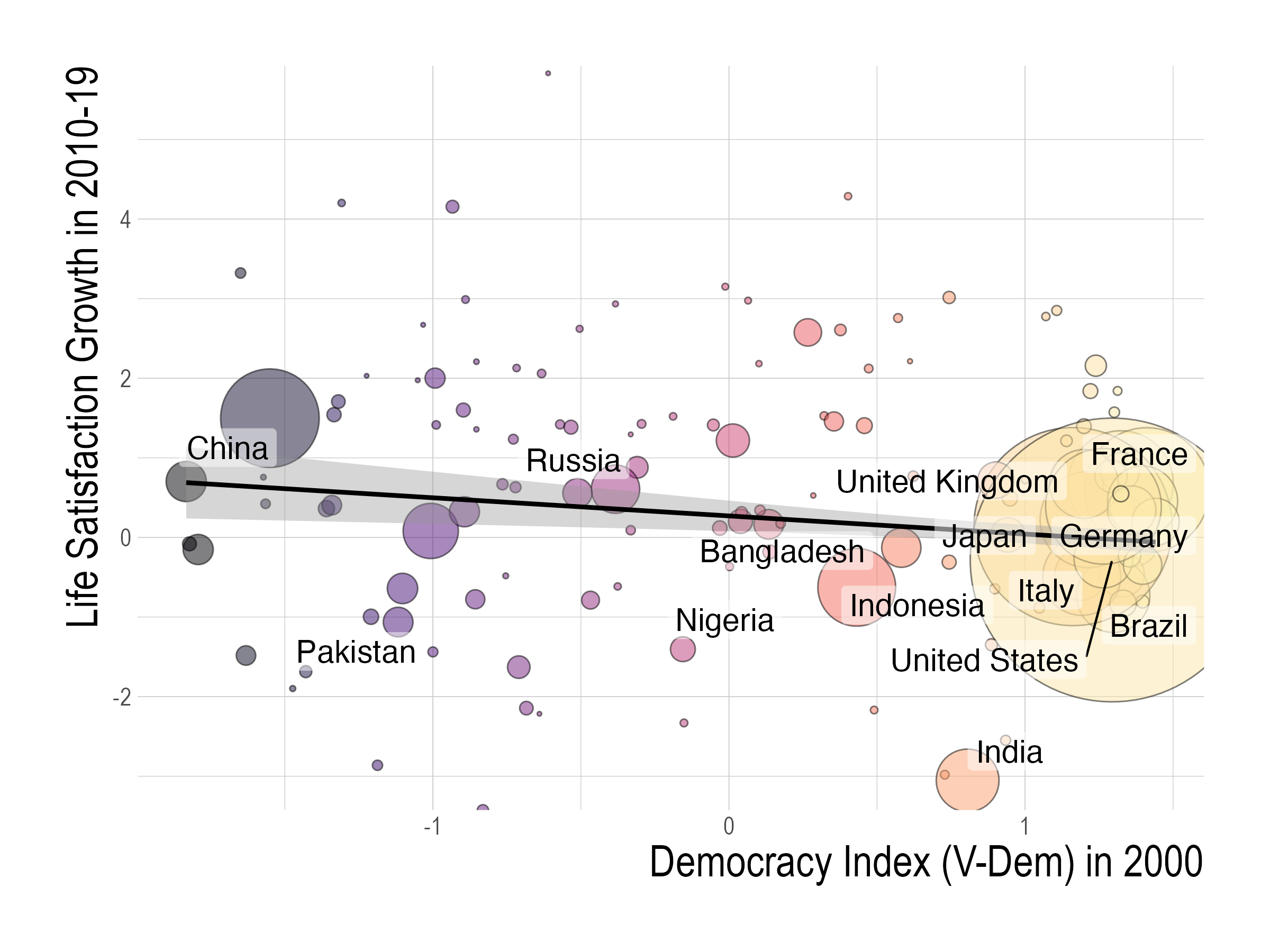}
  \end{subfigure}%
  \begin{subfigure}[c]{0.49\textwidth}
    \centering
    \caption{Mean Life Satisfaction Growth Rates in 2020-2022}
    \label{fig:democracy_outcomes_lf2022}
    \includegraphics[width=0.99\linewidth]{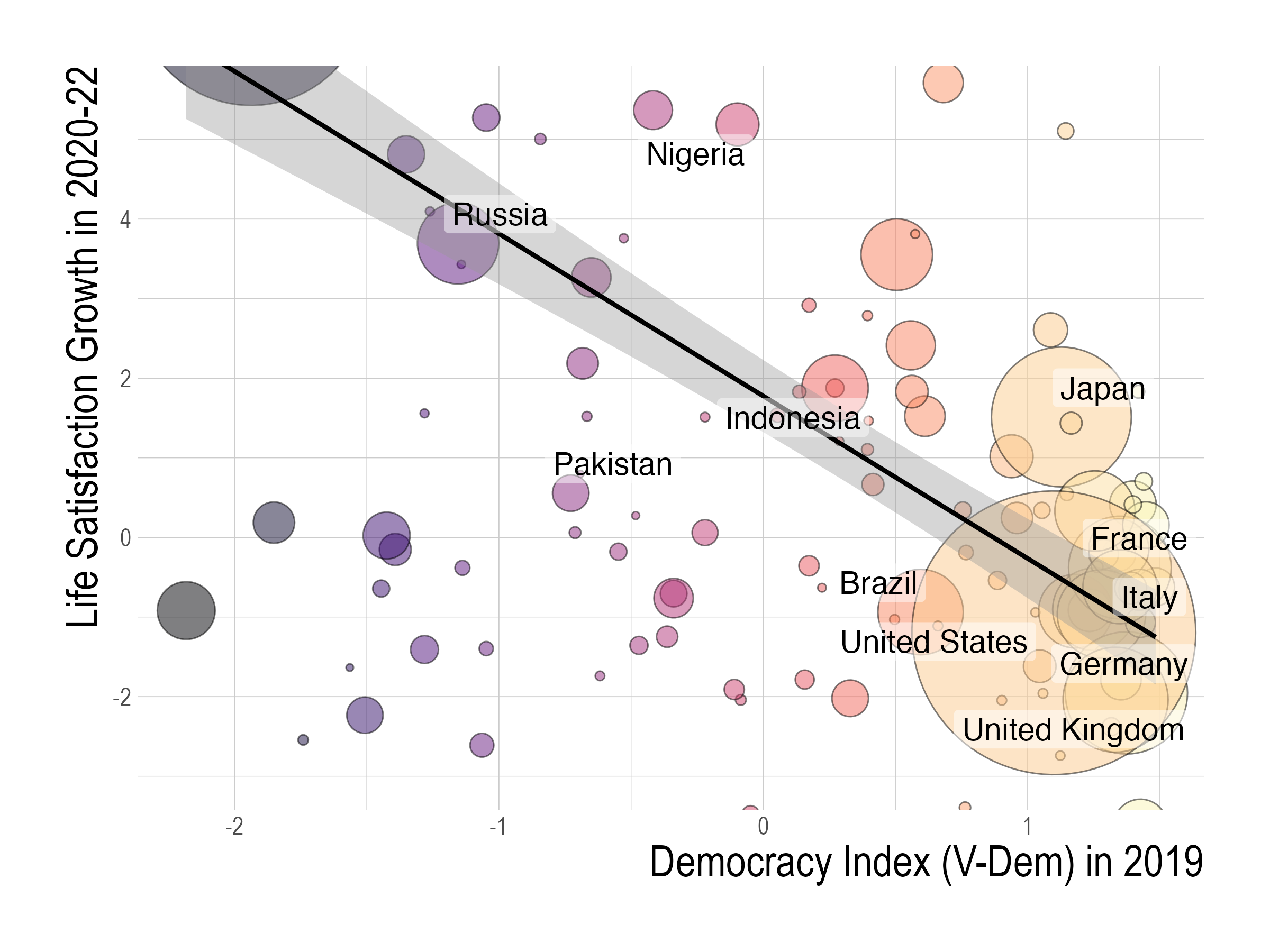}
  \end{subfigure}
  
  
  \begin{subfigure}[c]{0.49\textwidth}
    \centering
    \caption{Mean Top 1\% Income Share Growth Rates in 2001-2019}
    \label{fig:democracy_outcomes_in2019}
    \includegraphics[width=0.99\linewidth]{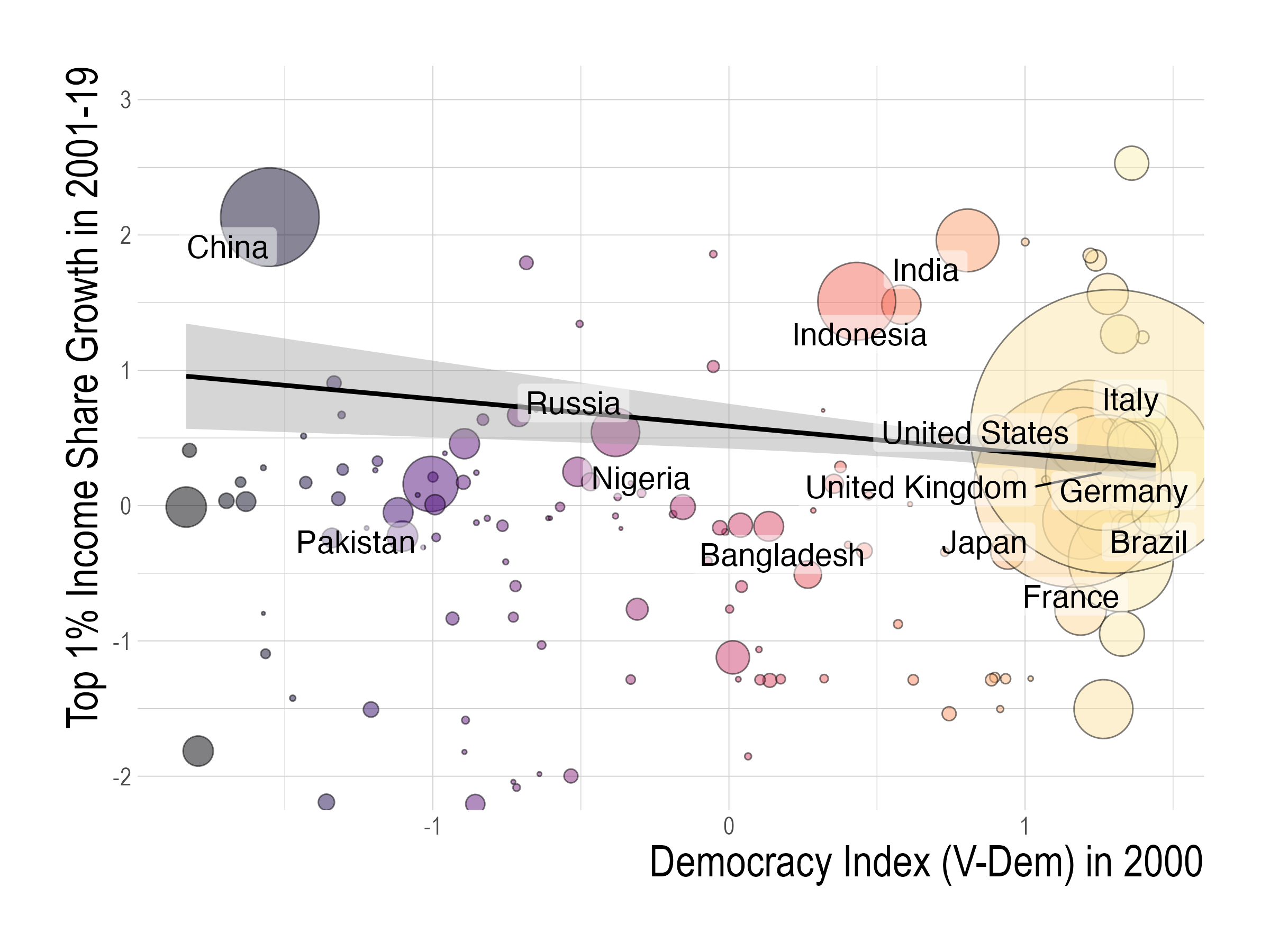}
  \end{subfigure}%
  \begin{subfigure}[c]{0.49\textwidth}
    \centering
    \caption{Mean Top 1\% Income Share Growth Rates in 2020-2022}
    \label{fig:democracy_outcomes_in2022}
    \includegraphics[width=0.99\linewidth]{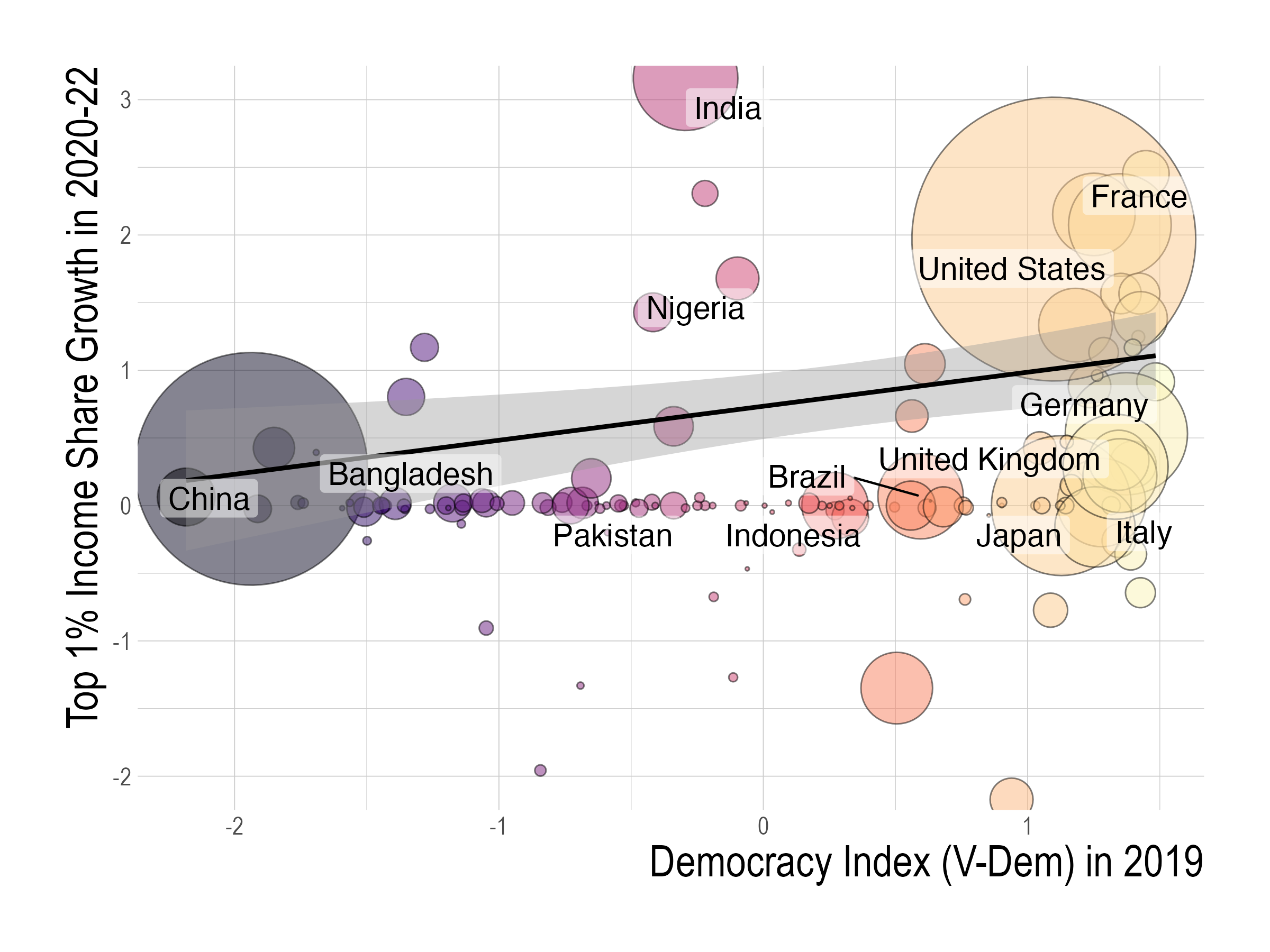}
  \end{subfigure}
\end{figure}

\begin{figure}[p]
  \ContinuedFloat
  \centering
  \captionsetup{width=0.99\textwidth}
  \caption{Continued}
  
  \begin{subfigure}[c]{0.49\textwidth}
    \centering
    \caption{Mean CO2 Emissions per Capita Growth Rates in 2001-2019}
    \label{fig:democracy_outcomes_co2019}
    \includegraphics[width=0.99\linewidth]{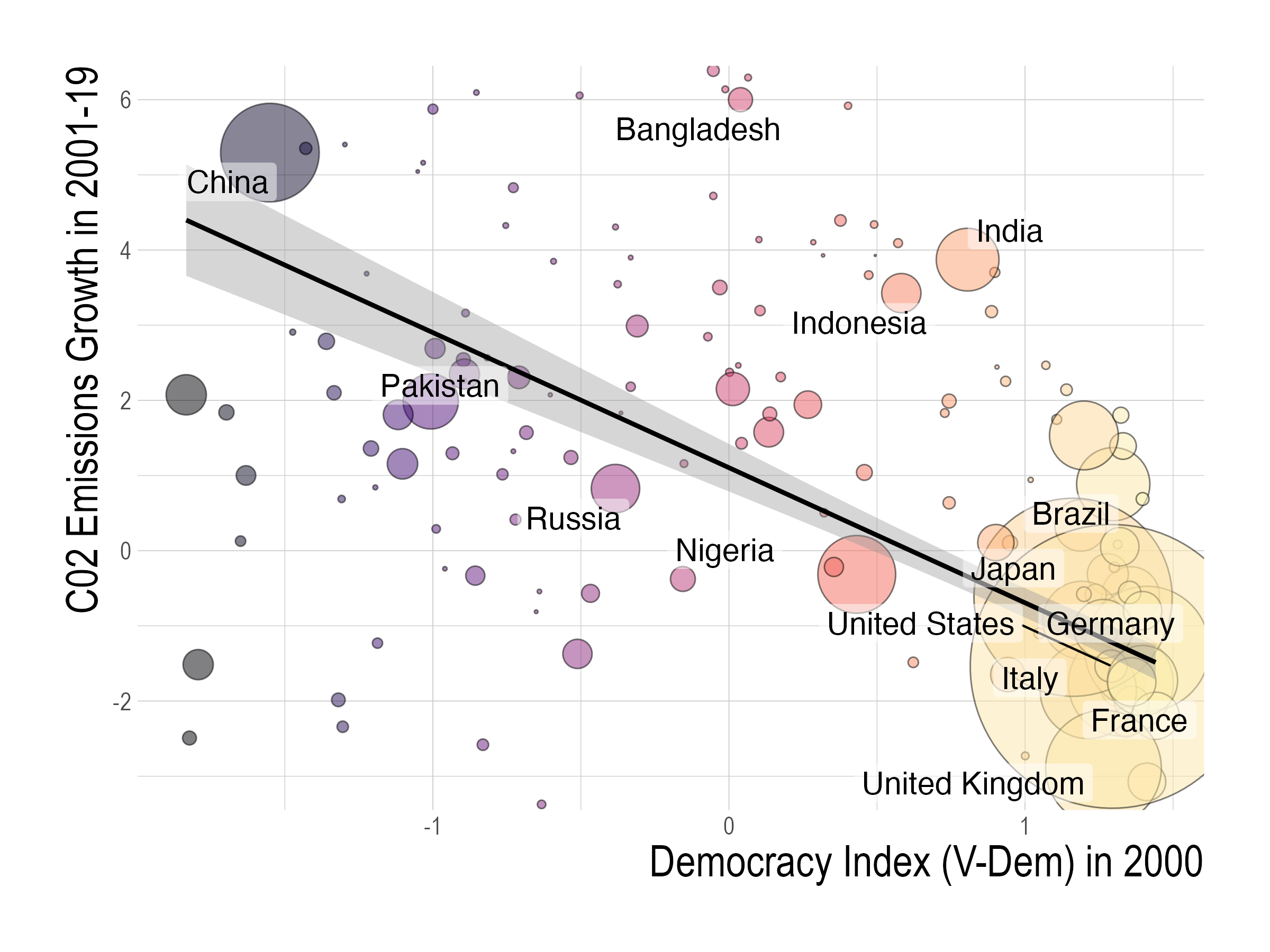}
  \end{subfigure}%
  \begin{subfigure}[c]{0.49\textwidth}
    \centering
    \caption{Mean CO2 Emissions per Capita Growth Rates in 2020-2022}
    \label{fig:democracy_outcomes_co2022}
    \includegraphics[width=0.99\linewidth]{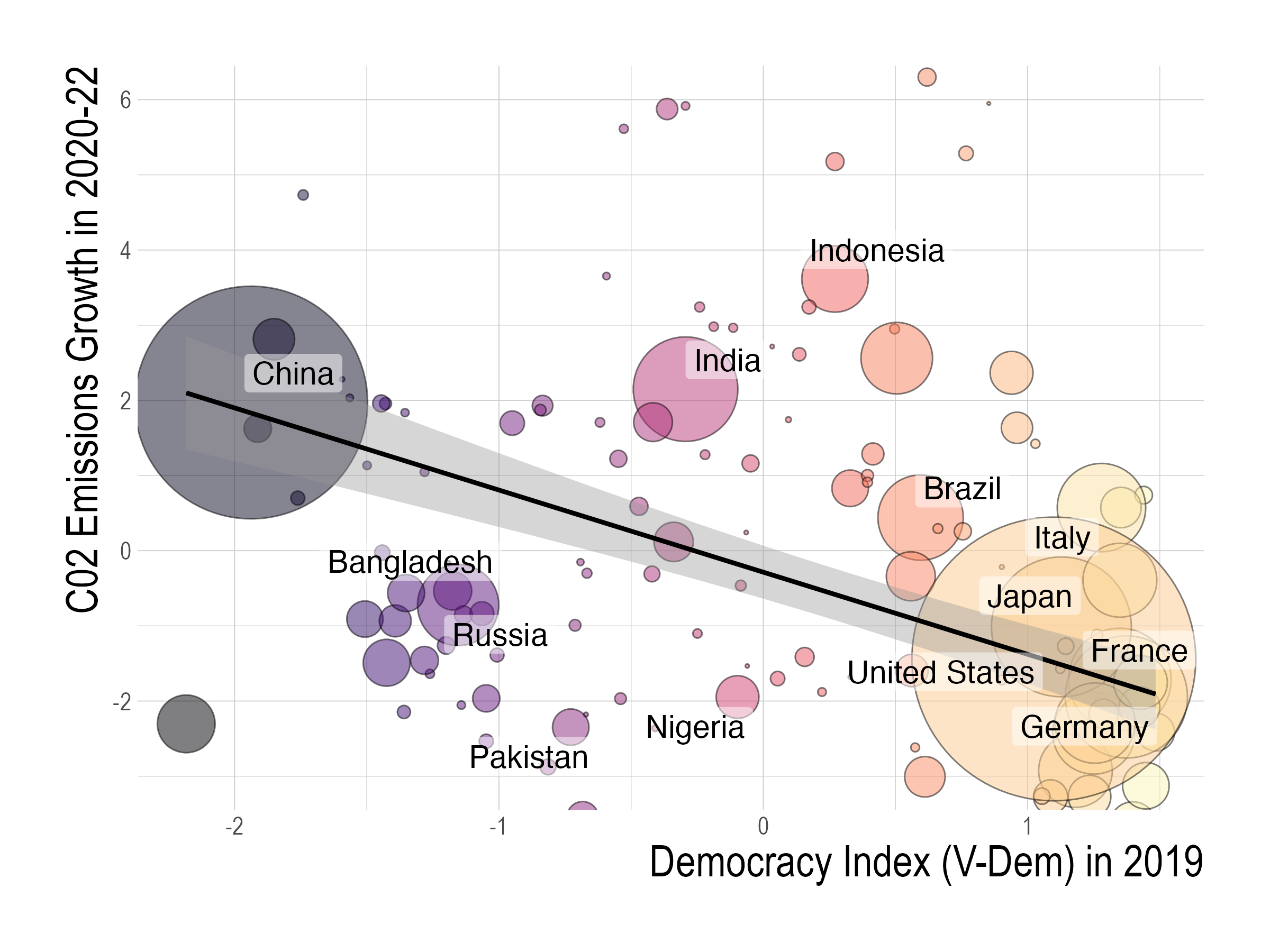}
  \end{subfigure}
  
  
  \begin{subfigure}[c]{0.49\textwidth}
    \centering
    \caption{Mean Energy Consumption per Capita Growth Rates in 2001-2019}
    \label{fig:democracy_outcomes_en2019}
    \includegraphics[width=0.99\linewidth]{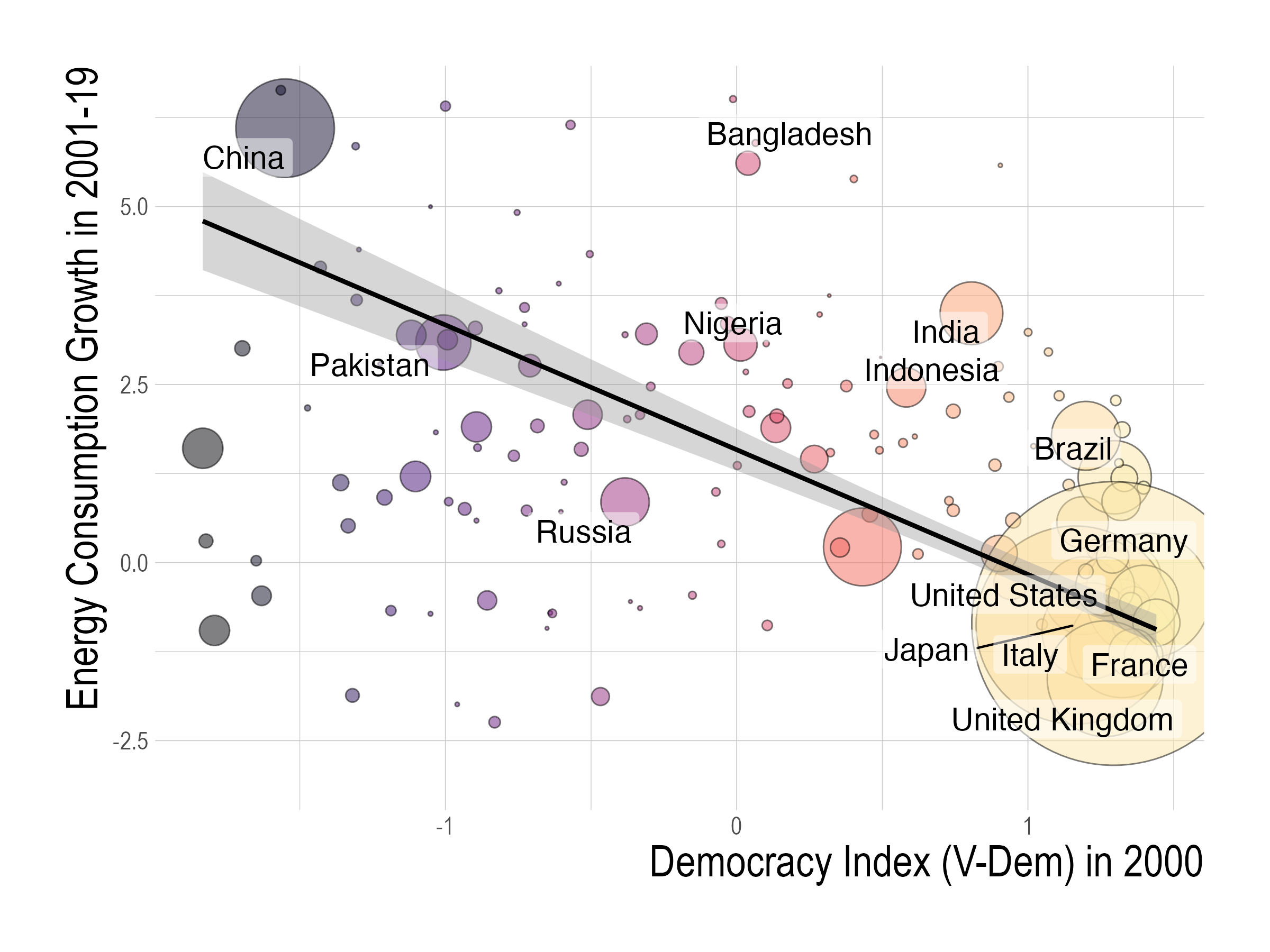}
  \end{subfigure}%
  \begin{subfigure}[c]{0.49\textwidth}
    \centering
    \caption{Mean Energy Consumption per Capita Growth Rates in 2020-2022}
    \label{fig:democracy_outcomes_en2022}
    \includegraphics[width=0.99\linewidth]{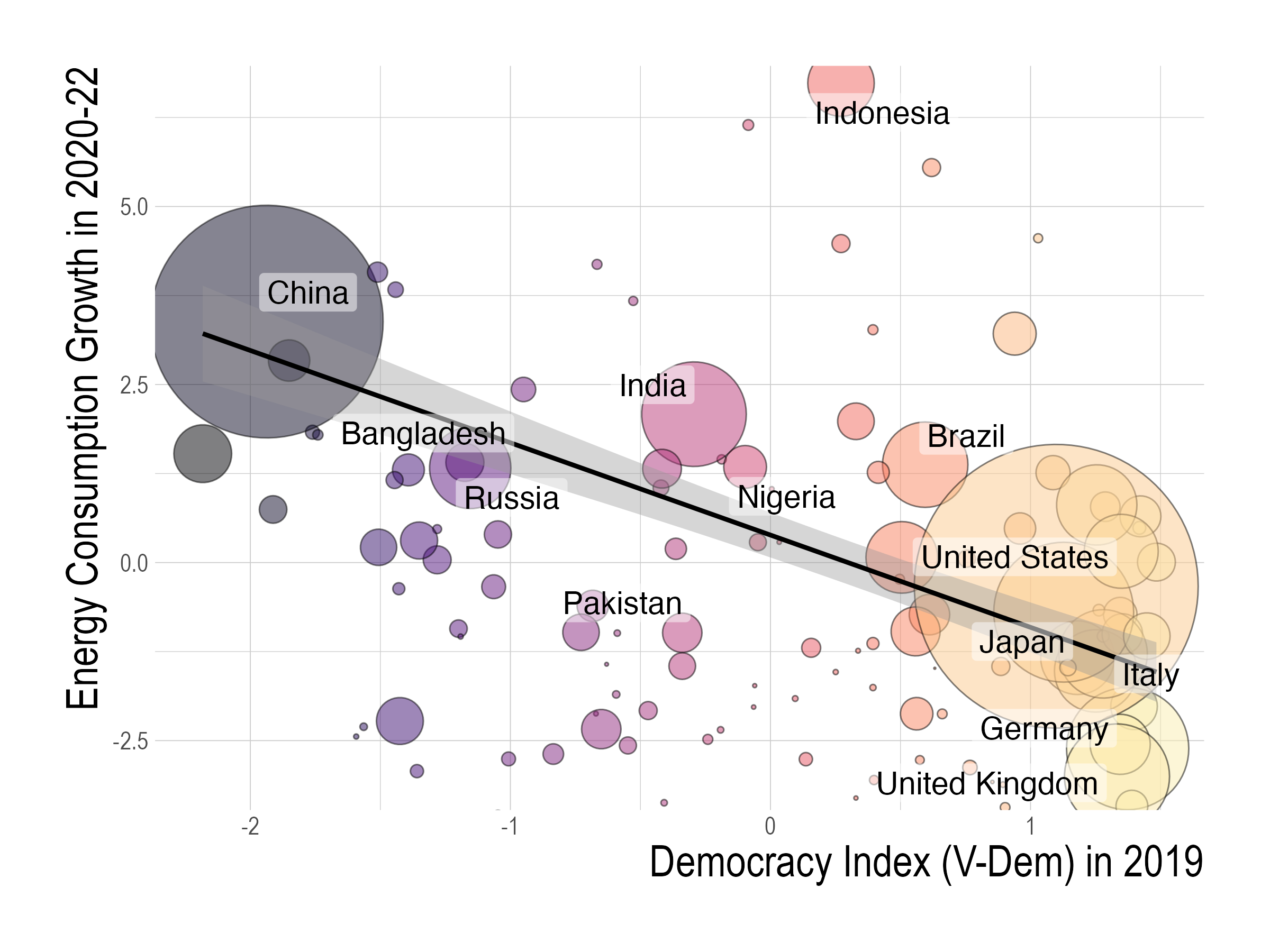}
  \end{subfigure}
  
  
  \caption*{\footnotesize{\textit{Notes:} This figure shows the correlation between democracy and additional outcomes.
  \input{supporting_files/coefs/explanation_dem} The size of each circle (country) is proportional to its baseline GDP. The colors depend on the level of the democracy index (warmer colors for democracy and darker colors for autocracies). The line is the fitted line from a univariate OLS regression of the outcome against the democracy index that weights observations by baseline GDP. The shaded area corresponds to the 95\% confidence interval.}}
\end{figure}

\restoregeometry

\clearpage
\begin{table}[!t]  
  \centering
  \caption{2SLS Regression Estimates of Democracy’s Effects on Additional Outcomes}
  \label{tab:2sls} 
  \fontsize{6}{7}\selectfont
  \begin{threeparttable}
  \begin{tabular}{@{\extracolsep{0pt}}lcccccccccccc} 
\\[-1.8ex]\hline 
\hline 
\\[-1.8ex] & (1) & (2) & (3) & (4) & (5) & (6) & (7) & (8) & (9) & (10) & (11) & (12)\\ 
\hline \\[-1.8ex] 
\multicolumn{13}{l}{\textbf{Panel A: Two-Stage Least Squares}} \\ 
& \multicolumn{12}{c}{Dependent Variable is Mean Excess Deaths per 100k People in 2020-2022} \\\cline{2-13}\\[-1.8ex]
Democracy Index (V-Dem, 2019)&        21.2&        30.2&        22.4&        26.7&        26.8&        25.9&        24.5&        20.1&        15.8&        24.9&        27.8&        29.7\\
&       (8.1)&       (6.3)&       (8.1)&       (3.4)&       (6.4)&       (4.9)&      (11.5)&      (11.8)&       (9.8)&       (7.0)&       (4.5)&       (2.9)\\
&        0.01&        0.00&        0.01&        0.00&        0.00&        0.00&        0.03&        0.09&        0.11&        0.00&        0.00&        0.00\\

& \multicolumn{12}{c}{Dependent Variable is Mean Life Satisfaction Growth Rate in 2010-2019} \\\cline{2-13}\\[-1.8ex]
Democracy Index (V-Dem, 2000)&         0.1&        -1.7&        -0.1&        -0.4&        -0.5&        -1.5&        -0.4&        -0.4&        -0.1&        -0.7&        -0.4&        -0.7\\
&       (0.5)&       (1.4)&       (0.4)&       (0.4)&       (0.2)&       (0.7)&       (0.3)&       (0.2)&       (0.3)&       (0.1)&       (0.2)&       (0.1)\\
&        0.85&        0.24&        0.76&        0.37&        0.04&        0.05&        0.21&        0.07&        0.79&        0.00&        0.09&        0.00\\

& \multicolumn{12}{c}{Dependent Variable is Mean Life Satisfaction Growth Rate in 2020-2022} \\\cline{2-13}\\[-1.8ex]
Democracy Index (V-Dem, 2019)&        -2.5&        -2.6&        -2.7&        -2.8&        -2.3&        -2.2&        -2.7&        -2.7&        -2.7&        -2.7&        -2.5&        -2.5\\
&       (0.2)&       (0.5)&       (0.4)&       (0.5)&       (0.4)&       (0.5)&       (0.4)&       (0.3)&       (0.4)&       (0.3)&       (0.2)&       (0.5)\\
&        0.00&        0.00&        0.00&        0.00&        0.00&        0.00&        0.00&        0.00&        0.00&        0.00&        0.00&        0.00\\

& \multicolumn{12}{c}{Dependent Variable is Mean Top 1\% Income Share Growth Rate in 2001-2019} \\\cline{2-13}\\[-1.8ex]
Democracy Index (V-Dem, 2000)&        -0.5&        -0.4&        -0.5&        -1.1&        -0.2&        -0.3&        -0.0&        -0.2&        -0.5&        -0.5&        -0.4&        -0.6\\
&       (0.2)&       (0.5)&       (0.2)&       (0.4)&       (0.3)&       (0.6)&       (0.3)&       (0.2)&       (0.2)&       (0.2)&       (0.1)&       (0.2)\\
&        0.02&        0.51&        0.01&        0.00&        0.40&        0.63&        0.93&        0.36&        0.01&        0.00&        0.00&        0.00\\

& \multicolumn{12}{c}{Dependent Variable is Mean Top 1\% Income Share Growth Rate in 2020-2022} \\\cline{2-13}\\[-1.8ex]
Democracy Index (V-Dem, 2019)&         0.5&         0.6&         0.4&         0.4&         1.0&         1.1&         0.5&         0.7&         0.1&         0.4&         0.4&         0.4\\
&       (0.2)&       (0.3)&       (0.2)&       (0.2)&       (0.4)&       (0.4)&       (0.3)&       (0.2)&       (0.2)&       (0.2)&       (0.1)&       (0.2)\\
&        0.01&        0.03&        0.03&        0.08&        0.02&        0.00&        0.13&        0.00&        0.53&        0.06&        0.00&        0.00\\

& \multicolumn{12}{c}{Dependent Variable is Mean CO2 Emissions per Capita in 2001-2019} \\\cline{2-13}\\[-1.8ex]
Democracy Index (V-Dem, 2000)&        -3.6&        -4.9&        -2.6&        -2.8&        -2.7&        -2.5&        -2.7&        -2.5&        -2.8&        -2.6&        -2.3&        -2.5\\
&       (0.9)&       (4.0)&       (0.5)&       (0.6)&       (0.5)&       (1.2)&       (0.5)&       (0.5)&       (0.5)&       (0.3)&       (0.2)&       (0.3)\\
&        0.00&        0.22&        0.00&        0.00&        0.00&        0.03&        0.00&        0.00&        0.00&        0.00&        0.00&        0.00\\

& \multicolumn{12}{c}{Dependent Variable is Mean CO2 Emissions per Capita in 2020-2022} \\\cline{2-13}\\[-1.8ex]
Democracy Index (V-Dem, 2019)&        -1.5&        -0.7&        -1.2&        -1.0&        -1.3&        -1.2&        -1.2&        -0.9&        -1.4&        -1.0&        -1.2&        -0.9\\
&       (0.4)&       (0.3)&       (0.2)&       (0.2)&       (0.3)&       (0.3)&       (0.2)&       (0.2)&       (0.2)&       (0.1)&       (0.2)&       (0.1)\\
&        0.00&        0.00&        0.00&        0.00&        0.00&        0.00&        0.00&        0.00&        0.00&        0.00&        0.00&        0.00\\

& \multicolumn{12}{c}{Dependent Variable is Mean Energy Consumption per Capita in 2001-2019} \\\cline{2-13}\\[-1.8ex]
Democracy Index (V-Dem, 2000)&        -3.2&        -5.0&        -2.4&        -2.7&        -2.5&        -2.5&        -2.0&        -2.0&        -2.7&        -2.3&        -2.3&        -2.4\\
&       (0.6)&       (3.8)&       (0.3)&       (0.5)&       (0.4)&       (1.0)&       (0.4)&       (0.4)&       (0.4)&       (0.3)&       (0.2)&       (0.3)\\
&        0.00&        0.19&        0.00&        0.00&        0.00&        0.01&        0.00&        0.00&        0.00&        0.00&        0.00&        0.00\\

& \multicolumn{12}{c}{Dependent Variable is Mean Energy Consumption per Capita in 2020-2022} \\\cline{2-13}\\[-1.8ex]
Democracy Index (V-Dem, 2019)&        -2.0&        -1.6&        -1.3&        -1.4&        -1.4&        -1.2&        -1.5&        -1.4&        -1.6&        -1.3&        -1.2&        -1.3\\
&       (0.5)&       (0.4)&       (0.1)&       (0.2)&       (0.2)&       (0.2)&       (0.4)&       (0.2)&       (0.2)&       (0.1)&       (0.1)&       (0.1)\\
&        0.00&        0.00&        0.00&        0.00&        0.00&        0.00&        0.00&        0.00&        0.00&        0.00&        0.00&        0.00\\
IVs & \multicolumn{2}{c}{settler mortality} &  \multicolumn{2}{c}{population density} & \multicolumn{2}{c}{legal origin} & \multicolumn{2}{c}{language} & \multicolumn{2}{c}{crops \& minerals} & \multicolumn{2}{c}{all IVs} \\
Number of IVs & 1 & 1 & 1 & 1 & 1 & 1 & 2 & 2 & 10 & 10 & 15 & 15 \\

\\[-1.8ex] 
\hline \\[-1.8ex] 

\multicolumn{13}{l}{\textbf{Panel B: Ordinary Least Squares}} \\ 
& \multicolumn{12}{c}{Dependent Variable is Mean Excess Deaths per 100k People in 2020-2022} \\\cline{2-13}\\[-1.8ex]
Democracy Index (V-Dem, 2019)&        24.2&        26.1&        28.6&        29.4&        28.6&        29.4&        15.2&        20.2&        15.0&        19.8&        28.9&        30.4\\
&       (5.5)&       (3.6)&       (4.1)&       (2.9)&       (4.1)&       (2.9)&       (9.5)&       (9.9)&       (9.5)&       (9.9)&       (3.9)&       (3.0)\\

& \multicolumn{12}{c}{Dependent Variable is Mean Life Satisfaction Growth Rate in 2010-2019} \\\cline{2-13}\\[-1.8ex]
Democracy Index (V-Dem, 2000)&        -0.4&        -0.8&        -0.4&        -0.7&        -0.4&        -0.7&        -0.2&        -0.6&        -0.2&        -0.6&        -0.5&        -0.7\\
&       (0.2)&       (0.1)&       (0.2)&       (0.1)&       (0.2)&       (0.1)&       (0.2)&       (0.2)&       (0.2)&       (0.2)&       (0.2)&       (0.1)\\

& \multicolumn{12}{c}{Dependent Variable is Mean Life Satisfaction Growth Rate in 2020-2022} \\\cline{2-13}\\[-1.8ex]
Democracy Index (V-Dem, 2019)&        -2.4&        -2.3&        -2.5&        -2.4&        -2.5&        -2.4&        -2.1&        -2.0&        -2.1&        -2.0&        -2.5&        -2.4\\
&       (0.2)&       (0.4)&       (0.2)&       (0.4)&       (0.2)&       (0.4)&       (0.3)&       (0.4)&       (0.3)&       (0.3)&       (0.2)&       (0.5)\\

& \multicolumn{12}{c}{Dependent Variable is Mean Top 1\% Income Share Growth Rate in 2001-2019} \\\cline{2-13}\\[-1.8ex]
Democracy Index (V-Dem, 2000)&        -0.4&        -0.4&        -0.4&        -0.4&        -0.4&        -0.4&        -0.2&        -0.2&        -0.2&        -0.2&        -0.4&        -0.4\\
&       (0.2)&       (0.2)&       (0.2)&       (0.2)&       (0.2)&       (0.2)&       (0.2)&       (0.2)&       (0.2)&       (0.2)&       (0.2)&       (0.2)\\

& \multicolumn{12}{c}{Dependent Variable is Mean Top 1\% Income Share Growth Rate in 2020-2022} \\\cline{2-13}\\[-1.8ex]
Democracy Index (V-Dem, 2019)&         0.3&         0.3&         0.4&         0.4&         0.4&         0.4&         0.3&         0.5&         0.3&         0.5&         0.4&         0.4\\
&       (0.2)&       (0.2)&       (0.1)&       (0.1)&       (0.1)&       (0.1)&       (0.2)&       (0.2)&       (0.2)&       (0.2)&       (0.1)&       (0.2)\\

& \multicolumn{12}{c}{Dependent Variable is Mean CO2 Emissions per Capita in 2001-2019} \\\cline{2-13}\\[-1.8ex]
Democracy Index (V-Dem, 2000)&        -2.2&        -1.7&        -2.1&        -2.0&        -2.1&        -2.0&        -1.8&        -1.5&        -1.8&        -1.5&        -2.1&        -2.0\\
&       (0.2)&       (0.4)&       (0.2)&       (0.4)&       (0.2)&       (0.4)&       (0.3)&       (0.4)&       (0.3)&       (0.4)&       (0.2)&       (0.4)\\

& \multicolumn{12}{c}{Dependent Variable is Mean CO2 Emissions per Capita in 2020-2022} \\\cline{2-13}\\[-1.8ex]
Democracy Index (V-Dem, 2019)&        -1.2&        -0.8&        -1.1&        -0.8&        -1.1&        -0.8&        -1.1&        -0.8&        -1.1&        -0.8&        -1.1&        -0.9\\
&       (0.1)&       (0.1)&       (0.1)&       (0.1)&       (0.1)&       (0.1)&       (0.1)&       (0.1)&       (0.1)&       (0.1)&       (0.2)&       (0.1)\\

& \multicolumn{12}{c}{Dependent Variable is Mean Energy Consumption per Capita in 2001-2019} \\\cline{2-13}\\[-1.8ex]
Democracy Index (V-Dem, 2000)&        -2.2&        -1.9&        -2.1&        -2.0&        -2.1&        -2.0&        -1.8&        -1.4&        -1.8&        -1.4&        -2.1&        -2.1\\
&       (0.2)&       (0.4)&       (0.2)&       (0.4)&       (0.2)&       (0.4)&       (0.3)&       (0.5)&       (0.3)&       (0.5)&       (0.2)&       (0.4)\\

& \multicolumn{12}{c}{Dependent Variable is Mean Energy Consumption per Capita in 2020-2022} \\\cline{2-13}\\[-1.8ex]
Democracy Index (V-Dem, 2019)&        -1.3&        -1.0&        -1.1&        -1.2&        -1.1&        -1.2&        -1.3&        -1.1&        -1.3&        -1.1&        -1.2&        -1.2\\
&       (0.2)&       (0.1)&       (0.1)&      (0.10)&       (0.1)&      (0.10)&       (0.2)&      (0.10)&       (0.2)&       (0.1)&      (0.08)&      (0.10)\\
Baseline Controls & \xmark & \cmark & \xmark & \cmark & \xmark & \cmark & \xmark & \cmark & \xmark & \cmark & \xmark & \cmark\\
N                   &          82&          82&          86&          86&          90&          90&         130&         130&         137&         137&          72&          72\\

\\[-1.8ex] 
\hline 
\hline \\[-1.8ex] 
\end{tabular} 

\begin{tablenotes} 
\item {\footnotesize {\textit{Notes:} 
Excess deaths refer to deaths occurring beyond the expected level, based on the typical, non-crisis mortality rate for the specific population in each country. Life Satisfaction values are based on answers to the main life evaluation questions asked in the poll. The poll asks respondents to think of a Satisfaction, with the best possible life for them being a 10 and the worst possible life being a 0. CO2 Emissions per Capita refers to the Carbon Dioxide emissions from fossil fuels and industry released into the atmosphere per person. Energy Consumption per Capita refers to the amount of primary energy consumed in kilowatt-hours per person.
 The Democracy Index (V-Dem) is normalized to have mean zero and standard deviation one in each year.\unskip 
For IVs, columns 1 and 2 use log European settler mortality, columns 3 and 4 use log population density in the 1500s, columns 5 and 6 use British legal origin, columns 7 and 8 use the fraction speaking English and the fraction speaking European, columns 9 and 10 use the ability to grow crops and mine minerals, and columns 11 and 12 use all the IVs together. \unskip 
Robust standard errors are in parentheses. \unskip 
The p-values are displayed as 0.00 if they are strictly smaller than the 0.005 threshold. The F-statistics are from the first-stage regressions of the IVs against the democracy index in 2019. Panel B reports the OLS estimates. 
Columns 1, 3, 5, 7, 9, and 11 have no controls, while columns 2, 4, 6, 8, 10, and 12 include the following baseline controls: absolute latitude, mean temperature, mean precipitation, population density, and median age. \unskip For outcomes in 2020-2022, we also control for diabetes prevalence.
N refers to the number of countries for which IVs, controls, and outcomes are available. \unskip
Variable definitions and data sources are in Appendix Table \ref{tab:sources}. \unskip}}
\end{tablenotes}
\end{threeparttable}
\end{table} 

\clearpage
\begin{center}
\begin{table}  \centering
  \caption {2SLS Regression Estimates of Democracy’s Effects on GDP per Capita Growth}
  \label{tab:2sls-gdppc} 
  \scriptsize
  \begin{threeparttable}

\begin{tabular}{@{\extracolsep{0pt}}lcccccccccccc} 
\\[-1.8ex]\hline 
\hline 
\\[-1.8ex] & (1) & (2) & (3) & (4) & (5) & (6) & (7) & (8) & (9) & (10) & (11) & (12)\\ 
\hline \\[-1.8ex] 
\multicolumn{13}{l}{\textbf{Panel A: Two-Stage Least Squares}} \\ 
& \multicolumn{12}{c}{Dependent Variable is Mean Per-Capita GDP Growth Rate in 2001-2019} \\\cline{2-13}\\[-1.8ex]
Democracy Index (V-Dem, 2000)&        -2.3&        -4.9&        -2.1&        -3.4&        -1.8&        -2.3&        -1.3&        -1.5&        -2.0&        -2.0&        -2.1&        -2.8\\
&       (0.4)&       (3.4)&       (0.5)&       (0.7)&       (0.6)&       (1.4)&       (0.7)&       (0.6)&       (0.5)&       (0.6)&       (0.4)&       (0.3)\\
&         0.0&         0.2&         0.0&         0.0&         0.0&         0.1&         0.0&         0.0&         0.0&         0.0&         0.0&         0.0\\

& \multicolumn{12}{c}{Dependent Variable is Mean Per-Capita GDP Growth Rate in 2020-2022} \\\cline{2-13}\\[-1.8ex]
Democracy Index (V-Dem, 2019)&        -0.9&        -1.0&        -1.0&        -1.2&        -0.9&        -0.9&        -0.9&        -0.8&        -1.0&        -1.0&        -1.0&        -1.1\\
&       (0.2)&       (0.2)&       (0.2)&       (0.1)&       (0.2)&       (0.1)&       (0.2)&       (0.2)&       (0.2)&       (0.2)&       (0.1)&       (0.1)\\
&         0.0&         0.0&         0.0&         0.0&         0.0&         0.0&         0.0&         0.0&         0.0&         0.0&         0.0&         0.0\\

& \multicolumn{12}{c}{Dependent Variable is Mean Per-Capita Nighttime Light Intensity Growth Rate in 2001-2013} \\\cline{2-13}\\[-1.8ex]
Democracy Index (V-Dem, 2000)&        -2.5&        -5.7&        -1.6&        -1.5&        -2.4&        -3.7&        -1.8&        -3.0&        -2.0&        -3.0&        -1.9&        -2.1\\
&       (0.6)&       (4.1)&       (0.4)&       (0.6)&       (0.5)&       (1.2)&       (0.9)&       (1.0)&       (0.5)&       (0.6)&       (0.3)&       (0.4)\\
&         0.0&         0.2&         0.0&         0.0&         0.0&         0.0&         0.0&         0.0&         0.0&         0.0&         0.0&         0.0\\
IVs & \multicolumn{2}{c}{settler mortality} &  \multicolumn{2}{c}{population density} & \multicolumn{2}{c}{legal origin} & \multicolumn{2}{c}{language} & \multicolumn{2}{c}{crops \& minerals} & \multicolumn{2}{c}{all IVs} \\
Number of IVs & 1 & 1 & 1 & 1 & 1 & 1 & 2 & 2 & 10 & 10 & 15 & 15 \\
Observations        &          81&          81&          86&          86&          90&          90&         130&         130&         136&         136&          71&          71\\

\\[-1.8ex]
\hline \\[-1.8ex] 

\multicolumn{13}{l}{\textbf{Panel B: Ordinary Least Squares}} \\ 
& \multicolumn{12}{c}{Dependent Variable is Mean Per-Capita GDP Growth Rate in 2001-2019} \\\cline{2-13}\\[-1.8ex]
Democracy Index (V-Dem, 2000)&        -2.0&        -2.1&        -1.9&        -2.3&        -1.9&        -2.3&        -1.4&        -1.3&        -1.4&        -1.3&        -2.0&        -2.3\\
&       (0.4)&       (0.5)&       (0.5)&       (0.5)&       (0.5)&       (0.5)&       (0.6)&       (0.7)&       (0.6)&       (0.7)&       (0.4)&       (0.5)\\

& \multicolumn{12}{c}{Dependent Variable is Mean Per-Capita GDP Growth Rate in 2020-2022} \\\cline{2-13}\\[-1.8ex]
Democracy Index (V-Dem, 2019)&        -1.0&        -1.0&        -1.1&        -1.1&        -1.1&        -1.1&        -0.9&        -0.9&        -0.9&        -0.9&        -1.1&        -1.1\\
&       (0.1)&      (0.10)&       (0.1)&      (0.08)&       (0.1)&      (0.08)&       (0.2)&       (0.2)&       (0.2)&       (0.2)&       (0.1)&      (0.08)\\

& \multicolumn{12}{c}{Dependent Variable is Mean Per-Capita Nighttime Light Intensity Growth Rate in 2001-2013} \\\cline{2-13}\\[-1.8ex]
Democracy Index (V-Dem, 2000)&        -1.9&        -2.0&        -1.8&        -1.9&        -1.8&        -1.9&        -1.2&        -1.4&        -1.2&        -1.4&        -1.8&        -1.9\\
&       (0.3)&       (0.4)&       (0.3)&       (0.4)&       (0.3)&       (0.4)&       (0.5)&       (0.6)&       (0.5)&       (0.6)&       (0.3)&       (0.5)\\
Baseline Controls & \xmark & \cmark & \xmark & \cmark & \xmark & \cmark & \xmark & \cmark & \xmark & \cmark & \xmark & \cmark\\
N                   &          81&          81&          86&          86&          90&          90&         130&         130&         136&         136&          71&          71\\

\\[-1.8ex]
\hline 
\hline \\[-1.8ex] 
\end{tabular} 

\begin{tablenotes} 
\item {\scriptsize {\textit{Notes:} Panel A reports the 2SLS estimates of democracy's effect on mean GDP per capita and nighttime light intensity per capita growth rates in 2001-2019 and 2020-2022, using five  different IV strategies. 
 The Democracy Index (V-Dem) is normalized to have mean zero and standard deviation one in each year.\unskip 
For IVs, columns 1 and 2 use log European settler mortality, columns 3 and 4 use log population density in the 1500s, columns 5 and 6 use British legal origin, columns 7 and 8 use the fraction speaking English and the fraction speaking European, columns 9 and 10 use the ability to grow crops and mine minerals, and columns 11 and 12 use all the IVs together. \unskip 
Robust standard errors are in parentheses. \unskip 
The p-values, presented under the standard errors, are displayed as 0.00 if they are strictly smaller than the 0.005 threshold. 
Panel B reports the OLS estimates. Columns 1, 3, 5, 7, 9, and 11 have no controls, while columns 2, 4, 6, 8, 10, and 12 have the following baseline controls: absolute latitude, mean temperature, mean precipitation, population density, and median age. For outcomes in 2020-22, we also control for diabetes prevalence.
N refers to the number of countries for which IVs, controls, and outcomes are available. \unskip 
Variable definitions and data sources are in Appendix Table \ref{tab:sources}. \unskip}}
\end{tablenotes}
\end{threeparttable}
\end{table} 
\end{center}

\clearpage
\begin{landscape}
\begin{table}\centering 
  \caption{2SLS Regression with Alternative Democracy Indices}
  \label{tab:2sls-compare-indices} 
    \begin{threeparttable}
    \scriptsize
\begin{tabular}{@{\extracolsep{0pt}}lcccccccccccc} 
\\[-1.8ex]\hline 
\hline 
& (1) & (2) & (3) & (4) & (5) & (6) & (7) & (8) & (9) & (10) & (11) & (12)\\ 
\hline \\[-1.8ex] 

\textbf{Panel A} & \multicolumn{12}{c}{Dependent Variable is Mean GDP Growth Rate in 2001-2019} \\ \cline{2-13} \\[-1.8ex] 

Democracy Index (V-Dem, 2000)&        -2.5&        -4.2&        -2.2&        -3.1&        -1.8&        -1.7&        -1.0&        -1.1&        -2.2&        -1.6&        -2.1&        -2.6\\
&       (0.3)&       (2.7)&       (0.4)&       (0.6)&       (0.5)&       (1.3)&       (0.9)&       (0.6)&       (0.5)&       (0.6)&       (0.2)&       (0.2)\\

Democracy Index (Polity, 2000)&        -2.8&        -3.6&        -2.6&        -3.4&        -2.0&        -1.5&        -1.3&        -1.4&        -2.8&        -2.0&        -2.3&        -2.5\\
&       (0.4)&       (1.3)&       (0.5)&       (0.8)&       (0.4)&       (1.2)&       (1.1)&       (0.6)&       (0.5)&       (0.5)&       (0.2)&       (0.3)\\

Democracy Index (Freedom House, 2003)&        -2.1&        -3.1&        -2.2&        -3.3&        -1.7&        -1.4&        -1.0&        -1.1&        -2.0&        -1.5&        -2.1&        -2.6\\
&       (0.3)&       (0.9)&       (0.4)&       (0.7)&       (0.5)&       (1.1)&       (1.0)&       (0.6)&       (0.4)&       (0.7)&       (0.3)&       (0.2)\\

Democracy Index (Economist Intelligence Unit, 2006)&        -2.7&        -3.3&        -2.5&        -4.2&        -1.9&        -1.7&        -1.3&        -1.4&        -2.3&        -1.6&        -2.4&        -2.9\\
&       (0.4)&       (1.0)&       (0.5)&       (1.1)&       (0.6)&       (1.5)&       (1.2)&       (0.8)&       (0.6)&       (0.8)&       (0.4)&       (0.5)\\

 \\[-1.8ex] 
\hline \\[-1.8ex] 

\textbf{Panel B} & \multicolumn{12}{c}{Dependent Variable is Mean GDP Growth Rate in 2020-2022} \\  \cline{2-13} \\[-1.8ex] 

Democracy Index (V-Dem, 2019)&        -0.9&        -0.8&        -0.9&        -1.0&        -0.8&        -0.7&        -0.7&        -0.6&        -1.0&        -0.8&        -0.9&        -1.0\\
&       (0.1)&       (0.2)&      (0.08)&      (0.09)&       (0.1)&       (0.1)&       (0.3)&       (0.2)&       (0.2)&       (0.2)&      (0.06)&      (0.07)\\

Democracy Index (Polity, 2018)&        -1.3&        -1.0&        -1.2&        -1.3&        -0.9&        -0.8&        -1.0&        -0.9&        -1.3&        -1.1&        -1.0&        -1.1\\
&       (0.2)&       (0.2)&       (0.2)&       (0.2)&       (0.2)&       (0.2)&       (0.3)&       (0.3)&       (0.2)&       (0.1)&      (0.06)&      (0.06)\\

Democracy Index (Freedom House, 2019)&        -1.1&        -1.0&        -1.1&        -1.2&        -0.8&        -0.8&        -0.8&        -0.7&        -1.1&        -0.9&        -1.0&        -1.1\\
&       (0.1)&       (0.2)&       (0.1)&       (0.2)&       (0.2)&       (0.2)&       (0.3)&       (0.3)&       (0.2)&       (0.2)&      (0.07)&      (0.09)\\

Democracy Index (Economist Intelligence Unit, 2019)&        -1.1&        -1.0&        -1.1&        -1.3&        -0.8&        -0.8&        -0.8&        -0.7&        -1.1&        -0.9&        -1.0&        -1.1\\
&       (0.1)&       (0.2)&       (0.1)&       (0.2)&       (0.2)&       (0.2)&       (0.3)&       (0.3)&       (0.2)&       (0.2)&      (0.08)&       (0.1)\\

 \\[-1.8ex] 
\hline \\[-1.8ex] 

\textbf{Panel C}  & \multicolumn{12}{c}{Dependent Variable is Mean Nighttime Light Intensity Growth Rate in 2001-2013} \\ \cline{2-13} \\[-1.8ex]

Democracy Index (V-Dem, 2000)&        -2.8&        -5.3&        -1.8&        -1.3&        -2.5&        -3.2&        -1.6&        -3.0&        -2.3&        -2.7&        -2.0&        -2.0\\
&       (0.7)&       (3.8)&       (0.4)&       (0.6)&       (0.6)&       (1.0)&       (1.1)&       (1.0)&       (0.5)&       (0.5)&       (0.2)&       (0.4)\\

Democracy Index (Polity, 2000)&        -3.1&        -4.5&        -2.1&        -1.5&        -2.7&        -2.8&        -2.2&        -4.0&        -2.9&        -3.0&        -2.2&        -2.1\\
&       (0.8)&       (2.7)&       (0.4)&       (0.5)&       (0.5)&       (0.7)&       (1.3)&       (1.3)&       (0.7)&       (0.6)&       (0.2)&       (0.4)\\

Democracy Index (Freedom House, 2003)&        -2.4&        -4.0&        -1.8&        -1.5&        -2.3&        -2.7&        -1.7&        -3.3&        -2.0&        -2.7&        -2.0&        -2.2\\
&       (0.4)&       (1.9)&       (0.4)&       (0.6)&       (0.3)&       (0.5)&       (1.1)&       (1.0)&       (0.4)&       (0.6)&       (0.2)&       (0.3)\\

Democracy Index (Economist Intelligence Unit, 2006)&        -3.1&        -4.0&        -2.1&        -1.8&        -2.6&        -3.2&        -2.1&        -3.9&        -2.1&        -2.4&        -2.2&        -2.3\\
&       (0.6)&       (1.8)&       (0.5)&       (0.7)&       (0.5)&       (0.8)&       (1.4)&       (1.6)&       (0.6)&       (0.6)&       (0.4)&       (0.5)\\
\hline \\[-1.8ex] IVs & \multicolumn{2}{c}{settler mortality} & \multicolumn{2}{c}{population density} & \multicolumn{2}{c}{legal origin} & \multicolumn{2}{c}{language} & \multicolumn{2}{c}{crops \& minerals} & \multicolumn{2}{c}{all IVs} \\ Baseline Controls & \xmark & \cmark & \xmark & \cmark & \xmark & \cmark & \xmark & \cmark & \xmark & \cmark & \xmark & \cmark\\
N                   &          77&          77&        82&        82&        86&        86&         120&         120&         124&         124&        68&        68\\

 \\[-1.8ex] 

\hline \hline \\[-1.8ex] 
\end{tabular} 
\begin{tablenotes}
\item {\footnotesize {\textit{Notes:} This table compares the results of 2SLS regressions on the mean GDP growth rate in 2001-2019 (Panel A), the mean GDP growth rate in 2020-2022 (Panel B), and the mean nighttime light intensity growth rate in 2001-2013 (Panel C) using four different democracy indices by V-Dem, Polity, Freedom House, and the Economist Intelligence Unit. When data for the democracy index does not exist for the baseline year, we use the value from the closest year. We normalize all indices to have mean zero and standard deviation one. 
Columns 1, 3, 5, 7, 9, and 11 have no controls, while columns 2, 4, 6, 8,  10, and 12 have the following baseline controls: absolute latitude, mean temperature, mean precipitation, population density, and median age. For the mean GDP growth rate in 2020-2022, we also control for diabetes prevalence. \unskip 
For IVs, columns 1 and 2 use log European settler mortality, columns 3 and 4 use log population density in the 1500s, columns 5 and 6 use British legal origin, columns 7 and 8 use the fraction speaking English and the fraction speaking European, columns 9 and 10 use the ability to grow crops and mine minerals, and columns 11 and 12 use all the IVs together. \unskip The estimates in this table are slightly different from those in Table \ref{tab:2sls_tab1} because this table uses only observations for which all of the democracy indices are available. 
N refers to the number of countries for which IVs, controls, and outcomes are available. \unskip
Robust standard errors are in parentheses. \unskip 
Variable definitions and data sources are in Appendix Table \ref{tab:sources}. \unskip}}
\end{tablenotes}
\end{threeparttable}
\end{table}
\end{landscape}

\clearpage
\begin{center}
\begin{table}  \centering
  \caption{2SLS Regression Estimates of Democracy's Effects Before, During, and After the Great Recession}
  \label{tab:2sls-recession} 
  \footnotesize
  \begin{threeparttable}
  
  \begin{tabular}{@{\extracolsep{0pt}}lcccccccccccc} 
\\[-1.8ex]\hline 
\hline 
\\[-1.8ex] & (1) & (2) & (3) & (4) & (5) & (6) & (7) & (8) & (9) & (10) & (11) & (12)\\ 
\hline \\[-1.8ex] 
\multicolumn{13}{l}{\textbf{Panel A: Two-Stage Least Squares}} \\ 
& \multicolumn{12}{c}{Dependent Variable is Mean GDP Growth Rate in 2001-2007} \\\cline{2-13}\\[-1.8ex]
Democracy Index (V-Dem, 2000)&        -2.8&        -5.5&        -2.3&        -3.2&        -2.2&        -2.6&        -1.5&        -1.7&        -2.4&        -2.1&        -2.4&        -2.9\\
&       (0.4)&       (3.9)&       (0.5)&       (0.7)&       (0.6)&       (1.3)&       (1.0)&       (0.6)&       (0.5)&       (0.6)&       (0.3)&       (0.4)\\
&        0.00&        0.17&        0.00&        0.00&        0.00&        0.06&        0.12&        0.00&        0.00&        0.00&        0.00&        0.00\\

& \multicolumn{12}{c}{Dependent Variable is Mean GDP Growth Rate in 2008-2009} \\\cline{2-13}\\[-1.8ex]
Democracy Index (V-Dem, 2007)&        -3.9&        -3.9&        -3.3&        -3.7&        -3.2&        -2.9&        -2.1&        -2.0&        -3.6&        -2.9&        -3.2&        -3.8\\
&       (0.6)&       (1.1)&       (0.5)&       (0.5)&       (0.6)&       (0.9)&       (0.9)&       (0.9)&       (0.6)&       (0.7)&       (0.2)&       (0.2)\\
&        0.00&        0.00&        0.00&        0.00&        0.00&        0.00&        0.02&        0.02&        0.00&        0.00&        0.00&        0.00\\

& \multicolumn{12}{c}{Dependent Variable is Mean GDP Growth Rate in 2010-2019} \\\cline{2-13}\\[-1.8ex]
Democracy Index (V-Dem, 2009)&        -1.7&        -1.6&        -1.8&        -2.3&        -1.4&        -1.2&        -1.1&        -0.9&        -1.8&        -1.3&        -1.6&        -1.9\\
&       (0.2)&       (0.2)&       (0.3)&       (0.4)&       (0.3)&       (0.5)&       (0.5)&       (0.4)&       (0.3)&       (0.4)&       (0.1)&       (0.1)\\
&        0.00&        0.00&        0.00&        0.00&        0.00&        0.01&        0.02&        0.04&        0.00&        0.00&        0.00&        0.00\\
IVs & \multicolumn{2}{c}{settler mortality} &  \multicolumn{2}{c}{population density} & \multicolumn{2}{c}{legal origin} & \multicolumn{2}{c}{language} & \multicolumn{2}{c}{crops \& minerals} & \multicolumn{2}{c}{all IVs} \\
Number of IVs & 1 & 1 & 1 & 1 & 1 & 1 & 2 & 2 & 10 & 10 & 15 & 15 \\
Observations        &          81&          81&          86&          86&          90&          90&         132&         132&         138&         138&          71&          71\\
\\[-1.8ex] 
\hline \\[-1.8ex] 

\multicolumn{13}{l}{\textbf{Panel B: Ordinary Least Squares}} \\ 
& \multicolumn{12}{c}{Dependent Variable is Mean GDP Growth Rate in 2001-2007} \\\cline{2-13}\\[-1.8ex]
Democracy Index (V-Dem, 2000)&        -2.3&        -2.2&        -2.2&        -2.4&        -2.2&        -2.4&        -1.9&        -1.5&        -1.9&        -1.5&        -2.3&        -2.4\\
&       (0.4)&       (0.5)&       (0.4)&       (0.5)&       (0.4)&       (0.5)&       (0.5)&       (0.6)&       (0.5)&       (0.6)&       (0.4)&       (0.5)\\

& \multicolumn{12}{c}{Dependent Variable is Mean GDP Growth Rate in 2008-2009} \\\cline{2-13}\\[-1.8ex]
Democracy Index (V-Dem, 2007)&        -2.9&        -2.7&        -2.9&        -3.1&        -2.9&        -3.1&        -2.4&        -1.8&        -2.4&        -1.7&        -3.0&        -3.2\\
&       (0.3)&       (0.5)&       (0.3)&       (0.4)&       (0.3)&       (0.4)&       (0.7)&       (0.9)&       (0.7)&       (0.9)&       (0.3)&       (0.4)\\

& \multicolumn{12}{c}{Dependent Variable is Mean GDP Growth Rate in 2010-2019} \\\cline{2-13}\\[-1.8ex]
Democracy Index (V-Dem, 2009)&        -1.6&        -1.5&        -1.5&        -1.6&        -1.5&        -1.6&        -1.4&        -1.0&        -1.3&        -1.0&        -1.6&        -1.7\\
&       (0.1)&       (0.2)&       (0.1)&       (0.2)&       (0.1)&       (0.2)&       (0.3)&       (0.4)&       (0.3)&       (0.4)&       (0.1)&       (0.1)\\
Baseline Controls & \xmark & \cmark & \xmark & \cmark & \xmark & \cmark & \xmark & \cmark & \xmark & \cmark & \xmark & \cmark\\
N                   &          81&          81&          86&          86&          90&          90&         132&         132&         138&         138&          71&          71\\

\\[-1.8ex] 
\hline 
\hline \\[-1.8ex] 
\end{tabular} 

\begin{tablenotes} 
\item {\footnotesize {\textit{Notes:} Panel A reports the 2SLS estimates of democracy's effect on mean GDP growth rates in 2001-7, 2008-9, and 2010-19, using five different IV strategies. 
 The Democracy Index (V-Dem) is normalized to have mean zero and standard deviation one in each year.\unskip 
For IVs, columns 1 and 2 use log European settler mortality, columns 3 and 4 use log population density in the 1500s, columns 5 and 6 use British legal origin, columns 7 and 8 use the fraction speaking English and the fraction speaking European, columns 9 and 10 use the ability to grow crops and mine minerals, and columns 11 and 12 use all the IVs together. \unskip 
Robust standard errors are in parentheses. \unskip 
The p-values are displayed as 0.00 if they are strictly smaller than the 0.005 threshold. Panel B reports the OLS estimates. 
Columns 1, 3, 5, 7, 9, and 11 have no controls, while columns 2, 4, 6, 8,  10, and 12 have the following baseline controls: absolute latitude, mean temperature, mean precipitation, population density, and median age.\unskip
N refers to the number of countries for which IVs, controls, and outcomes are available. \unskip
Variable definitions and data sources are in Appendix Table \ref{tab:sources}. \unskip}}
\end{tablenotes}
\end{threeparttable}
\end{table} 
\end{center}

\clearpage
\begin{landscape}
\begin{table}\centering 
  \caption{2SLS Regression Excluding the US and China}
  \label{tab:2sls-compare-samples}
  \begin{threeparttable}
  \scriptsize
\begin{tabular}{@{\extracolsep{0pt}}lcccccccccccc} 
\\[-1.8ex]\hline 
\hline  & (1) & (2) & (3) & (4) & (5) & (6) & (7) & (8) & (9) & (10) & (11) & (12)\\ 
\hline \\[-1.8ex] 
\textbf{Panel A} & \multicolumn{12}{c}{Dependent Variable is Mean GDP Growth Rate in 2001-2019}  \\ \cline{2-13} \\[-1.8ex]

Democracy Index (V-Dem, 2000)&        -2.5&        -4.1&        -2.2&        -3.1&        -1.8&        -1.8&        -1.0&        -1.1&        -2.2&        -1.6&        -2.1&        -2.6\\
&       (0.3)&       (2.7)&       (0.4)&       (0.6)&       (0.5)&       (1.3)&       (0.9)&       (0.6)&       (0.5)&       (0.6)&       (0.2)&       (0.2)\\
Include US \& China? & \cmark & \cmark  & \cmark & \cmark & \cmark & \cmark  & \cmark & \cmark & \cmark & \cmark & \cmark & \cmark\\
N                   &          82&          82&          87&          87&          91&          91&         133&         133&         139&         139&          72&          72\\

 \\

Democracy Index (V-Dem, 2000)&        -2.6&         4.0&        -1.9&        -3.9&         3.9&       -15.6&        -0.4&        -0.7&        -1.6&        -0.3&        -1.4&        -0.9\\
&       (0.8)&       (6.7)&       (1.0)&       (3.6)&       (8.6)&      (44.9)&       (1.2)&       (0.6)&       (0.5)&       (0.9)&       (0.4)&       (0.7)\\
Include US \& China? & \xmark & \xmark  & \xmark & \xmark   & \xmark & \xmark  & \xmark & \xmark  & \xmark & \xmark  & \xmark & \xmark \\
N                   &          80&          80&          85&          85&          89&          89&         131&         131&         137&         137&          70&          70\\

 \\

\hline \\[-1.8ex] 
 
\textbf{Panel B} & \multicolumn{12}{c}{Dependent Variable is Mean GDP Growth Rate in 2020-2022} \\ \cline{2-13} \\[-1.8ex] 

Democracy Index (V-Dem, 2019)&        -0.9&        -0.8&        -0.9&        -1.0&        -0.8&        -0.7&        -0.7&        -0.6&        -1.0&        -0.8&        -0.9&        -1.0\\
&       (0.1)&       (0.2)&      (0.08)&      (0.09)&       (0.1)&       (0.1)&       (0.3)&       (0.2)&       (0.2)&       (0.2)&      (0.06)&      (0.07)\\
Include US \& China? & \cmark & \cmark  & \cmark & \cmark & \cmark & \cmark  & \cmark & \cmark & \cmark & \cmark & \cmark & \cmark\\
N                   &          82&          82&          87&          87&          91&          91&         133&         133&         139&         139&          72&          72\\

 \\

Democracy Index (V-Dem, 2019)&        -0.9&        -0.6&        -1.0&        -1.7&         9.3&        17.6&        -0.3&        -0.3&        -0.6&        -0.3&        -1.0&        -1.5\\
&       (0.3)&       (1.4)&       (0.3)&       (1.0)&      (52.7)&      (74.9)&       (0.5)&       (0.4)&       (0.4)&       (0.6)&       (0.3)&       (0.7)\\
Include US \& China? & \xmark & \xmark  & \xmark & \xmark   & \xmark & \xmark  & \xmark & \xmark  & \xmark & \xmark  & \xmark & \xmark \\
N                   &          80&          80&          85&          85&          89&          89&         131&         131&         137&         137&          70&          70\\

 \\

 \hline \\[-1.8ex] 

 \textbf{Panel C} & \multicolumn{12}{c}{Dependent Variable is Mean Nighttime Light Intensity Growth Rate in 2001-2013} \\ \cline{2-13} \\[-1.8ex] 

Democracy Index (V-Dem, 2000)&        -2.8&        -5.2&        -1.8&        -1.4&        -2.5&        -3.2&        -1.6&        -3.0&        -2.3&        -2.7&        -2.0&        -2.0\\
&       (0.6)&       (3.7)&       (0.4)&       (0.6)&       (0.6)&       (1.0)&       (1.1)&       (1.0)&       (0.5)&       (0.5)&       (0.2)&       (0.4)\\
Include US \& China? & \cmark & \cmark  & \cmark & \cmark & \cmark & \cmark  & \cmark & \cmark & \cmark & \cmark & \cmark & \cmark\\
N                   &          81&          81&          86&          86&          90&          90&         130&         130&         136&         136&          71&          71\\

 \\[-1.8ex] 

Democracy Index (V-Dem, 2000)&        -3.5&        20.1&       -0.07&         0.4&        -1.5&         0.6&       -0.09&        -2.0&        -1.8&        -0.4&        -0.5&        -0.5\\
&       (1.3)&     (136.3)&       (0.5)&       (0.7)&       (3.6)&       (2.6)&       (2.4)&       (1.3)&       (0.8)&       (1.3)&       (0.5)&       (0.6)\\
Include US \& China? & \xmark & \xmark  & \xmark & \xmark   & \xmark & \xmark  & \xmark & \xmark  & \xmark & \xmark  & \xmark & \xmark \\
N                   &          79&          79&          84&          84&          88&          88&         128&         128&         134&         134&          69&          69\\
\hline \\[-1.8ex] IVs & \multicolumn{2}{c}{settler mortality} &  \multicolumn{2}{c}{population density} & \multicolumn{2}{c}{legal origin} & \multicolumn{2}{c}{language} & \multicolumn{2}{c}{crops \& minerals}  & \multicolumn{2}{c}{all IVs}  \\ Controls & \xmark & \cmark & \xmark & \cmark & \xmark & \cmark & \xmark & \cmark & \xmark & \cmark & \xmark & \cmark\\
 \\[-1.8ex] 
 

\hline \hline \\[-1.8ex] 
\end{tabular} 
\begin{tablenotes}
\item {\footnotesize {\textit{Notes:} This table compares the results of 2SLS regressions under two sample definitions (include the US and China vs. exclude the US and China). 
 The Democracy Index (V-Dem) is normalized to have mean zero and standard deviation one in each year.\unskip 
Columns 1, 3, 5, 7, 9, and 11 have no controls, while columns 2, 4, 6, 8,  10, and 12 have the following baseline controls: absolute latitude, mean temperature, mean precipitation, population density, and median age. For the mean GDP growth rate in 2020-2022, we also control for diabetes prevalence. \unskip
For IVs, columns 1 and 2 use log European settler mortality, columns 3 and 4 use log population density in the 1500s, columns 5 and 6 use British legal origin, columns 7 and 8 use the fraction speaking English and the fraction speaking European, columns 9 and 10 use the ability to grow crops and mine minerals, and columns 11 and 12 use all the IVs together. \unskip
N refers to the number of countries for which IVs, controls, and outcomes are available. \unskip
Robust standard errors are in parentheses. \unskip
Variable definitions and data sources are in Appendix Table \ref{tab:sources}. \unskip}}
\end{tablenotes}
\end{threeparttable}
\end{table} 
\end{landscape}

\clearpage
\begin{landscape}
    
\begin{table}\centering
\caption{2SLS Regression Excluding Outliers}\label{tab:2sls-outliers}
\begin{threeparttable}
\scriptsize
\begin{tabular}{l*{12}{c}}
\hline\hline
&\multicolumn{1}{c}{(1)}         &\multicolumn{1}{c}{(2)}         &\multicolumn{1}{c}{(3)}         &\multicolumn{1}{c}{(4)}         &\multicolumn{1}{c}{(5)}         &\multicolumn{1}{c}{(6)}         &\multicolumn{1}{c}{(7)}         &\multicolumn{1}{c}{(8)}         &\multicolumn{1}{c}{(9)}         &\multicolumn{1}{c}{(10)}      &\multicolumn{1}{c}{(11)}         &\multicolumn{1}{c}{(12)}       \\
\hline \\[-1.8ex]

& \multicolumn{12}{c}{Dependent Variable is Mean GDP Growth Rate in 2001-2019} \\\cline{2-13}\\[-1.8ex]
Democracy Index (V-Dem, 2000)&        -2.3&        -2.3&        -1.9&        -1.8&        -1.9&        -1.9&         0.3&        -0.8&        -1.9&        -2.0&        -2.0&        -2.0\\
&       (0.3)&       (0.3)&       (0.4)&       (0.4)&       (0.4)&       (0.4)&       (1.2)&       (1.0)&       (0.4)&       (0.4)&       (0.3)&       (0.3)\\
\hline
N                   &          80&          81&          81&          84&          85&          84&         123&         122&         128&         129&          69&          71\\

& \multicolumn{12}{c}{Dependent Variable is Mean GDP Growth Rate in 2020-2022} \\\cline{2-13}\\[-1.8ex]
Democracy Index (V-Dem, 2019)&        -0.9&        -0.9&        -0.9&        -0.9&        -0.8&        -0.8&        -0.7&        -0.7&        -1.0&        -1.0&        -0.9&        -0.9\\
&       (0.1)&       (0.1)&      (0.07)&      (0.08)&       (0.1)&       (0.1)&       (0.3)&       (0.3)&       (0.1)&       (0.1)&      (0.06)&      (0.06)\\
\hline
N                   &          81&          81&          84&          85&          88&          89&         128&         128&         133&         134&          71&          71\\

& \multicolumn{12}{c}{Dependent Variable is Mean Nighttime Light Intensity Growth Rate in 2001-2013} \\\cline{2-13}\\[-1.8ex]
Democracy Index (V-Dem, 2000)&        -2.8&        -2.8&        -1.7&        -1.7&        -2.4&        -2.4&        -1.5&        -1.5&        -2.2&        -2.2&        -1.9&        -1.9\\
&       (0.6)&       (0.6)&       (0.4)&       (0.4)&       (0.6)&       (0.6)&       (1.1)&       (1.1)&       (0.5)&       (0.5)&       (0.2)&       (0.2)\\
\hline
N                   &          80&          80&          82&          82&          86&          86&         125&         125&         130&         130&          70&          70\\
\hline \\[-1.8ex] IVs & \multicolumn{2}{c}{settler mortality} &  \multicolumn{2}{c}{population density} & \multicolumn{2}{c}{legal origin} & \multicolumn{2}{c}{language} & \multicolumn{2}{c}{crops \& minerals} & \multicolumn{2}{c}{all IVs}  \\ Baseline Controls & \xmark & \cmark & \xmark & \cmark & \xmark & \cmark & \xmark & \cmark & \xmark & \cmark & \xmark & \cmark\\
 \\[-1.8ex] 
\hline\hline \\[-1.8ex]
\end{tabular}
\begin{tablenotes}
\item {\footnotesize {\textit{Notes:} This table shows the 2SLS regression estimates on the effect of democracy on the mean GDP growth rate in 2001-2019, the mean GDP growth rate in 2020-2022, and the mean nighttime light intensity growth rate in 2001-2013 excluding countries with a standardized residual above 1.96 or below -1.96. For each 2SLS regression, we run the baseline specification, calculate the fitted values, use the fitted values to calculate the residual in the second stage regression, standardize the residuals to have mean zero and variance one, and finally rerun the 2SLS regression with the sample limited to countries that have a standardized residual between -1.96 and 1.96. 
 The Democracy Index (V-Dem) is normalized to have mean zero and standard deviation one in each year.\unskip 
Columns 1, 3, 5, 7, 9, and 11 have no controls, while columns 2, 4, 6, 8,  10, and 12 have the following baseline controls: absolute latitude, mean temperature, mean precipitation, population density, and median age. For the mean GDP growth rate in 2020-2022, we also control for diabetes prevalence. \unskip
For IVs, columns 1 and 2 use log European settler mortality, columns 3 and 4 use log population density in the 1500s, columns 5 and 6 use British legal origin, columns 7 and 8 use the fraction speaking English and the fraction speaking European, columns 9 and 10 use the ability to grow crops and mine minerals, and columns 11 and 12 use all the IVs together. \unskip
N refers to the number of countries for which IVs, controls, and outcomes are available. \unskip
Robust standard errors are in parentheses. \unskip
Variable definitions and data sources are in Appendix Table \ref{tab:sources}. \unskip}}
\end{tablenotes}
\end{threeparttable}
\end{table}
\end{landscape}

\clearpage

\begin{landscape}
\begin{table}\centering
\caption{2SLS Regression Excluding G7 Countries}\label{tab:2sls-remove-g7}
\footnotesize
  \begin{threeparttable}
\begin{tabular}{l*{12}{c}}
\hline\hline
                    &\multicolumn{1}{c}{(1)}         &\multicolumn{1}{c}{(2)}         &\multicolumn{1}{c}{(3)}         &\multicolumn{1}{c}{(4)}         &\multicolumn{1}{c}{(5)}         &\multicolumn{1}{c}{(6)}         &\multicolumn{1}{c}{(7)}         &\multicolumn{1}{c}{(8)}         &\multicolumn{1}{c}{(9)}         &\multicolumn{1}{c}{(10)}   
                    &\multicolumn{1}{c}{(11)}         &\multicolumn{1}{c}{(12)}  \\ \hline

& \multicolumn{12}{c}{Dependent Variable is Mean GDP Growth Rate in 2001-2019} \\\cline{2-13}\\[-1.8ex]
Democracy Index (V-Dem, 2000)&        -2.1&        -3.9&        -2.2&        -3.1&        -0.4&       -20.0&        -2.0&        -2.7&        -1.7&        -1.1&        -2.1&        -2.1\\
&       (0.5)&       (9.0)&       (0.5)&       (1.3)&       (2.3)&      (73.0)&       (0.6)&       (0.8)&       (0.6)&       (0.5)&       (0.3)&       (0.4)\\

& \multicolumn{12}{c}{Dependent Variable is Mean GDP Growth Rate in 2020-2022} \\\cline{2-13}\\[-1.8ex]
Democracy Index (V-Dem, 2019)&        -0.8&         0.2&        -1.1&        -1.2&        -0.7&         6.0&        -0.7&        -0.8&        -0.9&        -0.7&        -1.0&        -1.1\\
&       (0.3)&       (0.7)&       (0.2)&       (0.3)&       (0.4)&      (18.1)&       (0.3)&       (0.4)&       (0.3)&       (0.3)&       (0.1)&       (0.3)\\

& \multicolumn{12}{c}{Dependent Variable is Mean Nighttime Light Intensity Growth Rate in 2001-2013} \\\cline{2-13}\\[-1.8ex]
Democracy Index (V-Dem, 2000)&        -2.0&        -6.9&        -1.1&        -0.6&        -2.6&       -10.0&        -1.9&        -2.9&        -0.6&        -1.3&        -1.6&        -1.1\\
&       (0.5)&      (12.0)&       (0.5)&       (0.5)&       (1.5)&      (41.3)&       (0.7)&       (0.8)&       (0.7)&       (0.5)&       (0.4)&       (0.5)\\
\hline \\[-1.8ex] IVs & \multicolumn{2}{c}{settler mortality} &  \multicolumn{2}{c}{population density} & \multicolumn{2}{c}{legal origin} & \multicolumn{2}{c}{language} & \multicolumn{2}{c}{crops \& minerals} & \multicolumn{2}{c}{all IVs}  \\ Baseline Controls & \xmark & \cmark & \xmark & \cmark & \xmark & \cmark & \xmark & \cmark & \xmark & \cmark & \xmark & \cmark\\
N                   &          77&          77&          83&          83&          87&          87&         121&         121&         127&         127&          69&          69\\

 \\[-1.8ex] 
\hline\hline \\[-1.8ex]
\end{tabular}
\begin{tablenotes}
\item {\footnotesize {\textit{Notes:} This table shows the 2SLS regression estimates of democracy's effect on the mean GDP growth rate in 2001-2019, the mean GDP growth rate in 2020-2022, and the mean nighttime light intensity growth rate in 2001-2013 that exclude G7 countries (Canada, France, Germany, Italy, Japan, the United Kingdom, and the United States) from the sample. 
 The Democracy Index (V-Dem) is normalized to have mean zero and standard deviation one in each year.\unskip
Columns 1, 3, 5, 7, 9, and 11 have no controls, while columns 2, 4, 6, 8,  10, and 12 have the following baseline controls: absolute latitude, mean temperature, mean precipitation, population density, and median age. For the mean GDP growth rate in 2020-2022, we also control for diabetes prevalence. \unskip
For IVs, columns 1 and 2 use log European settler mortality, columns 3 and 4 use log population density in the 1500s, columns 5 and 6 use British legal origin, columns 7 and 8 use the fraction speaking English and the fraction speaking European, columns 9 and 10 use the ability to grow crops and mine minerals, and columns 11 and 12 use all the IVs together. \unskip
N refers to the number of countries for which IVs, controls, and outcomes are available. \unskip
Robust standard errors are in parentheses. \unskip
Variable definitions and data sources are in Appendix Table \ref{tab:sources}. \unskip}}
\end{tablenotes}
\end{threeparttable}
\end{table}

\end{landscape}

\clearpage
\begin{landscape}
\begin{table}\centering 
  \caption{2SLS Regression with Alternative Weightings}
  \label{tab:2sls-compare-weighting} 
  \small
  \begin{threeparttable}
\begin{tabular}{@{\extracolsep{0pt}}lcccccccccccc} 
\\[-1.8ex]\hline 
\hline 
& (1) & (2) & (3) & (4) & (5) & (6) & (7) & (8) & (9) & (10) & (11) & (12)\\ 
\hline \\[-1.8ex] 

\textbf{Panel A} & \multicolumn{12}{c}{Dependent Variable is Mean GDP Growth Rate in 2001-2019} \\ \cline{2-13} \\[-1.8ex]

Democracy Index (Weighting: GDP)&        -2.5&        -4.1&        -2.2&        -3.1&        -1.8&        -1.8&        -1.0&        -1.1&        -2.2&        -1.6&        -2.1&        -2.6\\
&       (0.3)&       (2.7)&       (0.4)&       (0.6)&       (0.5)&       (1.3)&       (0.9)&       (0.6)&       (0.5)&       (0.6)&       (0.2)&       (0.2)\\

Democracy Index (Weighting: Population)&        -1.7&        -2.0&        -3.4&        -4.0&        -1.1&        -1.2&        -2.7&        -2.5&        -2.9&        -1.5&        -1.6&        -1.9\\
&       (0.6)&       (0.8)&       (1.7)&       (1.3)&       (0.9)&       (0.8)&       (0.6)&       (0.6)&       (0.8)&       (0.6)&       (0.4)&       (0.4)\\

Democracy Index (Weighting: None)&        -1.0&        -0.2&        -1.2&        -1.3&        -2.2&        -3.8&        -1.1&        -1.1&        -1.2&        -0.4&        -1.3&        -1.5\\
&       (0.4)&       (4.0)&       (0.3)&       (0.5)&       (2.1)&       (5.8)&       (0.2)&       (0.4)&       (0.3)&       (0.6)&       (0.3)&       (0.5)\\

 \\[-1.8ex] 
\hline \\[-1.8ex] 

\textbf{Panel B} & \multicolumn{12}{c}{Dependent Variable is Mean GDP Growth Rate in 2020-2022} \\ \cline{2-13} \\[-1.8ex]

Democracy Index (Weighting: GDP)&        -0.9&        -0.8&        -0.9&        -1.0&        -0.8&        -0.7&        -0.7&        -0.6&        -1.0&        -0.8&        -0.9&        -1.0\\
&       (0.1)&       (0.2)&      (0.08)&      (0.09)&       (0.1)&       (0.1)&       (0.3)&       (0.2)&       (0.2)&       (0.2)&      (0.06)&      (0.07)\\

Democracy Index (Weighting: Population)&        -0.8&        -0.5&        -1.3&        -1.2&        -0.4&        -0.4&        -0.9&        -0.6&        -1.6&        -1.1&        -1.1&        -0.9\\
&       (0.4)&       (0.4)&       (0.3)&       (0.3)&       (0.7)&       (0.5)&       (0.2)&       (0.2)&       (0.4)&       (0.3)&      (0.10)&       (0.2)\\

Democracy Index (Weighting: None)&        -0.2&        18.3&         0.7&         1.8&         3.1&         3.5&         1.1&         2.3&        -0.6&         0.4&         0.8&         2.2\\
&       (1.5)&      (56.0)&       (1.5)&       (2.8)&       (4.8)&       (4.4)&       (1.3)&       (2.0)&       (0.5)&       (1.3)&       (1.5)&       (2.5)\\

 \\[-1.8ex] 
\hline \\[-1.8ex]

\textbf{Panel C} & \multicolumn{12}{c}{Dependent Variable is Mean Nighttime Light Intensity Growth Rate in 2001-2013} \\ \cline{2-13} \\[-1.8ex]

Democracy Index (Weighting: GDP)&        -2.8&        -5.2&        -1.8&        -1.3&        -2.5&        -3.2&        -1.6&        -3.0&        -2.3&        -2.7&        -2.0&        -2.0\\
&       (0.6)&       (3.7)&       (0.4)&       (0.6)&       (0.6)&       (1.0)&       (1.1)&       (1.0)&       (0.5)&       (0.5)&       (0.2)&       (0.4)\\

Democracy Index (Weighting: Population)&        -1.7&        -1.5&        -1.5&        -3.0&        -2.9&        -3.4&        -2.2&        -3.2&        -1.9&        -1.6&        -2.1&        -3.1\\
&       (0.8)&       (1.6)&       (0.8)&       (1.2)&       (1.2)&       (1.0)&       (0.5)&       (0.9)&       (0.6)&       (0.9)&       (0.5)&       (1.4)\\

Democracy Index (Weighting: None)&        -2.9&        -1.5&        -2.5&        -2.4&       -13.2&       -21.3&        -3.1&        -4.1&        -2.9&        -3.2&        -2.9&        -3.8\\
&       (1.1)&       (9.1)&       (1.1)&       (1.9)&      (10.0)&      (28.5)&       (0.6)&       (1.2)&       (0.9)&       (2.1)&       (0.8)&       (2.3)\\
\hline \\[-1.8ex] IVs & \multicolumn{2}{c}{settler mortality} &  \multicolumn{2}{c}{population density} & \multicolumn{2}{c}{legal origin} & \multicolumn{2}{c}{language} & \multicolumn{2}{c}{crops \& minerals}  & \multicolumn{2}{c}{all IVs} \\ Baseline Controls & \xmark & \cmark & \xmark & \cmark & \xmark & \cmark & \xmark & \cmark & \xmark & \cmark & \xmark & \cmark\\
N                   &          81&          81&          85&          85&          89&          89&         128&         128&         134&         134&          71&          71\\

 \\[-1.8ex] 

\hline \hline \\[-1.8ex] 
\end{tabular} 
\begin{tablenotes}
\item {\footnotesize {\textit{Notes:} This table compares the results of 2SLS regressions on the mean GDP growth rate in 2001-2019 (Panel A), the mean GDP growth rate in 2020-2022 (Panel B), and the mean nighttime light intensity growth rate in 2001-2013 (Panel C) with weighting of observations by baseline GDP, weighting by baseline population, and no weighting. 
 The Democracy Index (V-Dem) is normalized to have mean zero and standard deviation one in each year.\unskip 
Columns 1, 3, 5, 7, 9, and 11 have no controls, while columns 2, 4, 6, 8,  10, and 12 have the following baseline controls: absolute latitude, mean temperature, mean precipitation, population density, and median age. For the mean GDP growth rate in 2020-2022, we also control for diabetes prevalence. \unskip 
For IVs, columns 1 and 2 use log European settler mortality, columns 3 and 4 use log population density in the 1500s, columns 5 and 6 use British legal origin, columns 7 and 8 use the fraction speaking English and the fraction speaking European, columns 9 and 10 use the ability to grow crops and mine minerals, and columns 11 and 12 use all the IVs together. \unskip
N refers to the number of countries for which IVs, controls, and outcomes are available. \unskip
Robust standard errors are in parentheses. \unskip
Variable definitions and data sources are in Appendix Table \ref{tab:sources}. \unskip}}
\end{tablenotes}
\end{threeparttable}
\end{table} 
\end{landscape}

\clearpage
\begin{table}\centering
\def\sym#1{\ifmmode^{#1}\else\(^{#1}\)\fi}
\caption{Democracy's Effect on Economic Growth With Control for Baseline GDP}\label{tab:2sls-control-gdp} 
\scriptsize
  \begin{threeparttable}
\begin{tabular}{l*{12}{c}}
\hline\hline
                    &\multicolumn{1}{c}{(1)}         &\multicolumn{1}{c}{(2)}         &\multicolumn{1}{c}{(3)}         &\multicolumn{1}{c}{(4)}         &\multicolumn{1}{c}{(5)}         &\multicolumn{1}{c}{(6)}         &\multicolumn{1}{c}{(7)}         &\multicolumn{1}{c}{(8)}         &\multicolumn{1}{c}{(9)}         &\multicolumn{1}{c}{(10)} 
                    
                    &\multicolumn{1}{c}{(11)}         &\multicolumn{1}{c}{(12)}\\ \hline \\[-1.8ex]
 
 \multicolumn{12}{l}{\textbf{Panel A: No Control for Baseline GDP}} \\
 & \multicolumn{12}{c}{Dependent Variable is Mean GDP Growth Rate in 2001-2019} \\\cline{2-13}\\[-1.8ex]
Democracy Index (V-Dem, 2000)&        -2.5&        -4.2&        -2.2&        -3.1&        -1.8&        -1.8&        -1.0&        -1.1&        -2.2&        -1.6&        -2.1&        -2.6\\
&       (0.3)&       (2.7)&       (0.4)&       (0.6)&       (0.5)&       (1.3)&       (1.0)&       (0.6)&       (0.5)&       (0.6)&       (0.2)&       (0.2)\\

& \multicolumn{12}{c}{Dependent Variable is Mean GDP Growth Rate in 2020-2022} \\\cline{2-13}\\[-1.8ex]
Democracy Index (V-Dem, 2019)&        -0.9&        -0.8&        -0.9&        -1.0&        -0.8&        -0.7&        -0.7&        -0.6&        -1.0&        -0.8&        -0.9&        -1.0\\
&       (0.1)&       (0.2)&      (0.08)&      (0.09)&       (0.2)&       (0.1)&       (0.3)&       (0.2)&       (0.2)&       (0.2)&      (0.06)&      (0.07)\\

 \\[-1.8ex] 
 \hline \\[-1.8ex]

\multicolumn{12}{l}{\textbf{Panel B: Control for Baseline GDP Per Capita} } \\
& \multicolumn{12}{c}{Dependent Variable is Mean GDP Growth Rate in 2001-2019} \\\cline{2-13}\\[-1.8ex]
Democracy Index (V-Dem, 2000)&        -3.1&         1.9&        -2.1&        -3.0&         1.4&        19.1&        -1.6&        -0.8&        -2.1&        -1.1&        -2.1&        -2.1\\
&       (0.8)&       (6.5)&       (0.6)&       (0.9)&       (5.1)&      (86.0)&       (0.6)&       (0.5)&       (0.8)&       (0.6)&       (0.3)&       (0.3)\\

& \multicolumn{12}{c}{Dependent Variable is Mean GDP Growth Rate in 2020-2022} \\\cline{2-13}\\[-1.8ex]
Democracy Index (V-Dem, 2019)&        -1.3&        -1.3&        -1.2&        -1.5&        -0.8&         8.9&        -1.2&        -1.0&        -1.5&        -1.6&        -1.1&        -1.3\\
&       (0.2)&       (5.7)&       (0.1)&       (0.2)&       (0.4)&      (35.8)&       (0.3)&       (0.4)&       (0.2)&       (0.4)&      (0.10)&       (0.2)\\

 \\[-1.8ex] 
\hline \\[-1.8ex]

\multicolumn{12}{l}{\textbf{Panel C: Control for Baseline Total GDP} } \\ 
& \multicolumn{12}{c}{Dependent Variable is Mean GDP Growth Rate in 2001-2019} \\\cline{2-13}\\[-1.8ex]
Democracy Index (V-Dem, 2000)&        -2.6&        -7.1&        -2.2&        -3.1&        -1.3&        -0.3&        -1.2&        -0.9&        -2.4&        -1.6&        -2.1&        -2.5\\
&       (0.3)&      (12.7)&       (0.5)&       (0.6)&       (1.2)&       (2.6)&       (1.1)&       (0.6)&       (0.5)&       (0.8)&       (0.2)&       (0.3)\\

& \multicolumn{12}{c}{Dependent Variable is Mean GDP Growth Rate in 2020-2022} \\\cline{2-13}\\[-1.8ex]
Democracy Index (V-Dem, 2019)&        -1.0&        -0.9&        -1.0&        -1.1&        -0.9&        -0.8&        -0.8&        -0.9&        -1.0&        -1.0&        -1.0&        -1.0\\
&      (0.07)&       (0.2)&      (0.06)&      (0.09)&       (0.1)&       (0.2)&       (0.2)&       (0.2)&       (0.1)&       (0.1)&      (0.04)&      (0.08)\\
\hline \\[-1.8ex] IVs & \multicolumn{2}{c}{settler mortality} &  \multicolumn{2}{c}{population density} & \multicolumn{2}{c}{legal origin} & \multicolumn{2}{c}{language} & \multicolumn{2}{c}{crops \& minerals}  & \multicolumn{2}{c}{all IVs}\\ Baseline Controls Other Than Baseline GDP & \xmark & \cmark & \xmark & \cmark & \xmark & \cmark & \xmark & \cmark & \xmark & \cmark & \xmark & \cmark\\
N                   &          81&          81&          86&          86&          90&          90&         131&         131&         137&         137&          71&          71\\

 \\[-1.8ex] 
 
\hline\hline \\[-1.8ex]
\end{tabular}
\begin{tablenotes}
\item {\footnotesize {\textit{Notes:} This table compares the 2SLS regression estimates of democracy's effect on the mean GDP growth rate in 2001-2019 and the mean GDP growth rate in 2020-2022 without controls for baseline GDP (Panel A), with additional controls for baseline GDP per capita (Panel B), and with additional controls for baseline total GDP (Panel C). 
 The Democracy Index (V-Dem) is normalized to have mean zero and standard deviation one in each year.\unskip Columns 2, 4, 6, 8, 10, and 12 also have the following baseline controls: absolute latitude, mean temperature, mean precipitation, population density, and median age. For the mean GDP growth rate in 2020-2022, we also control for diabetes prevalence.
For IVs, columns 1 and 2 use log European settler mortality, columns 3 and 4 use log population density in the 1500s, columns 5 and 6 use British legal origin, columns 7 and 8 use the fraction speaking English and the fraction speaking European, columns 9 and 10 use the ability to grow crops and mine minerals, and columns 11 and 12 use all the IVs together. \unskip The sample sizes are slightly different from those in Table \ref{tab:2sls_tab1} because this table uses only observations for which all GDP per capita and total GDP data are available. 
N refers to the number of countries for which IVs, controls, and outcomes are available. \unskip
Robust standard errors are in parentheses. \unskip 
Variable definitions and data sources are in Appendix Table \ref{tab:sources}. \unskip}}
\end{tablenotes}
\end{threeparttable}
\end{table}



\clearpage

\begin{table}\centering
\caption{2SLS Regression with Continent Controls}\label{tab:2sls-control-continent} 
\scriptsize
  \begin{threeparttable}
\begin{tabular}{l*{12}{c}}
\hline\hline
                    &\multicolumn{1}{c}{(1)}         &\multicolumn{1}{c}{(2)}         &\multicolumn{1}{c}{(3)}         &\multicolumn{1}{c}{(4)}         &\multicolumn{1}{c}{(5)}         &\multicolumn{1}{c}{(6)}         &\multicolumn{1}{c}{(7)}         &\multicolumn{1}{c}{(8)}         &\multicolumn{1}{c}{(9)}         &\multicolumn{1}{c}{(10)}  
                    &\multicolumn{1}{c}{(11)}         &\multicolumn{1}{c}{(12)}  \\ \hline 
                    
& \multicolumn{12}{c}{Dependent Variable is Mean GDP Growth Rate in 2001-2019} \\\cline{2-13}\\[-1.8ex]
Democracy Index (V-Dem, 2000)&        -0.9&        -0.7&        -0.8&        -1.6&        -0.8&        -0.8&         1.5&        -6.4&        -2.8&        -1.8&        -1.1&        -0.8\\
&       (0.9)&       (3.1)&       (0.8)&       (0.5)&       (0.5)&       (0.5)&       (1.5)&       (6.4)&       (0.5)&       (0.6)&       (0.4)&       (0.3)\\

& \multicolumn{12}{c}{Dependent Variable is Mean GDP Growth Rate in 2020-2022} \\\cline{2-13}\\[-1.8ex]
Democracy Index (V-Dem, 2019)&        -0.4&        -0.3&        -0.9&        -1.0&        -0.4&       -0.02&         0.7&        -6.6&        -1.4&        -0.5&        -0.9&        -0.7\\
&       (0.7)&       (0.6)&       (0.9)&       (0.6)&       (0.4)&       (0.4)&       (1.8)&       (3.9)&       (0.2)&       (0.4)&       (0.2)&       (0.3)\\

& \multicolumn{12}{c}{Dependent Variable is Mean Nighttime Light Intensity Growth Rate in 2001-2013} \\\cline{2-13}\\[-1.8ex]
Democracy Index (V-Dem, 2000)&        -1.5&        -3.9&         0.6&        -0.1&        -1.6&        -2.0&        -8.7&         5.1&        -2.8&        -1.7&        -1.7&        -0.9\\
&       (0.9)&       (3.3)&       (1.2)&       (0.7)&       (0.6)&       (1.0)&       (9.8)&       (6.2)&       (0.5)&       (1.3)&       (0.4)&       (0.4)\\
\hline \\[-1.8ex] IVs & \multicolumn{2}{c}{settler mortality} &  \multicolumn{2}{c}{population density} & \multicolumn{2}{c}{legal origin} & \multicolumn{2}{c}{language} & \multicolumn{2}{c}{crops \& minerals} & \multicolumn{2}{c}{all IVs}  \\ Baseline Controls & \xmark & \cmark & \xmark & \cmark & \xmark & \cmark & \xmark & \cmark & \xmark & \cmark & \xmark & \cmark\\
N                   &          81&          81&          85&          85&          89&          89&         128&         128&         134&         134&          71&          71\\

 \\[-1.8ex]

\hline\hline \\[-1.8ex]
\end{tabular}
\begin{tablenotes}
\item {\footnotesize {\textit{Notes:} This table shows the 2SLS regression estimates of democracy's effect on the mean GDP growth rate in 2001-2019, the mean GDP growth rate in 2020-2022, and the mean nighttime light intensity growth rate in 2001-2013 that adds dummy variables for each continent (Africa, Asia, Europe, North America, Oceania, and South America) as controls. 
 The Democracy Index (V-Dem) is normalized to have mean zero and standard deviation one in each year.\unskip
Columns 1, 3, 5, 7, 9, and 11 only control for continents, while columns 2, 4, 6, 8, 10, and 12 also have the following baseline controls: absolute latitude, mean temperature, mean precipitation, population density, and median age. For the mean GDP growth rate in 2020-2022, we also control for diabetes prevalence.\unskip
For IVs, columns 1 and 2 use log European settler mortality, columns 3 and 4 use log population density in the 1500s, columns 5 and 6 use British legal origin, columns 7 and 8 use the fraction speaking English and the fraction speaking European, columns 9 and 10 use the ability to grow crops and mine minerals, and columns 11 and 12 use all the IVs together. \unskip
N refers to the number of countries for which IVs, controls, and outcomes are available. \unskip
Robust standard errors are in parentheses. \unskip
Variable definitions and data sources are in Appendix Table \ref{tab:sources}. \unskip}}
\end{tablenotes}
\end{threeparttable}
\end{table}

\clearpage
\newgeometry{left=0.3cm, right = 0.3cm, top = 1cm, bottom=1in}
\begin{figure}
\centering
\caption{Correlation Between Democracy and Economic Growth by Decade}\label{fig:ols-decade}
\captionsetup{width=0.99\textwidth}
\centering

\begin{subfigure}[c]{.99\textwidth}
  \subcaptionbox{1981-1990\label{fig:1980s}}{\includegraphics[width=.49\textwidth]{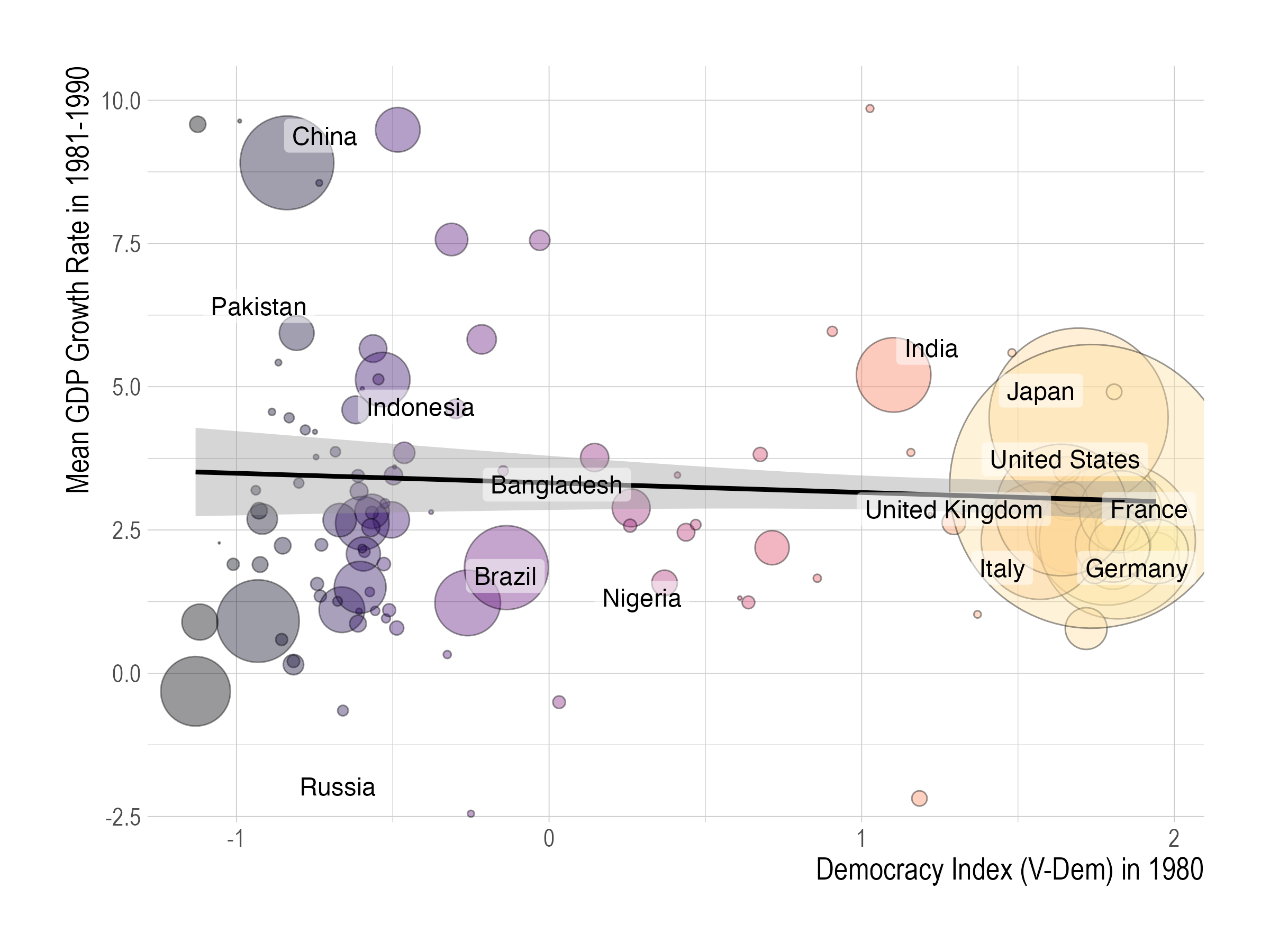}}
  \subcaptionbox{1991-2000\label{fig:1990s}}{\includegraphics[width=.49\textwidth]{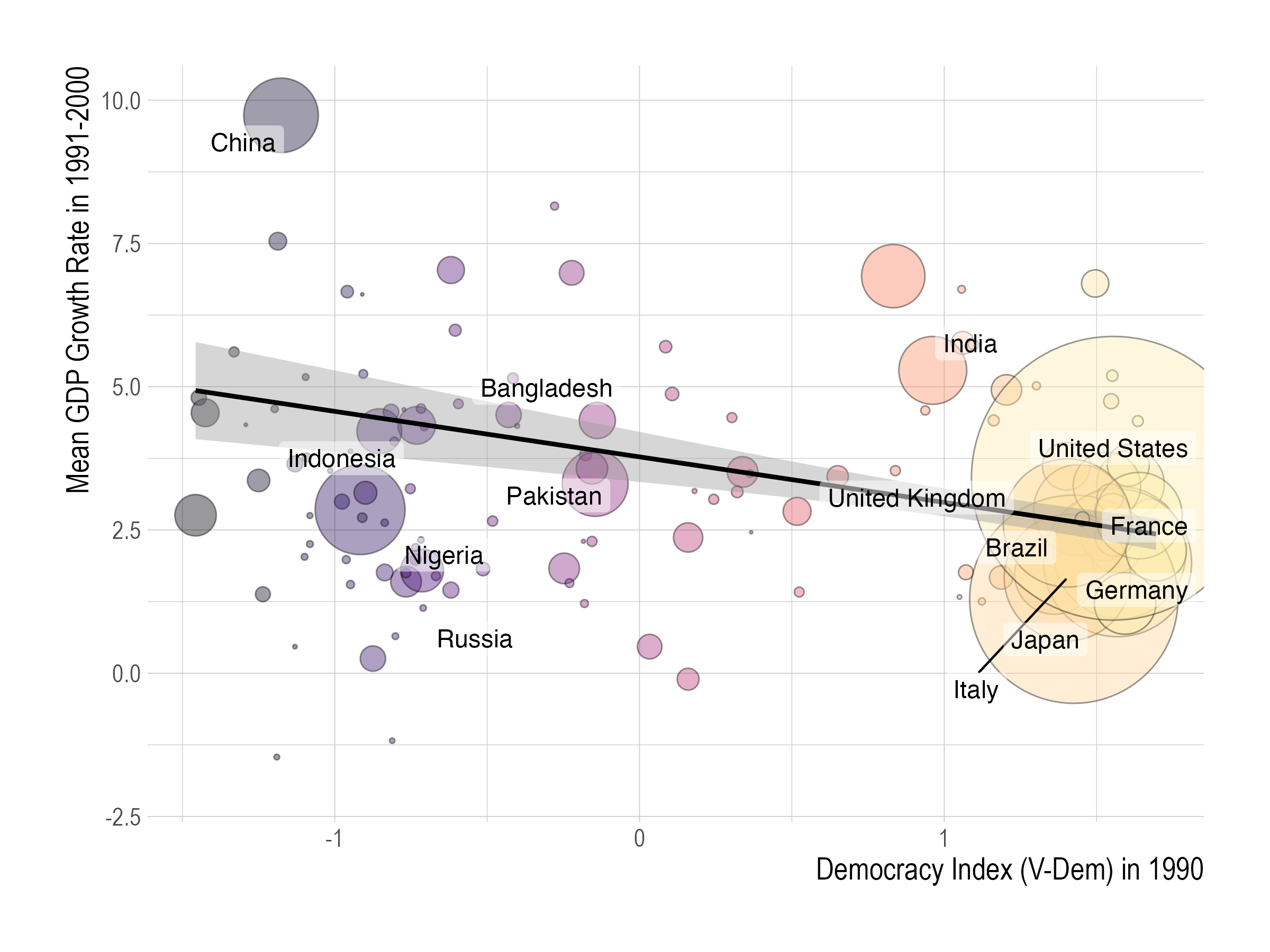}}
\end{subfigure}
\medskip
  
\begin{subfigure}[c]{.99\textwidth}
  \subcaptionbox{2001-2010 \label{fig:2000s}}{\includegraphics[width=.49\textwidth]{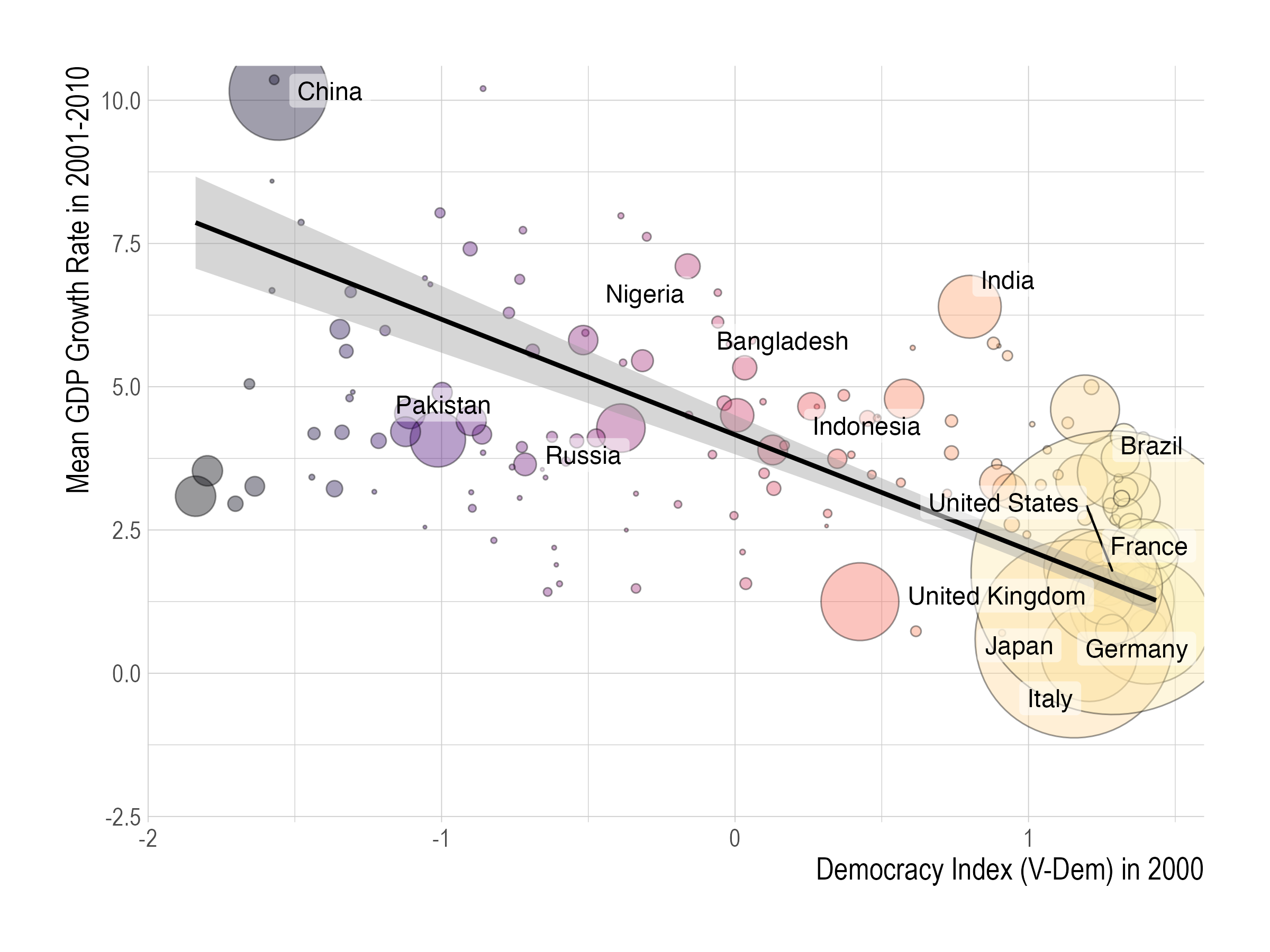}}
  \subcaptionbox{2011-2019 \label{fig:2010s}}{\includegraphics[width=.49\textwidth]{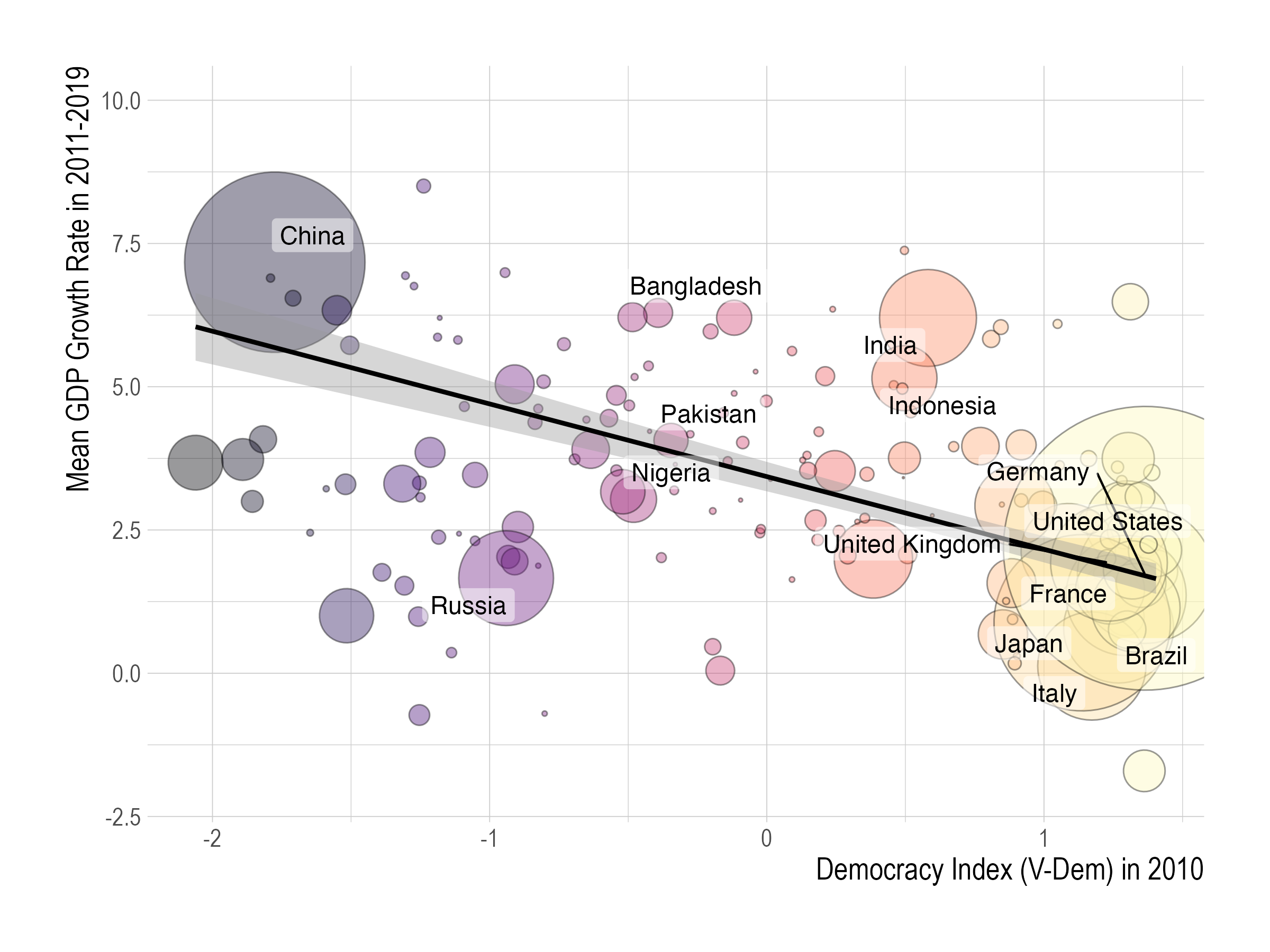}}
\end{subfigure}

\caption*{\footnotesize{\textit{Notes:} Panels (a)-(d) show the relationship between democracy and the mean GDP growth rates in four periods: 1981-1990, 1991-2000, 2001-2010, and 2011-2019. \input{supporting_files/coefs/explanation_dem} The size of each circle (country) is proportional to its baseline GDP. The colors depend on the level of the democracy index (warmer colors for democracy and darker colors for autocracies). The line is the fitted line from a univariate OLS regression of the outcome against the democracy index that weights observations by baseline GDP. The shaded area corresponds to the 95\% confidence interval. \input{supporting_files/coefs/explanation_vars}}}
\end{figure}

\restoregeometry

\clearpage
\begin{table}\centering
\def\sym#1{\ifmmode^{#1}\else\(^{#1}\)\fi}
\caption{Democracy's Effect on Economic Growth by Decade}\label{tab:2sls-by-decade} 
\scriptsize
\begin{threeparttable}
\begin{tabular}{l*{12}{c}}
\hline\hline
                    &\multicolumn{1}{c}{(1)}         &\multicolumn{1}{c}{(2)}         &\multicolumn{1}{c}{(3)}         &\multicolumn{1}{c}{(4)}         &\multicolumn{1}{c}{(5)}         &\multicolumn{1}{c}{(6)}         &\multicolumn{1}{c}{(7)}         &\multicolumn{1}{c}{(8)}         &\multicolumn{1}{c}{(9)}         &\multicolumn{1}{c}{(10)}  
                    &\multicolumn{1}{c}{(11)}         &\multicolumn{1}{c}{(12)} \\ 
                    \hline \\[-1.8ex]
\multicolumn{13}{l}{\textbf{Panel A: No Control for Baseline GDP}} \\
& \multicolumn{12}{c}{Dependent Variable is Mean GDP Growth Rate in 1981-1990} \\\cline{2-13}\\[-1.8ex]
Democracy Index (V-Dem, 1980)&        -0.6&        -0.7&        -1.1&       151.6&       -0.09&        0.09&        -0.7&        -0.6&        -0.8&        -0.2&        -0.2&       -0.10\\
&       (0.5)&       (1.3)&       (0.7)&    (2814.7)&       (0.6)&       (0.8)&       (0.8)&       (0.9)&       (0.5)&       (0.8)&       (0.5)&       (0.8)\\

& \multicolumn{12}{c}{Dependent Variable is Mean GDP Growth Rate in 1991-2000} \\\cline{2-13}\\[-1.8ex]
Democracy Index (V-Dem, 1990)&        -1.6&       -66.2&        -1.2&        -5.1&        -0.6&        -3.5&         0.3&         0.4&        -0.9&       0.007&        -1.0&        -1.9\\
&       (0.5)&     (554.8)&       (0.5)&       (2.6)&       (0.8)&       (5.6)&       (1.2)&       (0.9)&       (0.5)&       (1.0)&       (0.6)&       (0.7)\\

& \multicolumn{12}{c}{Dependent Variable is Mean GDP Growth Rate in 2001-2010} \\\cline{2-13}\\[-1.8ex]
Democracy Index (V-Dem, 2000)&        -3.2&        -5.7&        -2.6&        -3.2&        -2.5&        -3.0&        -1.5&        -1.7&        -2.8&        -2.4&        -2.5&        -3.0\\
&       (0.5)&       (4.1)&       (0.5)&       (0.7)&       (0.6)&       (1.5)&       (1.1)&       (0.7)&       (0.5)&       (0.6)&       (0.3)&       (0.4)\\

& \multicolumn{12}{c}{Dependent Variable is Mean GDP Growth Rate in 2011-2019} \\\cline{2-13}\\[-1.8ex]
Democracy Index (V-Dem, 2010)&        -1.7&        -1.5&        -1.8&        -2.1&        -1.2&        -1.1&        -1.2&        -1.1&        -1.9&        -1.5&        -1.6&        -1.7\\
&       (0.2)&       (0.1)&       (0.3)&       (0.3)&       (0.4)&       (0.5)&       (0.5)&       (0.3)&       (0.3)&       (0.2)&      (0.09)&       (0.1)\\

 \\[-1.8ex] 
\hline \\[-1.8ex] 

\multicolumn{13}{l}{\textbf{Panel B: Control for Baseline GDP Per Capita}} \\
& \multicolumn{12}{c}{Dependent Variable is Mean GDP Growth Rate in 1981-1990} \\\cline{2-13}\\[-1.8ex]
Democracy Index (V-Dem, 1980)&        -0.1&         0.6&         6.4&         4.4&         1.0&         1.0&        -0.2&        -0.1&       -0.05&         0.2&         0.9&         0.9\\
&       (1.2)&       (1.0)&       (6.5)&       (3.1)&       (0.6)&       (0.4)&       (1.2)&       (0.8)&       (0.6)&       (0.6)&       (0.3)&       (0.3)\\

& \multicolumn{12}{c}{Dependent Variable is Mean GDP Growth Rate in 1991-2000} \\\cline{2-13}\\[-1.8ex]
Democracy Index (V-Dem, 1990)&        -9.0&         3.0&        -3.0&        -4.6&         2.9&        16.2&       -0.07&         0.5&        -0.2&      0.0006&        -1.7&        -1.6\\
&       (7.0)&       (5.0)&       (1.4)&       (2.5)&      (17.2)&      (97.0)&       (1.1)&       (0.7)&       (0.6)&       (0.8)&       (0.8)&       (0.5)\\

& \multicolumn{12}{c}{Dependent Variable is Mean GDP Growth Rate in 2001-2010} \\\cline{2-13}\\[-1.8ex]
Democracy Index (V-Dem, 2000)&        -3.8&         2.0&        -2.0&        -2.7&         0.1&        36.7&        -2.0&        -1.1&        -2.5&        -1.7&        -2.2&        -1.9\\
&       (1.1)&       (5.6)&       (0.7)&       (0.8)&       (4.4)&     (368.2)&       (0.7)&       (0.4)&       (0.8)&       (0.6)&       (0.5)&       (0.3)\\

& \multicolumn{12}{c}{Dependent Variable is Mean GDP Growth Rate in 2011-2019} \\\cline{2-13}\\[-1.8ex]
Democracy Index (V-Dem, 2010)&        -2.2&        -1.9&        -2.6&        -4.3&         0.9&        25.0&        -1.4&        -0.9&        -2.2&        -1.6&        -1.7&        -1.8\\
&       (1.2)&       (2.7)&       (0.7)&       (1.6)&       (2.8)&     (115.5)&       (0.6)&       (0.4)&       (0.5)&       (0.4)&       (0.3)&       (0.6)\\

 \\[-1.8ex] 
\hline \\[-1.8ex]

\multicolumn{13}{l}{\textbf{Panel C: Control for Baseline Total GDP}} \\
& \multicolumn{12}{c}{Dependent Variable is Mean GDP Growth Rate in 1981-1990} \\\cline{2-13}\\[-1.8ex]
Democracy Index (V-Dem, 1980)&        -0.8&        -0.8&        -3.4&       -14.9&        -0.1&        0.03&        -4.0&        -2.9&        -0.7&        -0.4&        -0.4&        -0.2\\
&       (0.5)&       (1.3)&       (2.9)&      (22.0)&       (0.9)&       (0.9)&       (4.1)&       (5.0)&       (0.5)&       (1.1)&       (0.7)&       (0.8)\\

& \multicolumn{12}{c}{Dependent Variable is Mean GDP Growth Rate in 1991-2000} \\\cline{2-13}\\[-1.8ex]
Democracy Index (V-Dem, 1990)&        -1.8&      -114.9&        -1.7&        -5.1&        -0.7&        -3.2&        -0.6&         0.5&        -1.2&       -0.03&        -1.5&        -2.1\\
&       (0.5)&    (1645.4)&       (0.7)&       (2.2)&       (1.3)&       (6.0)&       (1.1)&       (0.9)&       (0.5)&       (1.1)&       (0.6)&       (0.7)\\

& \multicolumn{12}{c}{Dependent Variable is Mean GDP Growth Rate in 2001-2010} \\\cline{2-13}\\[-1.8ex]
Democracy Index (V-Dem, 2000)&        -3.2&        -9.4&        -2.4&        -3.2&        -2.1&        -1.4&        -1.5&        -1.0&        -2.9&        -2.3&        -2.4&        -2.7\\
&       (0.5)&      (17.8)&       (0.6)&       (0.7)&       (1.2)&       (2.9)&       (1.3)&       (0.7)&       (0.5)&       (0.8)&       (0.4)&       (0.4)\\

& \multicolumn{12}{c}{Dependent Variable is Mean GDP Growth Rate in 2011-2019} \\\cline{2-13}\\[-1.8ex]
Democracy Index (V-Dem, 2010)&        -1.7&        -1.8&        -2.0&        -2.3&        -1.1&        -0.9&        -1.4&        -1.3&        -2.0&        -2.0&        -1.7&        -2.0\\
&       (0.2)&       (0.3)&       (0.3)&       (0.4)&       (0.6)&       (0.7)&       (0.5)&       (0.5)&       (0.3)&       (0.2)&       (0.1)&       (0.2)\\
\hline \\[-1.8ex] IVs & \multicolumn{2}{c}{settler mortality} &  \multicolumn{2}{c}{population density} & \multicolumn{2}{c}{legal origin} & \multicolumn{2}{c}{language} & \multicolumn{2}{c}{crops \& minerals} & \multicolumn{2}{c}{all IVs}  \\ Baseline Controls Other Than Baseline GDP & \xmark & \cmark & \xmark & \cmark & \xmark & \cmark & \xmark & \cmark & \xmark & \cmark  & \xmark & \cmark\\
N                   &          73&          73&          75&          75&          78&          78&         109&         109&         113&         113&          65&          65\\

 \\[-1.8ex] 

\hline\hline \\[-1.8ex]

\end{tabular}
\begin{tablenotes}
\item {\footnotesize {\textit{Notes:} This table shows the 2SLS regression estimates of democracy's effect on mean GDP growth rates in 1981-1990, 1991-2000, 2001-2010, and 2011-2019. Panel A does not control for baseline GDP. Panel B controls for baseline GDP per capita. Panel C controls for baseline total GDP. 
 The Democracy Index (V-Dem) is normalized to have mean zero and standard deviation one in each year.\unskip Columns 2, 4, 6, 8, 10, and 12 also have the following controls: absolute latitude, mean temperature, mean precipitation, population density, and median age.
For IVs, columns 1 and 2 use log European settler mortality, columns 3 and 4 use log population density in the 1500s, columns 5 and 6 use British legal origin, columns 7 and 8 use the fraction speaking English and the fraction speaking European, columns 9 and 10 use the ability to grow crops and mine minerals, and columns 11 and 12 use all the IVs together. \unskip The sample size is slightly different from that in Table \ref{tab:2sls_tab1} because this table uses only observations for which all GDP per capita and total GDP growth rate data are available. 
N refers to the number of countries for which IVs, controls, and outcomes are available. \unskip
Robust standard errors are in parentheses. \unskip
Variable definitions and data sources are in Appendix Table \ref{tab:sources}. \unskip}}
\end{tablenotes}
\end{threeparttable}
\end{table}

\clearpage
\newgeometry{left=0.3cm, right = 0.3cm, top = 1cm, bottom=1in}
\begin{figure}
\centering
\caption{Correlation Between Democracy Quadratic and Economic Growth by Decade}\label{fig:ols-decade_quad}
\captionsetup{width=0.99\textwidth}
\centering

\begin{subfigure}[c]{.99\textwidth}
  \subcaptionbox{1981-1990\label{fig:1980s_q}}{\includegraphics[width=.49\textwidth]{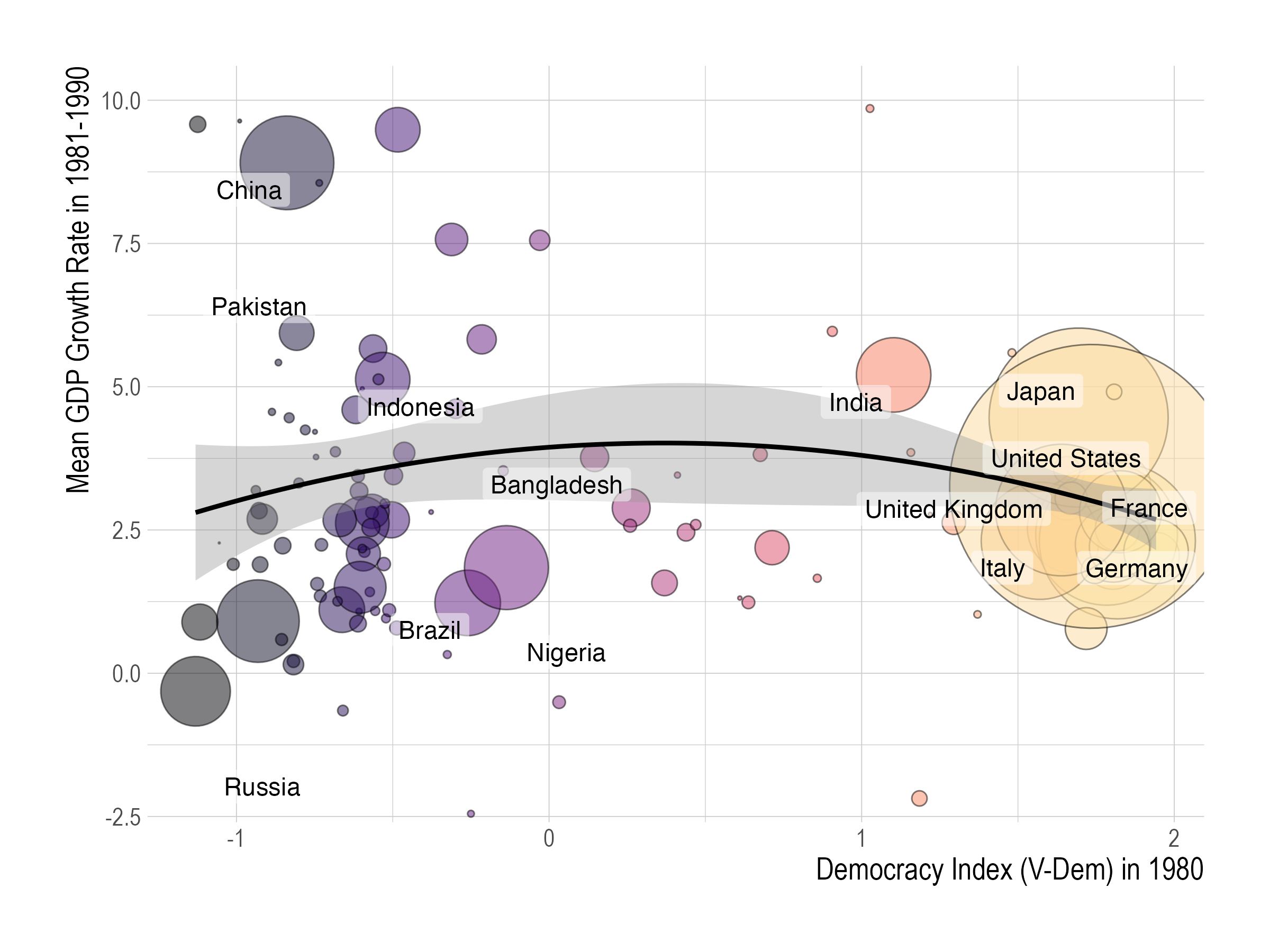}}
  \subcaptionbox{1991-2000\label{fig:1990s_q}}{\includegraphics[width=.49\textwidth]{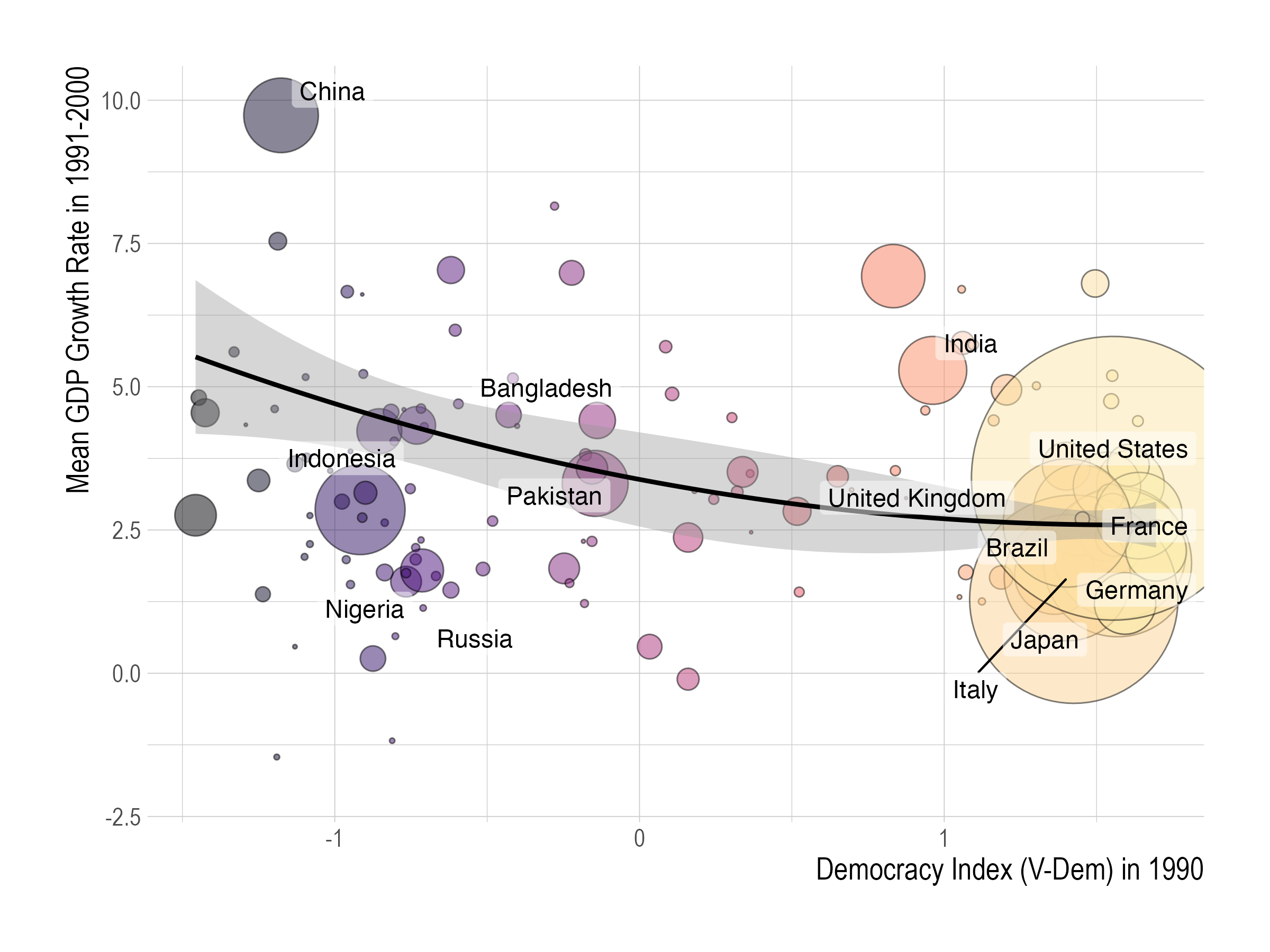}}
\end{subfigure}
\vspace{-0.1cm}
  
\begin{subfigure}[c]{.99\textwidth}
  \subcaptionbox{2001-2010 \label{fig:2000s_q}}{\includegraphics[width=.49\textwidth]{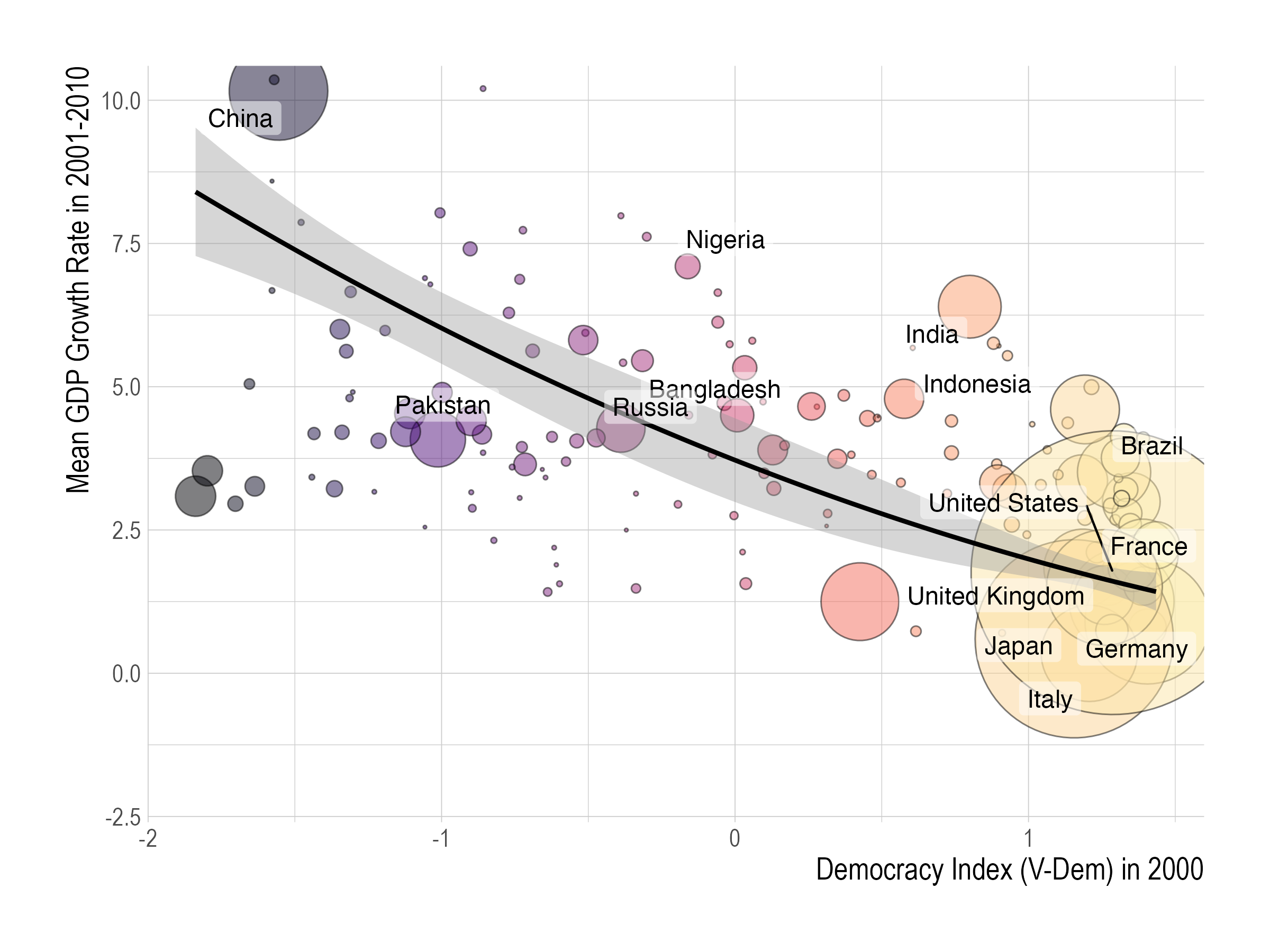}}
  \subcaptionbox{2011-2019 \label{fig:2010s_q}}{\includegraphics[width=.49\textwidth]{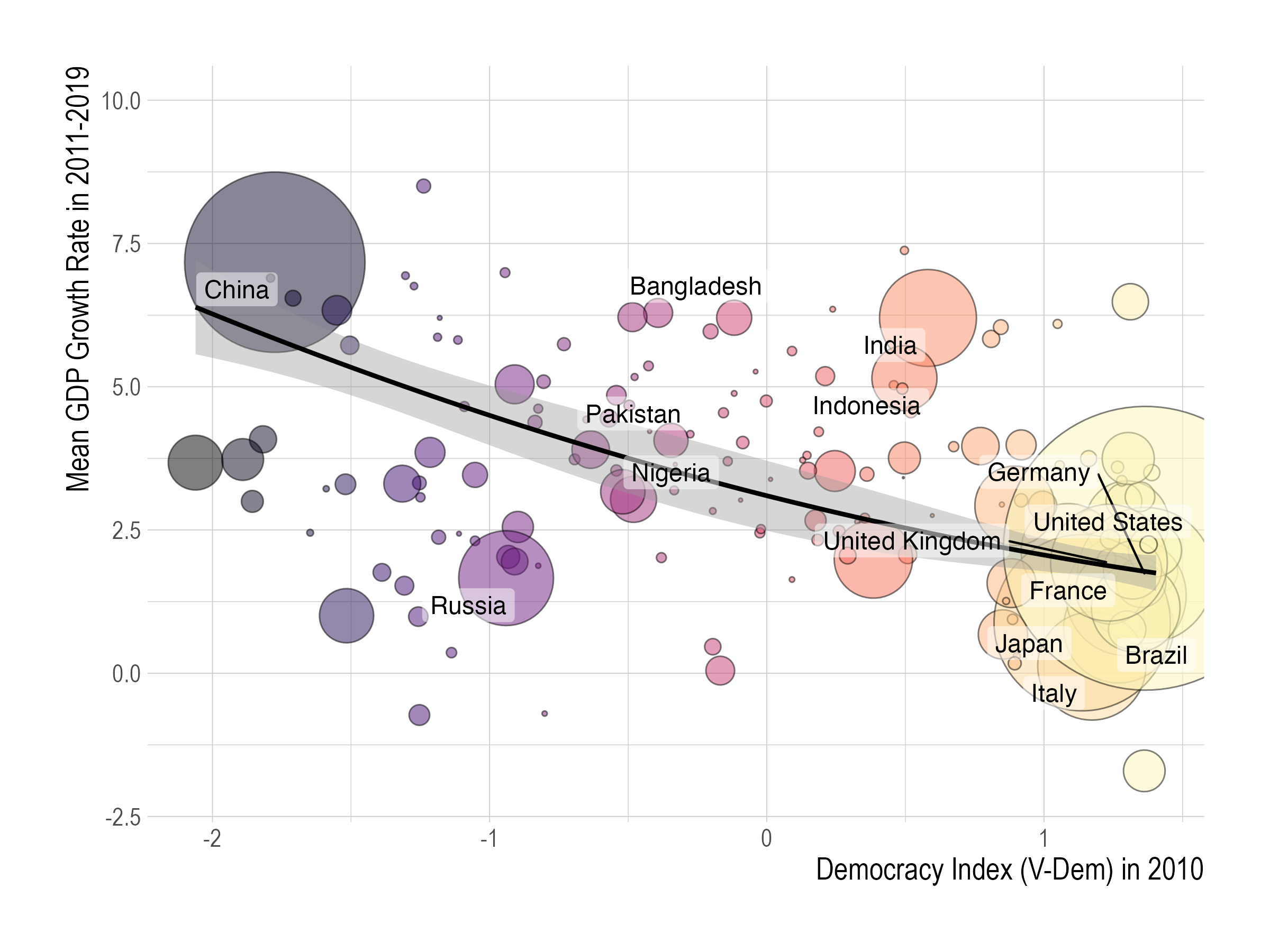}}
\end{subfigure}

\vspace{-0.1cm}
  
\begin{subfigure}[c]{.99\textwidth}
  \subcaptionbox{1993-2000 \label{fig:2000s_q_light}}{\includegraphics[width=.49\textwidth]{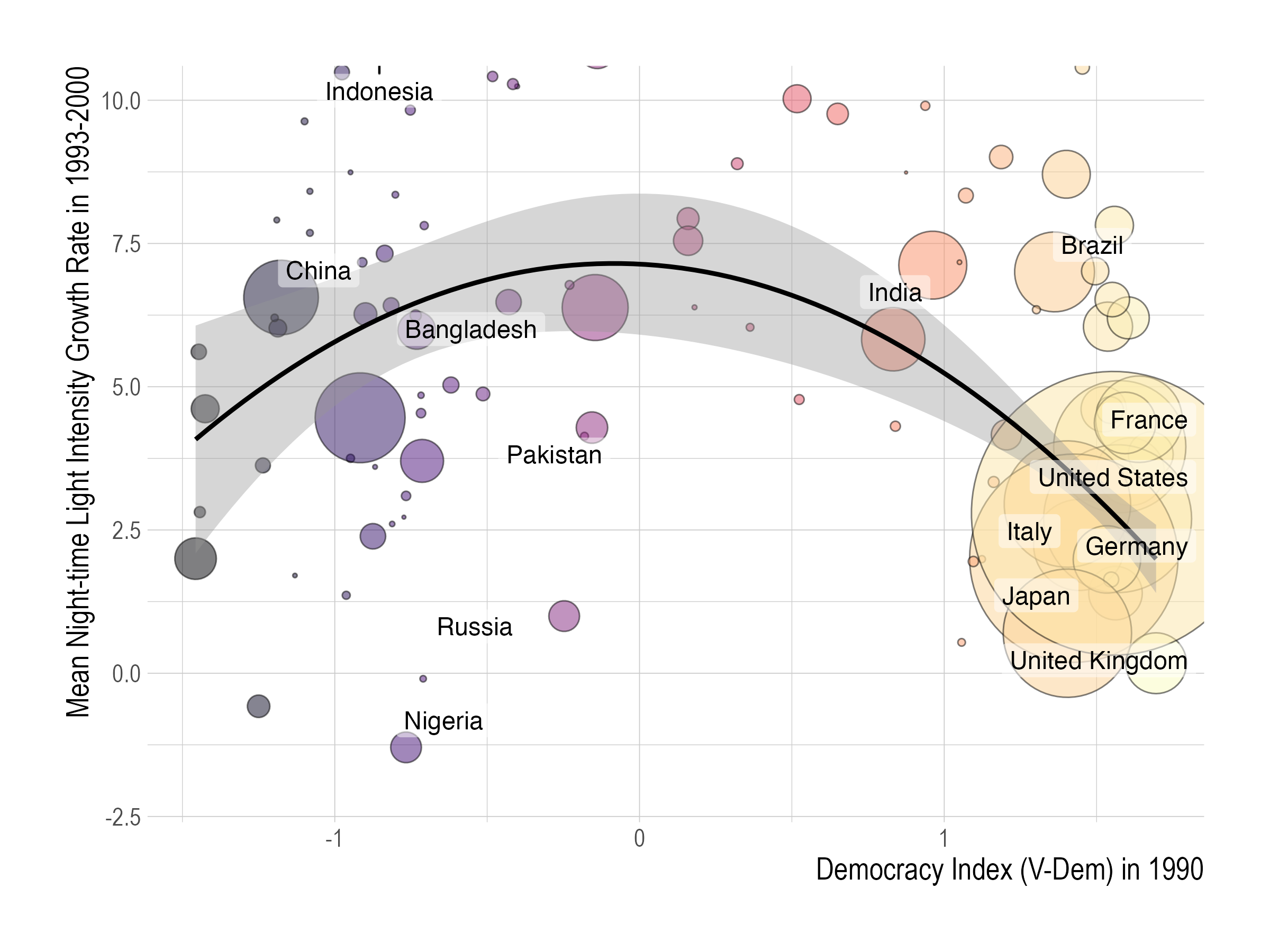}}
  \subcaptionbox{2001-2013 \label{fig:2010s_q_light}}{\includegraphics[width=.49\textwidth]{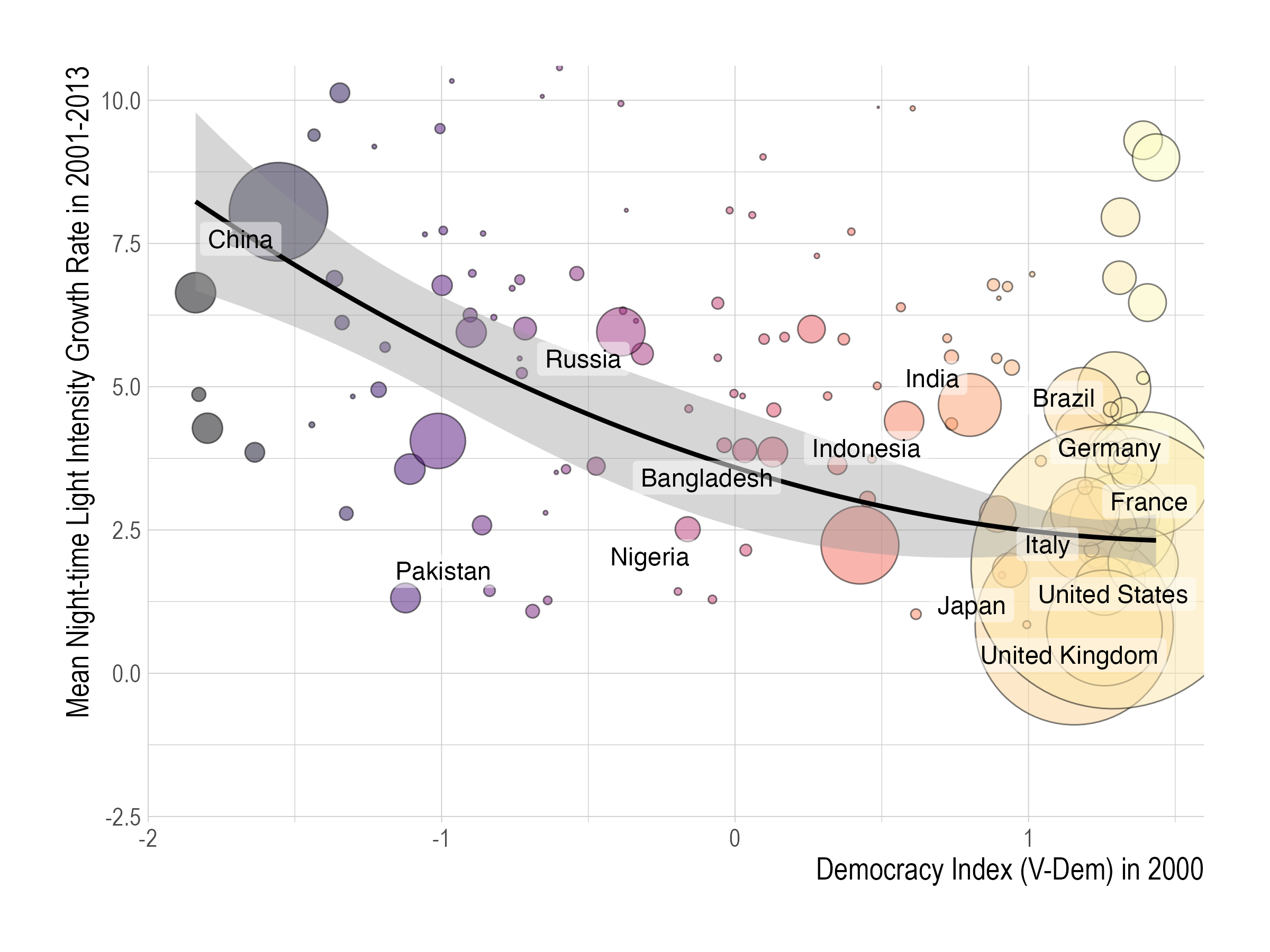}}
\end{subfigure}

\caption*{\footnotesize{\textit{Notes:} Panels (a)-(d) show the relationship between democracy and the mean GDP growth rates in four periods: 1981-1990, 1991-2000, 2001-2010, and 2011-2019. Panels (e)-(f) display the relationship between democracy and the mean nighttime light intensity growth rates in 1993-2000 and 2001-2013. \input{supporting_files/coefs/explanation_dem} The size of each circle (country) is proportional to its baseline GDP. The colors depend on the level of the democracy index (warmer colors for democracy and darker colors for autocracies). The line is the fitted line from an OLS regression of the outcome against the democracy index and its square, without controls, that weights observations by baseline GDP. The shaded area corresponds to the 95\% confidence interval. \input{supporting_files/coefs/explanation_vars}}}
\end{figure}

\restoregeometry

\clearpage
\newgeometry{left=0.3cm, right = 0.3cm, top = 1cm, bottom=1in}
\begin{figure}
\centering
\caption{Correlation Between Democracy Quadratic and Residualized Economic Growth by Decade}\label{fig:ols-decade_quad_res}
\captionsetup{width=0.99\textwidth}
\centering

\begin{subfigure}[c]{.99\textwidth}
  \subcaptionbox{1981-1990\label{fig:1980s_resid}}{\includegraphics[width=.49\textwidth]{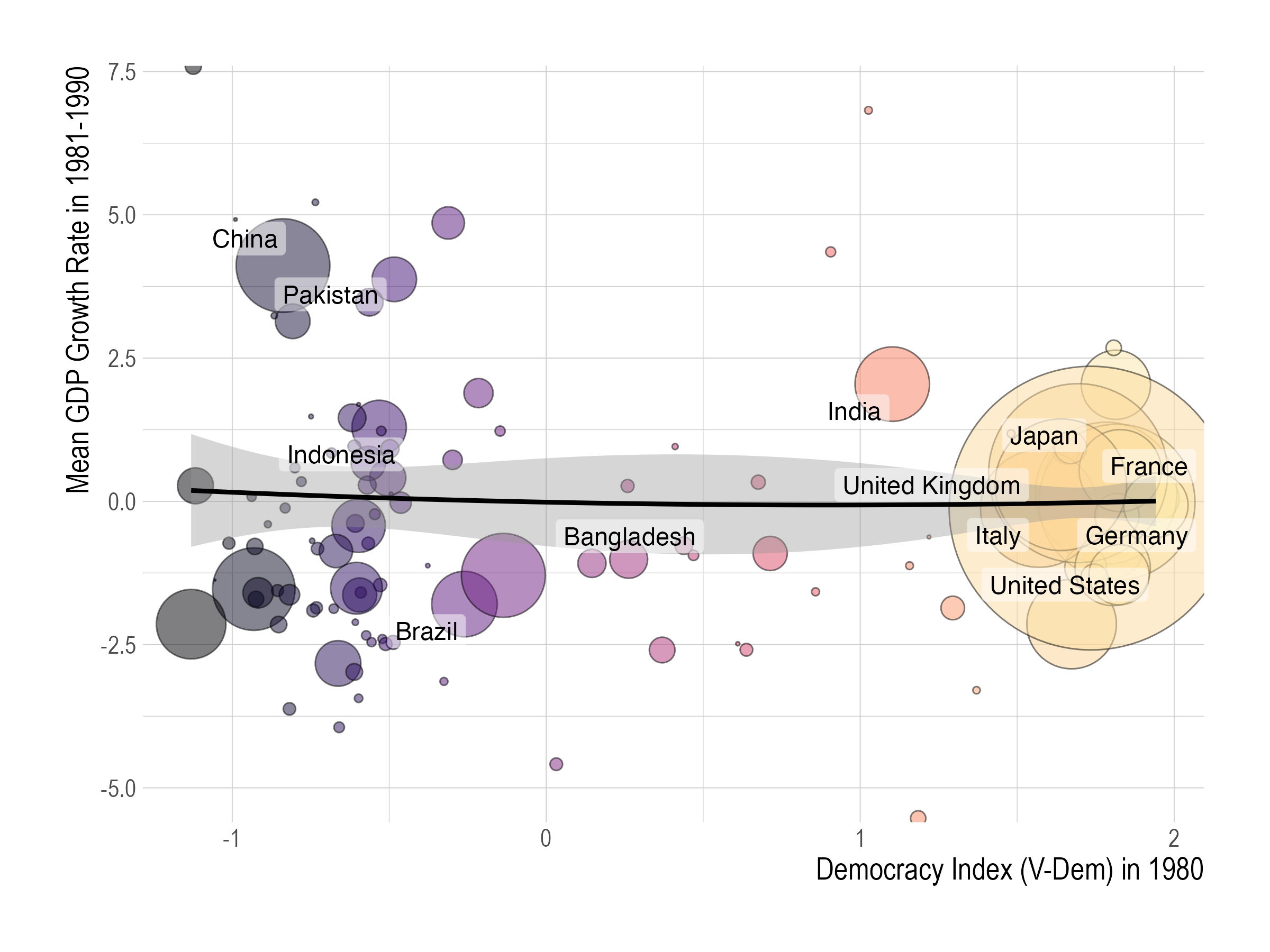}}
  \subcaptionbox{1991-2000\label{fig:1990s_resid}}{\includegraphics[width=.49\textwidth]{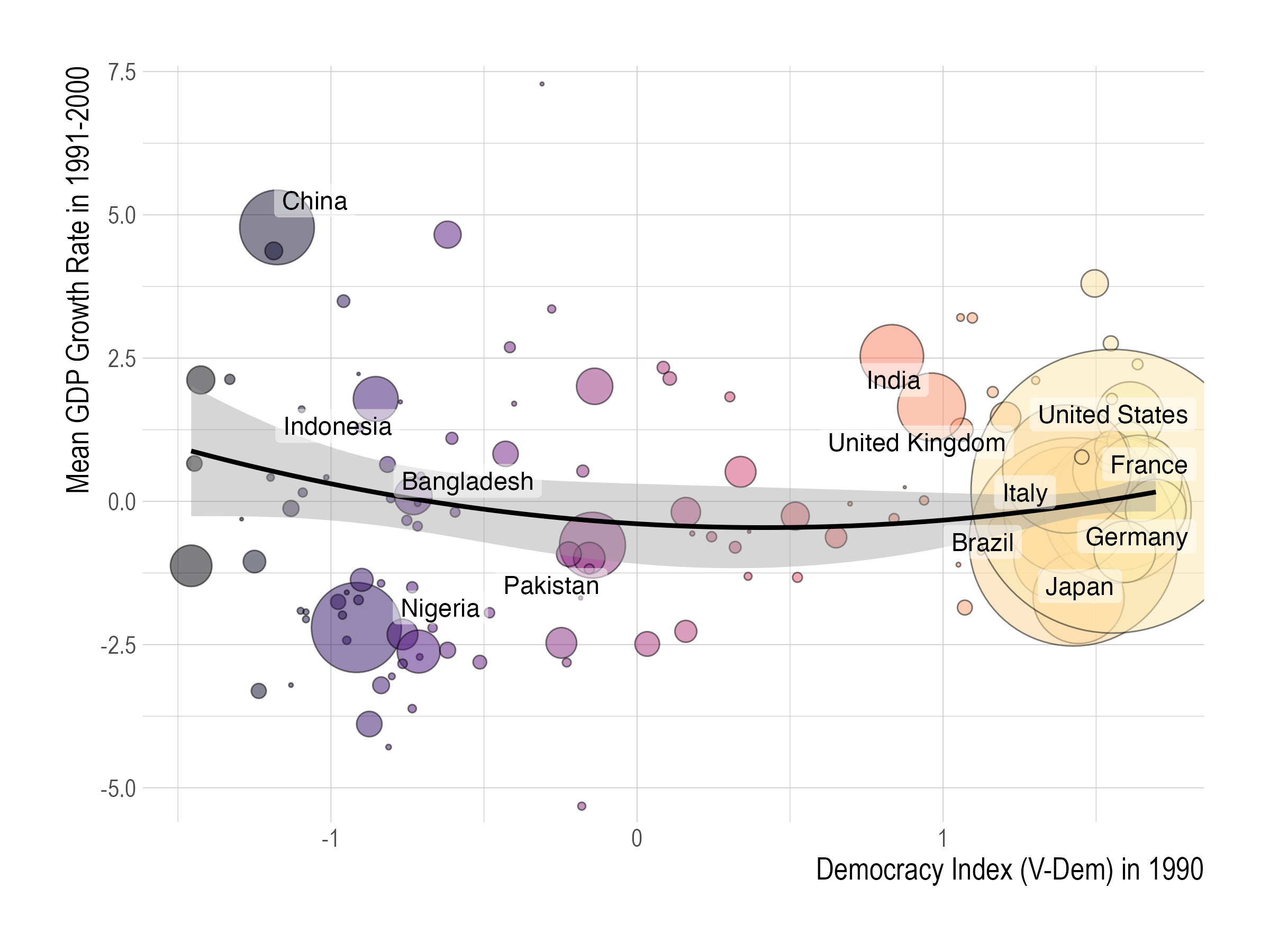}}
\end{subfigure}
\vspace{-0.1cm}
  
\begin{subfigure}[c]{.99\textwidth}
  \subcaptionbox{2001-2010 \label{fig:2000s_resid}}{\includegraphics[width=.49\textwidth]{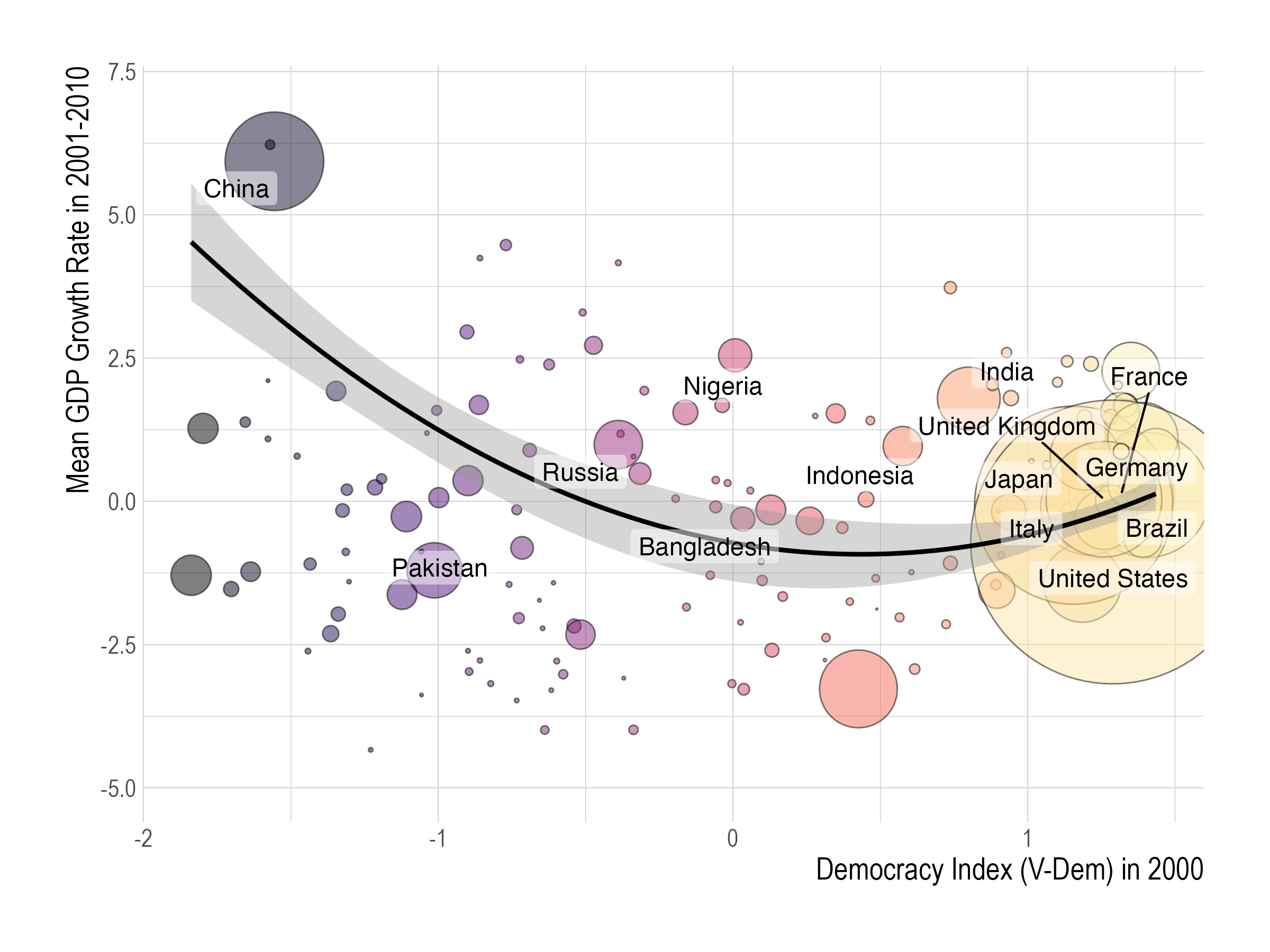}}
  \subcaptionbox{2011-2019 \label{fig:2010s_resid}}{\includegraphics[width=.49\textwidth]{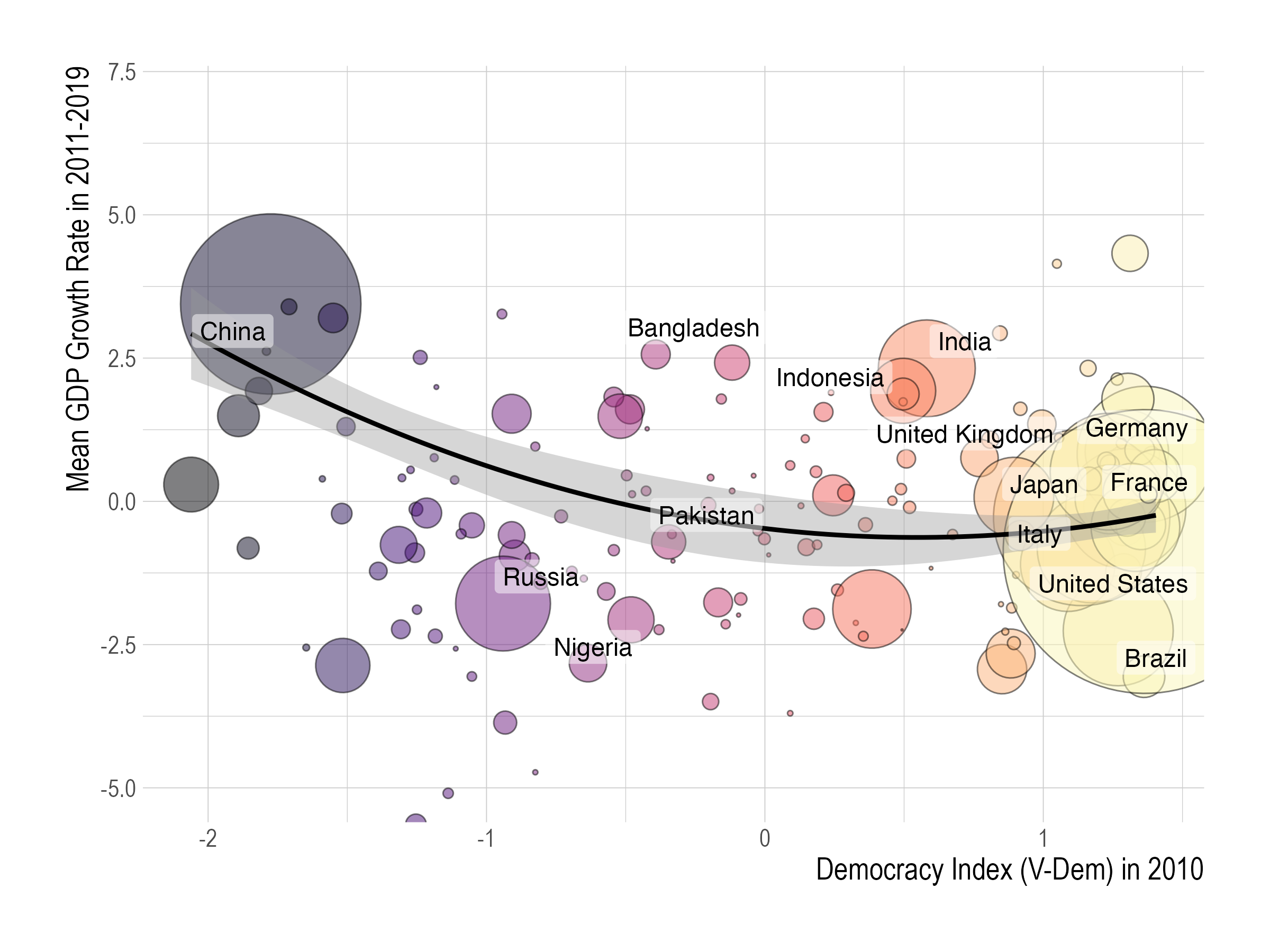}}
\end{subfigure}

=\vspace{-0.1cm}
  
\begin{subfigure}[c]{.99\textwidth}
  \subcaptionbox{1993-2000 \label{fig:2000s_resid_night}}{\includegraphics[width=.49\textwidth]{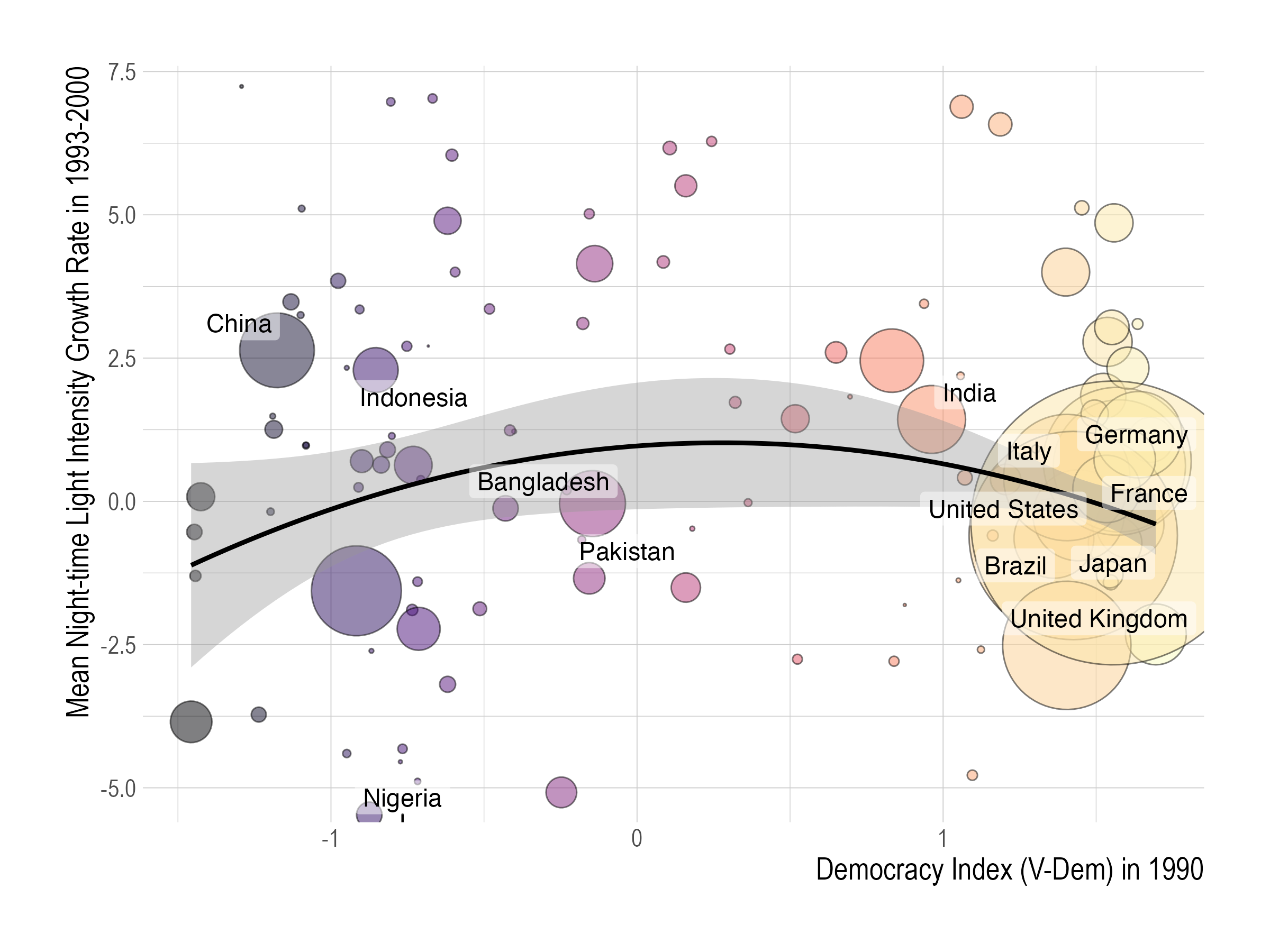}}
  \subcaptionbox{2001-2013 \label{fig:2010s_resid_night}}{\includegraphics[width=.49\textwidth]{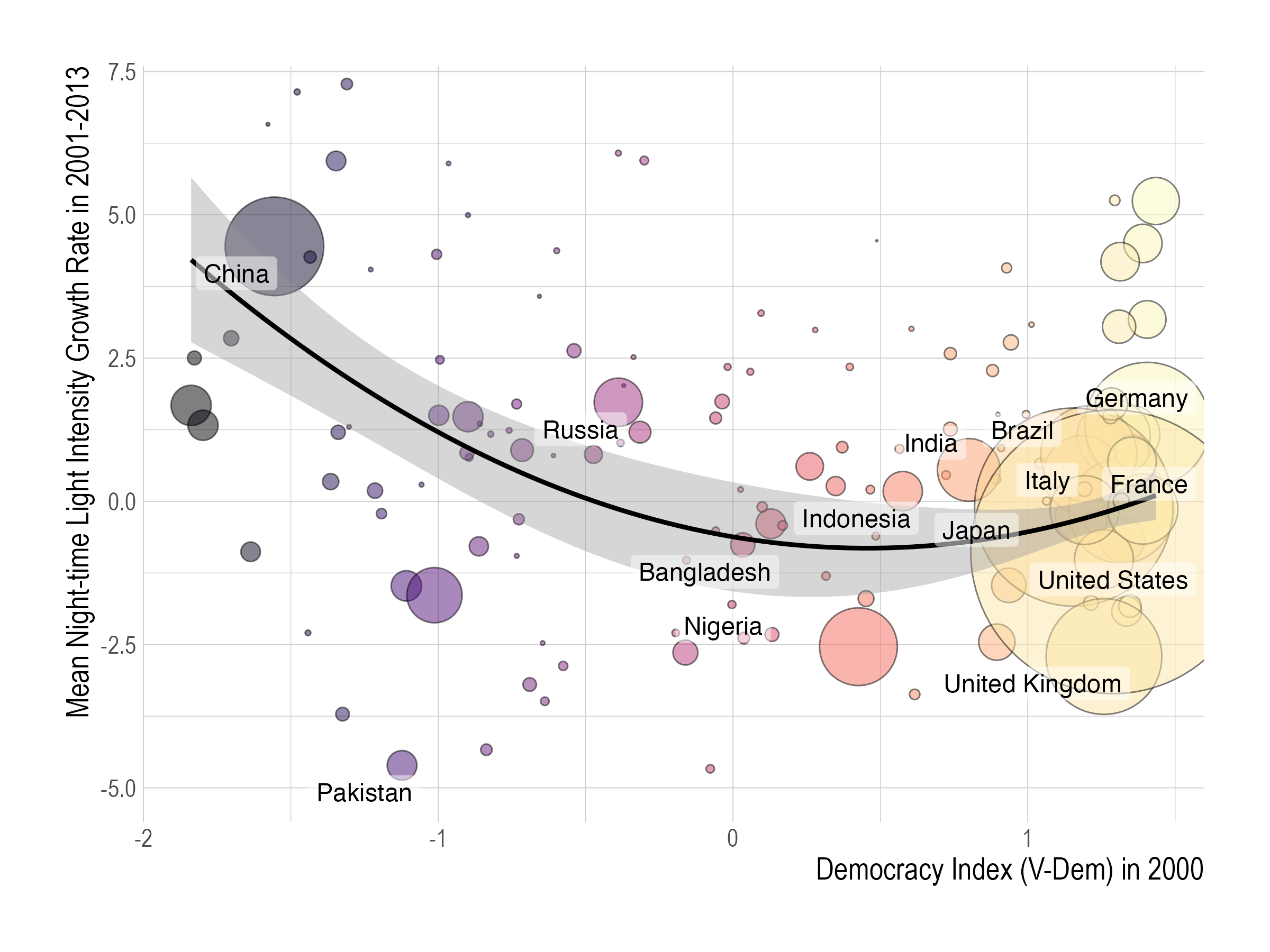}}
\end{subfigure}

\caption*{\footnotesize{\textit{Notes:} Panels (a)-(d) show the relationship between democracy and the residualized mean GDP growth rates in four periods: 1981-1990, 1991-2000, 2001-2010, and 2011-2019. Panels (e)-(f) display the relationship between democracy and the residualized mean nighttime light intensity growth rates in 1993-2000 and 2001-2013. Using a similar approach to \barroFixedCite, who residualize growth rates against covariates listed in Table 12.3 of their text, we residualize growth rates against our baseline controls: absolute latitude, mean temperature, mean precipitation, populationdensity, and median age.  \input{supporting_files/coefs/explanation_dem} The size of each circle (country) is proportional to its baseline GDP. The colors depend on the level of the democracy index (warmer colors for democracy and darker colors for autocracies). The line is the fitted line from an OLS regression of the outcome against the democracy index and its square, without controls, that weights observations by baseline GDP. The shaded area corresponds to the 95\% confidence interval. \input{supporting_files/coefs/explanation_vars}}}
\end{figure}

\restoregeometry

\clearpage
\newgeometry{left=0.3cm, right = 0.3cm, top = 1cm, bottom=1in}
\begin{figure}
\centering
\caption{Correlation Between Democracy Change and Outcomes}\label{fig:ols_change}
\captionsetup{width=0.99\textwidth}

    \begin{subfigure}[c]{0.49\textwidth}
        \centering
        \caption{Mean GDP Growth Rate in 2001-2019}
        \label{fig:ols_change1}
        \includegraphics[width=0.99\linewidth]{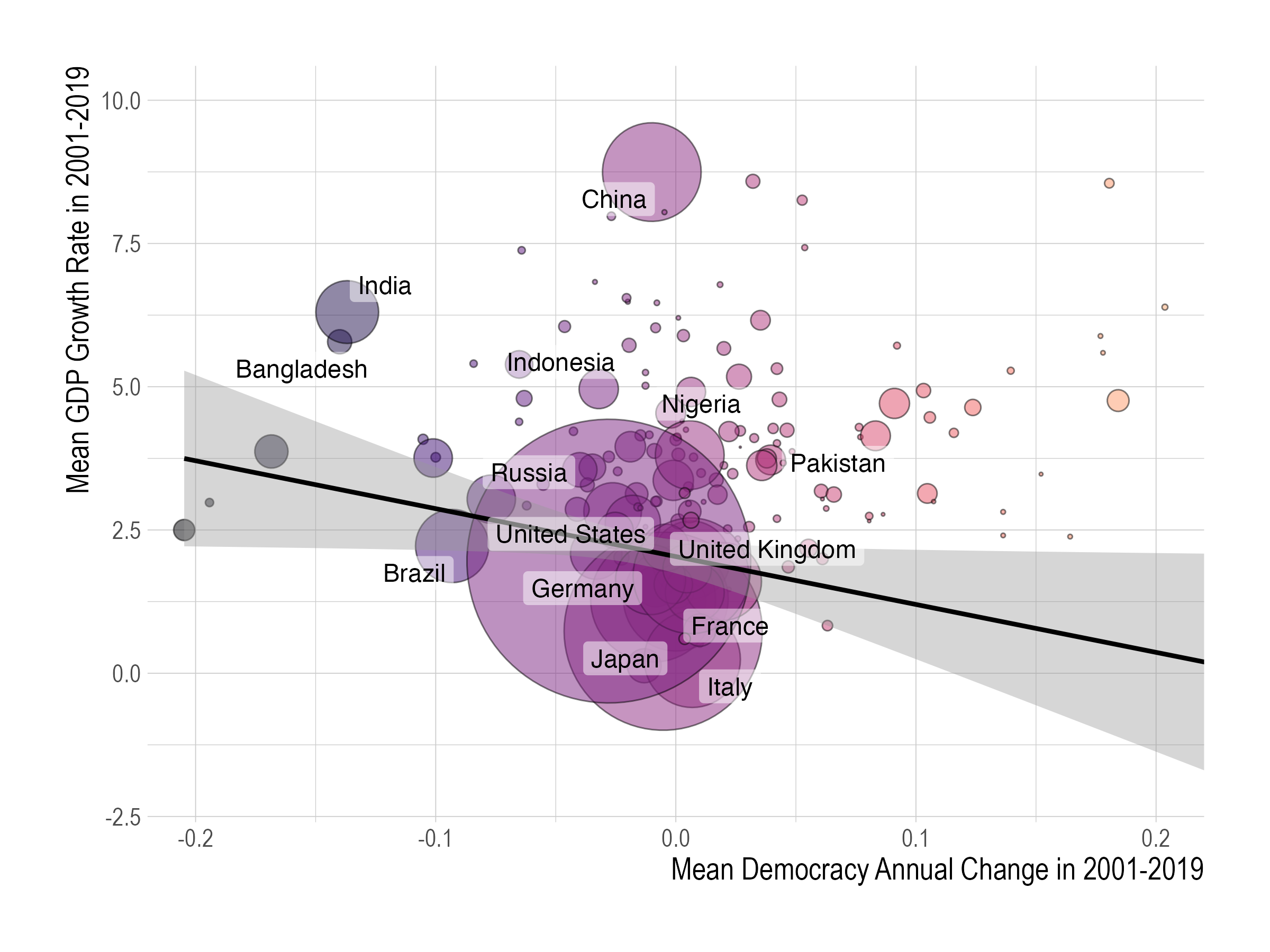}
    \end{subfigure}%
    \begin{subfigure}[c]{0.49\textwidth}
        \centering
        \caption{Mean Night-time Light Intensity Growth Rate in 2001-2013}
        \label{fig:ols_change2}
        \includegraphics[width=0.99\linewidth]{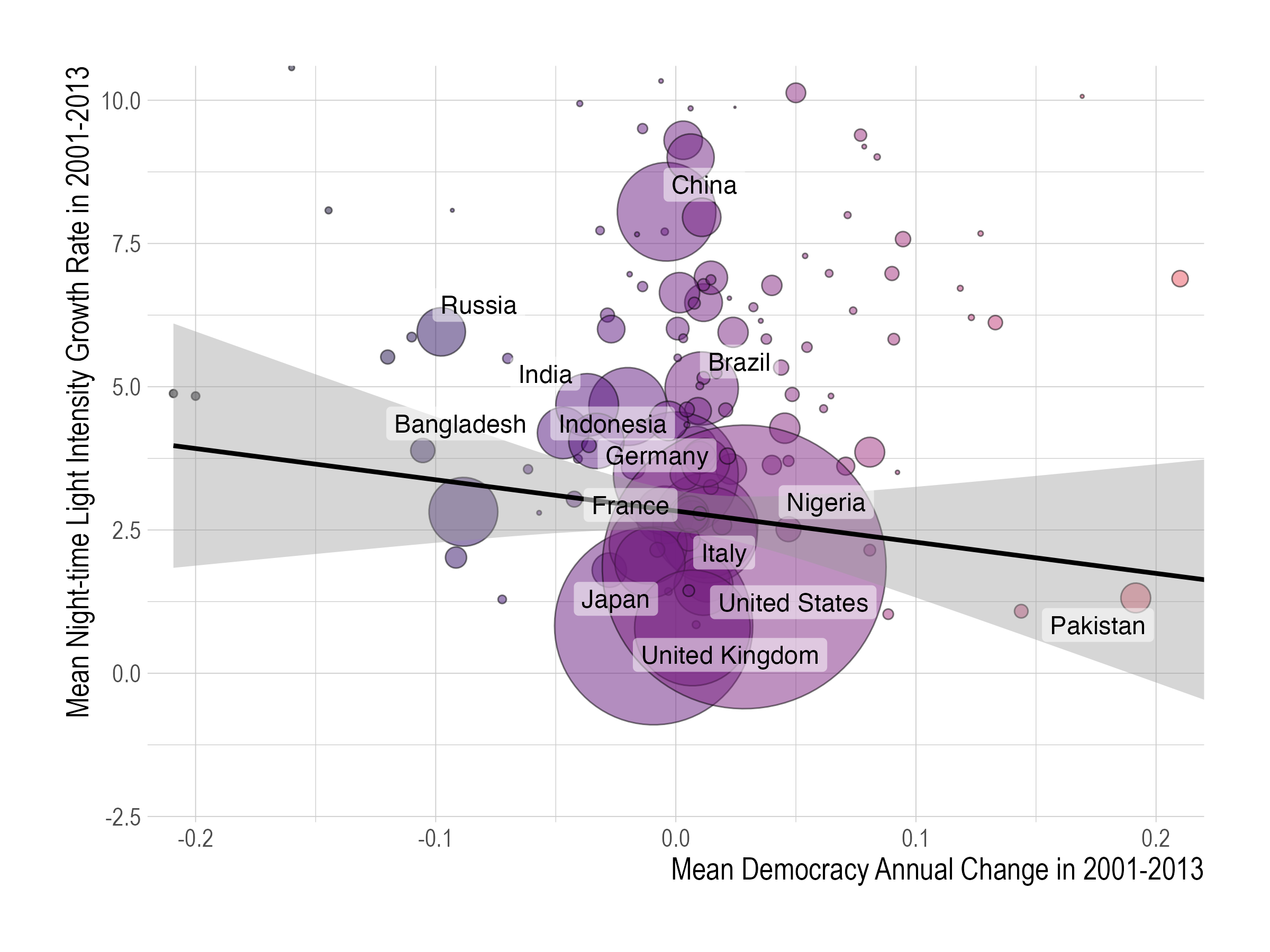}
    \end{subfigure}
    
    
    \begin{subfigure}[c]{0.49\textwidth}
        \centering
        \caption{Mean Democracy Annual Change in 2001-2019}
        \label{fig:ols_change3}
        \includegraphics[width=0.99\linewidth]{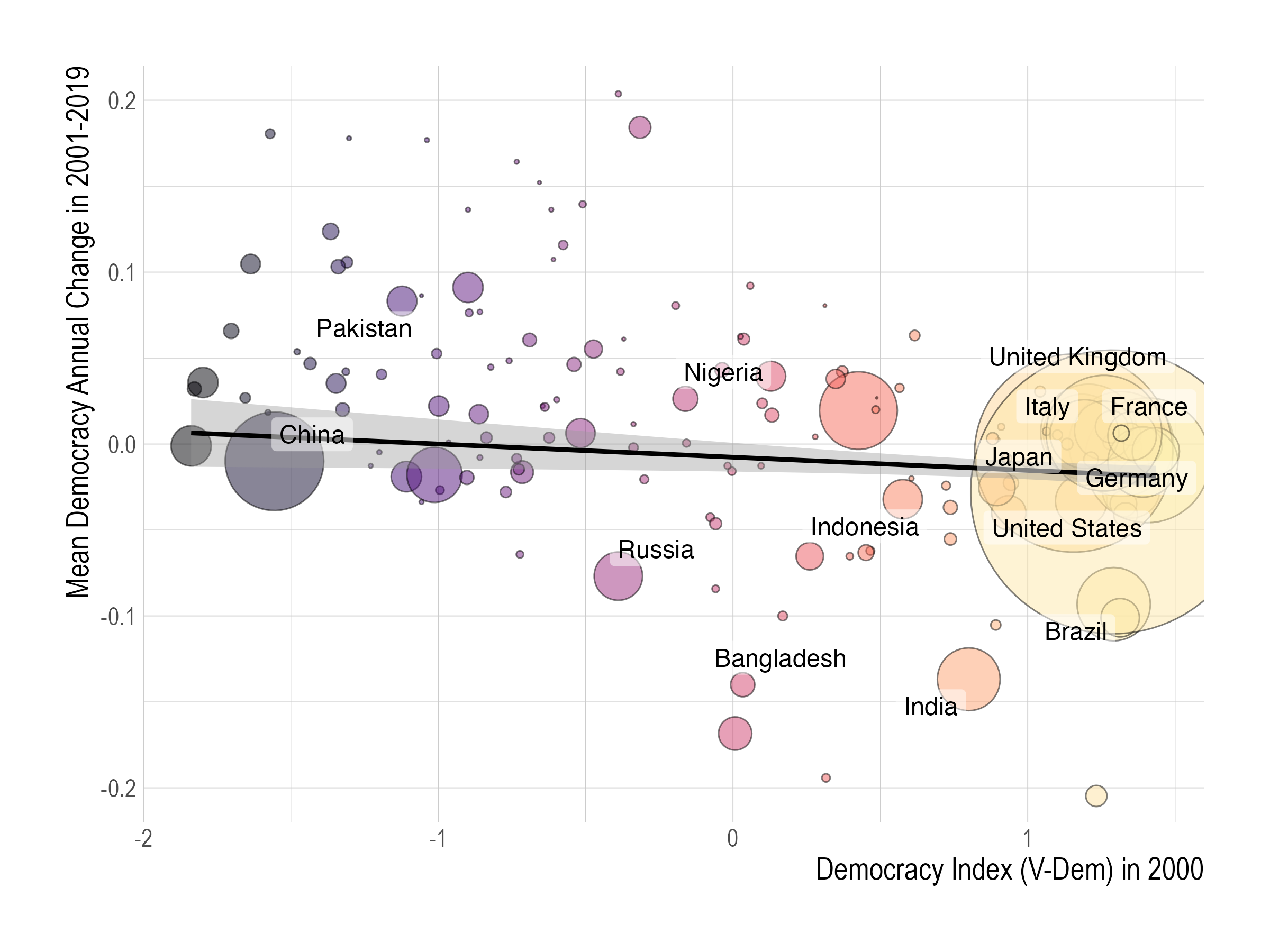}
    \end{subfigure}

\caption*{\footnotesize{\textit{Notes:} Panels (a) and (b) show the correlation between the mean annual change in democracy and outcomes (the mean GDP growth rate in 2001-19 and mean night-time light intensity growth rate in 2001-13). Panel (c) displays the correlation between the baseline democracy index and mean democracy index annual change in 2001-2019. Democracy change is scaled to $[-10, 10]$. The size of each circle (country) is proportional to its baseline GDP. The colors represent the level of democracy change, with warmer colors indicating higher levels of democracy change. The line represents the OLS regression fitted line without controls, weighted by countries' baseline GDP. The shaded area corresponds to the 95\% confidence interval. Variable definitions and data sources are in Appendix Table \ref{tab:sources}.}}
\end{figure}

\restoregeometry

\clearpage
\begin{center}
\begin{table}  
  \centering
  \caption{2SLS Regression Estimates of Democracy Change's Effects}\label{tab:2sls_change} 
  \tiny
  \begin{threeparttable}
  
  \begin{tabular}{@{\extracolsep{0pt}}lcccccccccccc} 
\\[-1.8ex]\hline 
\hline 
\\[-1.8ex] & (1) & (2) & (3) & (4) & (5) & (6) & (7) & (8) & (9) & (10) & (11) & (12)\\ 
\hline \\[-1.8ex] 
\multicolumn{13}{l}{\textbf{Panel A: Two-Stage Least Squares}} \\ 
& \multicolumn{12}{c}{Dependent Variable is Mean GDP Growth Rate in 2001-2019} \\\cline{2-13}\\[-1.8ex]
Mean Democracy Annual Change in 2001-2019&      -216.8&      -126.2&       606.8&      -340.2&       182.1&       119.2&        35.4&        91.5&       -50.9&       -13.5&         7.2&       -11.4\\
&     (274.1)&     (158.7)&    (2124.0)&     (668.0)&     (237.7)&     (374.3)&      (80.1)&      (89.2)&      (40.7)&      (16.0)&      (14.6)&       (8.9)\\
&        0.43&        0.43&        0.78&        0.61&        0.44&        0.75&        0.66&        0.31&        0.21&        0.40&        0.62&        0.20\\
F-Statistic (First stage)&         0.9&         0.8&         0.1&         0.2&         0.6&         0.2&         0.7&         0.9&         3.9&         0.8&        21.8&        10.5\\

& \multicolumn{12}{c}{Dependent Variable is Mean Nighttime Light Intensity Growth Rate in 2001-2013} \\\cline{2-13}\\[-1.8ex]
Mean Democracy Annual Change in 2001-2013&      -161.3&       -54.0&      -382.8&       167.9&      -230.6&       -87.5&       -52.7&      -130.2&       -26.4&       -38.5&       -19.8&       -17.0\\
&     (120.8)&      (25.9)&     (694.4)&     (324.5)&     (180.2)&      (43.1)&      (39.6)&      (53.8)&      (22.4)&      (21.0)&      (15.4)&      (12.8)\\
&        0.18&        0.04&        0.58&        0.60&        0.20&        0.04&        0.18&        0.02&        0.24&        0.07&        0.20&        0.19\\
F-Statistic (First stage)&         1.9&         7.1&         0.3&         0.4&         1.6&         4.1&         7.4&         3.7&        12.5&         3.8&    108262.3&       598.7\\

& \multicolumn{12}{c}{Dependent Variable is Mean GDP Growth Rate in 2001-2019} \\\cline{2-13}\\[-1.8ex]
Democracy Annual Change in 2000&        28.5&        16.4&        39.5&       -62.3&        11.6&         8.8&        -3.7&       -18.5&         7.5&         4.9&         2.8&        -0.8\\
&      (32.9)&      (16.8)&      (49.3)&      (96.2)&      (11.5)&      (14.9)&       (6.5)&      (18.4)&       (4.8)&       (3.0)&       (2.1)&       (1.8)\\
&        0.39&        0.33&        0.42&        0.52&        0.31&        0.55&        0.57&        0.32&        0.12&        0.10&        0.17&        0.65\\
F-Statistic (First stage)&         0.9&         1.6&         0.8&         0.4&         4.4&         1.4&         0.8&         1.3&         3.6&         2.1&        20.4&         5.6\\

& \multicolumn{12}{c}{Dependent Variable is Mean Nighttime Light Intensity Growth Rate in 2001-2013} \\\cline{2-13}\\[-1.8ex]
Democracy Annual Change in 2000&        31.9&        12.9&        32.7&        21.9&        16.1&        25.5&        20.9&       -58.1&         8.9&         7.0&         4.3&         1.4\\
&      (37.1)&      (10.7)&      (39.1)&      (36.4)&      (11.7)&      (27.3)&      (17.2)&      (71.0)&       (6.0)&       (5.7)&       (2.6)&       (1.9)\\
&        0.39&        0.23&        0.40&        0.55&        0.17&        0.35&        0.22&        0.41&        0.14&        0.22&        0.10&        0.47\\
F-Statistic (First stage)&         0.9&         3.2&         0.8&         0.5&         4.5&         1.0&         0.9&         0.4&         4.6&         2.2&        11.7&         3.7\\

& \multicolumn{12}{c}{Dependent Variable is Mean Democracy Annual Change in 2001-2019} \\\cline{2-13}\\[-1.8ex]
Democracy Index (V-Dem, 2000)&        0.01&        0.03&       -0.00&        0.01&       -0.01&       -0.01&       -0.01&       -0.01&        0.01&       -0.01&       -0.01&       -0.01\\
&      (0.02)&      (0.05)&      (0.01)&      (0.02)&      (0.01)&      (0.04)&      (0.01)&      (0.01)&      (0.01)&      (0.01)&      (0.01)&      (0.01)\\
&        0.45&        0.54&        0.77&        0.64&        0.38&        0.68&        0.46&        0.30&        0.46&        0.29&        0.05&        0.42\\
F-Statistic (First stage)&         5.6&         0.7&        14.3&        12.2&         5.9&         2.3&         2.7&         7.8&         4.2&         1.6&        84.1&        12.3\\
IVs & \multicolumn{2}{c}{settler mortality} &  \multicolumn{2}{c}{population density} & \multicolumn{2}{c}{legal origin} & \multicolumn{2}{c}{language} & \multicolumn{2}{c}{crops \& minerals} & \multicolumn{2}{c}{all IVs} \\
Number of IVs & 1 & 1 & 1 & 1 & 1 & 1 & 2 & 2 & 10 & 10 & 15 & 15 \\
 \\[-1.8ex] 
\hline \\[-1.8ex] 

\multicolumn{13}{l}{\textbf{Panel B: Ordinary Least Squares}} \\ 
& \multicolumn{12}{c}{Dependent Variable is Mean GDP Growth Rate in 2001-2019} \\\cline{2-13}\\[-1.8ex]
Mean Democracy Annual Change in 2001-2019&        -3.3&        -4.5&       -0.10&        -8.0&       -0.07&        -7.9&        -9.2&        -4.5&        -8.6&        -4.4&        -0.1&        -8.3\\
&       (8.2)&       (5.9)&       (9.7)&       (7.6)&       (9.7)&       (7.5)&       (6.3)&       (5.1)&       (6.2)&       (5.0)&      (10.3)&       (8.1)\\

& \multicolumn{12}{c}{Dependent Variable is Mean Nighttime Light Intensity Growth Rate in 2001-2013} \\\cline{2-13}\\[-1.8ex]
Mean Democracy Annual Change in 2001-2013&       -11.5&       -10.4&       -10.5&        -7.0&       -10.5&        -7.0&        -6.8&       -10.9&        -5.5&        -9.6&       -13.1&        -9.8\\
&       (8.4)&       (7.8)&       (8.4)&       (7.1)&       (8.4)&       (7.1)&       (8.0)&       (6.6)&       (7.2)&       (5.8)&      (10.2)&       (8.5)\\

& \multicolumn{12}{c}{Dependent Variable is Mean GDP Growth Rate in 2001-2019} \\\cline{2-13}\\[-1.8ex]
Democracy Annual Change in 2000&         0.2&        -1.0&         0.4&        -1.1&         0.4&        -1.1&         0.4&        -0.6&         0.4&        -0.6&         0.4&        -1.1\\
&       (1.3)&       (1.3)&       (1.2)&       (1.4)&       (1.2)&       (1.4)&       (1.0)&       (1.0)&       (1.0)&       (1.0)&       (1.2)&       (1.5)\\

& \multicolumn{12}{c}{Dependent Variable is Mean Nighttime Light Intensity Growth Rate in 2001-2013} \\\cline{2-13}\\[-1.8ex]
Democracy Annual Change in 2000&         0.4&        -1.3&         1.2&        -0.6&         1.1&        -0.6&        0.09&        -1.5&        0.07&        -1.7&         1.2&        -0.4\\
&       (1.3)&       (1.5)&       (1.1)&       (1.4)&       (1.1)&       (1.4)&       (1.2)&       (1.5)&       (1.2)&       (1.5)&       (1.1)&       (1.4)\\

& \multicolumn{12}{c}{Dependent Variable is Mean Democracy Annual Change in 2001-2019} \\\cline{2-13}\\[-1.8ex]
Democracy Index (V-Dem, 2000)&       -0.01&       -0.02&       -0.02&       -0.02&       -0.02&       -0.02&      -0.007&       -0.02&      -0.007&       -0.02&       -0.02&       -0.02\\
&     (0.006)&      (0.01)&     (0.007)&      (0.01)&     (0.007)&      (0.01)&     (0.004)&     (0.008)&     (0.004)&     (0.008)&     (0.007)&      (0.01)\\
Baseline Controls & \xmark & \cmark & \xmark & \cmark & \xmark & \cmark & \xmark & \cmark & \xmark & \cmark & \xmark & \cmark\\
N                   &          81&          81&          86&          86&          90&          90&         132&         132&         138&         138&          71&          71\\

 \\[-1.8ex] 
\hline 
\hline \\[-1.8ex] 
\end{tabular} 

\begin{tablenotes} 
\item {\footnotesize {\textit{Notes:} Panel A reports the 2SLS estimates of the effects of mean annual democracy change on the mean GDP growth rate in 2001-2019 and mean nighttime light intensity growth rate in 2001-2013. The panel also presents the 2SLS estimates of the effect of democracy on mean democracy annual change in 2001-2019. Democracy change is scaled to $[-10, 10]$. Democracy Index (V-Dem, 2000) is normalized to have mean zero and standard deviation one. 
For IVs, columns 1 and 2 use log European settler mortality, columns 3 and 4 use log population density in the 1500s, columns 5 and 6 use British legal origin, columns 7 and 8 use the fraction speaking English and the fraction speaking European, columns 9 and 10 use the ability to grow crops and mine minerals, and columns 11 and 12 use all the IVs together. \unskip 
Robust standard errors are in parentheses. \unskip 
The p-values, presented under the standard errors, are displayed as 0.00 if they are strictly smaller than the 0.005 threshold.
The high values of F-Statistics for the Mean Democracy Annual Change in 2001-2013 in Panel A (columns 11 and 12) are likely due to near-perfect multicollinearity among the instrumental variables. 
Columns 1, 3, 5, 7, 9, and 11 have no controls, while columns 2, 4, 6, 8, 10, and 12 include the following baseline controls: absolute latitude, mean temperature, mean precipitation, population density, and median age. \unskip 
N refers to the number of countries for which IVs, controls, and outcomes are available. \unskip Panel B reports the OLS estimates. 
Variable definitions and data sources are in Appendix Table \ref{tab:sources}. \unskip}}
\end{tablenotes}
\end{threeparttable}
\end{table} 
\end{center}

\clearpage
\newgeometry{left=2cm, right = 2cm, bottom = 1in, top =1in}
\begin{landscape}
\footnotesize
\begin{ThreePartTable}
\begin{TableNotes}
\item{\textit{Notes:} These tables present estimates of the effect of democratization on log GDP per capita for the four periods. The reported coefficient on democracy is multiplied by 100. Cols. 1--4 present results from the within estimator. Cols. 5--8 present results from Arellano and Bond's (1991) GMM estimator. The AR2 row reports the p-value for a test of serial correlation in the residuals of the GDP series. Cols. 9--12 present results from the HHK (Hahn et al., 2001) estimator. In all specifications, we control for a full set of country and year fixed effects. Cols. 4, 8, and 12 include eight lags of GDP per capita as controls, but we report only the p-value of a test for joint significance of lags 5--8. Standard errors robust against heteroskedasticity and serial correlation at the country level are reported in parentheses. See \citet{acemogluDemocracyDoesCause2018} for the details of the method.}
\end{TableNotes}

\begin{center}
\begin{spacing}{1}
\begin{longtable}{@{}lcccccccccccc}

\caption{ Effect of Democratization on (Log) GDP per Capita}
\label{tab:Table1_ext}
\\[-1.8ex] \hline 
\hline

\endfirsthead
\multicolumn{13}{c}{{\tablename} \thetable{} -- Continued}
\\ \hline 
\hline \\

\endhead

\endfoot

\begin{tabular}{@{\extracolsep{0.04pt}}lcccccccccccc}
&{(1)}&{(2)}&{(3)}&{(4)}&{(5)}&{(6)}&{(7)}&{(8)}&{(9)}&{(10)}&{(11)}&{(12)} \\
\hline \\
\multicolumn{13}{@{}l}{\makebox[0pt][l]{\textbf{Panel (a): 1960-2020}}} \\
Democracy   &       1.347   &       1.048   &       1.159   &       1.219   &       1.272   &       1.102   &       1.123   &       0.822   &      -0.025   &       0.405   &       0.576   &       1.050   \\
            &     (0.356)   &     (0.314)   &     (0.289)   &     (0.306)   &     (0.504)   &     (0.462)   &     (0.415)   &     (0.414)   &     (1.432)   &     (1.317)   &     (1.336)   &     (1.648)   \\
Log GDP Lag Controls &       1 lag   &       2 lags   &       4 lags   &       4 lags  &       1 lag  &       2 lags   &       4 lags   &       4 lags   &       1 lag  &       2 lags  &       14 lags  &       4 lags \\
    
p-value lags 5 to 8&&&& [  0.406] &&&& [  0.358] &&&& [  0.012] \\
Long-run effect of democracy&      45.877   &      30.526   &      28.632   &      25.073   &      25.717   &      19.970   &      18.445   &      11.623   &     128.034   &   $-$3616.231   &     442.944   &     107.518   \\
            &    (15.010)   &    (10.100)   &     (7.871)   &     (7.371)   &    (12.066)   &     (9.523)   &     (7.676)   &     (6.500)   &   (136.473)   & (91321.937)   &   (960.540)   &    (71.352)   \\
Effect of democracy after 25 years&      24.094   &      20.559   &      22.789   &      22.350   &      18.484   &      16.600   &      16.798   &      11.313   &      26.351   &      37.062   &      47.493   &      45.360   \\
            &     (6.560)   &     (6.268)   &     (5.789)   &     (5.860)   &     (7.872)   &     (7.469)   &     (6.688)   &     (6.117)   &     (9.913)   &    (11.468)   &    (11.046)   &    (13.486)   \\
Persistence of GDP process&       0.971   &       0.966   &       0.960   &       0.951   &       0.951   &       0.945   &       0.939   &       0.929   &       0.991   &       1.000   &       0.997   &       0.987   \\
            &     (0.005)   &     (0.005)   &     (0.006)   &     (0.007)   &     (0.008)   &     (0.007)   &     (0.008)   &     (0.010)   &     (0.009)   &     (0.009)   &     (0.007)   &     (0.008)   \\
AR2 test p-value&               &               &               &               &        0.14   &        0.00   &        0.95   &        0.78   &               &               &               &               \\
Observations&        8499   &        8335   &        7997   &        7287   &        8267   &        8103   &        7765   &        7067   &       11041   &       11041   &       11041   &       10797   \\
Countries in sample&         181   &         181   &         181   &         180   &         181   &         181   &         181   &         177   &               &               &               &               \\
 \\[-1.8ex] 
\hline \\
\multicolumn{13}{@{}l}{\makebox[0pt][l]{\textbf{Panel (b): 1980-2020}}} \\
Democracy   &       1.906   &       1.567   &       1.689   &       1.828   &       1.960   &       1.748   &       1.722   &       1.532   &      -0.025   &       0.405   &       0.576   &       1.050   \\
            &     (0.478)   &     (0.439)   &     (0.422)   &     (0.524)   &     (0.818)   &     (0.767)   &     (0.683)   &     (0.726)   &     (1.432)   &     (1.317)   &     (1.336)   &     (1.648)   \\
Log GDP Lag Controls&       1 lag   &       2 lags  &       4 lags   &       4 lags   &       1 lag   &       2 lags  &       4 lags   &       4 lags  &       1 lag   &       2 lags   &       4 lags   &      4 lags   \\

p-value lags 5 to 8&&&& [  0.060] &&&& [  0.015] &&&& [  0.032] \\
Long-run effect of democracy&      46.341   &      31.888   &      27.035   &      21.341   &      28.085   &      22.087   &      19.168   &      12.711   &       1.596   &     -18.678   &     -47.760   &      20.876   \\
            &    (13.652)   &     (9.655)   &     (7.148)   &     (6.618)   &    (15.210)   &    (11.503)   &     (8.630)   &     (6.547)   &    (90.860)   &    (64.375)   &   (138.601)   &    (30.194)   \\
Effect of democracy after 25 years&      30.122   &      25.607   &      25.366   &      22.108   &      23.484   &      20.410   &      18.935   &      12.942   &      -0.760   &      14.511   &      20.437   &      14.211   \\
            &     (7.835)   &     (7.330)   &     (6.425)   &     (6.707)   &    (11.502)   &    (10.157)   &     (8.390)   &     (6.732)   &    (43.478)   &    (46.778)   &    (46.424)   &    (20.864)   \\
Persistence of GDP process&       0.959   &       0.951   &       0.938   &       0.914   &       0.930   &       0.921   &       0.910   &       0.879   &       1.016   &       1.022   &       1.012   &       0.950   \\
            &     (0.006)   &     (0.006)   &     (0.008)   &     (0.012)   &     (0.014)   &     (0.012)   &     (0.012)   &     (0.015)   &     (0.015)   &     (0.014)   &     (0.015)   &     (0.033)   \\
 AR2 test p-value&               &               &               &               &        0.99   &        0.00   &        0.47   &        0.41   &               &               &               &               \\
Observations&        6558   &        6371   &        5995   &        5237   &        6326   &        6139   &        5763   &        5017   &        7421   &        7421   &        7421   &        7175   \\
Countries in sample&         181   &         181   &         181   &         180   &         181   &         181   &         181   &         177   &               &               &               &               \\
 \\[-1.8ex] 
\hline 
\end{tabular}\\

\begin{tabular}{@{\extracolsep{0.3pt}}lcccccccccccc}
\multicolumn{13}{@{}l}{\makebox[0pt][l]{\textbf{Panel (c): 1995-2020}}} \\
Democracy   &       0.196   &       0.382   &       0.450   &       0.350   &       1.817   &       2.010   &       1.781   &       1.301   &      -0.025   &       0.405   &       0.576   &       0.798   \\
            &     (0.694)   &     (0.680)   &     (0.638)   &     (0.822)   &     (1.356)   &     (1.373)   &     (1.295)   &     (1.491)   &     (1.432)   &     (1.317)   &     (1.336)   &     (1.613)   \\
Log GDP Lag Controls &       1 lag   &       2 lags  &       4 lags   &       4 lags   &       1 lag   &       2 lags  &       4 lags   &       4 lags  &       1 lag   &       2 lags   &       4 lags   &      4 lags   \\

p-value lags 5 to 8&&&& [  0.399] &&&& [  0.095] &&&& [  0.022] \\
Long-run effect of democracy&       3.569   &       6.357   &       6.112   &       3.471   &      18.814   &      20.435   &      16.316   &       9.414   &       1.596   &     -18.678   &     -47.760   &      15.683   \\
            &    (12.674)   &    (11.349)   &     (8.595)   &     (8.198)   &    (15.909)   &    (15.721)   &    (13.067)   &    (11.230)   &    (90.862)   &    (64.375)   &   (138.601)   &    (29.506)   \\
Effect of democracy after 25 years&       2.698   &       5.271   &       5.615   &       3.397   &      17.330   &      19.395   &      16.134   &       9.524   &      -0.760   &      14.511   &      20.437   &      10.622   \\
            &     (9.572)   &     (9.395)   &     (7.914)   &     (7.996)   &    (14.201)   &    (14.610)   &    (12.874)   &    (11.384)   &    (43.480)   &    (46.778)   &    (46.424)   &    (20.422)   \\
Persistence of GDP process&       0.945   &       0.940   &       0.926   &       0.899   &       0.903   &       0.902   &       0.891   &       0.862   &       1.016   &       1.022   &       1.012   &       0.949   \\
            &     (0.010)   &     (0.010)   &     (0.010)   &     (0.017)   &     (0.020)   &     (0.018)   &     (0.017)   &     (0.020)   &     (0.015)   &     (0.014)   &     (0.015)   &     (0.033)   \\
   AR2 test p-value&               &               &               &               &        0.35   &        0.00   &        0.33   &        0.93   &               &               &               &               \\
Observations&        4309   &        4118   &        3736   &        2974   &        4077   &        3886   &        3504   &        2754   &        4706   &        4706   &        4706   &        4550   \\
Countries in sample&         181   &         181   &         181   &         180   &         181   &         181   &         181   &         177   &               &               &               &               \\ \\[-1.8ex] 
\hline \\
\multicolumn{13}{@{}l}{\makebox[0pt][l]{\textbf{Panel (d): 2001-2020}}} \\
Democracy   &      0.197   &       0.210   &      -0.241   &       0.337   &       1.120   &       1.577   &       0.240   &      -0.717   &       0.140   &       1.150   &       0.528   &       0.913   \\
            &     (0.785)   &     (0.807)   &     (0.753)   &     (0.923)   &     (1.657)   &     (1.587)   &     (1.682)   &     (1.613)   &     (1.421)   &     (1.289)   &     (1.315)   &     (1.603)   \\
Log GDP Lag Controls&       1 lag   &       2 lags  &       4 lags   &       4 lags   &       1 lag   &       2 lags  &       4 lags   &       4 lags  &       1 lag   &       2 lags   &       4 lags   &      4 lags   \\
p-value lags 5 to 8&&&& [  0.020] &&&& [  0.003] &&&& [  0.014] \\
Long-run effect of democracy&       2.565   &       2.498   &      $-$2.139   &       1.391   &       9.982   &      12.494   &       1.529   &      $-$2.386   &      $-$7.612   &    $-$113.345   &     $-$41.301   &      16.057   \\
            &    (10.203)   &     (9.606)   &     (6.656)   &     (3.842)   &    (15.933)   &    (13.722)   &    (10.803)   &     (5.264)   &    (78.965)   &   (202.775)   &   (126.872)   &    (26.431)   \\
Effect of democracy after 25 years&       2.217   &       2.268   &      $-$2.079   &       1.387   &       9.473   &      12.188   &       1.526   &      $-$2.387   &       4.412   &      39.893   &      16.735   &      11.560   \\
            &     (8.823)   &     (8.723)   &     (6.469)   &     (3.828)   &    (14.924)   &    (13.286)   &    (10.781)   &     (5.265)   &    (44.214)   &    (38.934)   &    (44.537)   &    (19.359)   \\
Persistence of GDP process&       0.923   &       0.916   &       0.887   &       0.758   &       0.888   &       0.874   &       0.843   &       0.700   &       1.018   &       1.011   &       1.012   &       0.943   \\
            &     (0.007)   &     (0.009)   &     (0.011)   &     (0.025)   &     (0.021)   &     (0.019)   &     (0.018)   &     (0.029)   &     (0.015)   &     (0.014)   &     (0.014)   &     (0.030)   \\
AR2 test p-value&               &               &               &               &        0.32   &        0.02   &        0.22   &        0.19   &               &               &               &               \\
Observations&        3270   &        3079   &        2697   &        1936   &        3038   &        2847   &        2465   &        1718   &        3620   &        3620   &        3620   &        3500   \\
Countries in sample&         181   &         181   &         181   &         180   &         181   &         181   &         181   &         170   &               &               &               &               \\
 \\[-1.8ex] 

\end{tabular}\\

\hline \hline
\insertTableNotes
\end{longtable}
\end{spacing}
\end{center}
\end{ThreePartTable}
\end{landscape}
\restoregeometry

\clearpage
\newgeometry{left=2cm, right = 2cm, bottom = 1in, top =1in}
\begin{landscape}
\begin{footnotesize}
\vspace*{-2cm} 

\begin{ThreePartTable}
\begin{TableNotes}
\item{\textit{Notes:} These tables present IV estimates of the effect of democratization on log GDP per capita for the four periods. The reported coefficient of democracy is multiplied by 100. Within each period, Panel A presents 2SLS estimates instrumenting democracy with up to four lags of regional democracy waves and the p-value of a Hansen overidentification test. Panel B presents the corresponding first-stage estimates and the excluded (excl.)-instruments F statistic. Panel C presents results using the HHK (Hahn et al. 2001) estimator instrumenting democracy with up to four lags of regional democracy waves (except for col. 1, where we use only one lag). In all specifications we control for a full set of country and year fixed effects and four lags of GDP per capita. In addition, we control for the covariates specified in each column label and described in the text. Standard errors robust against heteroskedasticity and serial correlation at the country level are in parentheses. See \citet{acemogluDemocracyDoesCause2018} for the details of the method.}
\end{TableNotes}
\begin{center}
\begin{spacing}{0.85}
\begin{longtable}{@{}lccccccccc@{}}
\caption{IV Estimates of the Effect of Democratization on (Log) GDP per Capita}
\label{tab:Table2_ext}
\\[-1.8ex] \hline 
\hline\\

\endfirsthead
\multicolumn{10}{c}{{\tablename} \thetable{} -- Continued}
\\ \hline 
\hline \\

\endhead

\endfoot

\begin{tabular}{@{\extracolsep{0.1pt}}lccccccccc}
\multicolumn{10}{@{}l}{\makebox[0pt][l]{\textbf{Panel (a): 1960-2020}}} \\
& & &\multicolumn{7}{c}{Covariates Included} \\
 \cmidrule(lr){4-10} 
& & &GDP in 1960& & & &Regional& &Spatial Lags \tabularnewline
& & &Quintiles x Year&Soviet&Regional&Regional GDP&Unrest GDP&Spatial Lag&of GDP and \tabularnewline
& & &Effects&Dummies&Trends&and Trade&and Trade&of GDP&Democracy \tabularnewline
&{(1)}&{(2)}&{(3)}&{(4)}&{(5)}&{(6)}&{(7)}&{(8)}&{(9)} \tabularnewline

&\multicolumn{9}{c}{A. 2SLS Estimates with Fixed Effects} \\
 \cmidrule(lr){2-10} 
Democracy   &       3.012   &       3.098   &       1.125   &       3.132   &       1.697   &       4.042   &       3.689   &       3.055   &       3.021   \\
            &     (0.792)   &     (0.803)   &     (0.689)   &     (0.943)   &     (0.885)   &     (0.917)   &     (0.910)   &     (0.770)   &     (1.312)   \\
Long-run effect of democracy&      76.913   &      79.243   &      35.226   &      80.800   &      36.788   &      85.877   &      78.077   &      77.170   &      72.586   \\
            &    (25.172)   &    (25.684)   &    (23.846)   &    (28.928)   &    (20.657)   &    (24.084)   &    (23.069)   &    (24.356)   &    (35.601)   \\
Effect of democracy after 25 years&      59.974   &      61.723   &      25.618   &      62.522   &      32.051   &      72.245   &      65.669   &      60.578   &      54.747   \\
            &    (17.252)   &    (17.559)   &    (16.538)   &    (20.138)   &    (17.703)   &    (18.073)   &    (17.549)   &    (16.759)   &    (24.387)   \\

Observations&        7963   &        7960   &        5496   &        7960   &        6309   &        7960   &        7960   &        7769   &        7582   \\
Countries in sample&         180   &         180   &         148   &         180   &         174   &         180   &         180   &         177   &         176   \\

&\multicolumn{9}{c}{B. HHK Estimates} \\
 \cmidrule(lr){2-10} 
 Democracy  &       1.722   &       1.685   &       1.726   &       1.688   &       0.468   &       1.663   &       1.116   &       1.613   &       1.559   \\
            &     (0.584)   &     (0.542)   &     (0.597)   &     (0.553)   &     (0.469)   &     (0.445)   &     (0.382)   &     (0.478)   &     (0.453)   \\
Long-run effect of democracy&    1898.011   &     714.539   &      55.563   &     821.087   &       8.984   &      58.275   &      38.181   &     247.376   &     217.684   \\
            & (17105.111)   &  (2193.722)   &    (23.248)   &  (2962.032)   &     (9.467)   &    (25.498)   &    (17.762)   &   (328.407)   &   (243.348)   \\
Effect of democracy after 25 years&      56.487   &      54.610   &      37.806   &      54.686   &       7.753   &      36.758   &      24.354   &      48.708   &      46.258   \\
            &    (17.895)   &    (16.149)   &    (13.383)   &    (16.583)   &     (7.987)   &    (10.360)   &     (8.483)   &    (15.428)   &    (14.001)   \\

Observations&       11041   &       11041   &        9089   &       11041   &       10675   &       11041   &       11041   &       10858   &       10858   \\ \\[-1.8ex] 

\hline \\
\multicolumn{10}{@{}l}{\makebox[0pt][l]{\textbf{Panel (b): 1980-2020}}} \\

&\multicolumn{9}{c}{A. 2SLS Estimates with Fixed Effects} \\
 \cmidrule(lr){2-10} 
Democracy   &      10.589   &      10.783   &       8.393   &      14.012   &       5.294   &       9.957   &       8.962   &      10.080   &      11.231   \\
            &     (1.879)   &     (1.832)   &     (2.699)   &     (3.294)   &     (4.935)   &     (1.785)   &     (1.719)   &     (1.670)   &     (5.380)   \\
Long-run effect of democracy&     182.369   &     186.057   &     140.635   &     239.517   &      74.107   &     147.775   &     133.987   &     165.314   &     269.017   \\
            &    (43.396)   &    (43.536)   &    (56.457)   &    (68.835)   &    (79.319)   &    (35.129)   &    (31.746)   &    (36.974)   &   (242.820)   \\
Effect of democracy after 25 years&     165.078   &     168.257   &     128.381   &     215.716   &      72.918   &     138.666   &     125.338   &     152.242   &     240.240   \\
            &    (34.313)   &    (34.111)   &    (47.621)   &    (55.709)   &    (76.764)   &    (29.875)   &    (27.226)   &    (30.130)   &   (185.726)   \\

Observations&        5966   &        5966   &        5034   &        5966   &        4315   &        5966   &        5966   &        5860   &        5672   \\
Countries in sample&         180   &         180   &         148   &         180   &         174   &         180   &         180   &         177   &         176   \\

&\multicolumn{9}{c}{B. HHK Estimates} \\
 \cmidrule(lr){2-10} 
  Democracy  &       2.052   &       2.050   &       0.633   &       2.262   &       0.425   &       1.018   &       0.949   &       1.997   &       1.470   \\
            &     (0.856)   &     (0.791)   &     (0.905)   &     (0.797)   &     (0.798)   &     (0.638)   &     (0.516)   &     (0.733)   &     (0.717)   \\
Long-run effect of democracy&    $-$497.306   &    2264.355   &      72.719   &   $-$1481.879   &       4.290   &      18.761   &      20.072   &     399.691   &     161.145   \\
            &  (1322.104)   & (22204.012)   &   (126.368)   &  (9158.911)   &     (8.256)   &    (12.022)   &    (11.822)   &   (959.507)   &   (224.586)   \\
Effect of democracy after 25 years&      75.544   &      70.004   &      18.557   &      80.151   &       4.255   &      16.214   &      16.366   &      62.097   &      41.471   \\
            &    (29.748)   &    (24.535)   &    (26.329)   &    (25.864)   &     (8.172)   &     (9.981)   &     (8.903)   &    (25.795)   &    (21.608)   \\

  Observations&        7421   &        7421   &        7421   &        7421   &        7175   &        7421   &        7421   &        7298   &        7298   \\ \\[-1.8ex] 
\hline 
\end{tabular}\\

\begin{tabular}{@{\extracolsep{0.7pt}}lccccccccc}
\multicolumn{10}{@{}l}{\makebox[0pt][l]{\textbf{Panel (c): 1995-2020}}} \\
& & &\multicolumn{7}{c}{Covariates Included} \\
 \cmidrule(lr){4-10} 
& & &GDP in 1960& & & &Regional& &Spatial Lags \tabularnewline
& & &Quintiles x Year&Soviet&Regional&Regional GDP&Unrest GDP&Spatial Lag&of GDP and \tabularnewline
& & &Effects&Dummies&Trends&and Trade&and Trade&of GDP&Democracy \tabularnewline
&{(1)}&{(2)}&{(3)}&{(4)}&{(5)}&{(6)}&{(7)}&{(8)}&{(9)} \tabularnewline

&\multicolumn{9}{c}{A. 2SLS Estimates with Fixed Effects} \\
 \cmidrule(lr){2-10} 
Democracy   &      18.494   &       9.862   &      $-$1.099   &       9.862   &      $-$2.345   &       3.421   &       0.907   &      10.825   &       6.161   \\
            &    (11.535)   &     (6.857)   &    (11.964)   &     (6.857)   &    (13.126)   &     (5.657)   &     (6.520)   &     (6.891)   &     (6.715)   \\
Long-run effect of democracy&     262.797   &     136.101   &     $-$13.201   &     136.101   &     $-$15.725   &      39.196   &      10.121   &     141.110   &      68.471   \\
            &   (184.018)   &   (101.868)   &   (142.086)   &   (101.868)   &    (86.867)   &    (66.865)   &    (73.385)   &    (96.650)   &    (80.073)   \\
Effect of democracy after 25 years&     240.109   &     125.009   &     $-$12.504   &     125.009   &     $-$15.721   &      37.440   &       9.713   &     132.040   &      65.710   \\
            &   (162.783)   &    (91.538)   &   (134.822)   &    (91.538)   &    (86.848)   &    (63.537)   &    (70.330)   &    (88.935)   &    (75.906)   \\

Observations&        3714   &        3714   &        3132   &        3714   &        2063   &        3714   &        3714   &        3639   &        3451   \\
Countries in sample&         180   &         180   &         148   &         180   &         174   &         180   &         180   &         177   &         176   \\

&\multicolumn{9}{c}{B. HHK Estimates} \\
 \cmidrule(lr){2-10} 
 Democracy  &       0.296   &       0.179   &      $-$0.179   &       0.241   &      $-$0.032   &      $-$2.072   &      $-$0.949   &      $-$0.224   &      $-$0.011   \\
            &     (1.594)   &     (1.443)   &     (1.308)   &     (1.439)   &     (1.965)   &     (1.136)   &     (0.886)   &     (1.509)   &     (1.284)   \\
Long-run effect of democracy&     $-$15.499   &     $-$18.478   &      60.061   &     $-$25.402   &      $-$0.157   &     $-$41.487   &     $-$21.645   &      10.700   &       0.396   \\
            &    (86.364)   &   (153.651)   &   (497.578)   &   (158.773)   &     (9.772)   &    (31.882)   &    (24.729)   &    (73.982)   &    (44.440)   \\
Effect of democracy after 25 years&      14.462   &       7.013   &      $-$6.024   &       9.416   &      $-$0.157   &     $-$33.478   &     $-$16.455   &      $-$9.559   &      $-$0.555   \\
            &    (77.318)   &    (56.525)   &    (44.291)   &    (56.074)   &     (9.752)   &    (21.606)   &    (16.757)   &    (64.033)   &    (62.194)   \\

  Observations&        4706   &        4706   &        4706   &        4706   &        4550   &        4706   &        4706   &        4628   &        4628   \\ \\[-1.8ex] 
\hline \\

\multicolumn{10}{@{}l}{\makebox[0pt][l]{\textbf{Panel (d): 2001-2020}}} \\
&\multicolumn{9}{c}{A. 2SLS Estimates with Fixed Effects} \\
 \cmidrule(lr){2-10} 
Democracy   &      16.419   &       4.468   &      $-$4.187   &       4.468   &     $-$18.309   &       3.773   &       4.438   &       3.012   &       0.790   \\
            &    (10.892)   &     (6.555)   &    (11.708)   &     (6.555)   &    (22.670)   &     (6.924)   &     (6.934)   &     (5.658)   &     (6.931)   \\
Long-run effect of democracy&     151.853   &      39.663   &     $-$34.289   &      39.663   &     $-$71.669   &      31.484   &      37.022   &      26.263   &       5.467   \\
            &   (108.065)   &    (59.414)   &    (92.147)   &    (59.414)   &    (85.042)   &    (59.391)   &    (59.879)   &    (50.152)   &    (47.954)   \\
Effect of democracy after 25 years&     147.009   &      38.576   &     $-$33.654   &      38.576   &     $-$71.653   &      30.833   &      36.264   &      25.644   &       5.424   \\
            &   (103.574)   &    (57.644)   &    (90.765)   &    (57.644)   &    (85.032)   &    (58.018)   &    (58.468)   &    (48.908)   &    (47.567)   \\

Observations&        2681   &        2681   &        2255   &        2681   &        1030   &        2681   &        2681   &        2612   &        2424   \\
Countries in sample&         180   &         180   &         148   &         180   &         174   &         180   &         180   &         177   &         176   \\

&\multicolumn{9}{c}{B. HHK Estimates} \\
 \cmidrule(lr){2-10} 
 Democracy  &       0.254   &       1.045   &      $-$1.032   &       1.113   &      $-$0.302   &      $-$2.794   &      $-$1.876   &       0.900   &      $-$0.609   \\
            &     (2.198)   &     (1.689)   &     (1.549)   &     (1.702)   &     (3.047)   &     (1.504)   &     (1.286)   &     (1.940)   &     (1.550)   \\
Long-run effect of democracy&     $-$26.698   &    $-$108.637   &     168.169   &     $-$82.885   &      $-$1.516   &    $-$248.555   &     $-$62.557   &     $-$32.777   &      30.511   \\
            &   (254.597)   &   (248.661)   &   (641.086)   &   (157.804)   &    (15.325)   &   (673.599)   &    (70.136)   &    (67.863)   &    (93.297)   \\
Effect of democracy after 25 years&       9.523   &      38.217   &     $-$31.762   &      43.179   &      $-$1.510   &     $-$74.759   &     $-$37.516   &      39.697   &     $-$23.744   \\
            &    (81.435)   &    (63.510)   &    (48.189)   &    (68.032)   &    (15.263)   &    (50.397)   &    (28.385)   &    (90.685)   &    (58.016)   \\

Observations&        3620   &        3620   &        3620   &        3620   &        3500   &        3620   &        3620   &        3560   &        3560   \\ \\[-1.8ex] 
\end{tabular}\\

\hline \hline
\insertTableNotes
    
\end{longtable}
\end{spacing}
\end{center}
\end{ThreePartTable}
\end{footnotesize}
\end{landscape}
\restoregeometry

\clearpage
\newgeometry{left=0.3cm, right = 0.3cm, top = 1cm, bottom=1in}
\begin{figure}
        \centering
        \caption{Dynamic Panel Model Estimates of the Effects of Democratization on the Log of GDP Per Capita}\label{fig:Figure1_ext}
        \captionsetup{width=0.99\textwidth}
        \centering
        
         \begin{subfigure}[c]{.99\textwidth}
          \subcaptionbox{1960-2020}{\includegraphics[width=.49\textwidth]{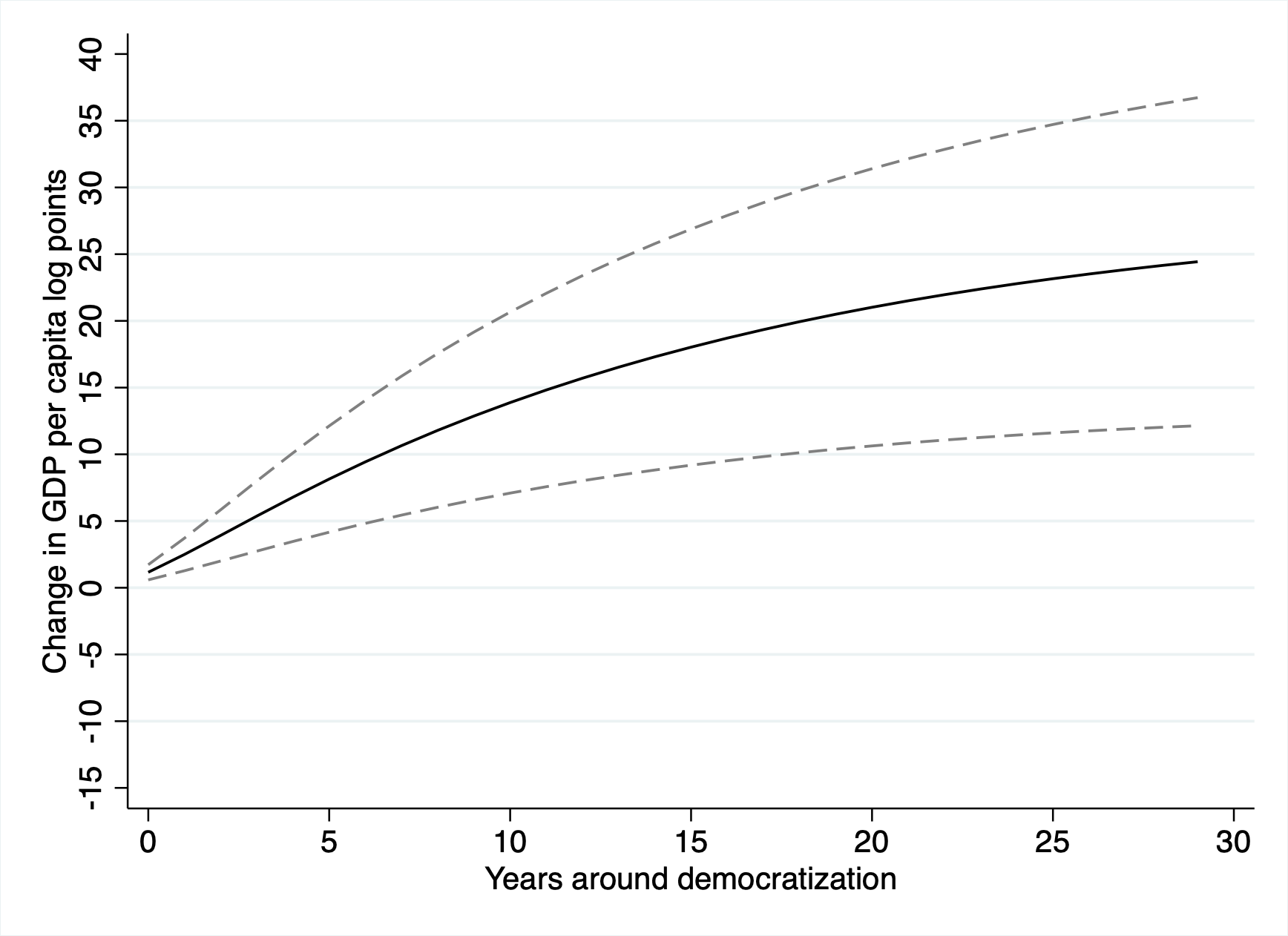}}
          \subcaptionbox{1980-2020}{\includegraphics[width=.49\textwidth]{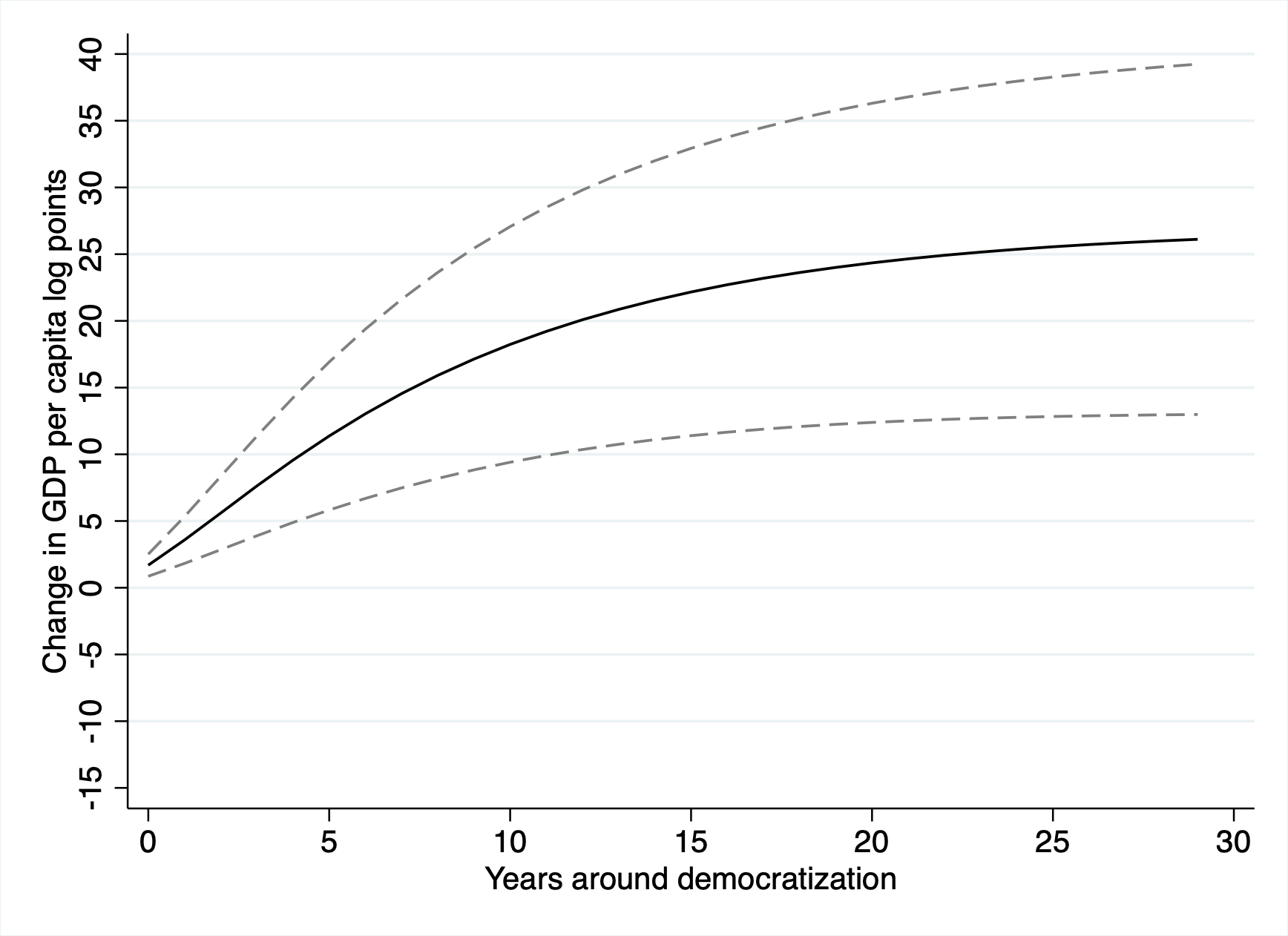}}
        \end{subfigure}
        
         \begin{subfigure}[c]{.99\textwidth}
          \subcaptionbox{1995-2020}{\includegraphics[width=.49\textwidth]{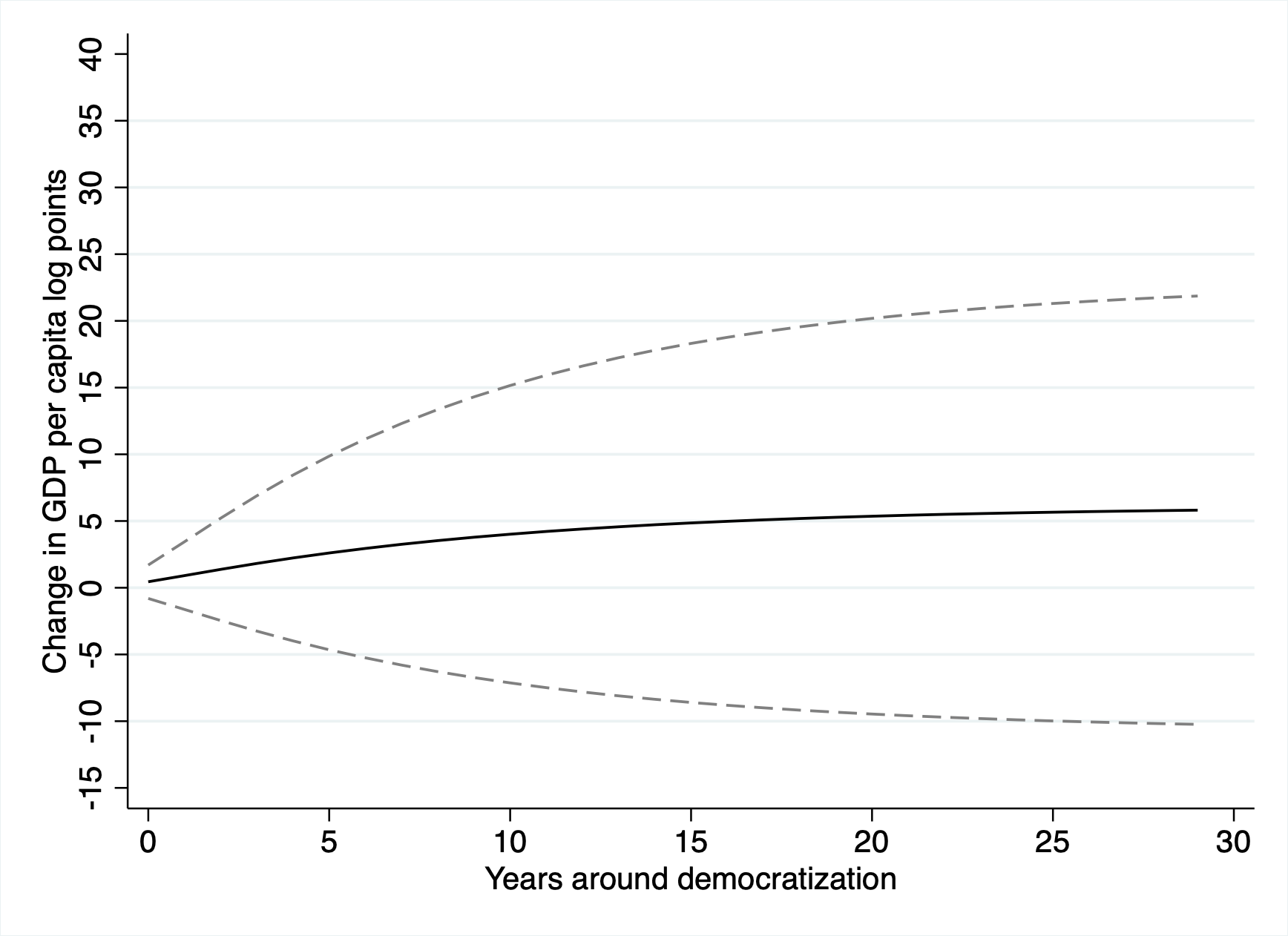}}
          \subcaptionbox{2001-2020}
          {\includegraphics[width=.49\textwidth]{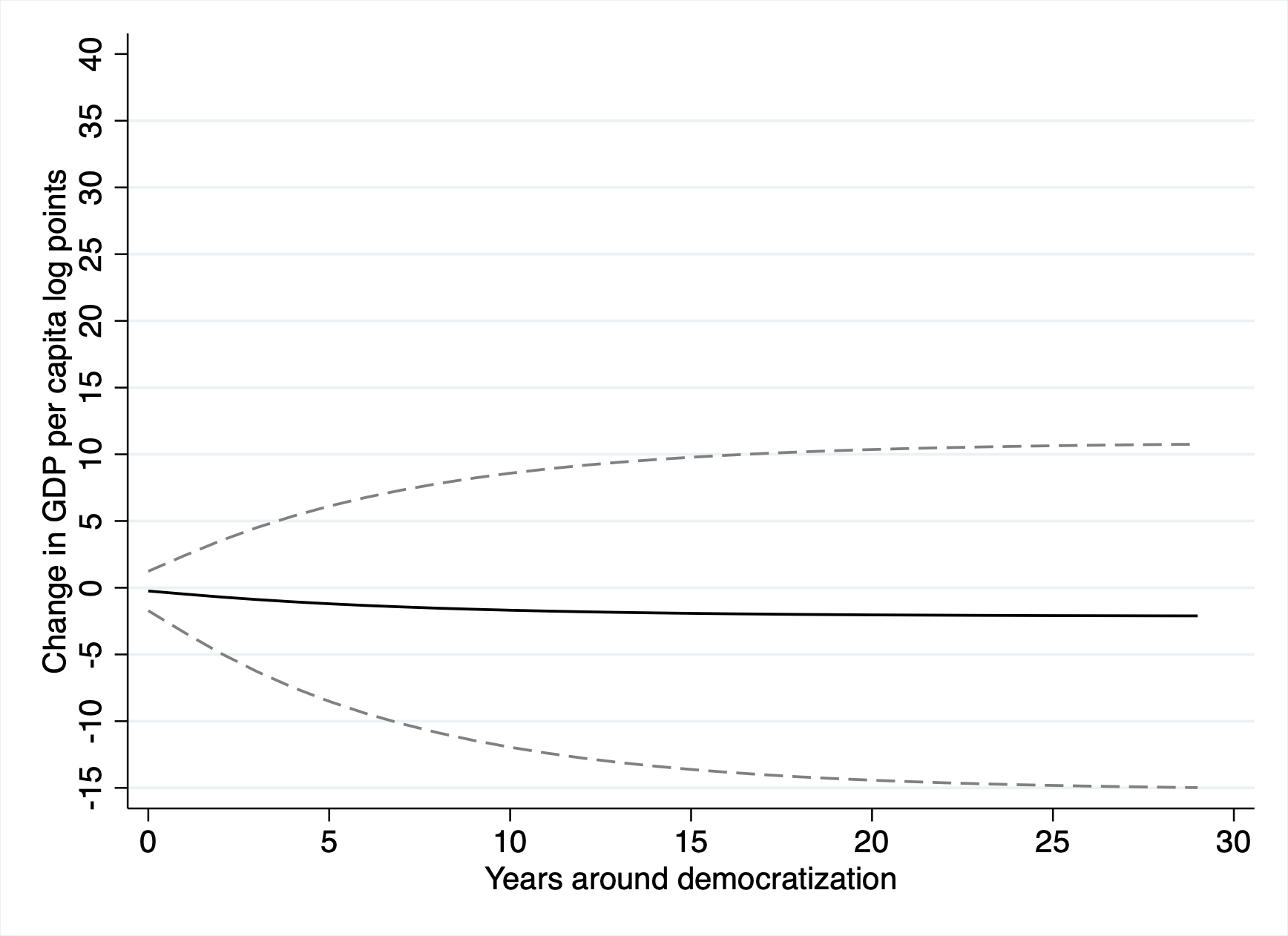}}
        \end{subfigure}
        
          \medskip
          
          \caption*{\footnotesize{\textit{Notes:} These figures plot the estimated change in the log of GDP per capita caused by a permanent transition to democracy. The effects are obtained by forward iteration of the estimated process for GDP. A 95 percent confidence interval obtained with the delta method is presented in dotted lines. Time (in years) relative to the year of democratization runs on the horizontal axis. See \citet{acemogluDemocracyDoesCause2018} for the details of the method.}}
    \end{figure}
\restoregeometry

\clearpage
\begin{table}\centering 
  \caption{Potential Policy Mechanisms Behind Democracy’s Effect in 2020}
  \label{tab:channels} 
  \scriptsize
  \begin{threeparttable}
\begin{tabular}{@{\extracolsep{0pt}}lcccccccccccc} 
\\[-1.8ex]\hline 
\hline 
 & (1) & (2) & (3) & (4) & (5) & (6) & (7) & (8) & (9) & (10) & (11) & (12)\\ 
\hline \\[-1.8ex] 
\textbf{ Panel A: Severity} & \multicolumn{12}{c}{ Dependent Variable is Containment Health Index at 10th Covid-19 Case (unit: std. deviation) } \\ \cline{2-13}  \\[-1.8ex]

Democracy Index (V-Dem, 2019)&        -0.4&        -0.3&        -0.4&        -0.4&        -0.4&        -0.3&        -0.4&        -0.3&        -0.3&        -0.4&        -0.4&        -0.4\\
&      (0.07)&      (0.07)&      (0.05)&      (0.03)&      (0.06)&      (0.04)&      (0.08)&      (0.05)&      (0.06)&      (0.04)&      (0.04)&      (0.02)\\

 \\[-1.8ex] 

 \textbf{ Panel B: Coverage} & \multicolumn{12}{c}{ Dependent Variable is Coverage of Containment Measures at 10th Covid-19 Case (unit: \%) } \\ \cline{2-13}  \\[-1.8ex]

Democracy Index (V-Dem, 2019)&        -8.1&        -7.1&       -10.2&        -9.2&        -9.4&        -7.8&        -9.3&        -8.8&        -8.2&        -9.0&        -9.7&        -9.5\\
&       (2.3)&       (1.9)&       (1.5)&       (0.5)&       (1.7)&       (1.3)&       (2.4)&       (1.1)&       (1.3)&       (0.7)&       (1.2)&       (0.5)\\

 \\[-1.8ex] 

 \textbf { Panel C: Speed}  & \multicolumn{12}{c}{Dependent Variable is Days Between 10th Covid-19 Case and Any Containment Measure (unit: days) } \\ \cline{2-13}  \\[-1.8ex]

Democracy Index (V-Dem, 2019)&         0.7&        -5.6&        -3.1&        -4.3&        -3.4&        -4.5&        -1.2&        -1.7&        -0.8&        -5.3&        -3.9&        -2.8\\
&       (3.5)&       (2.1)&       (2.2)&       (1.3)&       (2.7)&       (1.7)&       (2.7)&       (1.8)&       (2.3)&       (1.4)&       (1.7)&       (0.7)\\
\hline \\[-1.8ex] IVs & \multicolumn{2}{c}{settler mortality} &  \multicolumn{2}{c}{population density} & \multicolumn{2}{c}{legal origin} & \multicolumn{2}{c}{language} & \multicolumn{2}{c}{crops \& minerals}  & \multicolumn{2}{c}{all IVs} \\ Baseline Controls & \xmark & \cmark & \xmark & \cmark & \xmark & \cmark & \xmark & \cmark & \xmark & \cmark  & \xmark & \cmark\\
N                   &          80&          80&          84&          84&          88&          88&         129&         129&         132&         132&          70&          70\\

 \\[-1.8ex] 

\hline \hline \\[-1.8ex] 
\end{tabular} 
\begin{tablenotes} 
\item {\footnotesize {\textit{Notes:} This table reports the 2SLS estimates of democracy's effect on potential policy mechanisms behind democracy's negative impact in 2020, using five different IV strategies. Panel A reports the 2SLS estimates of democracy's effect on the containment health index at the 10th confirmed case of Covid-19. It is normalized to have standard deviation one. Panel B estimates the effect on the coverage of containment measures at the 10th confirmed case of Covid-19. Panel C estimates the effect on the number of days between the 10th confirmed case of Covid-19 and the introduction of any containment measure. 
 The Democracy Index (V-Dem) is normalized to have mean zero and standard deviation one in each year.\unskip
Columns 1, 3, 5, 7, 9, and 11 have no controls, while columns 2, 4, 6, 8, 10, and 12 have the following baseline controls: absolute latitude, mean temperature, mean precipitation, population density, median age, and diabetes prevalence.\unskip
For IVs, columns 1 and 2 use log European settler mortality, columns 3 and 4 use log population density in the 1500s, columns 5 and 6 use British legal origin, columns 7 and 8 use the fraction speaking English and the fraction speaking European, columns 9 and 10 use the ability to grow crops and mine minerals, and columns 11 and 12 use all the IVs together. \unskip
N refers to the number of countries for which IVs, controls, and outcomes are available. \unskip
Robust standard errors are in parentheses. \unskip
Variable definitions and data sources are in Appendix Table \ref{tab:sources}. \unskip}}
\end{tablenotes}
\end{threeparttable}
\end{table}

\clearpage
\section*{Data Sets (accessed March 2025)}

\bibliographystyle{chicago}
{\footnotesize 


}

\end{document}